\documentclass[12pt]{Jucthesis}
\usepackage{graphicx}
\usepackage{amssymb}
\usepackage{amsmath}
\usepackage[authoryear]{natbib}
\usepackage{color}


\begin{document}

\title{B-Machine Polarimeter: \\ A Telescope to Measure the Polarization of \\ the Cosmic Microwave Background}
\author{Brian Dean Williams}
\degreemonth{March} \degreeyear{2010} \degree{Doctor of Philosophy}
\approvalmonth{March} 
\chair{Professor Philip Lubin}  \experimentalmember{Professor Harry Nelson}\theorymember{Professor Omer Blaes}
\numberofmembers{3}

\field{Physics}
\campus{Santa Barbara}

\maketitle

\begin{frontmatter}

\thispagestyle{empty}\approvalpage

\copyrightpage

\begin{dedication}
\null\vfil
{\large
\begin{center}
For the courage and confidence my wife, Heather, has had in me. She moved to Santa Barbara, after living her entire life in Redding, Ca. for me.  As soon as she moved into some peoples guest house in some strange city, I left to go do field work for 2 weeks. Thanks Heather for putting up with this kind of behavior for far to long.
\end{center}}
\vfil\null
\end{dedication}

\begin{acknowledgements}
\ssp
There are a couple of people without whom I would most likely not be done yet. The primary driver and main knowledge base for the day to day questions and operations  is Dr. Peter Meinhold, I would like to thank and acknowledge him for always taking time out of his busy schedule to answer questions. Regardless of the quality of question or the frequency of questioning he always showed a great deal of patience and clarity in his answering. I would not have been able to make or test RF devices had it not been for the training from Dr. Jeff Childers. He showed me the techniques and attitude necessary to make good reliable devices.  Though Dr. Rodrigo Leonardi was only here for a couple of years, discussions with him on Cosmology, IDL code writing, and Latex type setting made the writing process easier.

Many people who have worked in the group, graduate students, undergraduates, and staff, made contributions both large and small towards making the B-Machine instrument a reality and have helped make the lab a fun and interesting place to work. These people in no particular order are Nate Stebor, Topher Mathews, Andrew Riley, Josh Zierton, John Billings, Nile Fairfield, Jared Martinez, Bernard Jackson, Hugh O'Neil, Ishai Rubin, Connor Wolf and the entire Staff at WMRS.

Support by the WMRS staff during the long months of observations were appreciated more than I can express. They went out of there way to make sure that our needs were meet, thanks again.

This research used resources from the National Energy Research Scientific Computing Center, which is supported by the Office of Science of the U.S. Department of Energy under Contract No. DE-AC02-05CH11231. 
\end{acknowledgements}

\begin{vitae}
 \newenvironment{vitaesectnodate}[1]{
\item {\bf #1}
\begin{list}{}{\leftmargin 0in \labelwidth 1.3in \labelsep .2in
\parsep 0in
\let\makelabel\vitaelabels}}{\end{list}}

\ssp

\hyphenation{Cryo-cooler}

\begin{vitaesection}{Education}

\item [2010] Doctor of Philosophy, Physics, University of California, Santa Barbara\\

\item [1999] Bachelor of Science, Physics, University of California, Santa Barbara\\

\end{vitaesection}

\begin{vitaesectnodate}{Teaching Experience}
\item Teaching Assistant in UCSB Physics Dept., 2001-2007\\
\item Physics 6A-6C Lab Manual editing and evaluation, at UCSB, 2006\\
\item Teacher of General Science Course at Brooks Institute of Photography, 2005-2006\\
\item Tutor at UCSB, Physics, 1996-1999\\
\item Tutor and Reader at Orange Coast College, Physics, Math, and Chemistry Depts., 1994-1995\\
\end{vitaesectnodate}

\begin{vitaesectnodate}{Publications}
\item Levy, A.~R. , Leonardi, R. , Ansmann, M. , Bersanelli, M., Childers, J., Cole, T.~D., D'Arcangelo, O. and Davis, G.~V. ,
	Lubin, P.~M., Marvil, J., Meinhold, P.~R., Miller, G., O`Neill, H., Stavola, F., Stebor, N.~C., Timbie, P.~T.,
	van der Heide, M. and Villa, F., Villela, T., Williams, B.~D., and Wuensche, C.~A., ''The White Mountain Polarimeter Telescope and
    an upper Limit on Cosmic Microwave Background Polarization,''  The Astrophysical Journal (2008),177:419-430.\\

\item Leonardi, R., Williams, B., Bersanelli, M., Ferreira, I., Lubin, P.~M., Meinhold, P.~R., O'Neill, H., Stebor, N.~C., Villa, F., Villela, T. and Wuensche, C.~A.,
 ''The Cosmic Foreground Explorer (COFE): A balloon-borne microwave polarimeter to characterize polarized foregrounds,'' New Astronomy Review (2006), 50:977-983.\\

\item Marvil, J., Ansmann, M., Childers, J., Cole, T., Davis, G.V., Hadjiyska, E., Halevi, D., Heimberg, G., Kangas, M., Levy, A.,
    Leonardi, R., Lubin, P., Meinhold, P., O'Neill, H., Parendo, S., Quetin, E., Stebor, N., Villela, T., Williams, B., Wuensche, C.
    A., and Yamaguchi, K.,  ``An Astronomical Site Survey at the Barcroft Facility of the White Mountain Research Station,'' New
    Astronomy (2006), 11:218-225.\\

\item Childers,  J., Bersanelli, M., Figueiredo, N., Gaier, T. C., Halevi, D., Kangas, M., Levy, A., Lubin, P. M., Malaspina, M.,
    Mandolesi, N., Marvil, J., Meinhold, P. R., Mejia, J., Natoli, P., O'Neill, H., Parendo, S., Seiffert, M. D., Stebor, N. C., Villa,
    F., Villela, T., Williams, B., and Wuensche, C. A.,  ``The Background Emission Anisotropy Scanning Telescope  (BEAST)
    Instrument Description and Performances,'' The Astrophysical Journal (2005), 158:124-138.\\

\item Meinhold, P. R., Bersanelli, Childers, J., M., Figueiredo, N., Gaier, T. C., Halevi, D., Huey, G. G., Kangas, M., Lawrence,
    C. R., Levy, A., Lubin, P. M., Malaspina, M., Mandolesi, N., Marvil, J., Mejia, J., Natoli, P., O'Dwyer, I., O'Neill, H.,
    Parendo, S., Pina, A., Seiffert, M. D., Stebor, N. C., Tello, C., Villa, F., Villela, T., Wade, L. A., Wandelt, B. D., Williams, B.,
    and Wuensche, C. A., ``A Map of the Cosmic Microwave Background from the BEAST Experiment,'' The Astrophysical Journal (2005),
    158:101-108.\\

\item O'Dwyer, I.~J. , Bersanelli, M. , Childers, J. ,Figueiredo, N. , Halevi, D. , Huey, G. , Lubin, P.~M. ,
	Maino, D. , Mandolesi, N. , Marvil, J. , Meinhold, P.~R. ,	Mej{\'{\i}}a, J. , Natoli, P. , O'Neill, H. , Pina, A. ,
	Seiffert, M.~D. , Stebor, N.~C. , Tello, C. , Villela, T. ,	Wandelt, B.~D. , Williams, B. and Wuensche, C.~A.,
    ''The Cosmic Microwave Background Anisotropy Power Spectrum from the BEAST Experiment,''The Astrophysical Journal (2005),
    158:93-100.\\

\item Figueiredo, N., Bersanelli, M., Childers, J., D'Arcangelo, O., Halevi, D., Janssen, M., Kedward, K., Lemaster, N., Lubin, P., Mandolesi, N., Marvil, J., Meinhold, P., Mej{\'{\i}}a, J., Mennella, A., Natoli, P., O'Neil, H., Pina, A., Pryor, M., Sandri, M., Simonetto, A., Sozzi, C., Tello, C. and Villa, F., Villela, T., Williams, B. and Wuensche, C.~A., ''The Optical Design of the Background Emission Anisotropy Scanning Telescope (BEAST),'' The Astrophysical Journal (2005),158:118-123.\\
\end{vitaesectnodate}

\begin{vitaesection}{Honors and Awards}
\item[2002, 2005-2009] UCSB California Space Grant Consortium Graduate Research Fellowship\\

\item[2004-2007] White Mountain Research Station Graduate Student Research Fellowship\\

\item[2002-2005] NASA Graduate Student Researchers Program (GSRP)\\

\item[1995] Platinum Tutoring Award Orange Coast College\\

\end{vitaesection}

\end{vitae}

\begin{abstract}
  The B-Machine Telescope is the culmination of several years of development, construction, characterization and observation. The telescope is a departure from standard polarization chopping of correlation receivers to a half wave plate technique. Typical polarimeters use a correlation receiver to chop the polarization signal to overcome the $1/f$ noise inherent in HEMT amplifiers. B-Machine uses a room temperature half wave plate technology to chop between polarization states and measure the polarization signature of the CMB. The telescope has a demodulated $1/f$ knee of 5 mHz and an average sensitivity of 1.6 $\mathrm{mK}\sqrt{\mathrm{s}}$. This document examines the construction, characterization, observation of astronomical sources, and data set analysis of B-Machine. Preliminary power spectra and sky maps with large sky coverage for the first year data set are included.
\end{abstract}

\addcontentsline{toc}{chapter}{Contents}
\tableofcontents

\listoffigures

\listoftables

\end{frontmatter}

\numberwithin{equation}{chapter}

\pagestyle{plain} 
\ssp
\chapter{Introduction}
From the beginning of human history man has looked to the sky for answers about our origins. Why are we here? Where did we come from? These questions have for the most part been in the realm of philosophy and religion.  With the evolution of Cosmology, science can start to address some of these questions. Plato and his student Aristotle created cosmologies to search for higher meaning, focusing on the Earth/Sun system. Their ideas were exemplified by the Ptolemaic Earth centered system which dominated western thinking for over 2000 years. Not until the sixteenth century did a different Heliocentric Copernican school of thought emerge. At the time this was ground breaking not just changing the position of the Earth and Sun, but also declaring that the Earth was not the center of the Universe.  During this time period observational Astronomy was beginning to blossom. Galileo Galilei turned his telescope to the sky and saw multiple satellites orbiting Jupiter, the roughness of the Moons surface and sunspots. These observations directly challenged standard dogma. Tycho Brahe was gathering unprecedented measurements of the heavens, though he still believed in an Earth centered Universe, his data eventually led to Kepler's discovery of elliptical orbits and descriptions of planetary motion. By the 18th century basic foundations of gravity and physics were being laid down by Newton, Euler, and Laplace. This marked the beginning of a truly cosmological type of thinking expanding our Universe to the edges of our Galaxy. For the first time, the Universe was more than just the Earth/Sun system. Herschel presented evidence of a vast network of stars that laid between 2 planes and stretched out a large distance and proposed a method to find our location in this stratum of stars. Man has moved from the center of the Universe to some unspecified position on the edge of our Galaxy amongst many galaxies in a vast sea of space.  This is really the start of empirical Astronomy and Cosmology with the advancements in photometry and spectroscopy, chemical properties of celestial objects could be found. Though many of the conclusions were questionable until the early twentieth century, Cosmology truly separated from philosophy into an observational science. Like the Universe our understanding of it had an inflationary epoch from 1915 to 1930. The size of the Universe in human understanding increased exponentially from our galaxy to a possibly infinite space and time Universe. Hubble's discovery that everything was moving away from us in every direction and our new understanding of energy, matter, gravity, space and time, from Einstein, made it possible to realize that in the distant past things were densely packed. The Universe was packed so close that everything was all in one place and was followed by a BIG BANG.

\section{Hot Big Bang}
The Hot Big Bang or standard cosmological model consists of a homogeneous and isotropic Universe whose development is described by the Friedman equations derived from Einstein's field equations of gravitation (General Relativity).
\begin{eqnarray}\label{eqn:FriedmanEquations}
    \mathrm{H}^{2}=\left(\frac{\dot{a}}{a}\right)^{2}= \frac{8 \pi \mathrm{G}}{3}  \rho -\frac{k c^{2}}{a^{2}} + \frac{\Lambda c^{2}}{3} \nonumber\\
    \dot{\mathrm{H}}+ \mathrm{H}^{2}=\frac{\ddot{a}}{a}=  - \frac{4\pi \mathrm{G}}{3}\left( \rho + \frac{3p}{c^{2}} \right) + \frac{\Lambda c^{2}}{3},
\end{eqnarray}
where $a$ is the expansion parameter, the dot represents a derivative with respect to proper time $\tau$, G is the gravitation constant, $\rho$ is the mass density, p is the pressure, c is the speed of light in a vacuum, k the curvature parameter and H is the Hubble parameter. The Universes expansion is parameterized by the Hubble constant, $H_0$, where $v = H_0 d$ gives the relationship between the recessional velocity, $v$, of a galaxy and its distance, $d$, from Earth. $H_0$ is the Hubble parameter now and has been measured by the ``Hubble Space Telescope Key Project'' to have a value of $72\pm8$~km~s$^{-1}$~Mpc$^{-1}$ \citep{freedman01}.

\begin{figure}[p]
\begin{center}
\includegraphics[width =\textwidth]{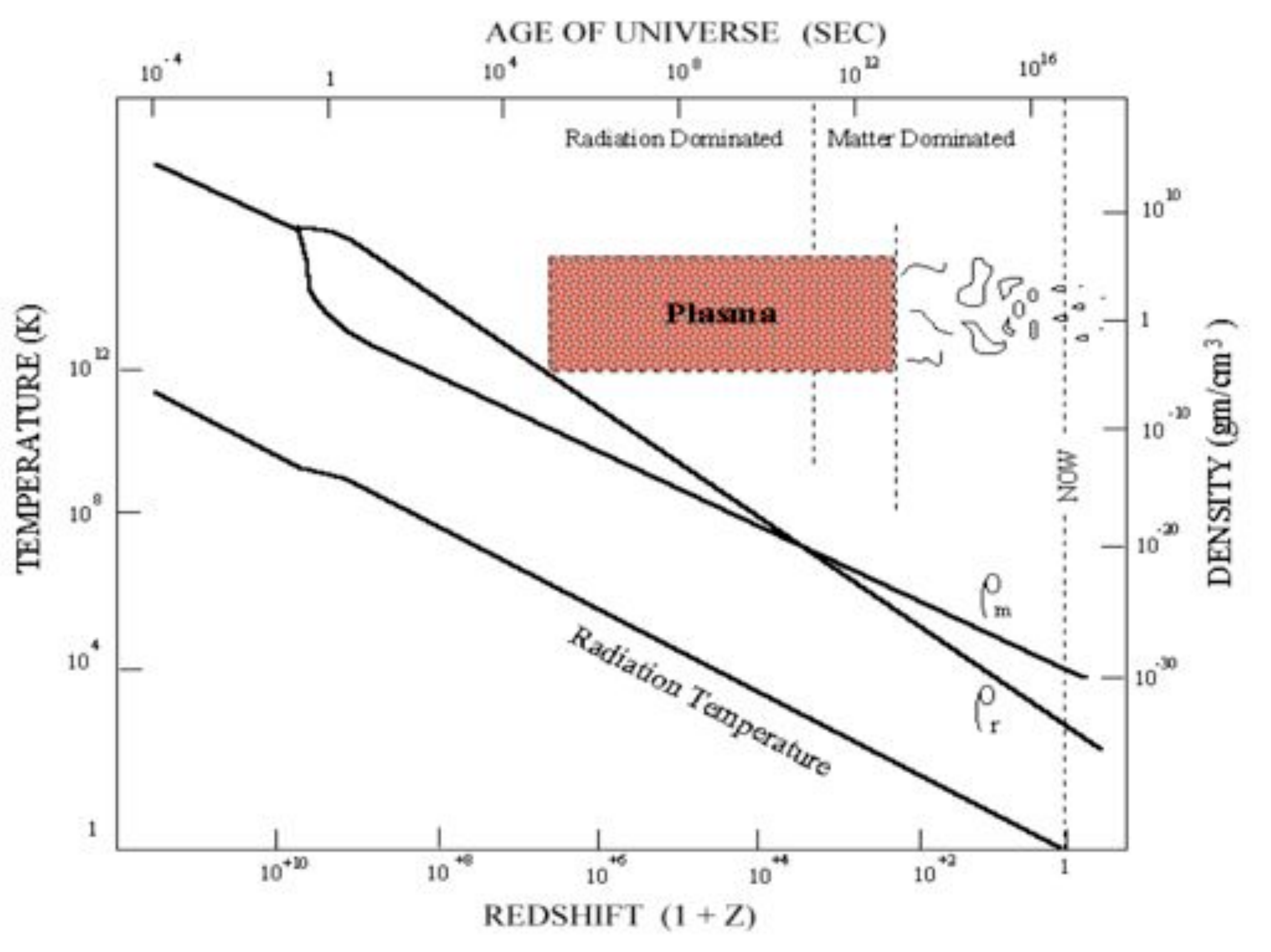}
\caption[Radiation temperature and density history of the Universe]{Thermal and density history of the Universe. CMB observations originate from the end of the period when the Universe was a plasma. The top set of lines are the energy densities, matter $\rho_{m}$ and radiation $\rho_{r}$. The lowest line is the CMB temperate and has only one major feature from the slight increase in temperature when the Universe experienced reheating from electron positron annihilation.  \label{fig:ThermalHistory}}
\end{center}
\end{figure}

The Big Bang did not occur at a single point in space but rather simultaneously everywhere in the Universe. Our understanding of the Universe starts shortly after the big bang. Before the Planck time,  $10^{-43}$ s, General Relativity needs to be modified to take into account quantum corrections which become significant at these scales. From the Heisenberg uncertainty principle in the form
\begin{equation}
\Delta E \Delta t \simeq \hbar,
\end{equation}
it can be seen that at very early times the energy levels would require masses or energies of a black hole. Post Planck time as the Universe cooled, neutrinos decoupled from the primordial plasma, particles froze out, and dark matter began to coalesce. At about 10 s the Universe experienced a brief moment of reheating when the temperature dropped below the threshold energy of the electrons and positrons which began to annihilate releasing energy.  As the Universe continued to expand, matter began to dominate the energy density.  The cooling continued until nucleosytheis created nuclei up to Lithium and Beryllium, all other heavier elements had to wait millions of years  till the formation of the first stars and their eventual death in supernova. When the temperature was sufficiently low the electrons combined with nuclei and created neutral Hydrogen or Helium. With a neutral Universe the mean free path of the typical photon increased to longer than the size of the Universe. The transition from a photon-baryon fluid to a neutral gas marked the time of last scattering of the primordial photons.

\subsection{Surface of Last Scattering}
\begin{figure}[p]
\includegraphics[width=\textwidth]{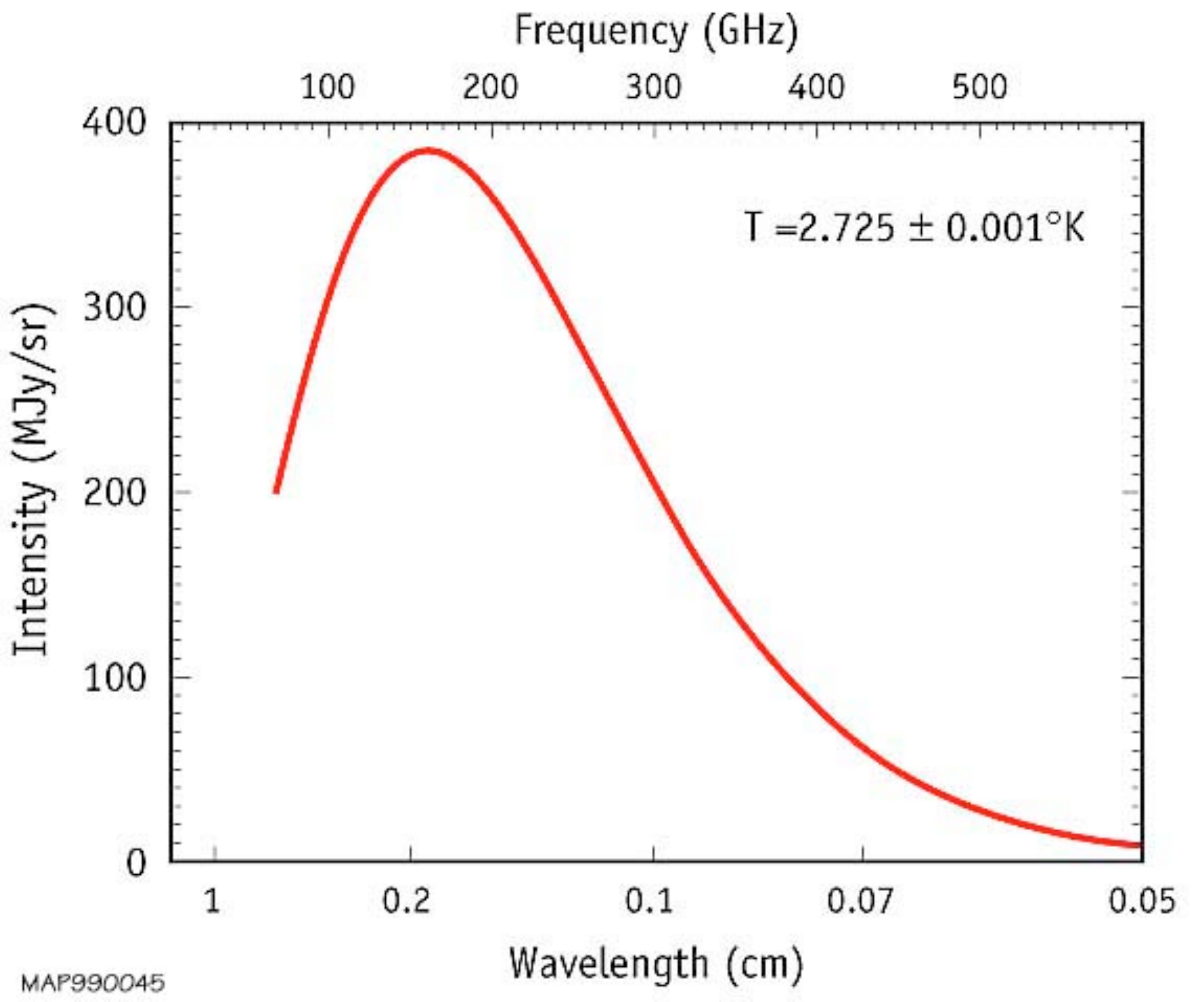}
\caption[COBE CMB temperature spectrum]{COBE temperature spectrum showing a blackbody spectrum with peak radiation consistent with a  2.72 K blackbody. Errors are smaller than the width of the line. Source: lambda.gsfc.nasa.gov. \label{fig:CobeBlackbody}}
\end{figure}

The surface of last scattering was the last time most of the primordial photons directly scattered off of matter, embedding information about this time into the remanent photon field known as the Cosmic Microwave Background (CMB). The CMB was first discovered by Penzias and Wilson in 1965 \citep{penzias65} and was found to be a uniform blackbody over the entire sky (see Appendix~\ref{app:blackbody} for a brief explanation of blackbody temperature and antenna temperature), Figure~\ref{fig:CobeBlackbody}. The initial radiation field has cooled to the point where the radiation is now in the microwave bands ($\sim$2.72 K). Not until the launch of the COBE satellite \citep{cobe92} where variations found in the background temperature, with the best measurements, to date, of the non-uniformities coming from the WMAP (Wilkinson Microwave Anisotropy Probe \citep{bennett03a}) space mission.

\begin{figure}[p]
\begin{center}
\begin{tabular}{c}
\includegraphics[width=\textwidth]{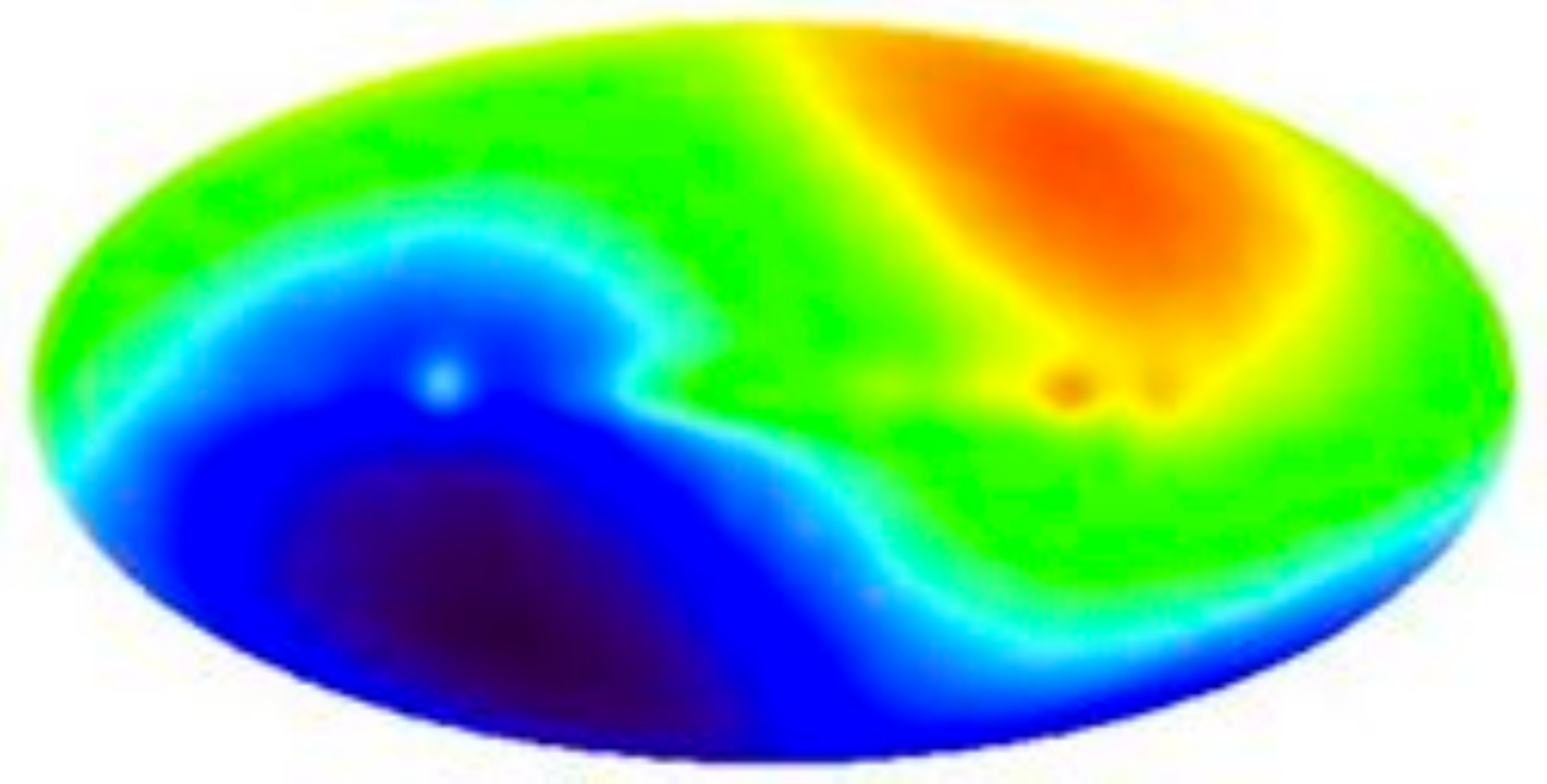}\\
\includegraphics[width=\textwidth]{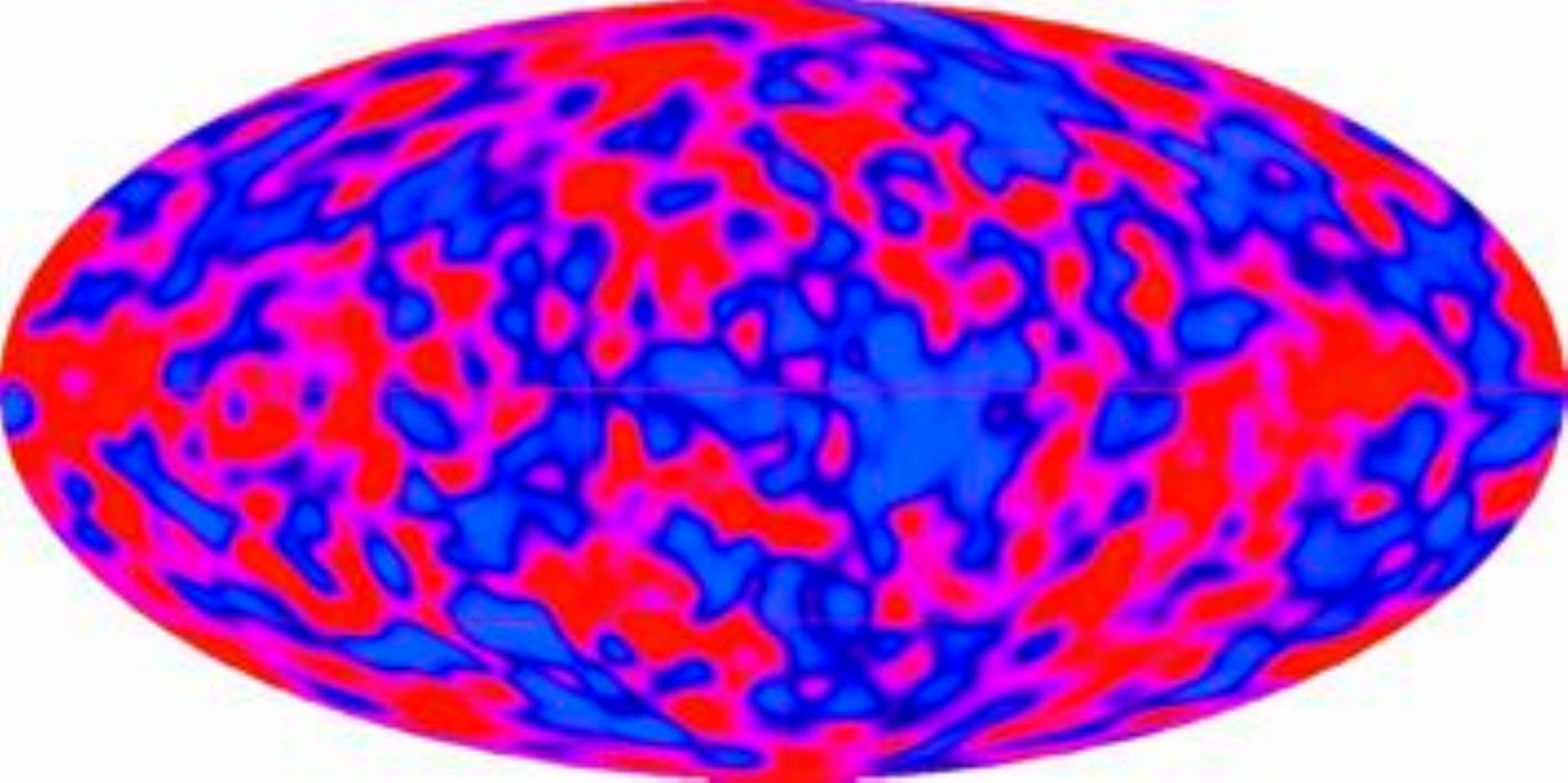}\\
\end{tabular}
\caption[COBE sky temperature map]{Top, picture of the dipole as seen by COBE where red is hotter and blue is cooler where the magnitude of the dipole is $~3.353\pm0.024\mathrm{~mK}$. Bottom, picture of the fluctuations in the mean CMB temperature with the dipole and galactic foregrounds removed. Source: lambda.gsfc.nasa.gov  \label{fig:CobeWmapSky}}
\end{center}
\end{figure}

\begin{figure}[p]
\includegraphics[width=\textwidth]{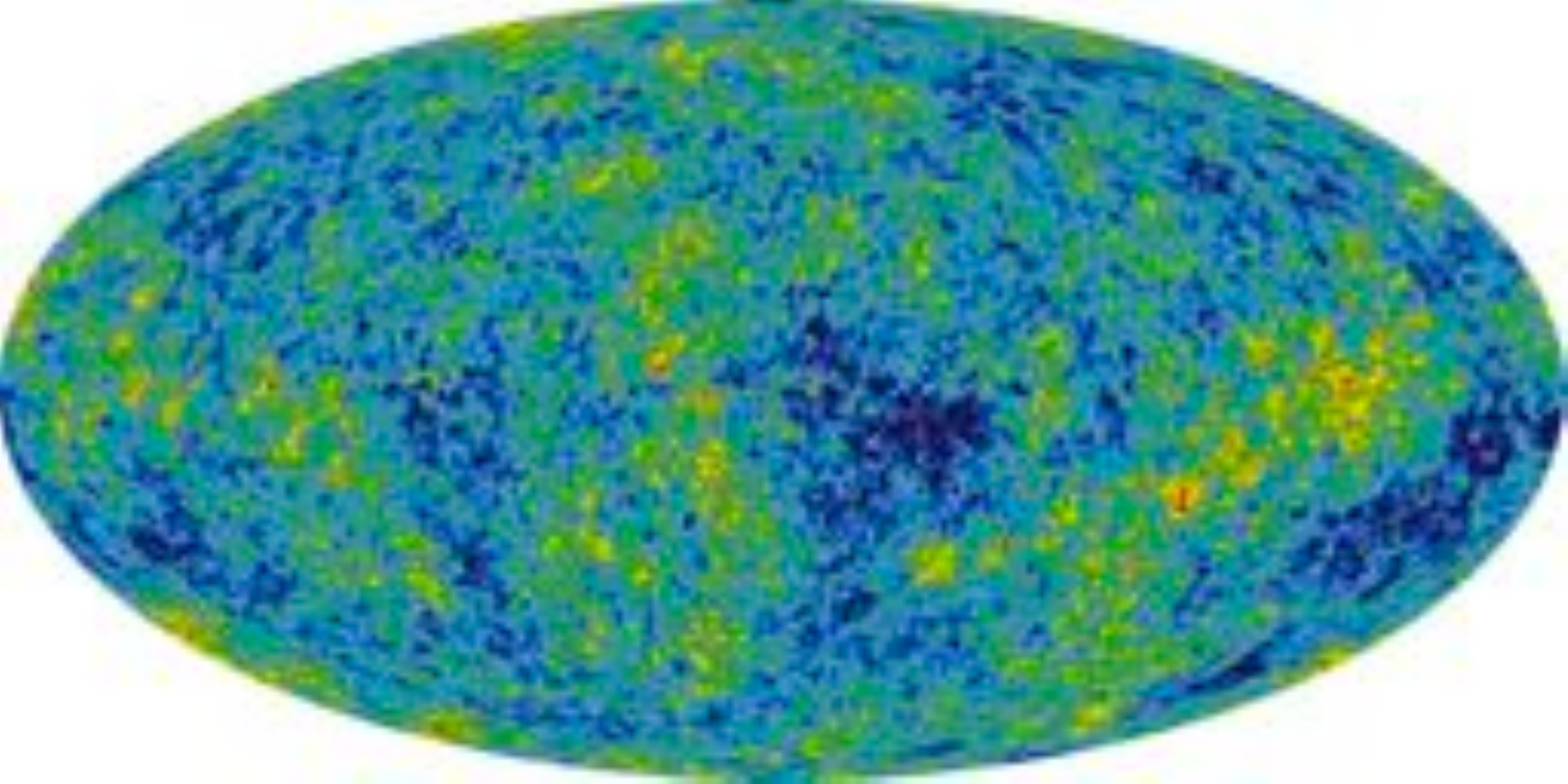}
\caption[WMAP CMB temperature spectrum]{The CMB fluctuations as seen by the WMAP space mission \citep{bennett03a} which also measured a dipole temperature consistent with COBE. The northern galactic pole is at the top of the map. Source: lambda.gsfc.nasa.gov  \label{fig:wmapsky}}
\end{figure}

The temperature fluctuations come from the oscillation of the primordial plasma, caused by quantum fluctuations expanded by inflation. The oscillatory behavior of the perturbed plasma can be described as a forced harmonic oscillator,
\begin{equation}\label{eqn:OscillatorEq}
    \ddot{ \Theta } + c_{s}^{2} k^{2} \Theta = - \frac{k^{2}}{3} \Psi -\ddot{ \Phi },
\end{equation}
with $c_{s}$ being the sound speed and $k$ the wave number. Perturbations, where $\Theta$ are perturbations in the metric and $\Psi$ are perturbations in the spatial curvature, manifest themselves as small temperature fluctuations in the background temperature. The wealth of data that we get from the anisotropies comes from the primordial temperature differences with changing $\theta$ and $\phi$ described by,
\begin{equation}
 \mathrm{T}\left( \vec{x},\theta,\phi,\eta \right) =\mathrm{T}\left(\eta\right) \left[1 + \Theta \left(\vec{x},\theta,\phi,\eta \right) \right],
 \end{equation}
where $\eta$ is the conformal time defined by,
\begin{equation}
\eta=\int\frac{dt}{a(t)}.
\end{equation}
The temperature differences originated from 3 effects described by,
\begin{equation}
 \frac{\Delta T}{T}=\frac{\phi}{c^{2}}-\frac{\hat{r} \cdot \vec{v}}{c}+\frac{1}{3}\delta.
\end{equation}
The first term on the right corresponds to the depth of the potential well, second the velocity of the fluid relative to the observer, and the final term the intrinsic temperature of the region. Getting at the information embedded in the measurements of these small, $\frac{\Delta \mathrm{T}}{\mathrm{T}}\approx 10^{-5}$, temperature differences contained in a map, see Figure~\ref{fig:CobeWmapSky}, is done by expanding the temperature anisotropies into their spherical harmonic components,
\begin{equation}
 \Theta \left(\vec{x},\theta,\phi,\eta \right)=\sum_{l=1}^{\infty} \sum_{m=-l}^{l} a_{lm}\left(\vec{x},\eta \right) Y_{lm}\left(\theta,\phi\right).
\end{equation}
No predictions of any particular $a_{lm}$ is possible, but information from the distribution from which they are drawn can be made, as long as the fluctuations that generated the parent distribution of the $a_{lm}$'s are described by a Gaussian random process. If so the angular power spectrum is given by,
\begin{equation}
 \langle a_{lm}^{\mathrm{T}} a_{l^{'}m^{'}}^{\mathrm{T}*} \rangle = \delta_{l l^{'}} \delta_{m m^{'}}C_{l}^{\mathrm{TT}}.
\end{equation}
The T superscript denotes the cross correlation of the temperature, more on this in Section~\ref{subsec:CMBpol}. It is common to plot $C_{l}$ in a way that removes the monopole, dipole, and corrects for the scale invariance of the power of each $l$ such that,
\begin{equation}
  \frac{l(l+1)C_l}{2\pi} = \left(\frac{\Delta T}{T}(\theta)\right)^2.
\end{equation}
Once angular power spectra and maps are generated from a given data set, see Figure~\ref{fig:WMAPPSclTTandTE}, the power spectra and map obtained can be compared to theoretical models to determine which cosmological model has the most likely parameter correspondence. 
Many Cosmological parameters have been found with unprecedented accuracy by the WMAP space mission, see Table~\ref{tab:parameters}, and will soon be refined by an order of magnitude by the Planck Space Mission \citep{Planck06}. For Table~\ref{tab:parameters},  BAO is the Baryon Acoustic Oscillations,  which searches for the distribution of galaxies in 3 dimensions, and SN is supernovae data. In addition to the temperature anisotropies the CMB is polarized at the microKelvin level and maps of the polarization provide a complimentary data set.

\begin{table}[p]
\caption{Key Cosmological Parameters from WMAP + BAO + SN  \label{tab:parameters}}
\begin{center}
\begin{tabular}{l|c|c}
\hline
Parameter & Symbol &  Value \\
\hline
\hline
Total density & $\Omega_0$ & $1.0052 \pm 0.0064$  \\
Dark energy density & $\Omega_{\Lambda}$ & $0.721 \pm 0.015$ \\
Matter density & $\Omega_m$ & $0.27 \pm 0.04$\\
Baryon density  &  $\Omega_B$ & $0.0462 \pm 0.0015$\\
Hubble constant & $H_{0}$ & $70.1\pm 1.3$ km/s/Mpc \\
Age of the Universe & $t_0$ & $13.73 \pm 0.12$ Gyr\\
Age at decoupling &  $t_{dec}$ & $375938^{+3148}_{-3115}$ yr \\
Redshift of Reionization & $z_{reion}$ & $10.8 \pm 1.4$ Myr \\
\hline
\end{tabular}
\end{center}
\end{table}

\section{CMB Polarization}\label{subsec:CMBpol}

\begin{figure}[p]
\begin{center}
\includegraphics[width = 13.5cm]{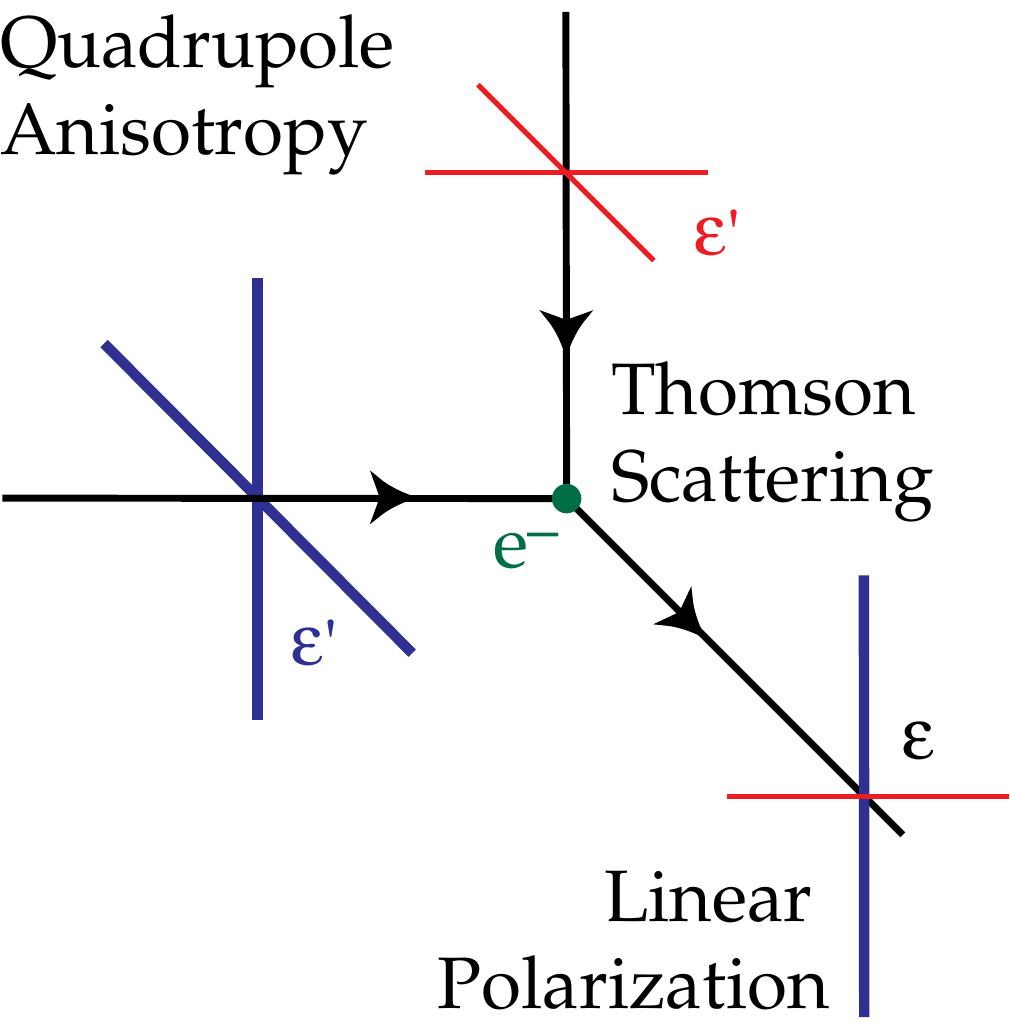}
\caption[Quadrupole illumination of electron]{Geometry of Thomson scattering for generation of polarized signal from unpolarized quadrupole illumination (adapted from \citet{hw97b}, also see website at http://background.uchicago.edu/$\sim$whu/). The incoming unpolarized radiation on the left/right (thick blue lines) is scattered by the free electron either up and down or into and out of the page (depending on the polarization). Similarly the radiation from above/below the free electron is scattered up and down or into and out of the page. The end result when viewing radiation emitted is that the horizontal polarization, comes from a cooler region than the vertical polarization giving rise to a polarized signal.\label{fig:cmbquadpol}}
\end{center}
\end{figure}

If a charged particle is illuminated by a quadrupole pattern, such as that in the CMB anisotropies (Figure~\ref{fig:cmbquadpol}), a polarized signal is generated, even if the illuminating radiation is not intrinsically polarized. A polarized electromagnetic wave of the form,
\begin{equation}
\vec{E} = E_x (t) \cos(kz - \omega t + \phi_x)\hat{i} + E_y (t) \cos(kz - \omega t + \phi_y)\hat{j},
\end{equation}
can be completely characterized by its stokes parameters. The parameters are given by
\begin{eqnarray}
    I \equiv \langle E^2_x\rangle + \langle E^2_{y}\rangle \\
    Q \equiv \langle E^2_x\rangle - \langle E^2_{y}\rangle \\
    U \equiv \langle 2 E_x E_y \cos(\phi_y - \phi_x) \rangle \\
    V \equiv \langle 2 E_x E_y \sin(\phi_y - \phi_x) \rangle
\end{eqnarray}
where the brackets denote a time average. The Stokes parameter $I$ is the total intensity of the radiation with $I^2 \geq Q^2 + U^2 + V^2$. $Q$ and $U$ describe the linear polarization of the wave and $V$ describes the circular polarization, these are equal to zero for unpolarized radiation. The angle of polarization is defined as,
\begin{equation}
\alpha \equiv\frac{1}{2}\tan^{-1}\left(\frac{U}{Q}\right),
\end{equation}
and the total polarization fraction, $P$, is
\begin{equation}
P \equiv\frac{\sqrt{Q^2 + U^2 + V^2}}{I}.
\end{equation}
$I$ and $V$ are rotationally invariant but $Q$ and $U$ transform under rotation by
\begin{equation}
 Q^{\prime} = Q\cos(2\varphi) + U\sin(2\varphi),
\end{equation}
\begin{equation}
 U^{\prime} = -Q\sin(2\varphi) + U\cos(2\varphi)
\end{equation}
where $\varphi$ is the rotation angle. However, it is clear that the quantity $Q^2 + U^2$ is rotationally invariant.
\begin{figure}[p]
\begin{center}
\includegraphics[width=\textwidth]{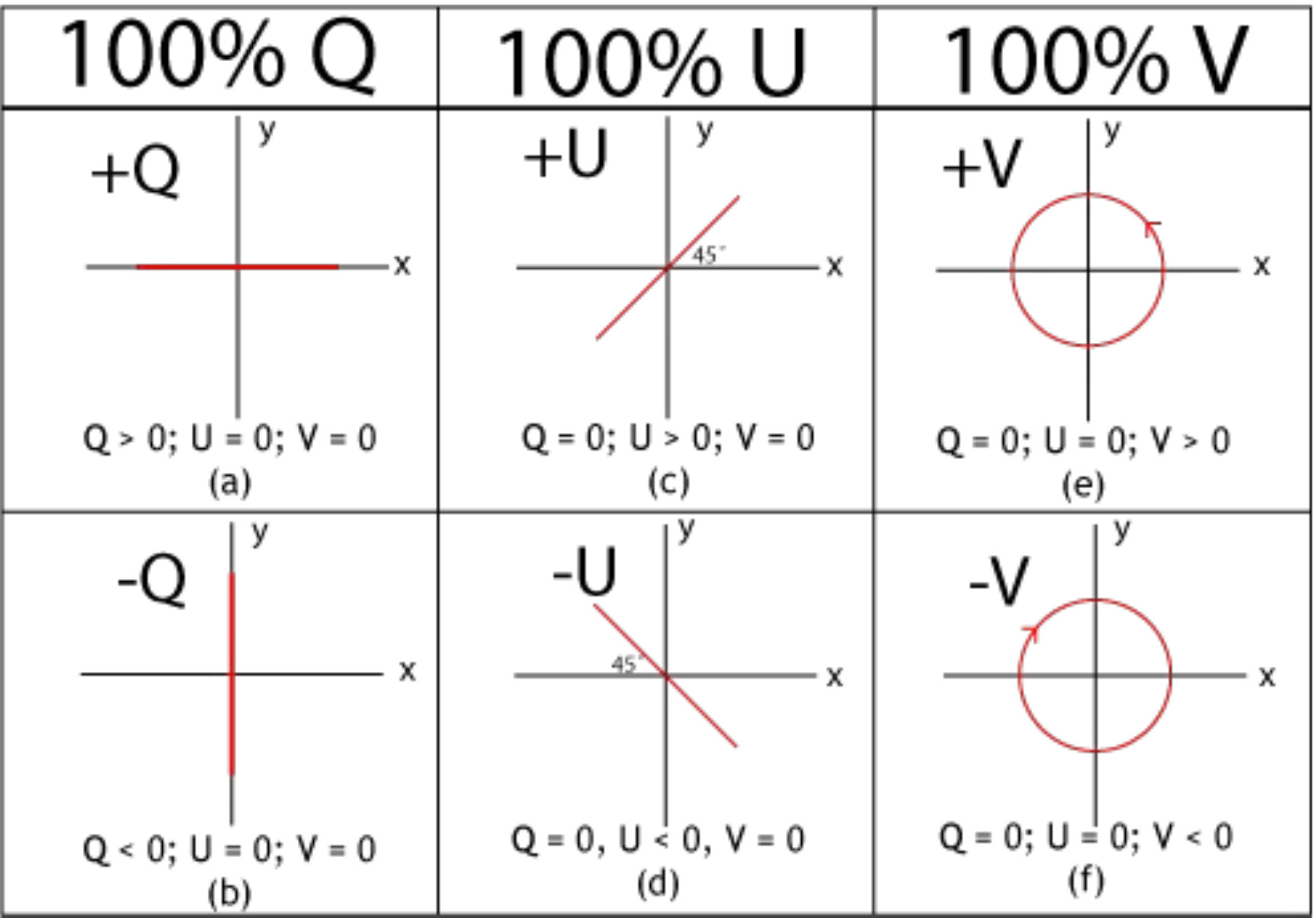}
\caption[Stokes parameters in degenerate states]{Examples of Stokes parameters in degenerate states retrieved May 15, 2009 from $\mathrm{en.wikipedia.org} / \mathrm{wiki} / \mathrm{StokesVector}$. \label{fig:stokesparameters}}
\end{center}
\end{figure}

To standardize measurements a polarization convention was defined by the International Astronomical Union in 1973 and is summarized by \citet{Hamaker96}. At each point on the celestial sphere a cartesian coordinate system with the x and y axes pointing respectively toward the North and East, and the z axis along the line of sight pointing toward the observer (inwards) for a right-handed system.  Though to confuse the issue slightly following the mathematical and CMB literature tradition, HEALPix (the most common pixelization scheme for CMB anisotropy maps) defines a cartesian referential with the x and y axes pointing respectively toward the South and East, and the z axis along the line of sight pointing away from the observer (outwards). This difference introduces a minus sign in U that has to be kept track of for power spectra generation and comparisons.

The recombination of the Universe at the surface of last scattering is not instantaneous but rather takes a finite amount of time. This leaves some fraction of charged particles to interact, through Thomson scattering, with the background anisotropies. This process is expected to give the CMB a polarization signature. Only the quadrupole moments and above generate polarization anisotropies, Figure~\ref{fig:cmbquadpol}, shows how the quadrupole moment causes a polarization from an unpolarized signal. Thomson scattering is only expected to polarize the CMB by $\sim10\%$ and will not generate any circular polarization, hence $V$ is expected to be zero.  Though some circular polarization might be generated from gravitational lensing and  galactic magnetic fields, this signal will be significantly smaller than that of the linear polarization signature.

Observations of the CMB polarization signature generate maps of the various Stokes parameters. Holding to the Helmholtz's decomposition any sufficiently smooth, rapidly decaying vector field (the Universe was/is finite in extent) can be decomposed into a divergence-free vector field (gradient) and a curl-free (divergence) vector field. The typical nomenclature for CMB, analogous to electromagnetic notation, is an E field (divergence) and a B field (gradient).

\begin{figure}[p]
\includegraphics[width=13.5cm]{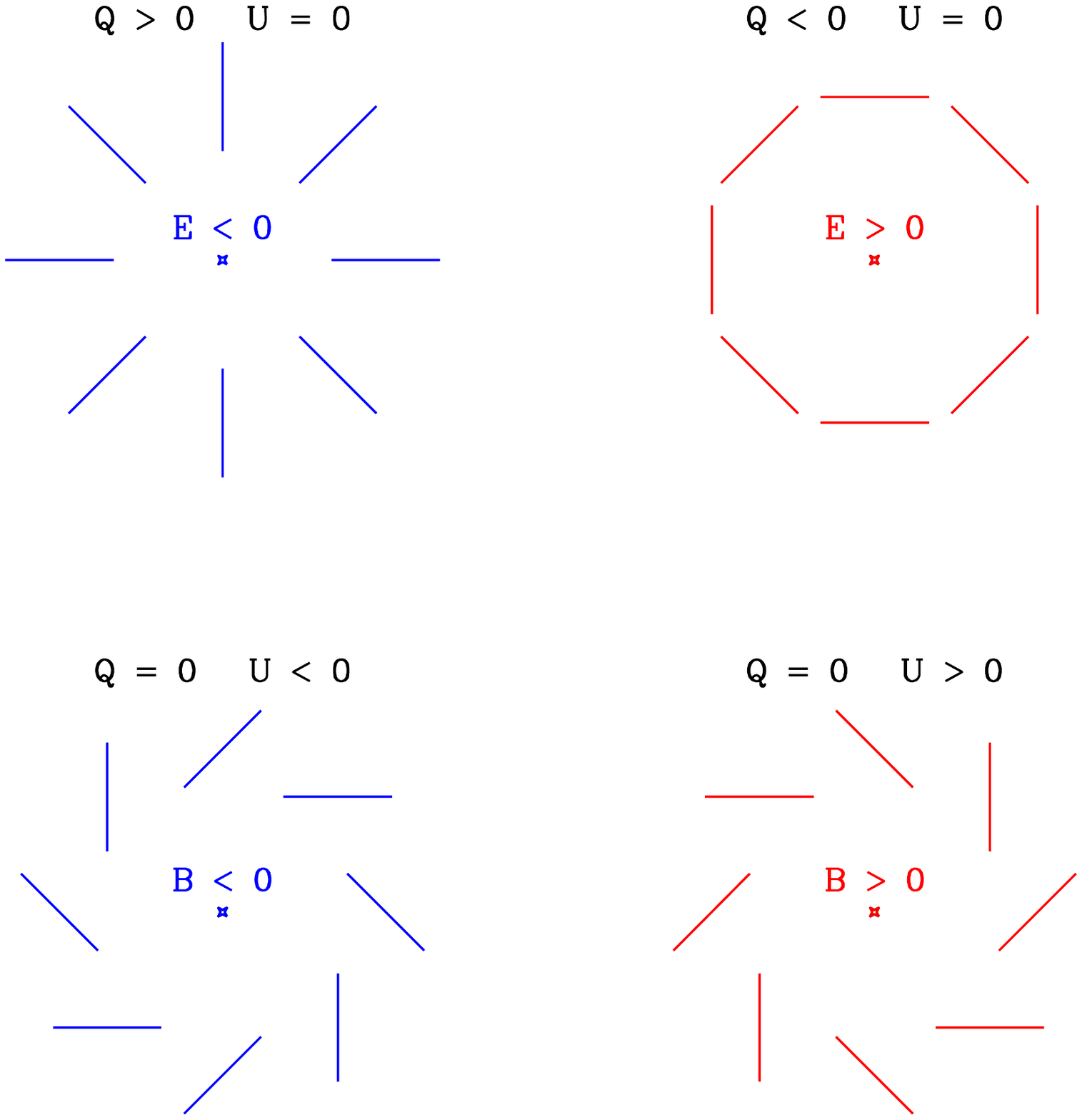}
\caption[E and B patterns for different Stokes parameter values]{E and B patterns for different Stokes parameter values (from \citet{zal04}). Note the parity differences in the E and B vector fields.\label{fig:EandBpatterns}}
 \end{figure}
\begin{figure}[p]
\includegraphics[width=13.5cm]{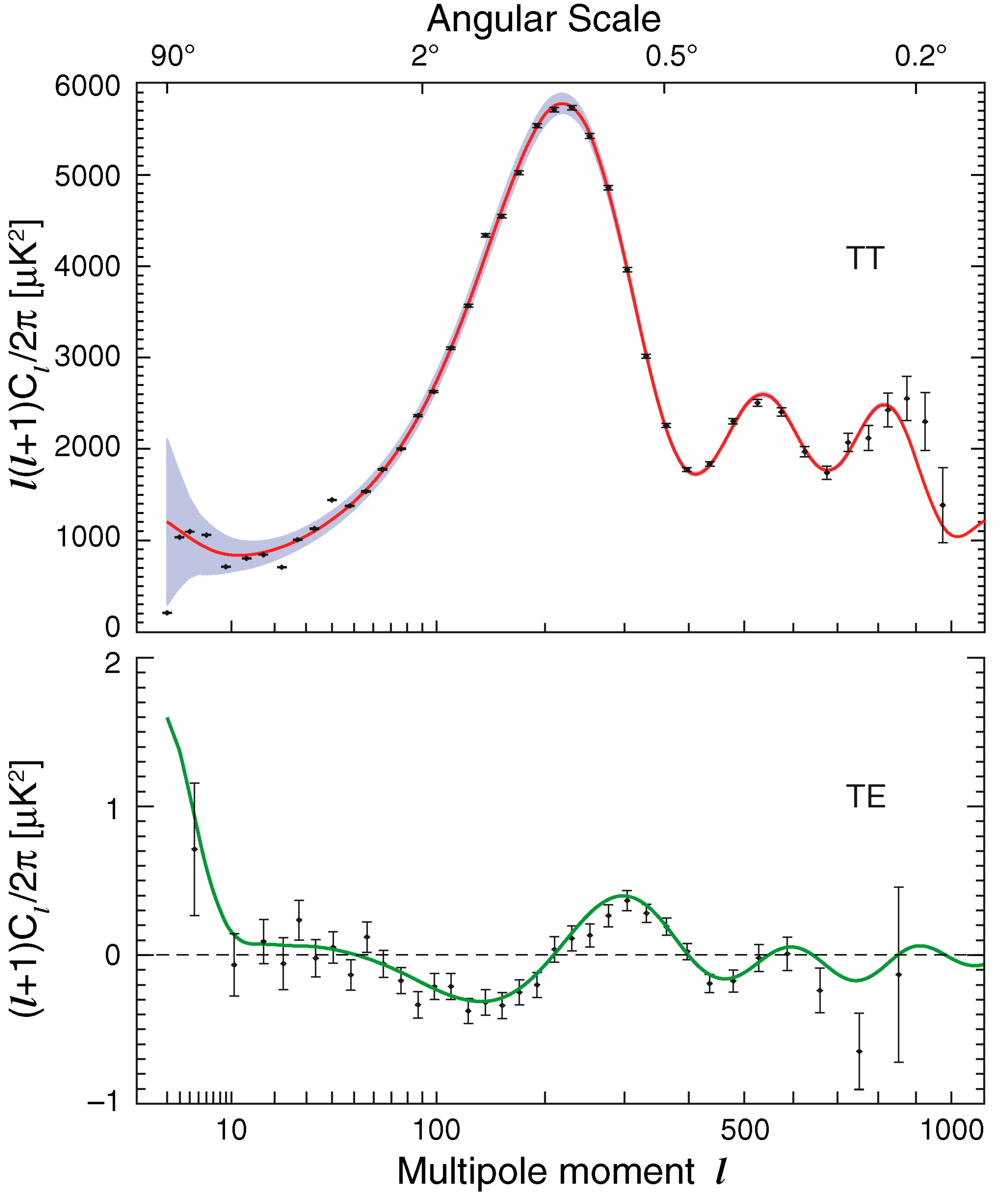}
\caption[WMAP TT and TE angular power spectra]{WMAP TT and TE power spectra. Solid curves represent the best-fit theory spectrum from $\Lambda$CDM \citep{d09}. Grey area on the left represents the cosmic variance limit and the increase in error bars on the right side are caused by the finite beam size of WMAP.  Source: lambda.gsfc.nasa.gov.  \label{fig:WMAPPSclTTandTE}}
\end{figure}

Transforming into E-modes and B-modes (E and B from here out) lets us take advantage of the fact that E and B are scalar spin-0 quantities like temperature and the maps can be interpreted similar to that of temperature. Congruent with the temperature expansions E and B can be expanded into spherical harmonics:

\begin{equation}
E(\theta,\phi) = \sum_{l,m} a^E_{lm}Y_{lm}(\theta,\phi)
\end{equation}

\begin{equation}
B(\theta,\phi) = \sum_{l,m} a^B_{lm}Y_{lm}(\theta,\phi).
\end{equation}
giving rise to angular power spectra that are defined by

\begin{equation}
 C^{XX^{\prime}}_l \equiv \langle a^{X\ast}_{lm}a^{X^{\prime}}_{lm}\rangle
\end{equation}
where $X$ and $X^{\prime}$ can be $T$, $E$, or $B$ resulting in six possible power spectra $C^{TT}_l$, $C^{TE}_l$, $C^{EE}_l$, $C^{BB}_l$, $C^{TB}_l$, and $C^{EB}_l$. $C^{TT}_l$ denotes the temperature anisotropy angular power spectrum which has been previously discussed , $C^{TE}_l$ is the temperature polarization cross-power spectrum (see Figure~\ref{fig:WMAPPSclTTandTE}), and $C^{EE}_l$ and $C^{BB}_l$ are the $E$-mode and $B$-mode angular power spectra.  E and T have an even parity while B has an odd parity, as seen in Figure~\ref{fig:EandBpatterns} and this property reduces the number of power spectra under cross-correlation from $6$ to $4$ since $C^{TB}_l$ and $C^{EB}_l$ are expected to be zero.

\begin{figure}[p]
\includegraphics[width=13.5cm]{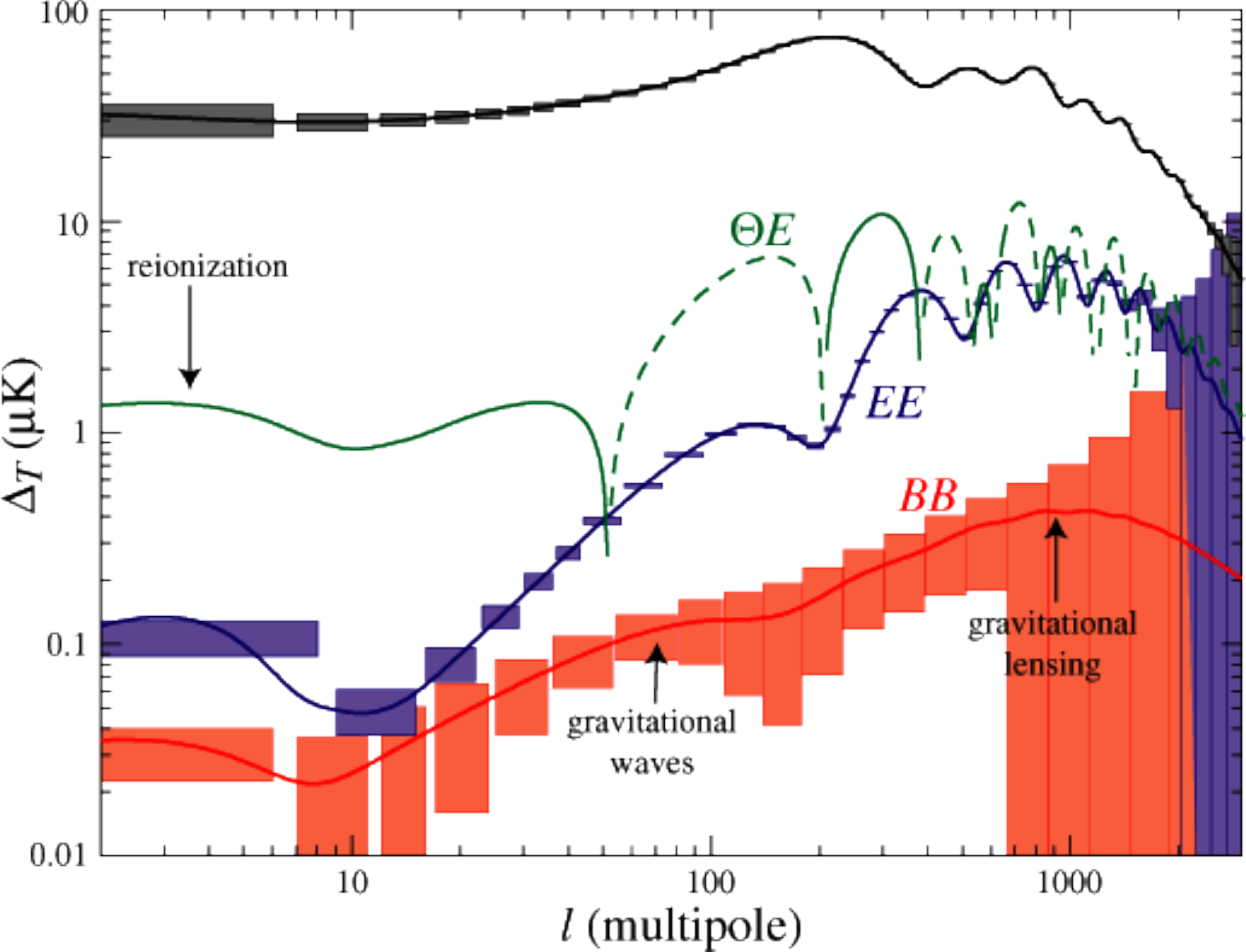}
\caption[Planck satellite power spectra estimates]{Temperature and polarization spectra for $\Omega_{tot}=1$, $\Omega_{\Lambda}= 2/3$, $\Omega_{b} h^{2}= 0.02$, $\Omega_{m} h^{2}= 0.16$, $n=1 $, $z_{ri}= 7$, and $E_{i}= 2.2\times 10^{16}$GeV. The dashed lines indicate negative cross correlation and the boxes are the statistical errors of the Planck satellite. Plot courtesy of \citet{Hudodelson02}. \label{fig:PlanckPowerSpectra}}
\end{figure}

The expected EE power spectrum has extremes that correspond to scales where the fluid is in motion, maximizing the quadrupole of the CMB temperature. The motion of the fluid induces a quadrupole moment that correlates to the maximum velocity fields causing the maxima of the EE spectrum to correspond to the minimum of the TT spectrum with maximum correlation in the middle as seen in Figure~\ref{fig:WMAPPSclTTandTE}.  Measurement of the polarization power spectra gives an independent confirmation of the temperature results. With the addition of the cross correlated (TE) spectrum the two additional pieces of information can lead to better constrained cosmological parameters and the breaking of degeneracies in different cosmological models. The reionization history of the Universe has a slight effect on the smallest scales of the TT spectrum, but leaves a drastic signature on the EE and TE spectra. The final spectrum as of yet undetected is the BB power spectrum and is one of the very few direct probes of inflation that exists. No direct detection of the BB spectrum has been made up to this point, but efforts to determine the scalar-to-tensor ratio are in the works and more sensitive instruments are constantly being developed.  A detection of the BB spectrum would give constraints on the energy scale of inflation and help theorists confine the class of theories for inflation. The currently favored theory for inflation is a  single parameter slow role model.  The interested reader is encouraged to read through \citet{samtleben07}, \citet{samtleben07}, and \citet{Hudodelson02} for more information on the Cosmic Microwave Background temperature and polarization spectra.

\subsection{Foregrounds}
Unfortunately, observations of both the temperature and polarization signatures of the Big Bang are polluted by material between us and the surface of last scattering, known as foregrounds. The foregrounds have been well characterized for the temperature maps, and template subtractions from these maps have been successful. The difficulty arises when trying to understand the polarization of the foreground sources. The sources include reionization, gravitational lensing, synchrotron radiation, free-free emission, extragalactic point sources, atmosphere, and spinning dust grains. Of the 7 sources only 3 of them present severe problems for polarization observations. Reionization is expected to effect low-$l$ measurements of the EE spectrum, gravitational lensing will effect the B-mode measurements, extragalactic point sources are good for calibration and can be masked out fairly easily and the atmosphere (for ground based telescopes only) is not expected to have any polarization effects, but may contribute other systematic effects (see Section~\ref{subsec:barcroft}). This leaves synchrotron, free-free emission, and spinning dust grains as the primary obstacles. Synchrotron radiation is caused by relativistic charged particles interacting with the Galactic magnetic field and can be highly polarized. While free-free emission is due to electron-ion scattering and is expected to be unpolarized, but through Thomson re-scattering by electrons at the edges of the HII regions will become polarized tangentially to the edges of the clouds up to $\sim10\%$. Spinning dust radiation is not well understood, it is thought that the radiation is generated by electric dipole radiation from small rapidly rotating dust particles and has the potential to be significantly polarized. The signal from spinning dust grains is likely to peak at or around 20 GHz at $100~\mu$K and role off rapidly up to $\sim60$ GHz.  Cosmic signals can be distinguished from foregrounds by their frequency dependence and their spatial power spectra.  Using polarimeters that cover a wide range of frequencies, large sky coverage, and correlations with other lower frequency observations can generate significant information about polarized foregrounds. A more in depth analysis of foregrounds and their effects can be found in \citep{tegmark00}.

\begin{figure}[p]
\includegraphics[width=\textwidth]{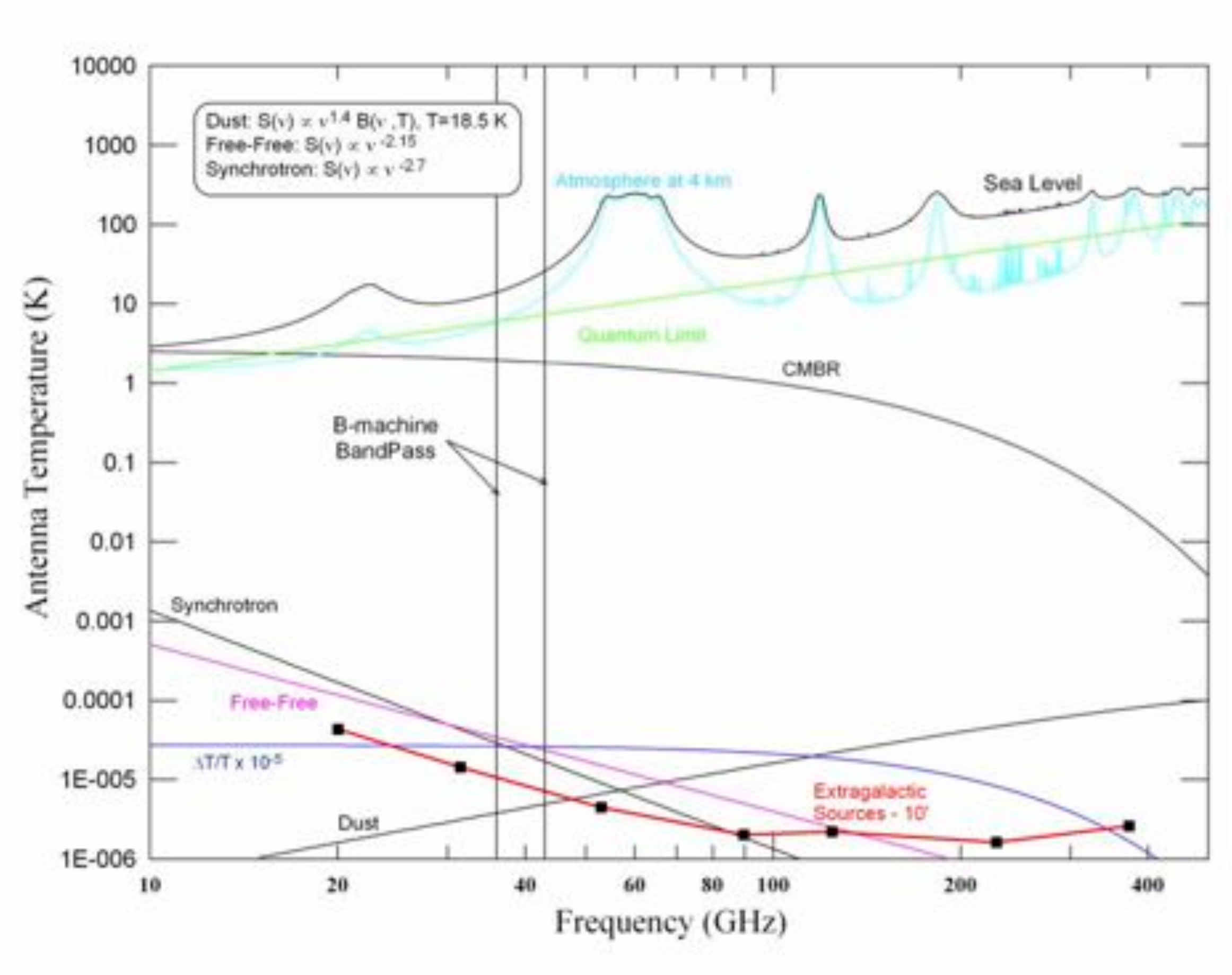}
\caption[Atmospheric and foreground emission]{Atmospheric and foreground emissions, generated by ATMOS32. Sky temperature is a sum of $H_{2}O$, $O_{2}$, and $O_{3}$ using a sea level water vapor density of $10 \mathrm{ ~g/m}^{3}$ and foreground spectral indices from \citet{bennett03b}. \label{fig:Atmosphere}}
\end{figure}
An interesting comprehensive treatment of the pertinent foregrounds is treated in excruciating detail in \citet{bennett03b}.

\section{B-Machine at White Mountain Research Station, Barcroft}
With this abridged overview of the CMB it is clear that maps with large sky coverage and widely separated frequency bands will play a role in discovering information about the origins and the current state of our local Universe. I have endeavored to build, field, operate, and analyze data from a telescope that is dedicated to mapping the E-modes and B-modes. B-Machine (named for eventually detecting B-modes not for Brian) has been placed at a high altitude site (see Subsection~\ref{subsec:barcroft}) and has been observing for several months. An in depth description of the instrument and its systems can be found in Chapter~\ref{chap:instrument} and Chapter~\ref{chap:polarizationrotator}. Characterizing the instrument and fielding it has been described in Chapter~\ref{chap:telescopechar} and finally an explanation of the preliminary data set and sky maps is presented in Chapter~\ref{chap:analysis}.

\subsection{Barcroft} \label{subsec:barcroft}
Atmospheric loading plays a significant role in both design and use of a telescope. When testing a telescope at sea level (Santa Barbara, Ca.) the typical sky zenith temperature is around 30 K versus about 10 K at a high altitude site (White Mountain Research Station at Barcroft, Ca.).  This is mostly due to colder air temperatures and waters scale height of 2 km. Integrated precipitable water vapor (IPWV) for moderate latitudes at sea level varies from $1-2\mathrm{~cm}$, while at high altitude sites, at appreciable latitudes, only varies from $1-2\mathrm{~mm}$, see \citet{marvil05}. It is critical to field ground based telescopes at high altitude sites because rapid changes of the IPWV can mimic sky signals in either temperature or polarization and noise scales directly with antenna temperature (see Appendix~\ref{app:blackbody}). 
We have had a great deal of experience in fielding telescopes at a high altitude site that is a reasonable driving distance from Santa Barbara, California. White Mountain Research Station, Barcroft (referred to as WMRS) has been developed into a reliable site over the past decade.  Power and personnel issues have been all but eliminated and with the knowledge and experience gained by the previously 2 fielded telescopes (BEAST \citep{childers05} and WMPol \citep{levy08}) it was an easy decision to place B-Machine at WMRS. A comprehensive site survey was done early on, comparing WMRS to other high altitude sites, and it was found to be akin to others, see \citet{marvil05} for full results. The only major  preparation needed at WMRS for the installation of B-Machine was the construction of a building with a fully retractable roof. The vast majority of the work and credit for the successful design and construction of the building goes to Andrew Riley. Construction of a building at a high altitude site is much more difficult than normal construction and Andrew went through some heroics to get the concrete foundation laid and the building made in time for B-Machine to be fielded.

\begin{figure}[p]
\begin{tabular}{cc}
\includegraphics[width=6.75cm,height=6.75cm]{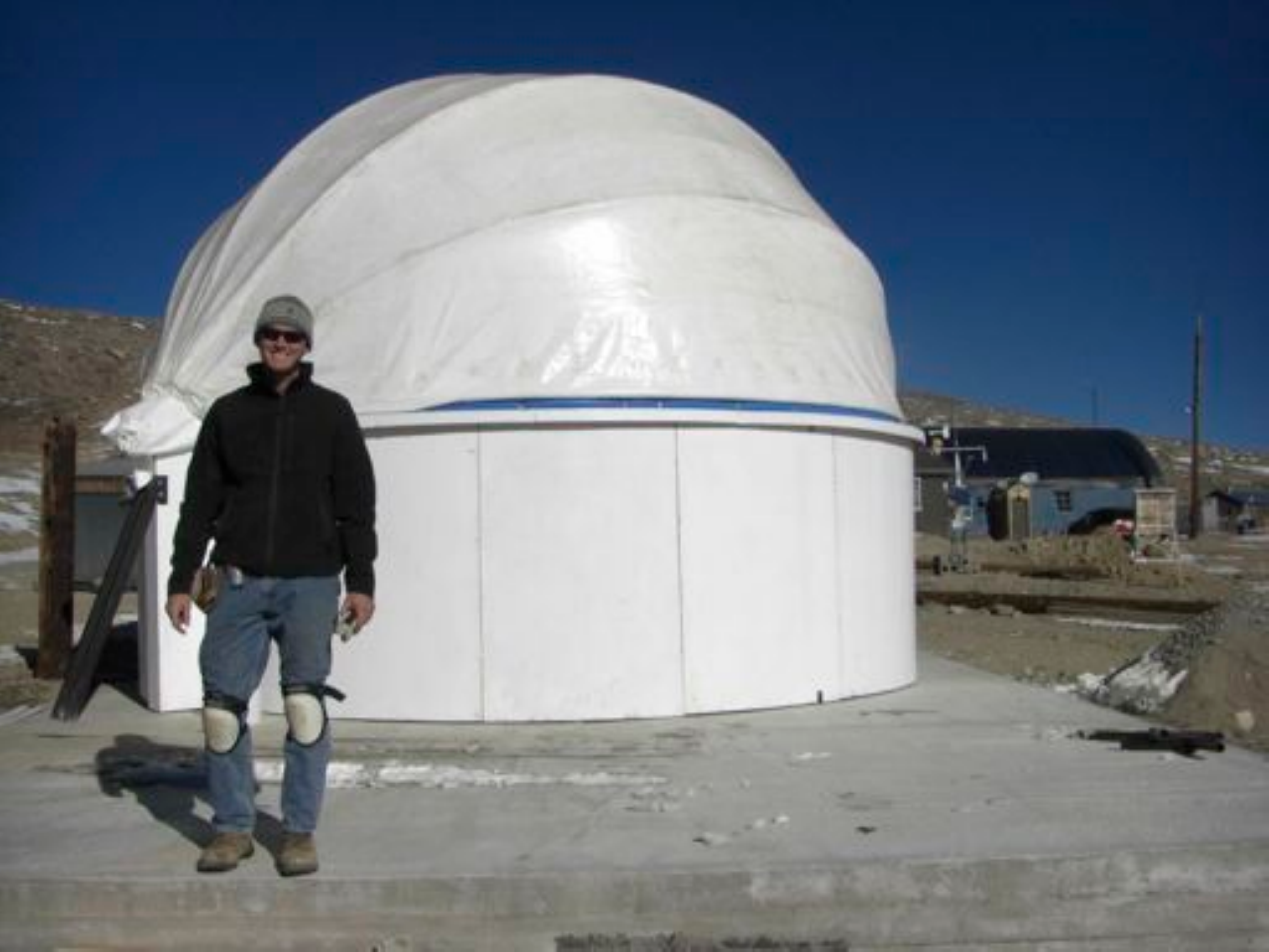}&
\includegraphics[width=6.75cm,height=6.75cm]{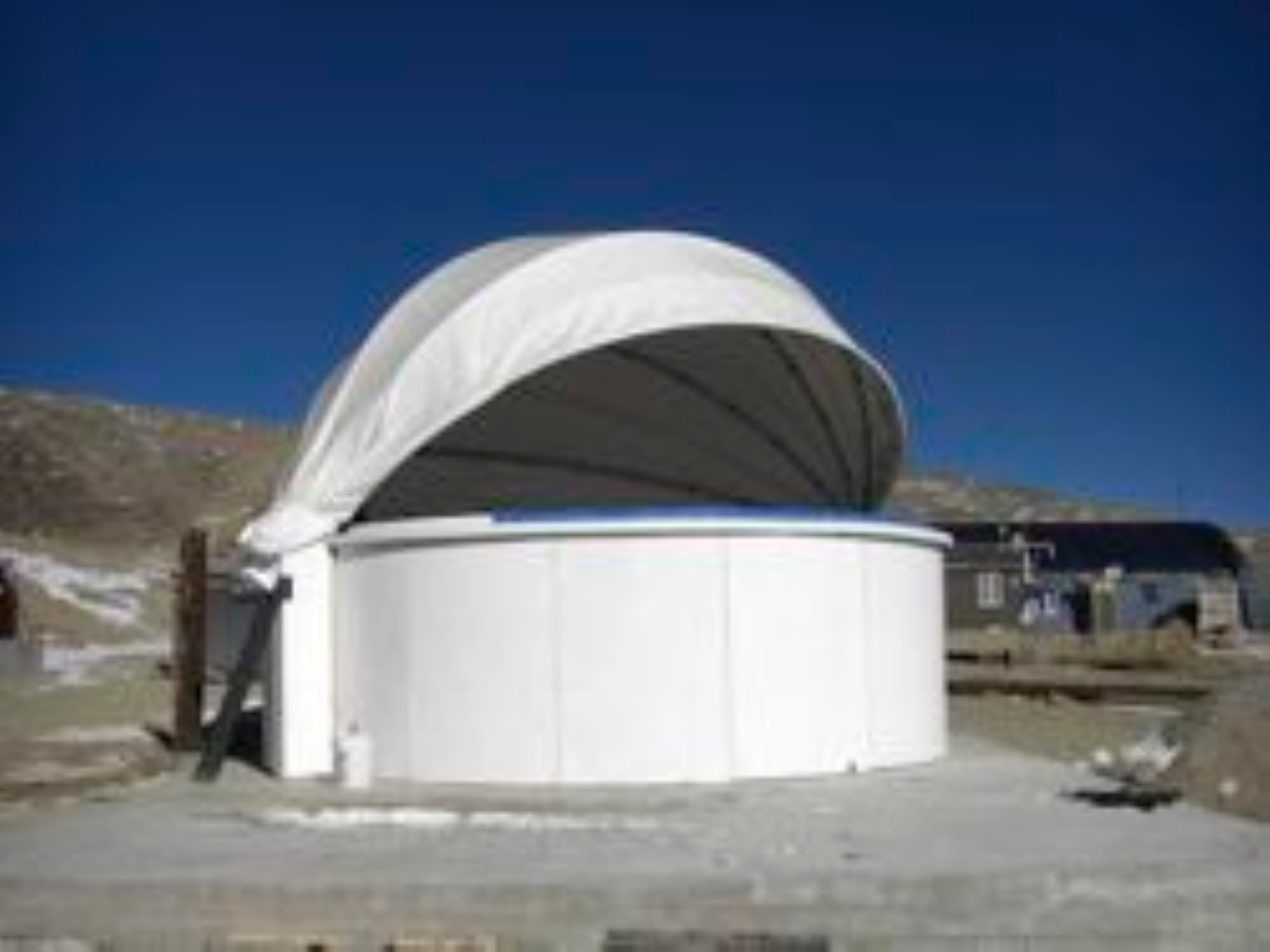}\\
\includegraphics[width=6.75cm,height=6.75cm]{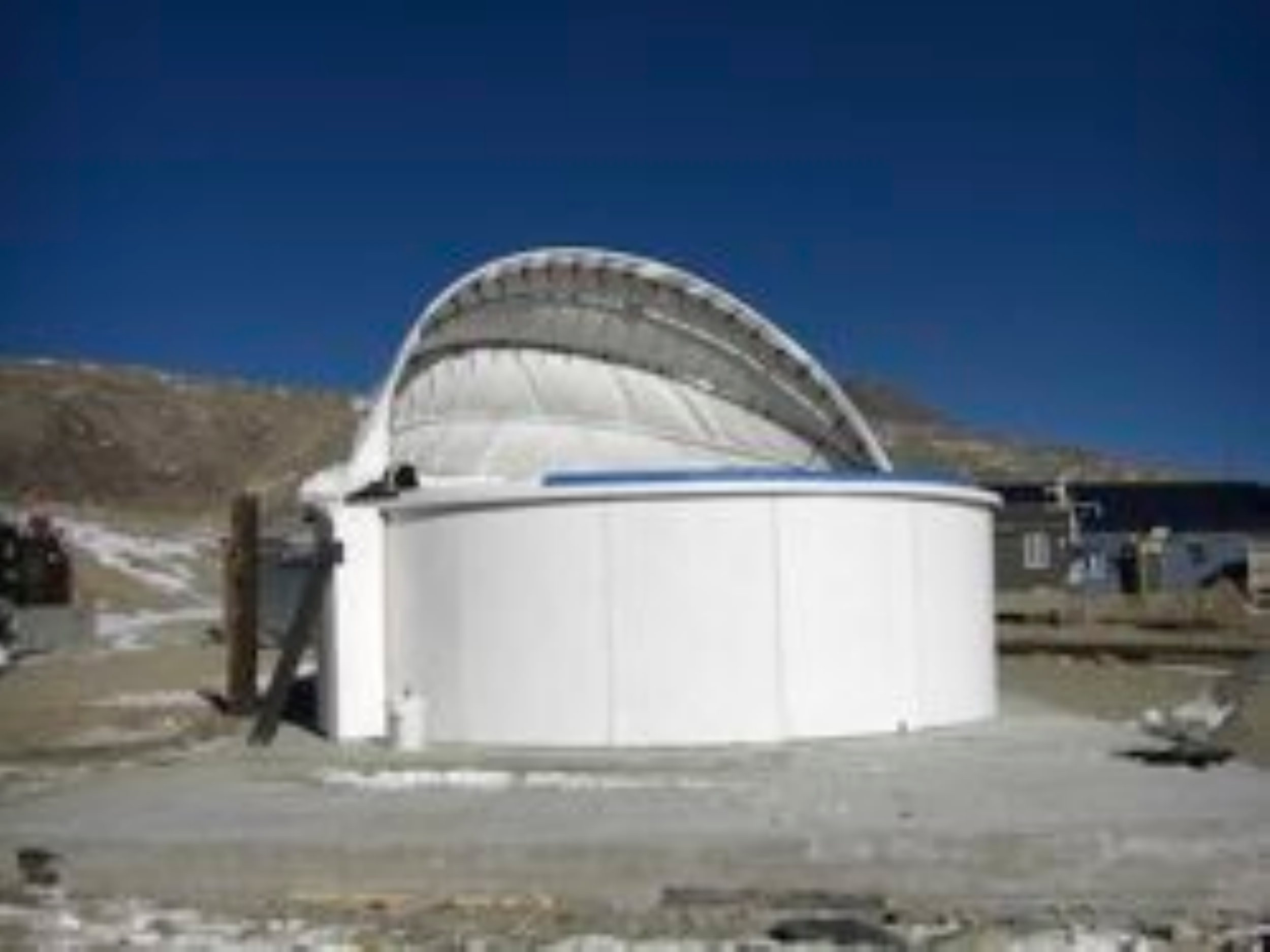}&
\includegraphics[width=6.75cm,height=6.75cm]{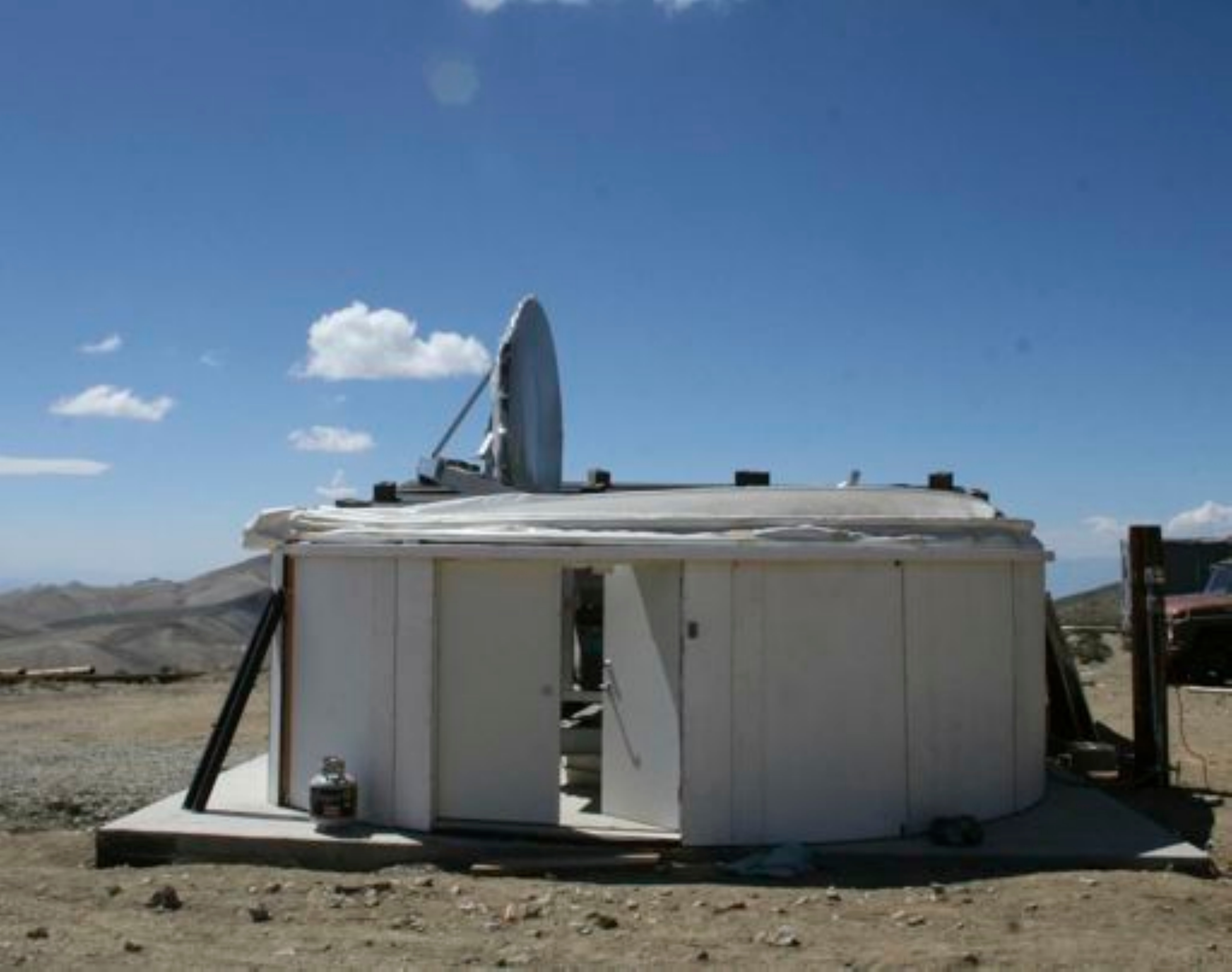}\\
\end{tabular}
\caption[B-Machine dome opening sequence]{Opening sequence for B-Machine dome, notice the roof is completely retractable. Andrew Riley is seen in the top left image standing on the 20 ft x 20 ft concrete pad that he laid to construct the dome on. The bottom right picture is from a different angle and has B-Machine in the dome before the baffling had been added. \label{fig:DomeOpening}}
\end{figure}

\chapter{Description of the B-Machine Instrument}\label{chap:instrument}

The B-Machine telescope was designed to test a new technique in CMB polarization detection (see Section~\ref{chap:polarizationrotator}) and to measure CMB polarization from a previously established site (White Mountain Research Station, Barcroft, henceforth referred to as WMRS). The construction of the telescope has been an on going process for the last several years. Each of the telescopes subsystems was constructed and tested at UCSB prior to full integration and deployment to WMRS. The majority of the work constructing the telescope was performed by me, with general design and construction help coming from lab personnel including Peter Meinhold, Jared Martinez, Hugh O'Neil, and Andrew Riley. There are also a handful of undergraduates and others that deserve some thanks and a list of them can be found in the acknowledgements section.

\section{Telescope}\label{sec:telescope}

\begin{figure}[p]
\begin{center}
\includegraphics[width = \textwidth]{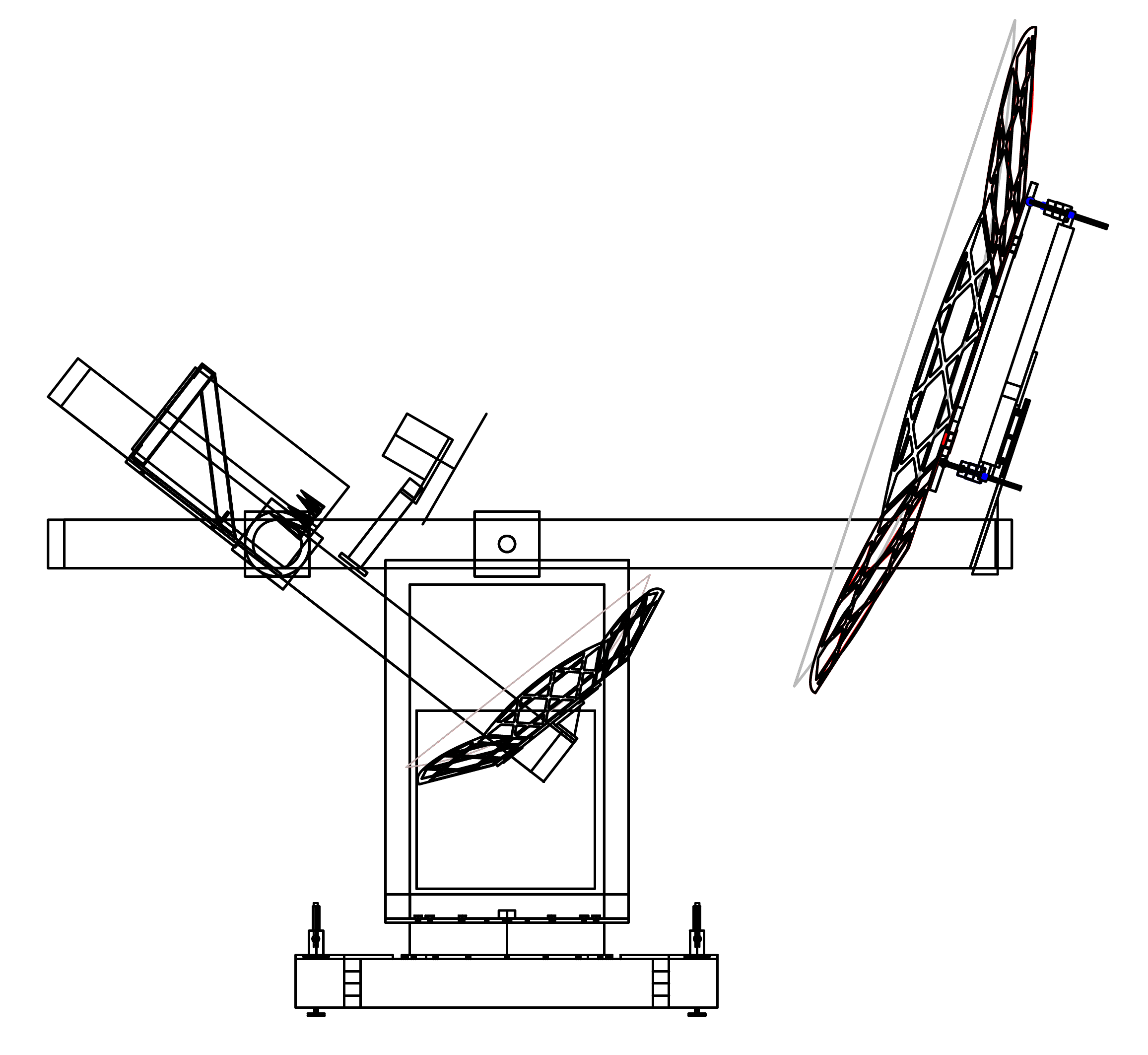}
\end{center}
\caption[B-Machine line drawing]{Line Drawing of B-Machine telescope including Polarization Rotator, dewar, optics, and table. \label{fig:bmachinedrawing}}
\end{figure}

\begin{figure}[p]
\begin{center}
\includegraphics[width = \textwidth]{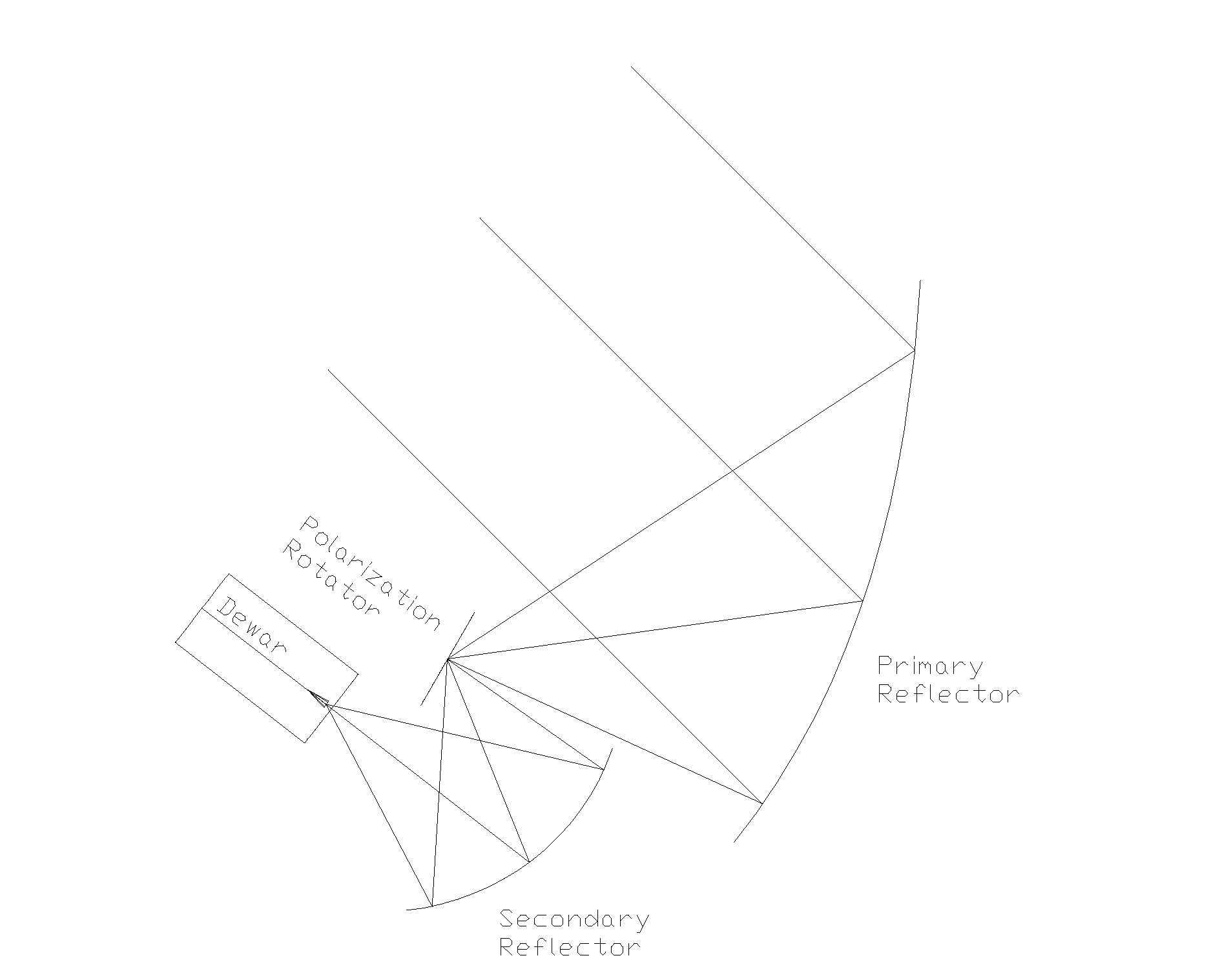}
\end{center}
\caption[Optical design of B-Machine telescope]{Optical design of B-Machine Telescope. The parabolic primary reflector, plane mirror Polarization Rotator, ellipsoidal secondary reflector, and the dewar are shown. Radiation from the sky is focused by the primary onto the common focal point of the primary and secondary where the plane mirror Polarization Rotator is located and then focused by the secondary to the phase center of the central corrugated feed horn.\label{fig:opticaldesign}}
\end{figure}

\subsection{Optical Design}
B-Machine is a modified off-axis Gregorian telescope with a reflecting half wave plate polarization modulator at the confocal point. This design is a slight modification of the BEAST and WMPOL optical design \citep{childers05,figueiredo05, meinhold05, mejia05, odwyer05}.  The optics consist of a primary $2.2$ m off-axis parabolic reflector, a $0.9$ m ellipsoidal secondary reflector and a reflecting polarization modulator as seen in Figure~\ref{fig:opticaldesign}. The telescope meets the Dragone-Mizugutch condition for minimal cross-polarization contamination and maximum focal plane area \citep{Dragone78,Mizugutch76}.

\begin{figure}[p]
\includegraphics[width = \textwidth]{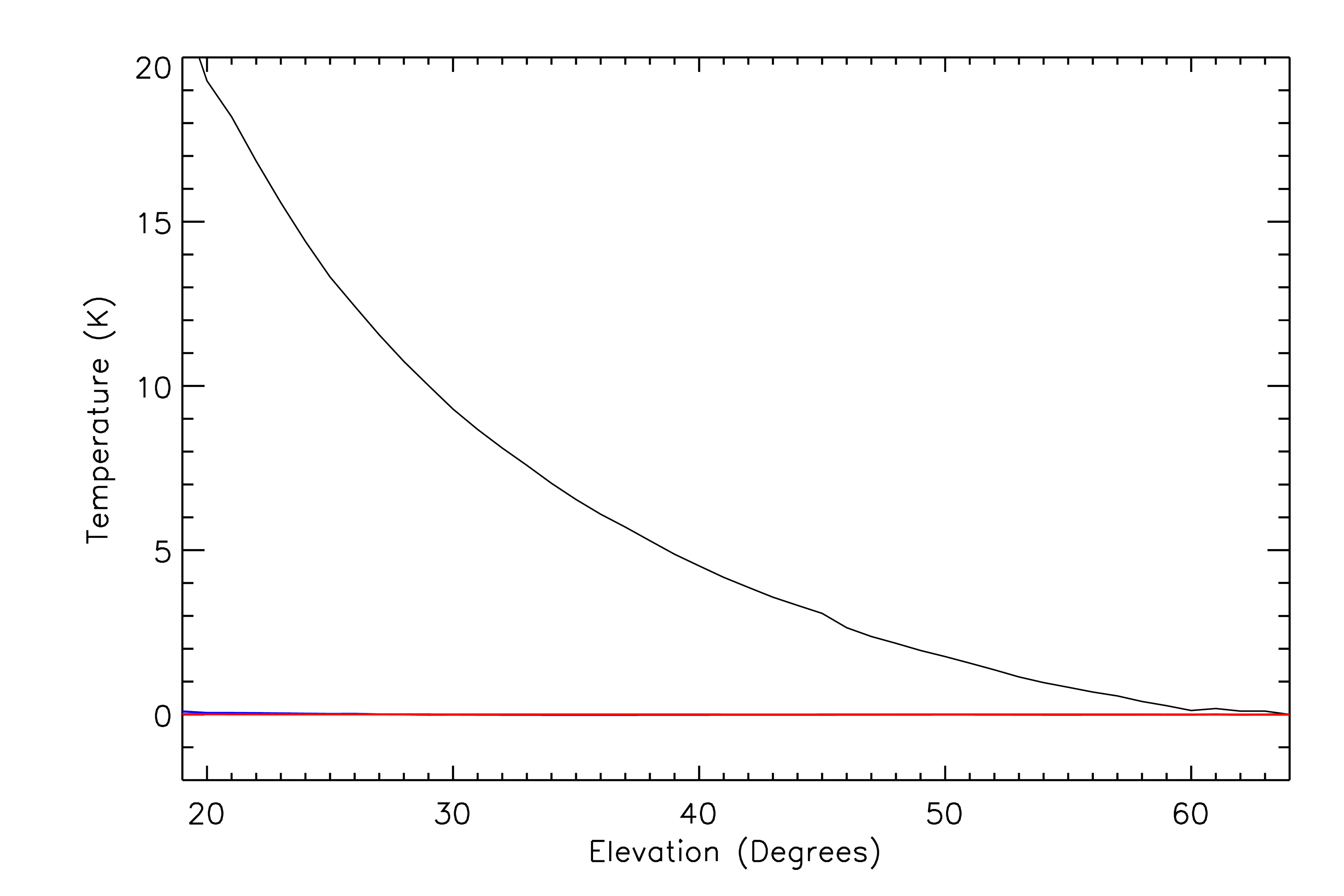}
\caption[Cross polarization isolation using sky dip]{Sky dip with $\mathrm{T}_{\mathrm{sys}}$ removed, black line is the temperature change as a function of azimuth ($90^{\circ}$ corresponds to zenith), blue and red lines are $Q$ and $U$, respectively. The slope of $Q$ and $U$ is over 20 dB lower than that of $T$ indicating a isolation of $>20$ dB. The Zenith sky temperature calculated from these data is 10 K, consistent with atmospheric models of the site. \label{fig:crosspolarization}}
\end{figure}
The Dragone-Mizugutch condition for an off-axis Gregorian telescope is shown to be
\begin{equation}\tan{\alpha} = \left(\frac{e+1}{1-e}\right)
\tan{\beta},
\end{equation}
by \citet{Dragone78}, where the parameters $\alpha$, $\beta$, and $e$ can be found in Table~\ref{tab:optics}. The Reflecting Polarization Modulator is positioned at the confocal point of the primary and secondary reflectors. The optical system is folded about this point so that the overall beam paths have the same lengths as the Beast \citep{childers05} and WMPOL \citep{levy08} optical design.

\begin{table}[p]
\begin{center}
\caption[Optics Parameters]{Design Parameters of B-Machine Optics \label{tab:optics}}
\begin{tabular}{lc}
\hline\hline\\
Parameter & Value\\\\ \hline \\
Primary focal length (mm)                 & 1250.0 \\
Primary max. physical dimension (mm)      & 2200.0 \\
Primary min. physical dimension  (mm)     & 1966.1 \\\\
\hline \\
Secondary semimajor axis (mm)             & 886.7 \\
Secondary semiminor axis (mm)             & 853.4 \\
Secondary focal length (mm)               & 240.7 \\
Secondary eccentricity, e                 & 0.2714 \\\\
\hline\\
Feed angle, $2\alpha$ (degrees)           & 58.2 \\
Angle between axes, $2\beta$ (degrees)    & 35.4 \\\\
\hline
\end{tabular}
\end{center}
\end{table}

\begin{figure}[p]
\begin{center}
\includegraphics[width = \textwidth]{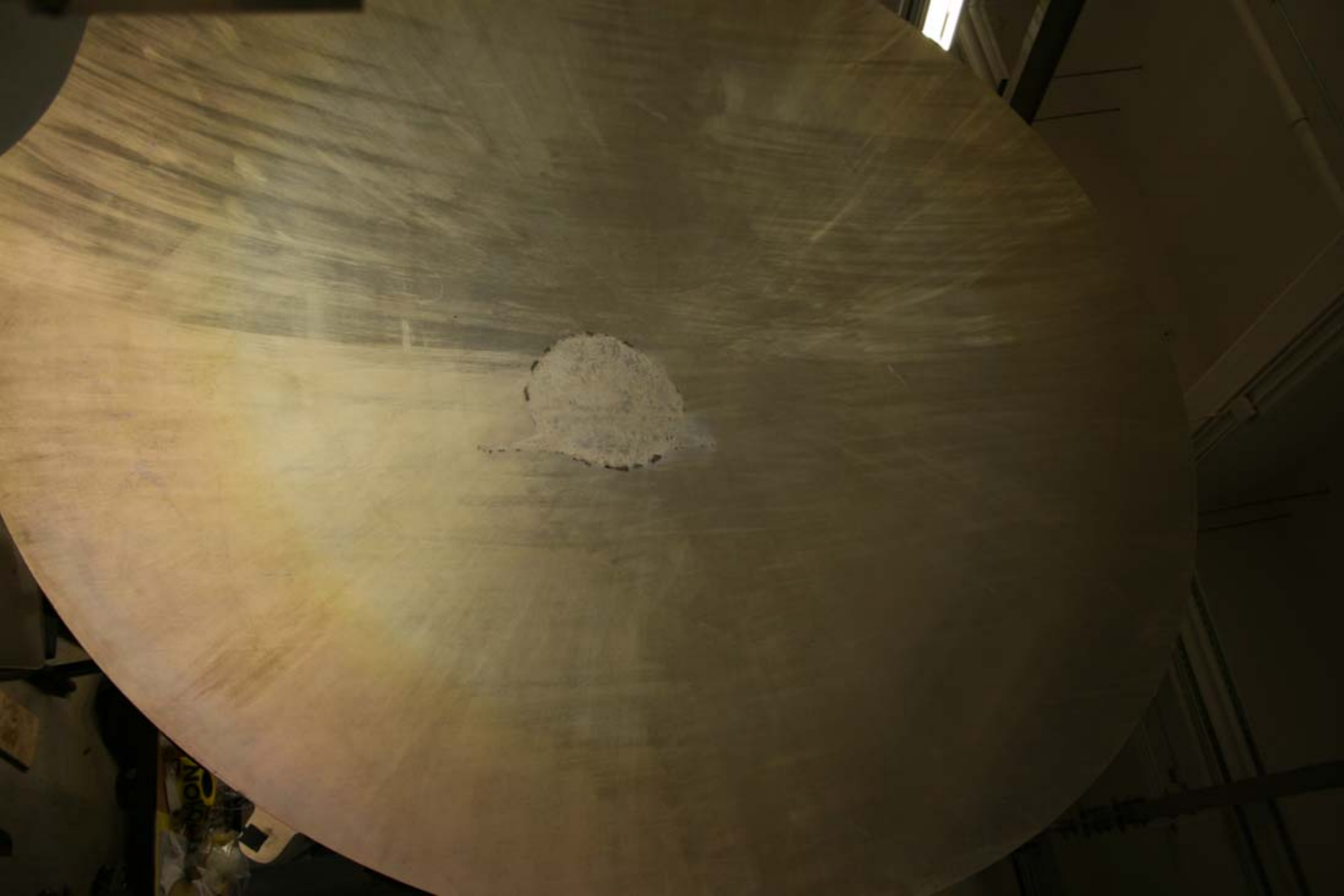}
\caption[Blemish on primary mirror from water evaporation]{Large discoloration in the middle of the primary mirror is oxidation of the Aluminum layer from water that was allowed to pool during long term storage. This blemish caused an $\sim5$ K polarized offset. \label{fig:PrimaryBlemish}}
\end{center}
\end{figure}

The same mirrors were used on both the Beast telescope and the B-Machine telescope and were stored in a large crate between observing campaigns. When unpacked from storage there was a large blemish in the middle of the primary reflector, as seen in Figure~\ref{fig:PrimaryBlemish}. It is believed that a small amount of water sat on the mirror and degraded the AL surface during storage. The mirrors were installed and used for testing in spite of the damage. While testing, a large polarization offset was observed when viewing the sky. The blemish appeared to be the cause of the offset and the mirrors were sent to Surface Optics Incorporated and resurfaced. The polarization offset was in fact eliminated following the re-coat of the mirrors.

\subsection{Table}
The mount for B-Machine is a rotating table with a stationary base, the majority of the instrumentation is mounted on the top of the rotating section of the table. The Azimuth drive system for the table consists of a Galil, planetary drive, motor, relative encoder, slip ring, cone bearings, absolute encoder, and control software. The table was retrofitted from a drive cone system \citep{levy08} to a gear reducing direct drive system. A DC motor\footnote{Amtek 40 V} (relative encoder attached to base of motor) was attached via belt to the lower drive cog of the planetary drive gearbox\footnote{Sipco Mechanical Linkage 105}. The upper side of the planetary drive gearbox, which drives the main pulley (19.05") and gives a 127:1 gear reduction between the output of the motor and the table rotation rate. With this gearing the table can rotate anywhere between 3 rotations per minute down to $0.1$ rotations per minute (see Chapter~\ref{chap:telescopechar} Section~\ref{subsec:scanstrategy} for more details). The system is controlled by a servo code originally written for WMPOL and modified, by Marcus Ansman, for use with B-Machine. The code communicates with a multi-axis motion controller\footnote{Galil DMC-2140}, that uses feedback from the relative encoders and the absolute encoders\footnote{Gurley Precision Instruments A25S 16-Bit} to move the table in both azimuth and elevation. The servo computer (the computer that controls the motion, as opposed to the DAQ computer which collects the scientific data) needs constant contact with the Galil for precise motion of the telescope. This communication channel is achieved through a wireless router which enables the moving servo computer to communicate with the stationary Galil (the Galil is mounted to the bottom of the stationary section of the table).  The Galil is unable to source sufficient current to power the motors for direct motion control so a Linear Servo Amplifier\footnote{Western Servo Design Inc. LDH-A1-4/15} is used on both axes enabling a small control voltage, $\pm10$ V, to control the high current motors. The elevation drive system is similar in that the same Galil is used, and an absolute and relative encoder are used in tandem to control its motion. The elevation drive uses a linear actuator to drive the experiment up and down. The elevation drive is not used during normal operation; it is fixed (bolted down) to minimize jitter in this axis.

To properly control and power the experiment it is important that the stationary base of the experiment can communicate with the rotating platform. Though wireless communications are the most sensible, they are not suited for higher current or constant voltage applications. Since, it is necessary to route the power for the entire telescope through the connection a slip ring assembly was chosen. The slip ring allows 12 connections between the moving and stationary platforms. Of the 12 possible connections only 10 were used, 5 for the 220/120 V AC power and the remaining 5 for the elevation encoder. These lines consisted of the A incremental encoder phase, 5 V power from Galil to the relative encoder, Galil ground, and $\pm$ Control lines for the elevation linear amplifier.

\subsection{Leveling}
\begin{figure}[p]
\includegraphics[width = \textwidth]{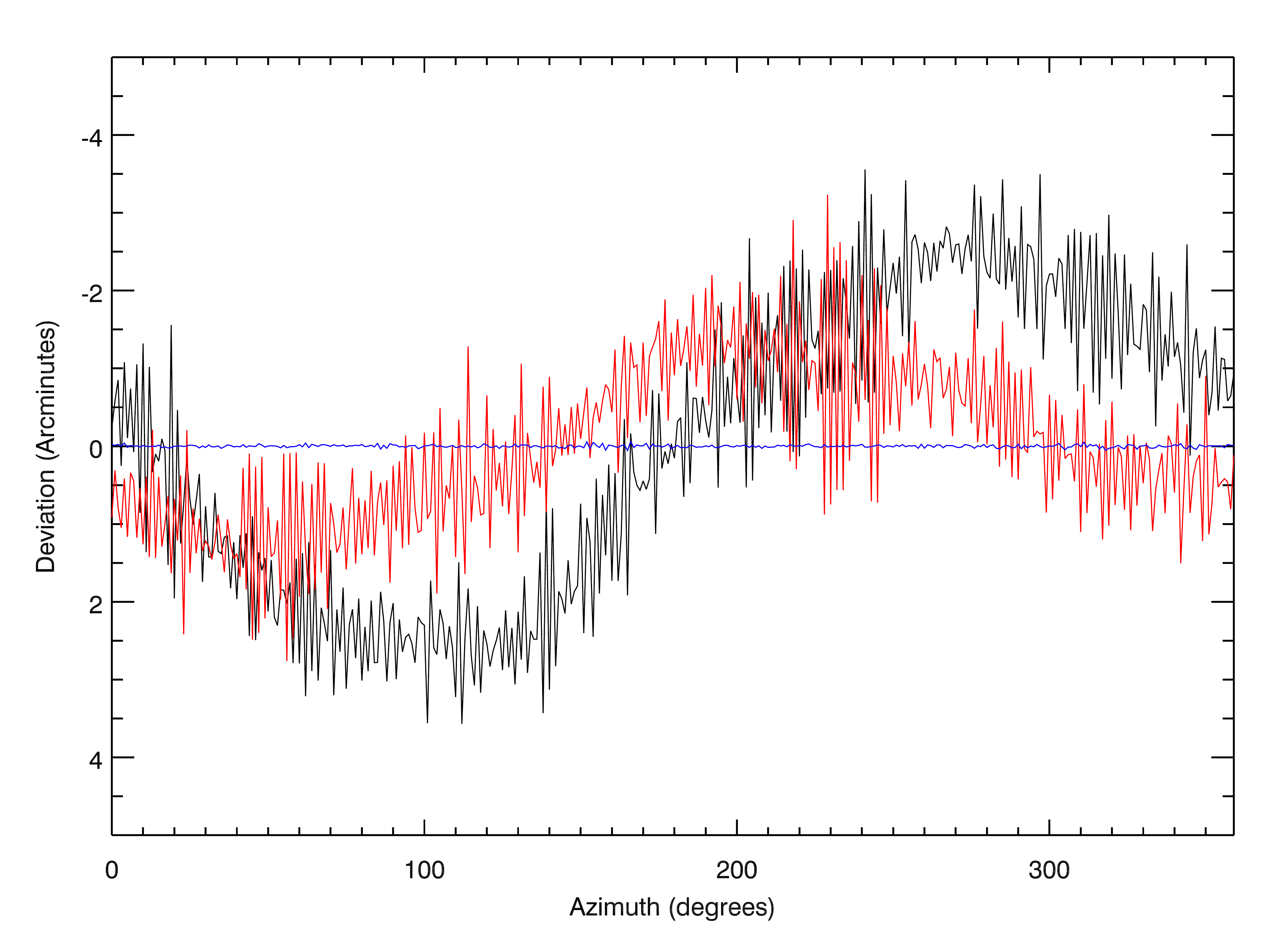}
\caption[X tilt]{Tilt readings from the X axis of the clinometer. Black is the tilt of B-Machine just after fixing of the drive system, red is the tilt after adjusting slightly (current tilt) and blue is tilt measurements with the experiment stationary binned into $360^{\circ}$ bins. Both black and red are averaged over 10 rotations and binned into $1^{\circ}$ sections. The stationary data set is looking at $\sim31^{\circ}$ azimuth and binned into 360 bins for ease of plotting (number of samples per bin similar for all curves).\label{fig:xtilt}}
\end{figure}

\begin{figure}[p]
\includegraphics[width = \textwidth]{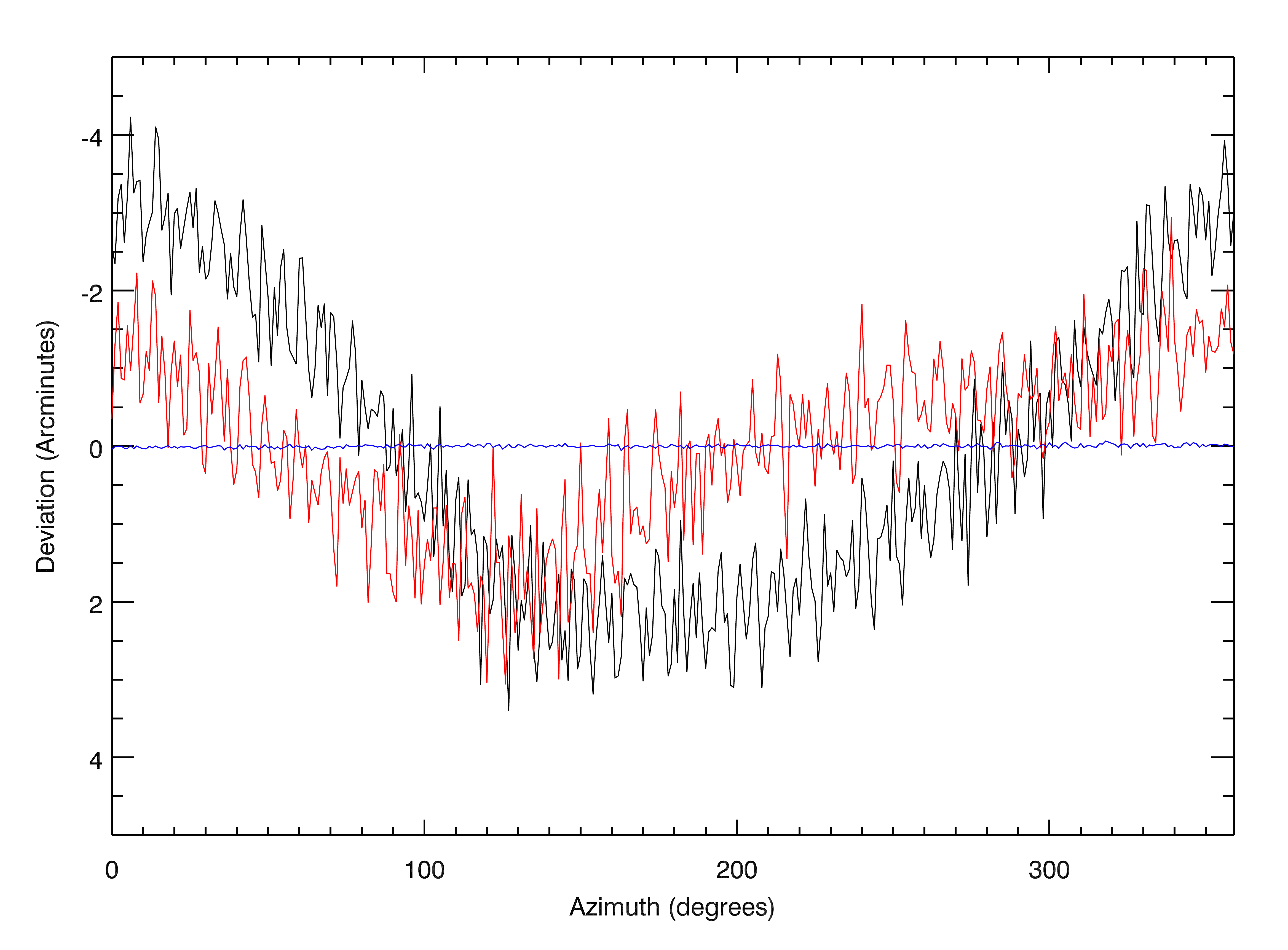}
\caption[Y tilt]{Tilt readings from the Y axis of the clinometer. Black is the tilt of B-machine just after fixing of the drive system, red is the tilt after adjusting slightly (current tilt) and blue is tilt measurements with the experiment stationary binned into $360^{\circ}$ bins. Both black and red are averaged over 10 rotations and binned into $1^{\circ}$ sections. The stationary data set is looking at $\sim31^{\circ}$ azimuth and binned into 360 bins for ease of plotting (number of samples per bin similar for all curves).\label{fig:ytilt}}
\end{figure}

When the experiment was installed into the dome care was taken to level the experiment. Each of the 3 corners of the table rests on 2 6" Aluminum I beams, on top of each of the I beams is multiple thicknesses of shim material (thin pieces of brass) for fine adjustment of the height. A 3 ft long bubble level was used for a rough level of the table and a clinometer\footnote{Applied Geomechanics Inc. Model 904-TH} with a $\pm10^{\circ}$ range was used to level the experiment to operational tolerances, see Figures~\ref{fig:xtilt} and~\ref{fig:ytilt}. The clinometer was read throughout the observing campaign, with each servo file containing X tilt , Y tilt and the temperature of the clinometer. When initially inspecting the clinometer data at UCSB it was found that no small variations in the signal level could be seen due to the input noise level of the data acquisition board. To solve this the signal was run into a times 10.76 amplifying board and then routed into the servo computer.

When inspecting the experiment after reassembly and precursory testing at WMRS it was noticed that 2 of the support/bearing cones between the moving and stationary parts of the table were not always in contact. While adjusting the cones it was found that 3 of the 4 bolts that hold the table top to the drive system were broken, sheared in half, presumably from the 20 miles of unpaved road that the experiment was shipped over. After removing and replacing all of the broken bolts, the level of the experiment was rechecked. The level hadn't changed significantly from previous measurements, but it was re-leveled again as a precaution, see Table~\ref{tab:levelingext}.

\begin{table}[p]
\begin{center}
\caption[Leveling Extremes Before and After Re-leveling]{Leveling Extremes Before and After Re-leveling \label{tab:levelingext}}
\begin{tabular}{|c|c|c|c|}
  \hline
  Tilt Axis & Min & Max & Total Deviation\\
  \hline
  Y before & -4.24' & 3.40' & 7.64'\\
  \hline
  Y After & -2.95' & 3.06' & 6.01'\\
  \hline
  Y Stationary & -0.07' & 0.06'& 0.13'\\
  \hline
  X before & -3.56' & 3.57' & 7.13'\\
  \hline
  X After & -3.23' & 2.76' & 5.99'\\
  \hline
  X Stationary & -0.07' & 0.06' & 0.13'\\
  \hline
\end{tabular}
\end{center}
\end{table}
Leveling monitored through out the campaign shows no significant change in the overall leveling of the instrument from the beginning to the end of observations.

\subsection{Pointing} \label{sec:pointing}
Determination of the pointing of the experiment is achieved through the use of 2 16-bit absolute encoders\footnote{Gurely Precision Instruments Model A25S} and precise leveling of the experiment. In addition to the pointing encoders several other pointing checks were used, Moon crossings, CCD images, and point sources. The Moon was observed once or twice a day (depending on the day) for about 50$\%$ of the observing days. Each moon crossing was used to align the beam by fitting all of the data to a moon model and taking topological corrections for the position of the experiment on the surface of the earth.  A CCD \footnote{Electrophysics Corp. Model WAT-902} equipped with a motorized zoom lens\footnote{Computer Model V10Z1618} and aligned with the beam, captured images of star patterns sporadically during the data campaign. Upon further inspection the CCD images were inconsistent enough that the pointing corrections found from these images were not used. The final pointing evaluation tool was to find bright point sources and correct using the known position of the source. The only source bright enough to see real time was Tau A (Crab Nebula). Evaluating each day for a Tau A crossing and adjusting pointing gave 9 days of additional data to the Moon data. When comparing days with both Tau A and Moon crossings the pointing was consistent to within a beam size. Due to some mechanical difficulties the Moon and Tau A crossings were used for all of the pointing reconstruction. See Chapter~\ref{chap:analysis} Section~\ref{sec:pointreconstruct} for an in depth discussion of the pointing reconstruction.

\subsection{Data Acquisition}
To keep up with the required data rates B-Machine uses 2 separate data acquisition computers:  one computer to collect the scientific data (called the DAQ computer) and another computer for housekeeping data and servo control (called the servo computer). Housekeeping is a generic term used to describe the various pieces of information that are needed to turn the science data into useful information. The servo computer uses a PCI based board\footnote{Measurement Computing Corporation model PCI-DAS6402/16} to read position, tilt, temperatures, time, cryogenic temperatures, and status of gain and calibrator. In tandem with this board the servo computer also incorporates a PCI-DIO24 board to read in a 24 bit synchronization number. The DAQ computer reads in 10 channels of scientific data, synchronization number and time using a USB based board\footnote{IOtech model DaqBoard/3005USB}. Of the remaining 6 science channels 2 are used for the Thermopile and polarization calibrator and 4 are blank though functional.

To recombine the complimentary data sets a synchronization number is generated by using the index pulse from the Polarization Rotator encoder to count each revolution. Each line of the data for both computers has a synchronization number associated with it allowing for recombination of the data sets at a later time. It is essential that the synchronization number is unique on multi day time scales. Given our sample rate of $33.4\mathrm{~Hz}$ and 16 hours of data per day the number will repeat itself every 6 days, giving sufficient time to avoid errors in the recombination process.

\section{Radiometer}\label{sec:radiometer}
\begin{figure}[p]
\includegraphics[width = 13.5cm]{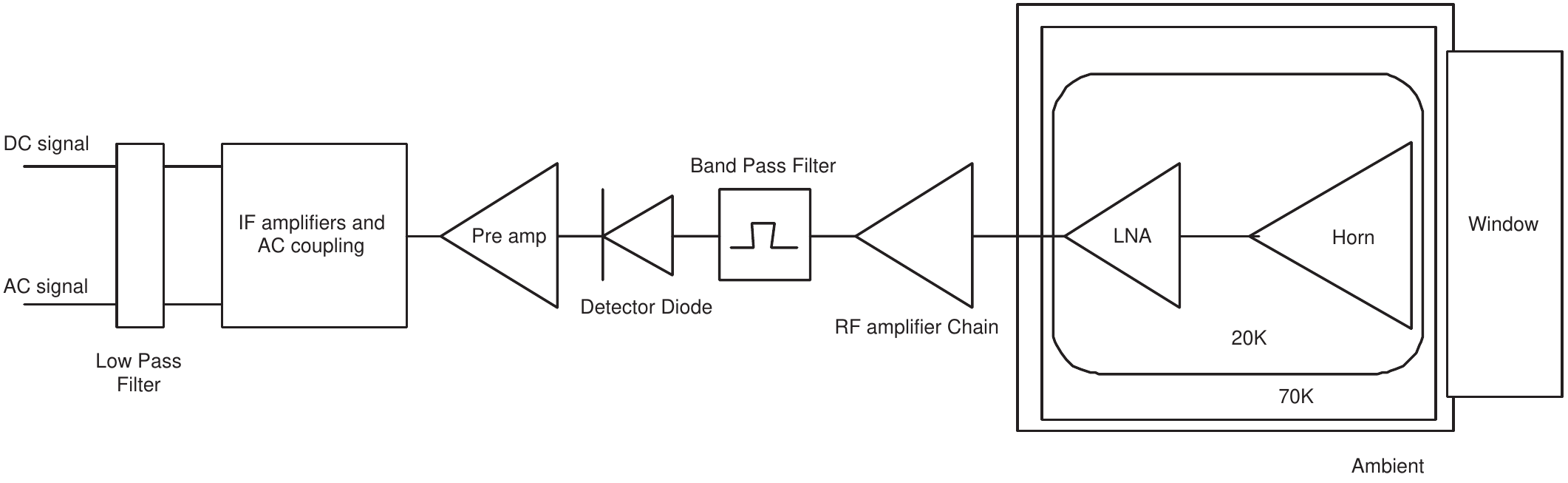}
\caption[Radiometer outline drawing]{Outline drawing of a single radiometer chain. Radiation enters from the right through a microwave transparent window and is coupled into the first LNA by a corrugated feed horn. Low loss coaxial cable carries the signal from the cryogenic LNA to the room temperature LNA's (front end to back end amplifiers). The signal is then passed through a band-pass filter and rectified using a square law diode. Directly attached to the diode is an IF (Intermediate Frequency) amplifier where it is split off into an AC coupled and DC signal. Each signal is passed through a 1.7 kHz low pass filter before entering the A/D converter. \label{fig:radiometeroutline}}
\end{figure}

Much of the main guts (basic wiring and overall structure) of the radiometer were salvaged and reconstituted from the BEAST experiment \citep{childers05}. Very few of the BEAST RF chains survived the punishment of time and static discharge to be used in B-Machine, but the basic signal path is the same.

The microwave signal enters at a sealed window, a low loss extruded polystyrene material that provides a vacuum seal and a first layer of infrared blocking. Between the window and the $20$ K detector array lies a multilayered IR blocking window. This window consists of several (10-20) thin sheets of low loss extruded polystyrene material, each layer is separated by a small gap and attached to a cooling shield. This allows each layer to run slightly cooler than the one above it enabling the horn array to view an IR source that is significantly colder than ambient temperature. Without IR blocking the thermal radiative load would cause the horns to run significantly warmer than the rest of the array giving rise to thermal load dependent signal which could fluctuate significantly from day to day or hour to hour.  The corrugated scalar feed horns and the front end amplifiers are housed in a cryogenically cooled dewar and kept at $\sim20$ K. The RF signal comes out of the vacuum vessel through low loss coaxial cable where it is further amplified, filtered, and rectified by temperature regulated back end RF chains. It then ultimately is saved via analog to digital (A/D) conversion on hard disk for post processing.

\subsection{Feed Horns}
\begin{figure}[p]
\begin{center}
\includegraphics[width = 13.5cm]{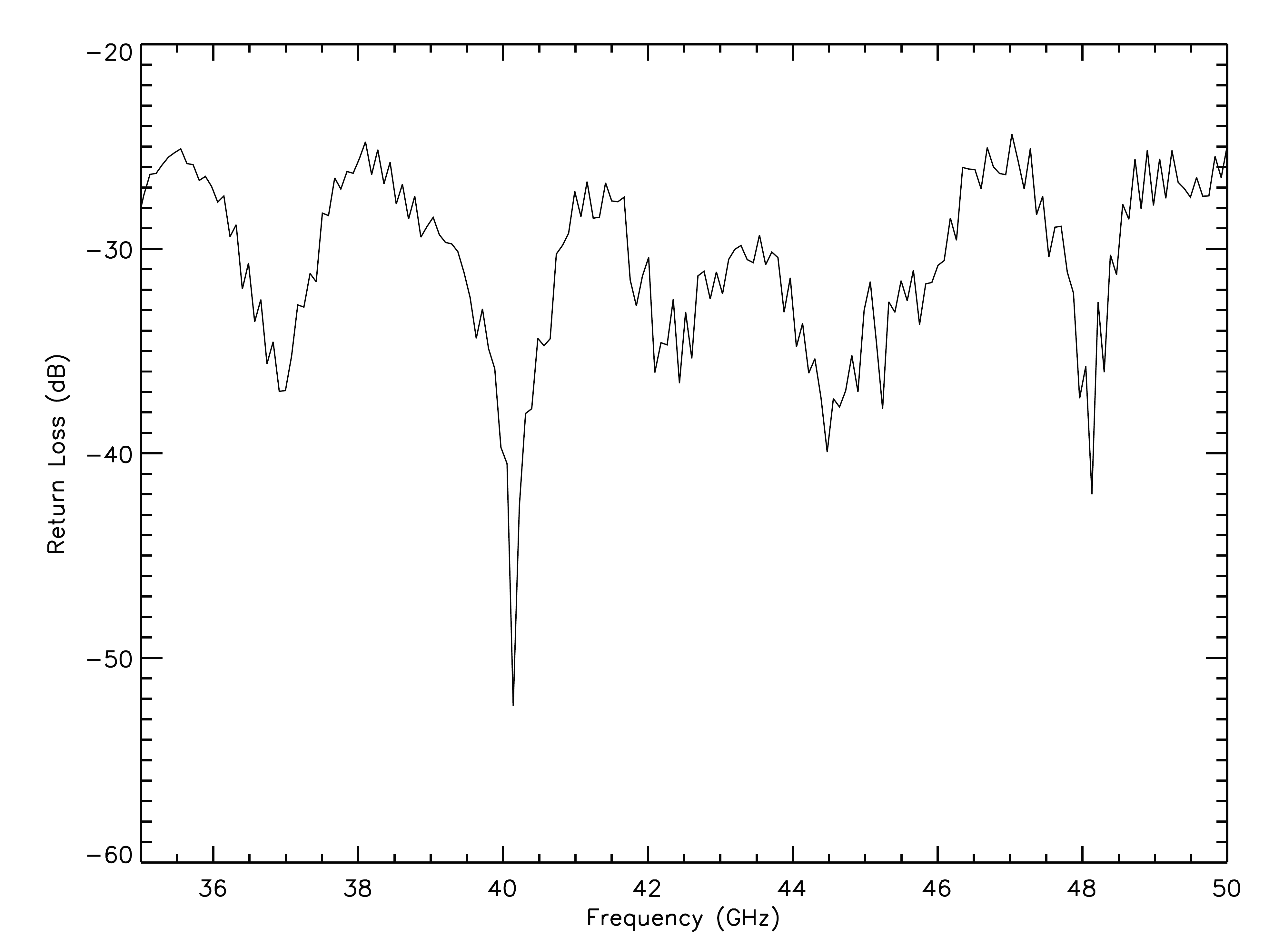}
\end{center}
\caption[Measured corrugated feed horn return loss (S11) ]{Measured input return loss (S11) of a Q-band feed horn normalized so that 0 dB is all power reflected. \label{fig:hornretloss}}
\end{figure}

As shown in Figure~\ref{fig:radiometeroutline}, conical corrugated scalar feed horns \citep{villa97} couple the microwave radiation from the sky to the telescope. Figure~\ref{fig:hornretloss} shows the return loss of a horn from data taken on a Vector Network Analyzer (VNA)\footnote{HP8510C Vector Network Analyzer from 45 MHz to 50 GHz}. Designed specifically for CMB experiments the full details of the horn design and testing can be found in \citep{villa98,villa97}.

\subsection{Amplifiers}
\begin{figure}[p]
\includegraphics[width = 13.5cm]{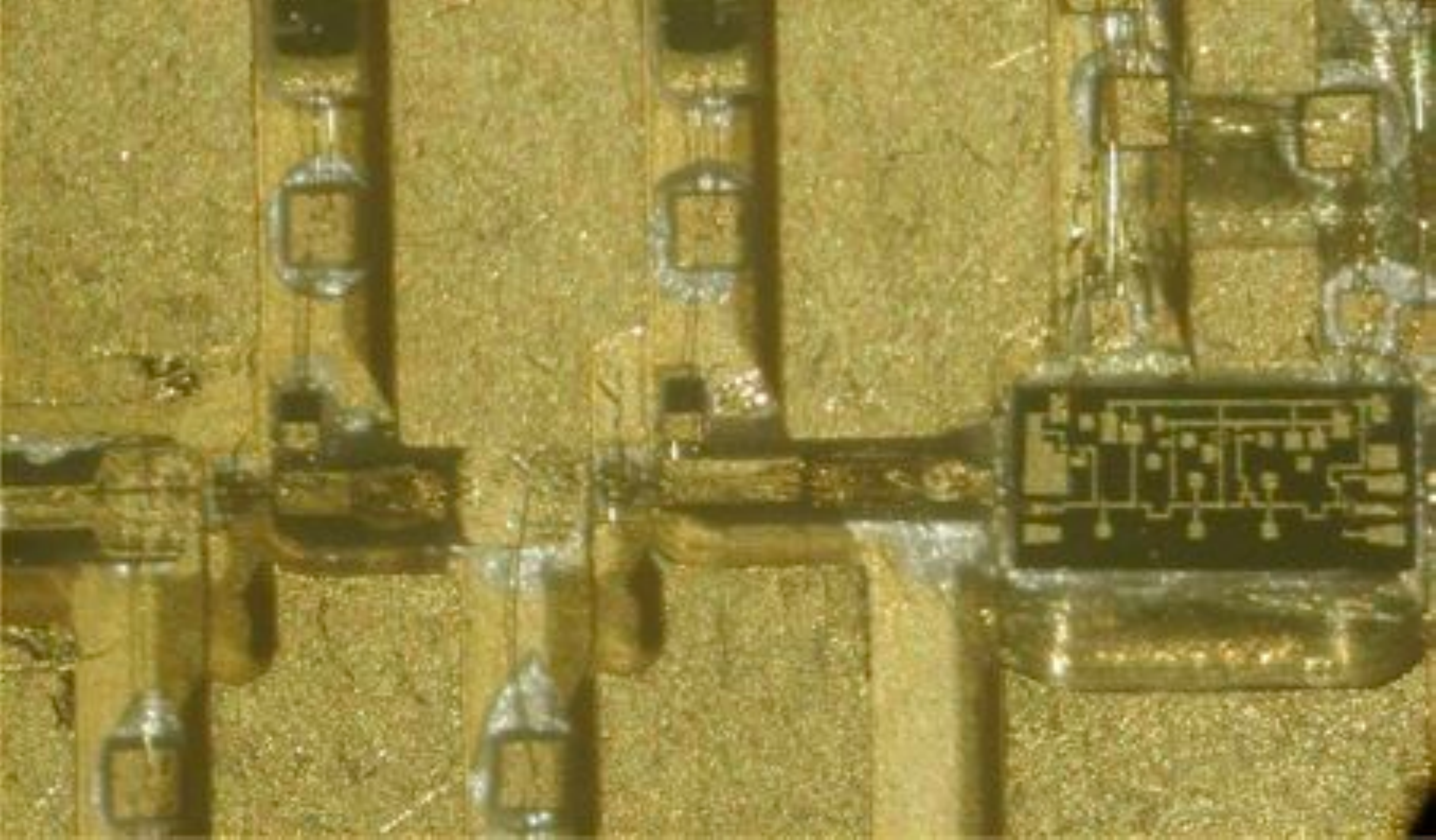}
\caption[MIC and MMIC amplifier]{A hybrid test amplifier that contains both MIC amplifiers (left side) and MMIC amplifier (right side). On the MIC side the discrete amplifiers can be seen with the bias and bias protection circuits running in from both sides, the top is the drain side and the bottom is the gate side. The 2 MICs in the circuit give $\sim10$ dB of gain while the MMIC gives $\sim25$ dB of gain. \label{fig:MMICVsMIC}}
\end{figure}

The detector array on B-Machine is equipped with 5 Low Noise Amplifiers (LNA's) that use FET (field effect transistor) technology, 3 of which are microwave integrated circuits (MIC) based and the 2 remaining are monolithic microwave integrated circuits (MMIC). A MMIC has the majority of the bias network coalesced into one small chip, as opposed to the MIC which requires additional bias electronics and tuning. The 2 different types of circuits can be seen in Figure \ref{fig:MMICVsMIC} in a single package. Each of the LNA stages has approximately 25 dB of gain and has been optimized for the lowest noise temperature in our band-pass (38-45 GHz). Each LNA is followed by a back end module that consists of several room temperature MMIC's, with roughly 60 dB of gain, a band-pass filter and a detector diode. With the exception of the detector diode all of the RF chains were designed, assembled, tuned, and tested at UCSB by either Jeff Childers or myself.

\begin{figure}[p]
\includegraphics[width = 13.5cm]{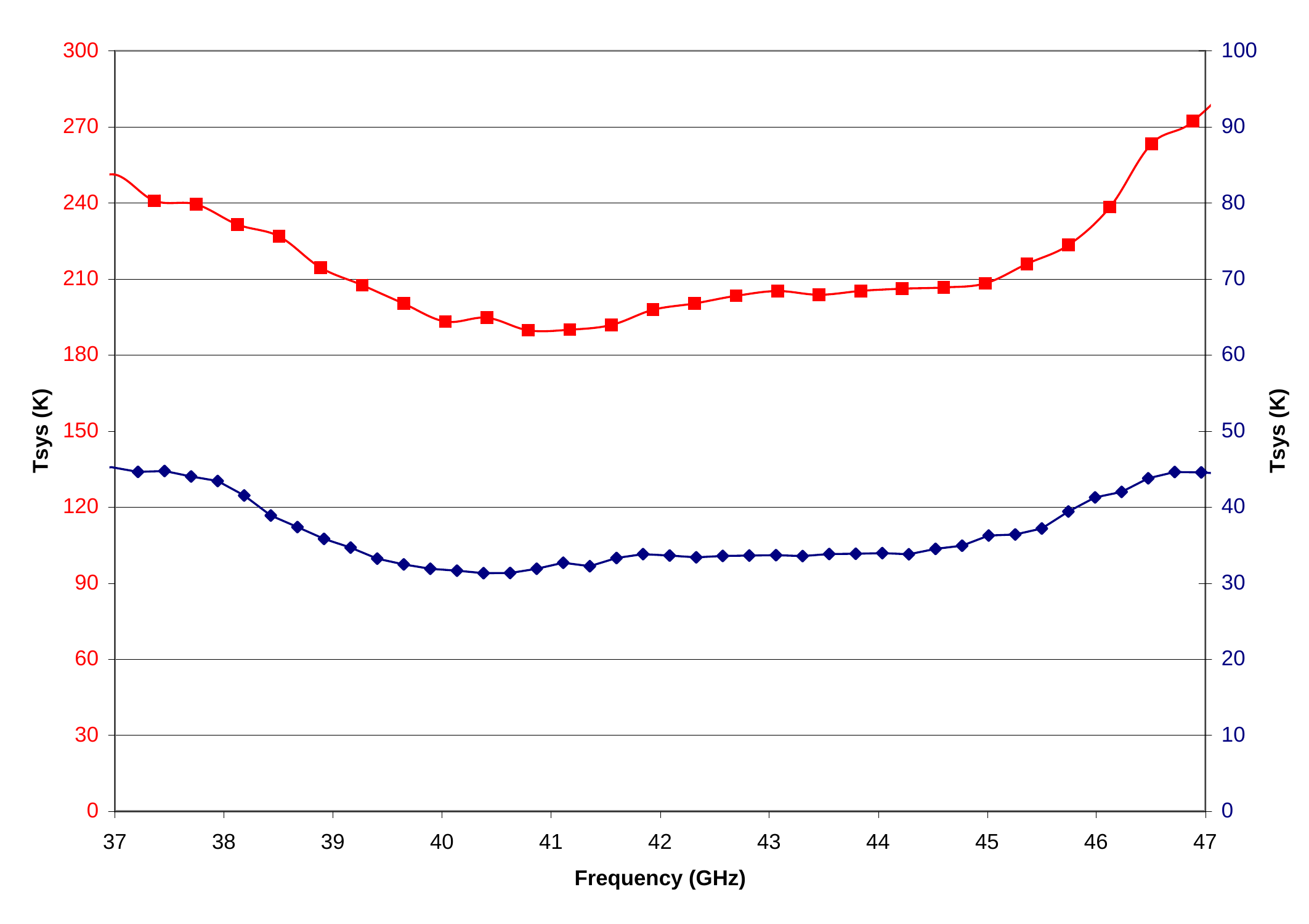}
\caption[45LN1 noise temperature at ambient and 20 K]{The noise temperature of a 45LN1 MMIC chip at ambient temperature left axis in red and 20 K right axis in blue. Chip fabricated at HRL using a Sandy Weinreb design. Amplifier assembled and tested at UCSB. \label{fig:NTVstemp}}
\end{figure}

The 2 MMIC based amplifiers were test chips from JPL wafer runs and when testing a stability problem associated with a feedback capacitor on the gate of the first stage was found. The problem is outside of our band pass and presents some minor stability problems for our measurements, but a new wafer run of the chip was too far off to try and implement any new designs. Each of the LNA's  is cooled to  $\sim20$ K using a CTI Cryogenic Cryodyne refrigerator\footnote{Helic Company Model SC Compressor and Model 350CP Cold Head}.  The noise temperature of these LNA's drops approximately linearly with temperature from 300 K to $\sim20$ K, see Figure~\ref{fig:NTVstemp}, greatly reducing the effective integration time necessary to achieve desired error bars.  Further cooling yields little to no improvement on the noise temperature and requires significant work implementing.

The LNA's output is coupled via stainless steel coax (for thermal isolation) and copper coax to a back end amplification block. The coax cable provides both thermal and RF isolation between the front end and back end amplifiers and carries the signal with minimal loss to the room temperature section of RF gain, see Table~\ref{tab:backends}. Each individual component in the back end modules is attached together using a gold plated brass carrier that allows for the replacement and testing of the individual components prior to assembly. Though each back end is much more massive than is necessary, the mass provides thermal stability for the unit. All of the back ends are bolted to a large Aluminum plate that is insulated and thermally regulated. Thermal regulation is achieved through a temperature sensor feed back loop that uses an AD590 temperature sensor and power resistors for heating. The back ends are set to run at 305 K, to stay in thermal regulation during both night and day cycling. Three of the 5 back ends used were from the BEAST experiment with only minor alterations and the additional 2 were the same design but complete rebuilds.  The three amp chains from BEAST needed tuning and gain adjustment before use.

\begin{table}[p]
\begin{center}
\caption[Back End Amplifier Blocks]{Components of Back End Amplifier Blocks \label{tab:backends}}
\begin{tabular}{|c|c|c|c|c|c|}
  \hline
  Channel & Amp 1 & Amp 2 & Amp 3 & Filter (GHz) & Diode\\
  \hline
  \hline
  0  &   44LNA1$\_$80  &   ALH192C    &  HMMC-5040 & 38-45 & 75KC50\\
  \hline
  1   &   44LNA1$\_$80  &   ALH192C    &  HMMC-5040 & 38-45 & 75KC50\\
  \hline
  2   &   ALH244   &   ALH244     &   ALH244  & 37-44 & 75KC50\\
  \hline
  3  &   ALH376    & ALH386    &  ALH386   &   37-44 & 75KC50\\
  \hline
  6  &   44LNA1$\_$80  &   ALH192C  &  HMMC-5040 & 38-45 & 75KC50\\
  \hline
\end{tabular}
\end{center}
\end{table}
With the exception of the 44LNA1$\_$80 all of the back end chips are commercially available. The 44LNA1$\_$80 was a test chip from JPL and is no longer made due to the availability of mass produced devices. The band-pass for all of the amplifier chains is quite large and it is necessary to define the band-pass with an external filter. The first round of filters (38-45 GHz) from BEAST had a band-pass that was dictated by the minimal in the noise temperature of the front end amplifiers. With the second pass of filters shifted down slightly in frequency to avoid the $45\mathrm{~GHz}$ oxygen line in the atmosphere, see Subsection~\ref{subsect:filters} for complete design details.
\begin{figure}[p]
\includegraphics[width = 13.5cm]{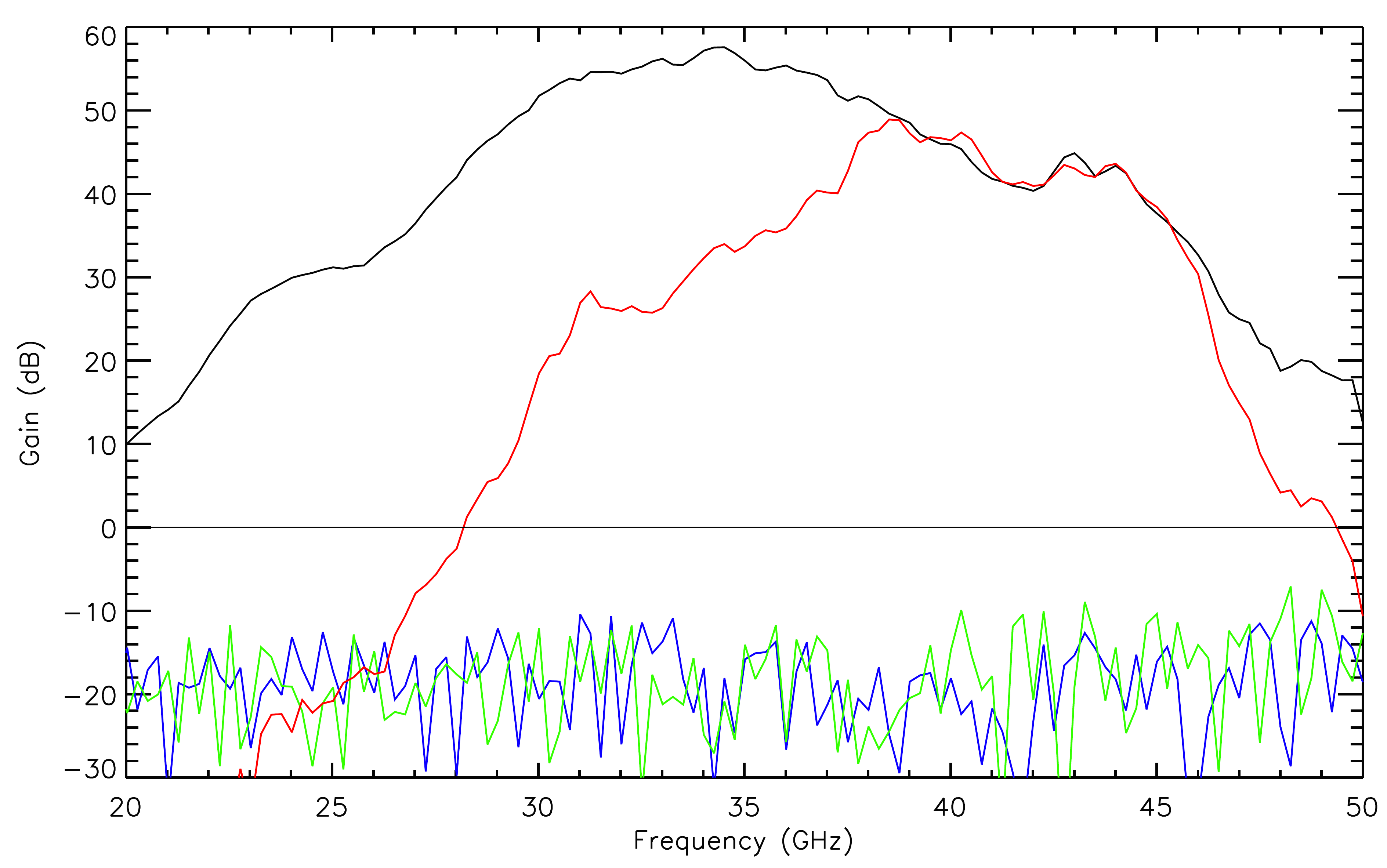}
\caption[Gain profile of typical back end]{Back end gain profile for two back ends used in B-Machine. Neither back end contained a band-pass filter and a 10 dB attenuator was attached to the output of the back end to avoid damaging the test equipment. \label{fig:qbe8}}
\end{figure}

\subsection{Filters}\label{subsect:filters}
Each RF chain has significant gain both above and below the desired band-pass. As a result a coupled line filter was produced to significantly shrink the band-pass of the instrument. The filters reflect all of the out of band power, hence, attenuation is necessary between the output of the final amplification stage and the input of the filter. For every 1 dB of attenuation, 2 dB of isolation is gained. There is typically 6-8 dB of attenuation between the 2 elements and the S22 parameter for the output amps is typically around 10 dB giving an isolation of at least 20 dB. The technique to produce the filters was developed jointly by Jeff Childers, Alan Levy and myself.  

\begin{figure}[p]
\includegraphics[width = 13.5cm]{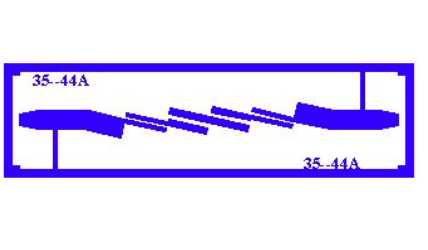}
\caption[Filter mask outline]{Mask drawing of band-pass filter with simulated response from 35-45 GHz and tested response of 37-44 GHz . The large leads on either side of the filter are meant to couple the line to 50 ohm transmission line or glass bead feedthru. The filter is $0.5$ inches long by $\sim0.1$ inches wide. \label{fig:FilterDrawing}}
\end{figure}

The filters are first modeled and optimized for the desired band in Libra/Eesof (a software tool that is the pre-cursor to ADS RF EDA software). Libra supplied the dimensions for the inductive fins and the spacing which was then used to draw a template in AUTOCAD, Figure~\ref{fig:FilterDrawing}. From the template a positive photo mask was produced\footnote{CadArt Services Poway, CA}.  Standard substrate material consisting of $1/2$ oz. copper separated from the bottom $2$ oz copper layer by 5 mils of Quflon (Teflon)  was etched using standard photo-lithography and Ferric Chloride etching techniques and packaged in a generic housing for use in the back end amplifier blocks.

\begin{figure}[p]
\includegraphics[width = 13.5cm]{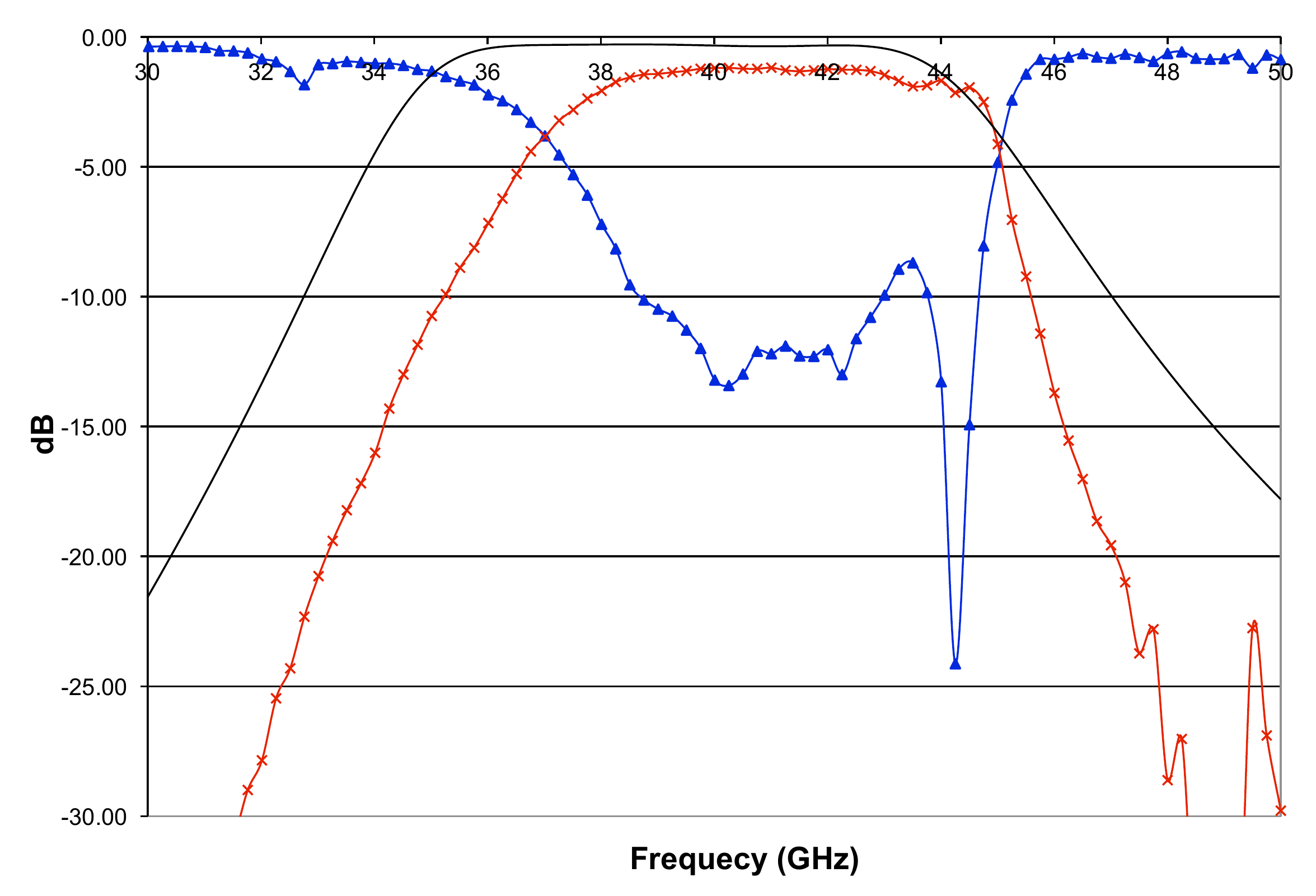}
\caption[Band-pass filter response]{Blue (S11) and red (S21) lines are measured responses from the fabricated filter and the black line is the simulated response from Libra. The discrepancy was known before hand and planned on for final filter results.\label{fig:Filterresponse}}
\end{figure}

Early in the fabrication development processing it was found that the bandwidth of the filters shifted slightly up in frequency and narrowed, shown in Figure~\ref{fig:Filterresponse}. This effect was seen uniformly throughout many different filter designs and was compensated for to achieve the desired band-pass for the filters.

\subsection{Data Input}\label{Subsect:datainput}

\begin{figure}[p]
\includegraphics[width = 13.5cm]{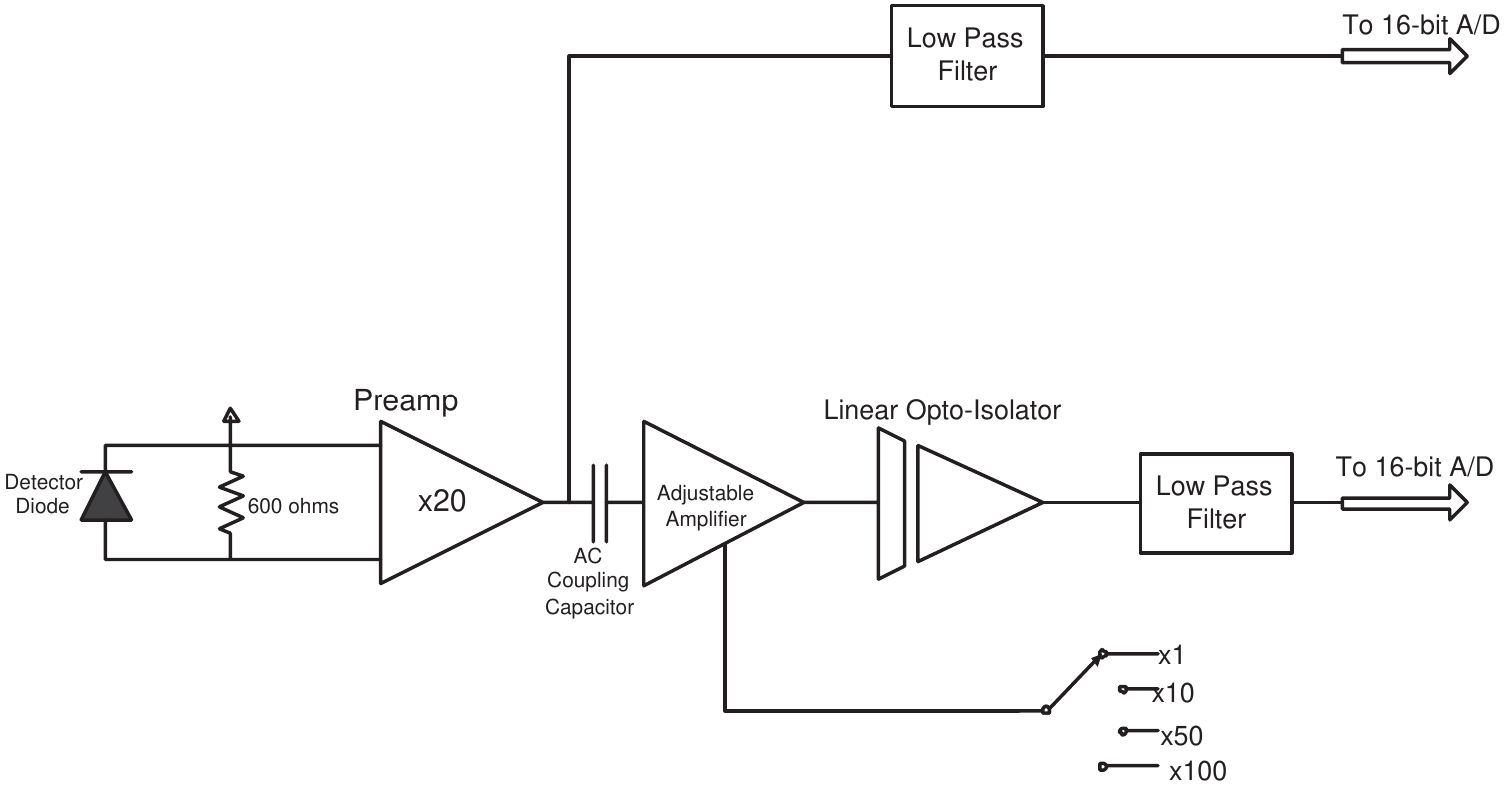}
\caption[DAQ layout drawing]{Simplified layout drawing of the data acquisition system components from the end of  the RF gain section (square law detector diode) to the input of the computer (16-bit A/D converter). The pre-amp is attached to the detector diode directly and connected to the remainder of the electronics (located in a shielded Mu Metal box) via coaxial cabling. \label{fig:DAQIFoutline}}
\end{figure}

Following the band-pass filter the signal is rectified using a linear rectifying diode\footnote{Anritsu Company 75K50 Microwave Detector Diode}, see Section~\ref{Sec:Calibration} for response characteristics, that converts the RF power incident on it to a proportional DC voltage. All gain after the diode is below 1 MHz and referred to as audio amplification. First, in line is a 20 times audio amplifier that is connected directly to the diode; this configuration reduces sensitivity to systematic noise from external sources such as ground loops and radiation from wireless devices. The first stage audio gain also provides the appropriate loading for the diode (600 ohms). Following the first stage the signal is routed to the shielded Planck Acquisition box, called this because it was originally used for testing prototype cards for the Planck Satellite Mission. The signal into the Planck box goes to an audio amplifier (x10), from here the signal is split into 2 paths, see Figure~\ref{fig:DAQIFoutline}. The upper path is referred to as the DC signal, since the voltage is proportional to the absolute power that hits the diode. The alternate path is first AC coupled, taking any offset voltages out, and ran through another adjustable amplifier gain stage, and then to an opto-isolator out of the box. Each of the front end LNA's is associated with 2 data channels, a DC channel and an AC coupled channel (primary science data). The AC coupled channel has a switchable gain setting on it that can be switch between x1, x10, x50, and x100. From the output of the Planck Acquisition Box each channel is run through a 1.7 kHz low pass filter and then to the input of the data acquisition board\footnote{Iotech, Inc. DaqBoard/3005USB} on the DAQ computer.

Each of the DC channels is used to calibrate the corresponding AC channel, see Chapter~\ref{chap:telescopechar} Section ~\ref{Sec:Calibration}. A complete understanding of the gain difference (from the adjustable gain amp) between the AC and the DC channels is critical for the proper calibration of the system. To measure the gain of the system each channel had a small (0.00052 V) sine wave input to the x20 pre-amp. Data were taken for approximately one minute for each of the gain settings. The sine wave data for each channel was fit using IDL. The amplitude from the fit data for each sine wave was divided by the input fit for the next gain level. For each of the channels there are 3 possible gain divisions x10$/$x1, x50$/$x10, and x100$/$x50. In the x1 setting there is also the preceding audio gain of $\sim$x200, this is referred to as the front-end gain here. Also, the DC channels were compared to the AC x1 gain channels to confirm no gain difference between AC and DC channels with this setting.

\begin{table}[p]
\begin{center}
\caption[IF Gain Measurements]{IF Gain Measurements~\label{tab:ifgain}}
\begin{tabular}{|c|c|c|c|c|c|}
 \hline
 AC & Front-end & x10 & x50 & x100 & Total Gain \\
     channel    &           &     &     &      &   at x100 \\
 \hline
 \hline
 Channel 1 & 199.112 &   9.983  &  50.247   & 100.827 & 20,075.87 \\
 \hline
 Channel 2  & 196.588   &   9.957  &  50.334  &  100.782 & 19,812.53 \\
 \hline
 Channel 3   &  195.244  &   9.958  &    49.901  & 100.312 & 19,585.32 \\
 \hline
 Channel 6  &    195.825    &  9.991    &  50.413 & 100.648 & 19,709.39 \\
 \hline
\end{tabular}
\end{center}
\end{table}
In the standard operating mode the adjustable gain switch is set to x100. The total gain is the gain from the output of the rectifying diode to the input of the IOTech data acquisition board. The x100 column in Table~\ref{tab:ifgain} is the multiplication factor for calibration between the AC and the DC channels in standard observing mode.

\section{Electronics}
The telescope runs on the back of multiple electronic subsystems, ranging from simple power distribution to the digital 24-bit synchronization number.

\subsection{Power Distribution}

\begin{figure}[p]
\includegraphics[width = 13.5cm]{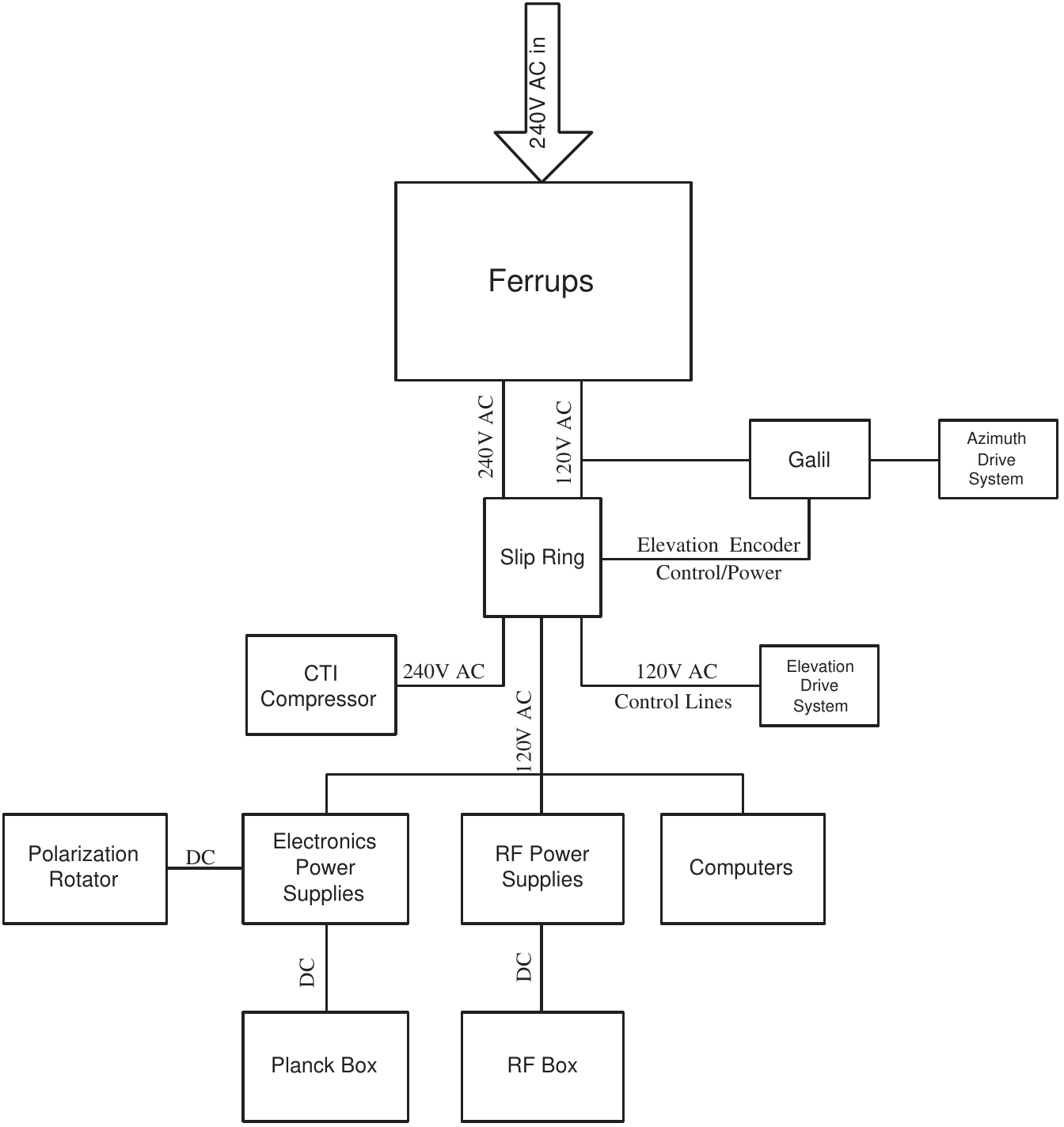}
\caption[Schematic of power distribution]{Layout drawing of the power distribution of B-Machine. Input power is sourced from an independent power grid that gets its power from a battery bank that is charged by either a diesel generator or a bank of solar panels. \label{fig:PowerSchematic}}
\end{figure}

All telescope power is filtered by a Ferro Resonant Uninterrupted Power System\footnote{Ferrups, Eaton Corp. Model FE2.1 kVA}. This system uses 4 12 Volt 35 AH batteries as backup power, allowing sufficient time to shut down all critical systems before a hard shutdown occurs in case of power outages.  The input power coming from the station is sourced from either a solar/battery powered inverter or a diesel generator. The power was switched between these two sources depending on the battery status, load, and weather conditions. Management of the station power was a split effort between telescope personnel and WMRS facility employees.

Each day started with the facility staff checking the status of the battery bank first thing in the morning (around 6 am).  At this point if the battery status was above $90\%$ the generator would be switched off and power would be sourced from the solar/battery inverters.  With the generator off several criteria needed to be evaluated every couple of hours. First, were the solar panels getting direct sunlight,  if not what state were the batteries in. Second, how long had the facility been on batteries and under what kind of load. This depended on the number of people that were using the WMRS facilities at any given time. The generator was typically left off until about 6 pm.  At this point the generator would be turned on for a couple of hours for night time showering and dinner. Following dinner the generator was turned off and it was the telescope personnel's responsibility to turn the generator on at night before bed. At any point if the charge level of the batteries reached the low $60\%$ level the generator was immediately turned on. The station's power requirements have been greatly reduced by upgrading existing systems to more power friendly devices.  Also, the crew will typically conserve power when possible by shutting off overhead lighting and turning off the water heater during solar/battery operation. Even with all of this the power failed due to low battery levels several times. The telescope power is very stable and is buffered by multiple pieces of equipment before getting to the experiment.  This arrangement worked well and I must thank the WMRS staff in doing a job that wasn't there responsibility. 
Power from the generator or solar/battery arrays is run from the main Pace building out to the telescope buildings via 240 V lines. The power comes into the building and goes directly into the UPS and is split into 1 240 V line and 6 120 V lines. The 240 V AC power that is routed through the slip ring goes directly to the CTI cooler and is not used for anything else. This line has its own ground and the cooler is isolated from the rest of the experiment. One of the 120 V AC lines from the UPS is routed through the slip ring and powers all of the systems on the moving part of the table. There are 7 linear power supplies that convert the 120 V AC power to DC power for the assorted electronics (see sections below).  The stationary part of the table uses several of the 120 V lines from the ferro resonant power system for power; this includes the Galil, all azimuth drive systems, and the wireless router that is used for Galil communications.

\subsection{Amplifier Bias}
\begin{center}
\begin{figure}[p]
\centering
\includegraphics[width = 19cm, height=5.5cm, angle=90]{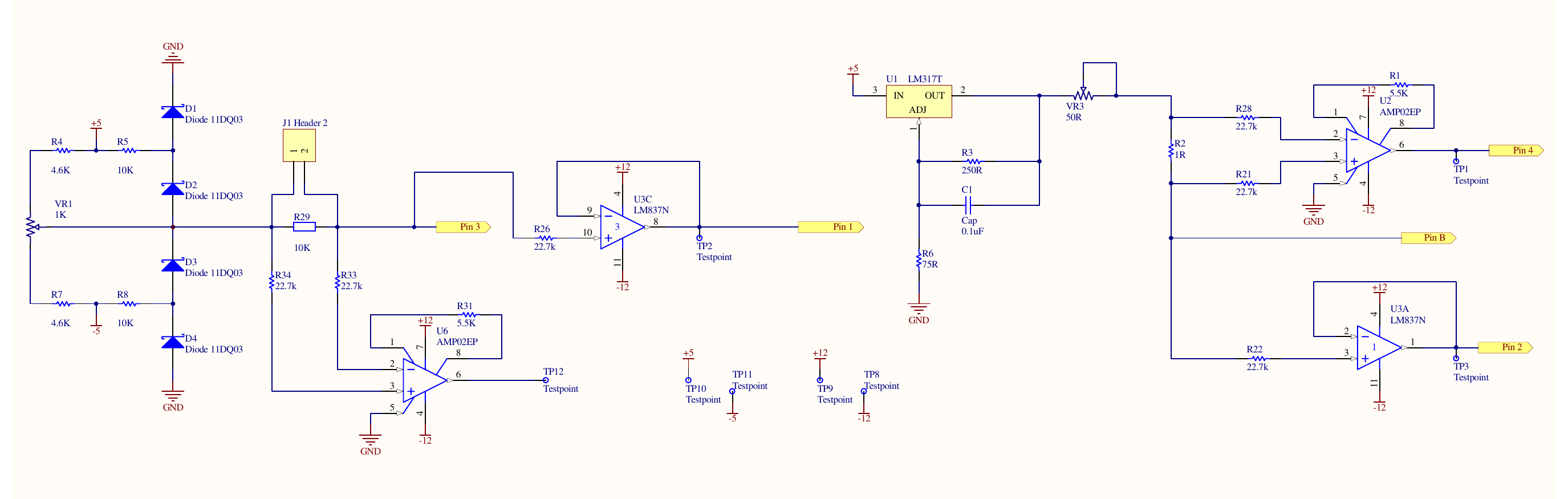}
\caption[Schematic of MMIC bias circuit]{Schematic of front end MMIC Bias board. Originally designed for room temperature back end bias and modified to have readout circuitry compatible with MIC bias board pin outs.\label{fig:MMICBiasSchematic}}
\end{figure}
\end{center}

\begin{center}
\begin{figure}[p]
\centering
\includegraphics[width = \textwidth]{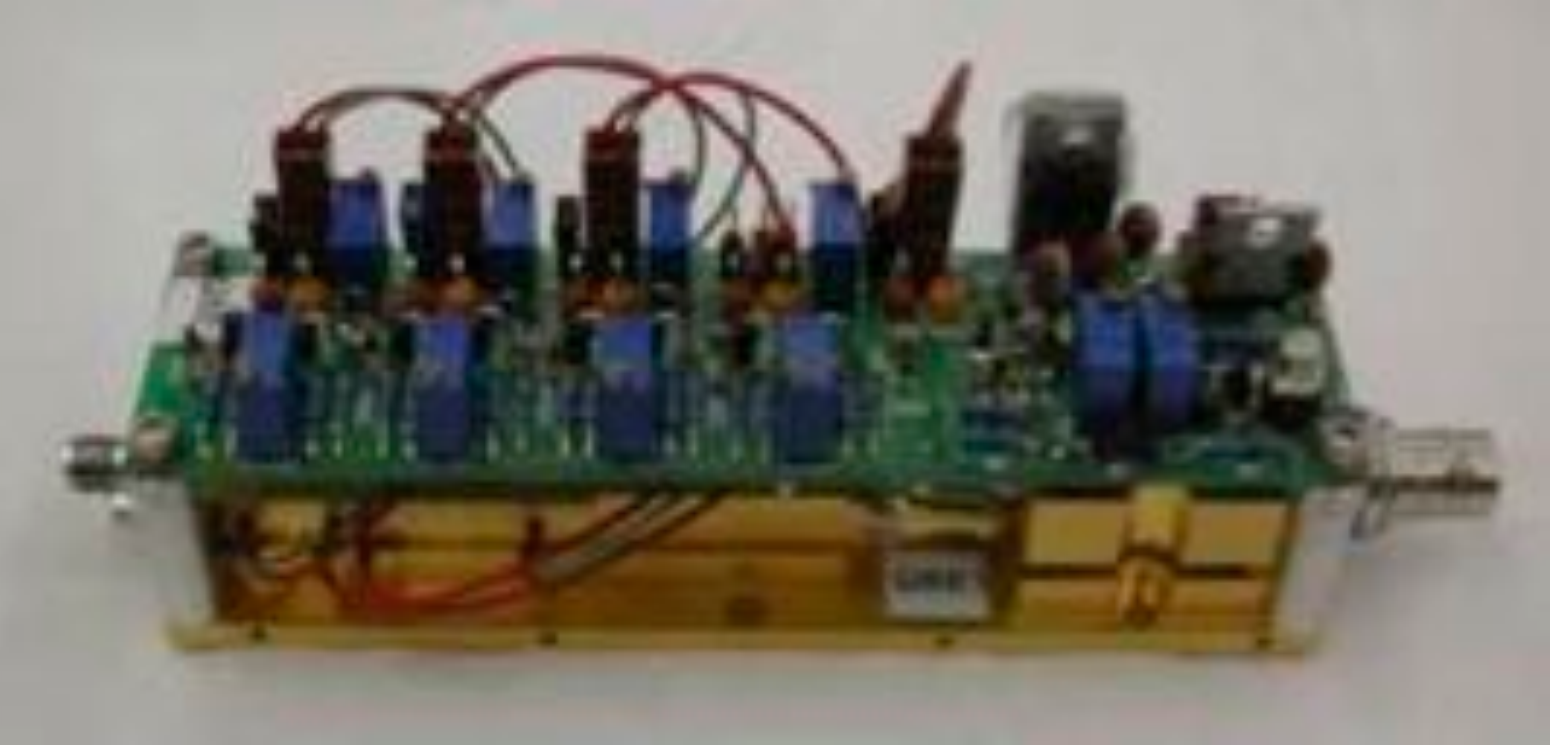}
\caption[Picture of back end module]{Picture of back end module. The RF input is on the left and the diode output is on the right. Stood off above the RF components is the bias board.  \label{fig:backendmodule}}
\end{figure}
\end{center}

There are 2 basic amplifier bias schemes used on B-Machine, one for each class of amplifier (MIC and MMIC). Each of the front end LNA's with discrete amplifiers needs separate bias lines for each FET. The circuit that was used was developed at UCSB by Jeff Cook. This circuit allows for 2 different biasing possibilities, constant current or constant voltage. The normal operational state of the bias is the constant current mode. In this mode the drain voltage and gate current is set to the desired points and the gate voltage is servoed until the drain current is within an appropriate tolerance of the set point. In the time ordered data power spectrum a broad bump can be seen around 800 Hz. This bump is the servo rate of the gate control circuit. In addition to biasing, this board also provides some over voltage and over current protection; it limits the maximum and minimum values that the amplifiers can be supplied to, 1.75 V on the drain, 10 mA drain current or $\pm0.4$ V on the gates.  These boards are not suited to bias the MMIC's due to the current requirements of the larger chips.
To bias the MMICs, 2 front end chips and all of the back end modules, a constant voltage bias scheme was used. The front end MMICs use a board that is the same bias design as the back end boards with minor modifications for voltage and current readouts.  Using the pin outs from the front end MIC bias boards, a front end MMIC bias board was developed that supplied the appropriate readouts to the front panel BNC connectors, see Figure~\ref{fig:MMICBiasSchematic}. Each back end board is directly attached to the back end module, as shown in Figure~\ref{fig:backendmodule}.  The back end bias boards provided 3 important elements, bias power, voltage protection, and sequencing.  The voltage protection on the gates is provided by diode protection and ensures that any potentially dangerous spikes in the voltages are not conducted to the gates but rather shunted to ground and limits the gate voltages to $\pm0.4$ V. Drain/gate sequencing is required for MMICs in general, otherwise they have the possibility of burning out on power cycling.  If the gate voltage is not present when the drain voltage comes on the device can draw excessive amounts of current and burn out the FETs on the chip.

\subsection{24-bit Synchronization Number}
One of the most important subsystems in B-Machine is the synchronization number (sync number), which allows the data from the 2 different computers, DAQ and servo, to be recombined. Unfortunately, in my haste to get the telescope working there is very little documentation for this board. I will attempt to outline the basic operation of this board and include the fundamental elements (IC chips) that are used.
The essential components of the board are 3 8-bit binary counters with output registers (54LS590\footnote{Texas Instruments}) and a hex inverter chip (74LS04~\footnote{Motorola}) which contains 6 independent inverters. The binary counter chips are cascaded in series such that the previous chip triggers the following chip.  The input to the first counter is the index pulse that is generated either from the polarization modulator encoder or the encoder eliminator board, see subsection~\ref{subsec:encodereliminator}. It is necessary to invert the index pulse before it is feed into the first of the 8-bit binary counter chips.  The 8-bit chips are divided into 2 4-bit counters, and each 4-bit counter uses the previous binary word to trigger it. From here it is just as one would expect, skipping the gory details of getting the chips to cascade appropriately, with all of the 4-bit words in series being read out by each of the data acquisition boards. One feature of the chip is a clock clear pin; this allows for the entire 24-bit sync number to be reset to zero with the use of a small button built into the board. By pulling the pin high ($+5$ V) the entire word is reset.

\subsection{Encoder Eliminator}\label{subsec:encodereliminator}
On occasion it was necessary to run the experiment without the Polarization Rotator running; this only occurred during testing.  For this reason an additional board was installed on the experiment to simulate the encoder output. A Programmable Crystal Oscillator (PXO-600\footnote{Statek Co.}) was used to generate a square wave at either 10 Hz or 30 Hz.  The square wave signal was split into 2 waves with one of them getting a 1/4 wave phase shift. The unshifted signal was the A phase and the shifted signal was the B phase of the encoder output. An additional counter chip was used to count 128 pulses of the A phase and send out an index pulse. This setup closely mimics the output of the encoder on the Polarization Rotator encoder.

\subsection{Temperature Sensors}
The temperature is monitored by one of 3 systems. All cryogenic temperature sensors use a biased silicon temperature sensor\footnote{Lake Shore, Inc.} that gets a constant $10$ $\mu$A.  The voltage across the diode is temperature dependent and readout via the servo computer. Ambient temperature sensors are mounted to several components on the experiment which include the calibrator, primary and secondary mirrors, polarization rotator, tilt sensor, and frame temperature. These sensors use an AD590\footnote{Analog Devices, Inc.} two-terminal IC temperature transducer that has been calibrated prior to use for temperature readout. The third and final temperature sensors are active and utilize an AD590 for the temperature readout in tandem with a set of power resistors mounted for the heating elements. A control voltage can be set to raise the temperature of an insulated system above ambient.
\begin{figure}[p]
\begin{center}
\includegraphics[width = 13.5cm]{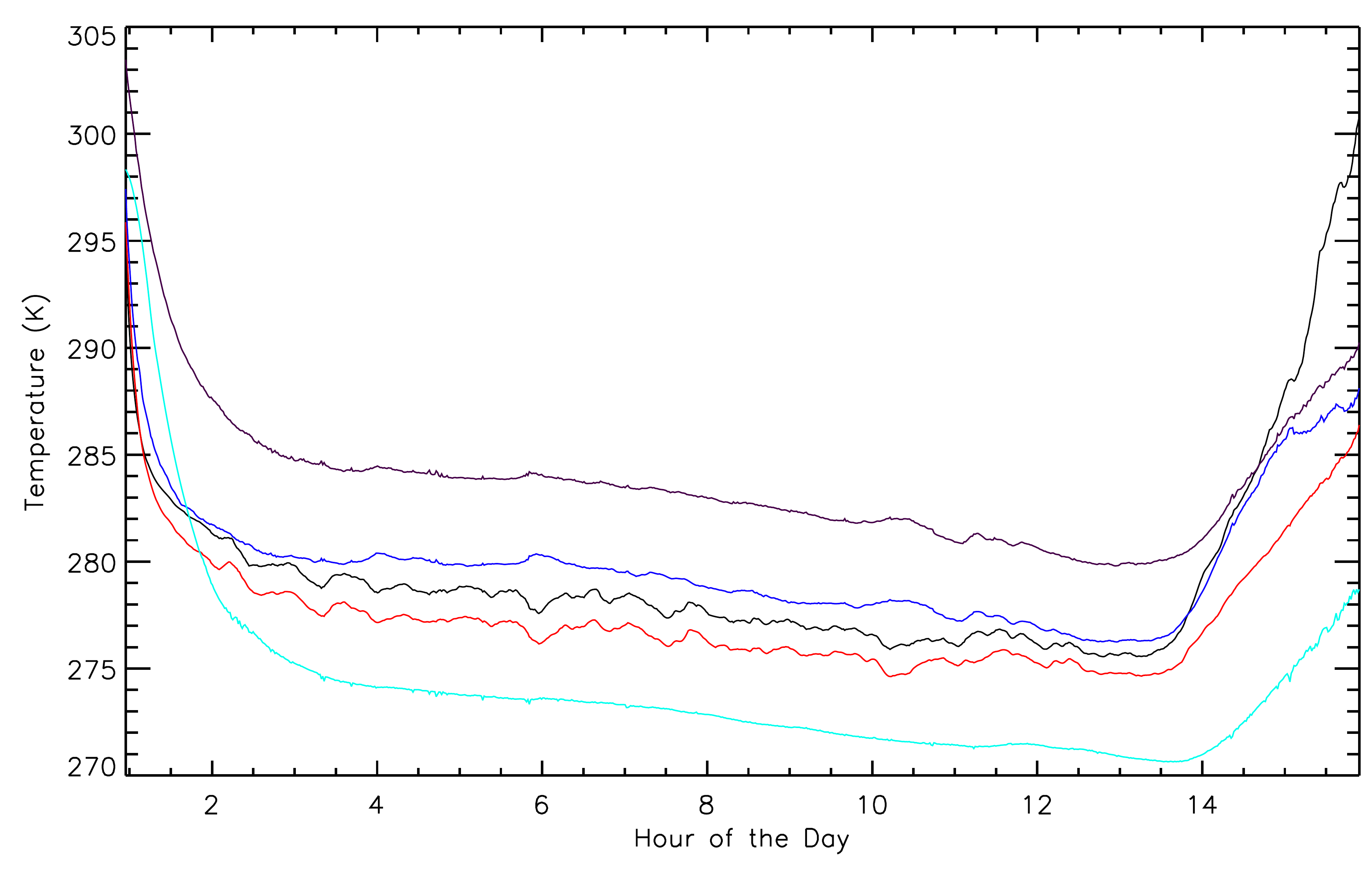}
\caption[Ambient temperatures for multiple sensors over one observing day]{Ambient temperatures for multiple sensors over one observing day. Temperatures binned into 1 minute averages for primary reflector (black), secondary reflector (red), frame (blue), calibrator (light blue), and Polarization Rotator (purple). \label{fig:AmbientTemperatures}}
\end{center}
\end{figure}

\chapter{Polarization Rotator}\label{chap:polarizationrotator}
\begin{figure}[p]
\begin{center}
\includegraphics[width=\textwidth]{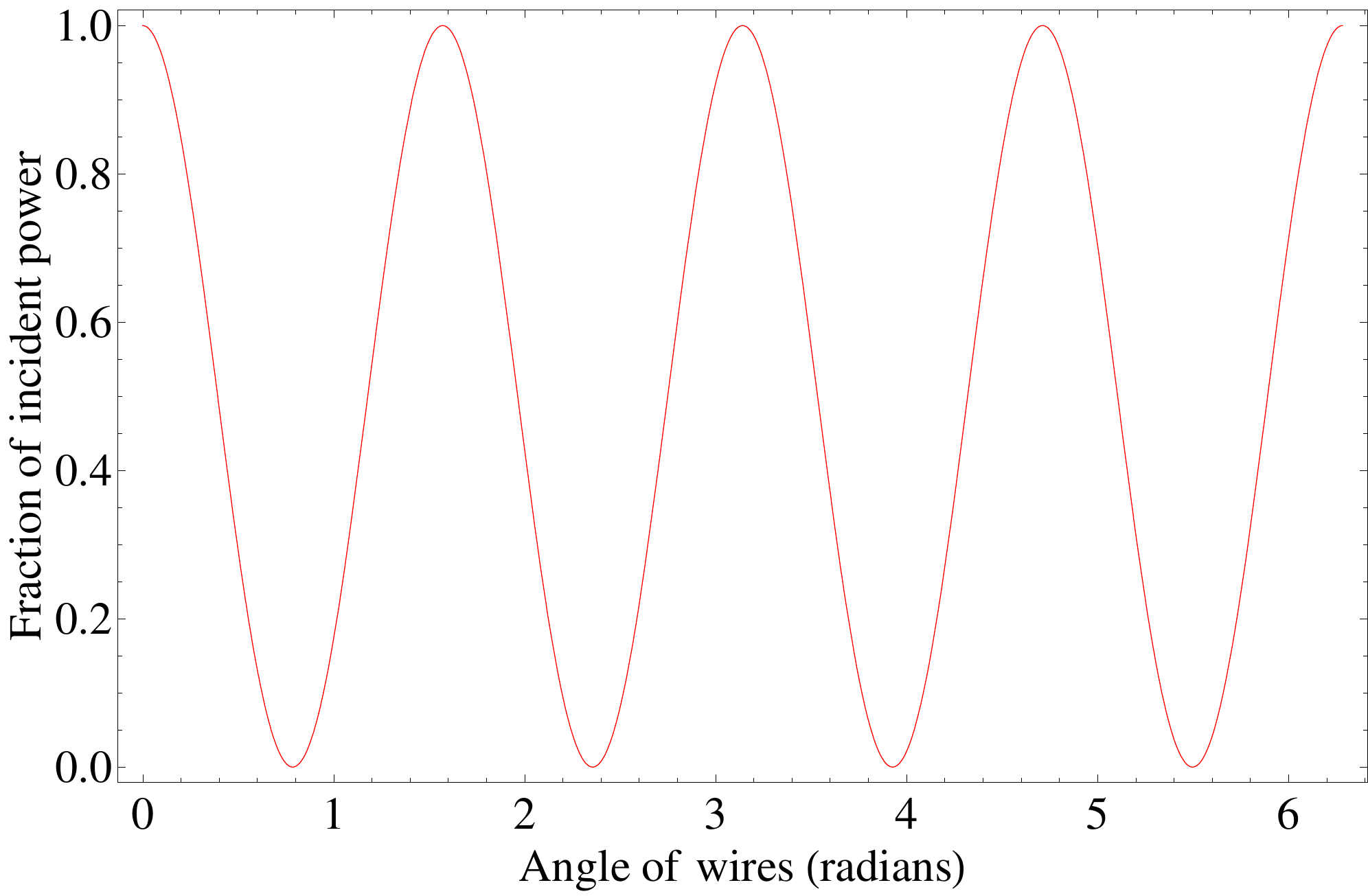}
\caption[Model of fractional power as a function of wire angle]{Fraction of the polarized incident power that reaches the detector as a function of Polarization Rotator wire angle . The wire angle is referenced to the horizontal. \label{fig:signal}}
\end{center}
\end{figure}

To overcome the effects of $1/f$ noise contributed to the data stream by the HEMT amplifiers, some sort of chopping is required. For temperature experiments a Dicke switch radiometer is typically used, which chops rapidly between 2 temperatures. In B-Machine a new technique, that works in a similar fashion to a half wave plate, to chop between polarization states is being used. The Polarization Rotator consists of a linear polarizing wire grid mounted in front of a plane reflecting mirror (polished Aluminum plate). The wire grid decomposes the input radiation into its 2 polarization components, parallel and perpendicular to the wires. The parallel component is reflected off of the wire grid surface while the perpendicular component passes through the wire grid, where it is reflected off of the plane mirror, passing through the grid again and combining with the parallel component. The spacing between the plane mirror and the grid introduces a phase shift between the two polarization components, effectively rotating the plane of polarization of the input wave. A schematic to illustrate the operation of the polarization modulator is shown in Figure~\ref{fig:WireGridExplained}B. By rotating the grid the incident polarization can be rotated 2 times per revolution, as shown in Figure~\ref{fig:signal}, giving the single polarization sensitive receiver a chop between the 2 polarizations 4 times per revolution.

\begin{figure}[p]
\begin{center}
\includegraphics[width = \textwidth]{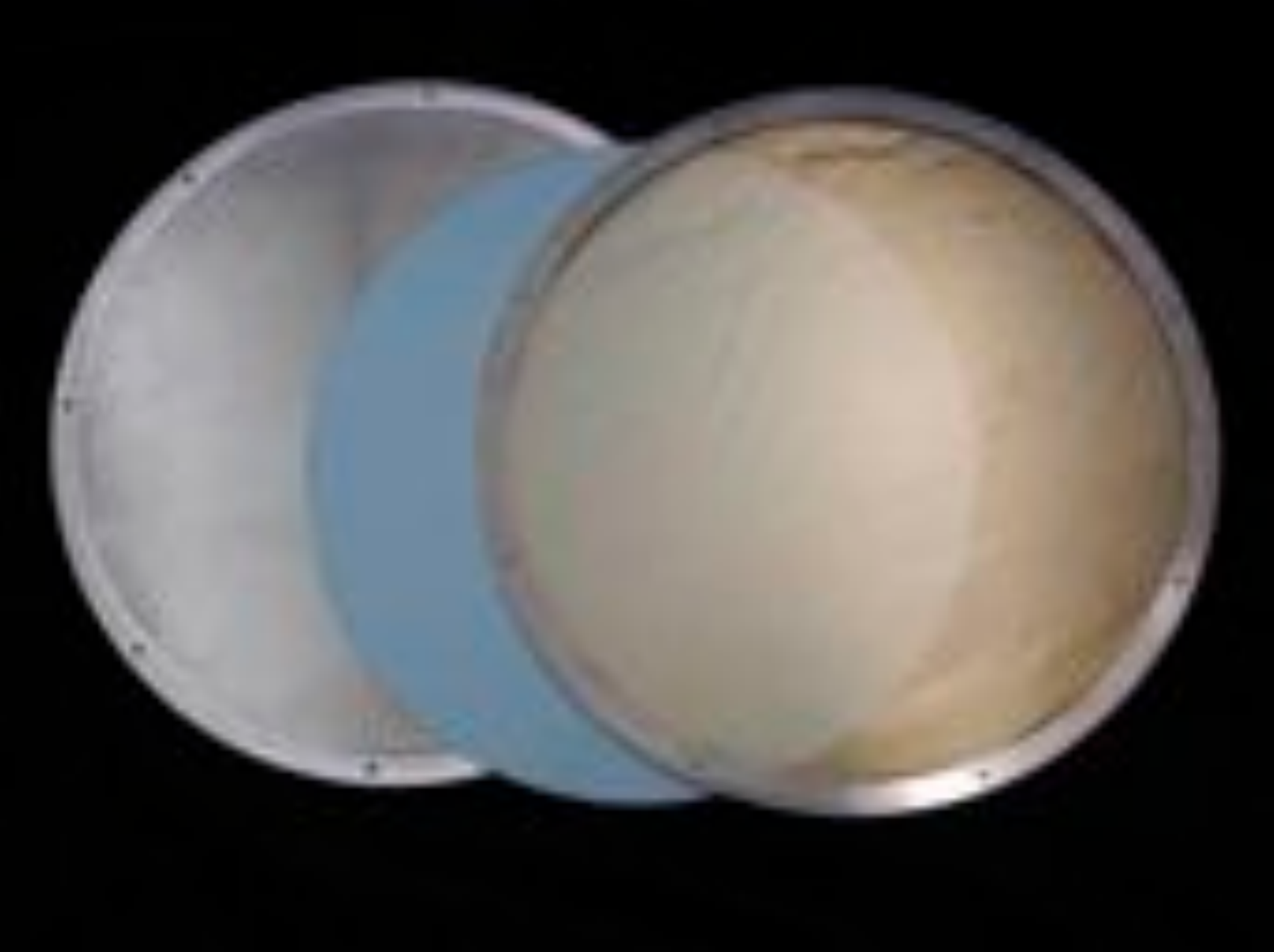}
\caption[Wire grid, blue foam, and plane mirror picture]{Polarization Rotator assembly with plane mirror on bottom, blue foam spacer (transparent at our frequencies), and wire grid upside down so wires and support ring can be seen.
\label{fig:PolGridFoamPlate}}
\end{center}
\end{figure}

\section{Theory}
Reducing the operation of the Polarization Rotator to a simple example of a polarized signal incident on the polarization modulator is the easiest way to examine its workings. Polarized radiation of the form,

\begin{equation}\label{eqn:polrotIncident}
   E_{incident}=E_{0}\cos(\kappa y + \omega t)\hat{j},
\end{equation}
is incident on the top surface of the wire grid with magnitude $E_{0}$, angular frequency $\omega$, and the polarization vector makes an angle of $\theta$ with the wires. Transforming the incident radiation basis into the wire grid basis (assuming wires are in the $\hat{j}'$ direction), see Figure~\ref{fig:WireGridExplained}A, and adding a phase shift, $\delta$, to the polarization that passes through the wire grid and is reflected off of the plane mirror backing plate gives,

\begin{equation}\label{eqn:polrotTransformed}
  E_{transformed}=E_{0}\sin(\theta)\cos(\kappa y + \omega t+\delta)\hat{i}'+E_{0}\cos(\theta)\cos(\kappa y + \omega t)\hat{j}'.
\end{equation}
Combining the 2 radiation paths and converting back into the original basis gives,

\begin{eqnarray}\label{eqn:polrotback}
  \lefteqn{E_{out}=}\nonumber \\
    & E_{0}\sin(\theta)\cos(\theta)[\cos(\kappa y + \omega t+\delta)- \cos(\kappa y + \omega t)]\hat{i}+ \nonumber \\
    & E_{0}[\sin^{2}(\theta)\cos(\kappa y + \omega t+\delta) + \cos^{2}(\theta)\cos(\kappa y + \omega t)]\hat{j},
\end{eqnarray}
see Figure~\ref{fig:WireGridExplained}B. Since the detector is sensitive to power it is easier to look at the power in each of the polarization states as a function of angle of the wire grid rather than electric field strength. When averaging over time it is assumed that the detector is at $y=0$ and $P\propto\langle E^{2} \rangle_{t}$, which yields,

\begin{eqnarray}\label{eqn:PowerinE}
  P_{\hat{i}}& = & 2E_{0}^{2}\sin^{2}(\theta)\cos^{2}(\theta)\sin^{2}\left(\frac{\delta}{2}\right),\\
  P_{\hat{j}}& =& \frac{1}{2}E_{0}^{2}[1-4\sin^{2}(\theta)\cos^{2}(\theta)\sin^{2}\left(\frac{\delta}{2}\right)],
\end{eqnarray}
where $\theta$ is the angle the wires on the wire grid make with the incident polarized signal's polarization angle and $\delta$ is the phase shift.  Keeping in mind that the detector is only sensitive to either $P_{\hat{i}}$ or $ P_{\hat{j}}$, a wire angle can be found using the above framework to rotate any arbitrary incident polarization into the single polarization angle that the detector is sensitive too. The reflection and transmission losses have little effect on the outcome but do complicate the calculation significantly. As a result of this, they have been omitted from the calculations here.

\subsection{Wire Grid Plane Mirror Spacing}
\begin{figure}[p]
\begin{tabular}{cc}
\includegraphics[width=7cm]{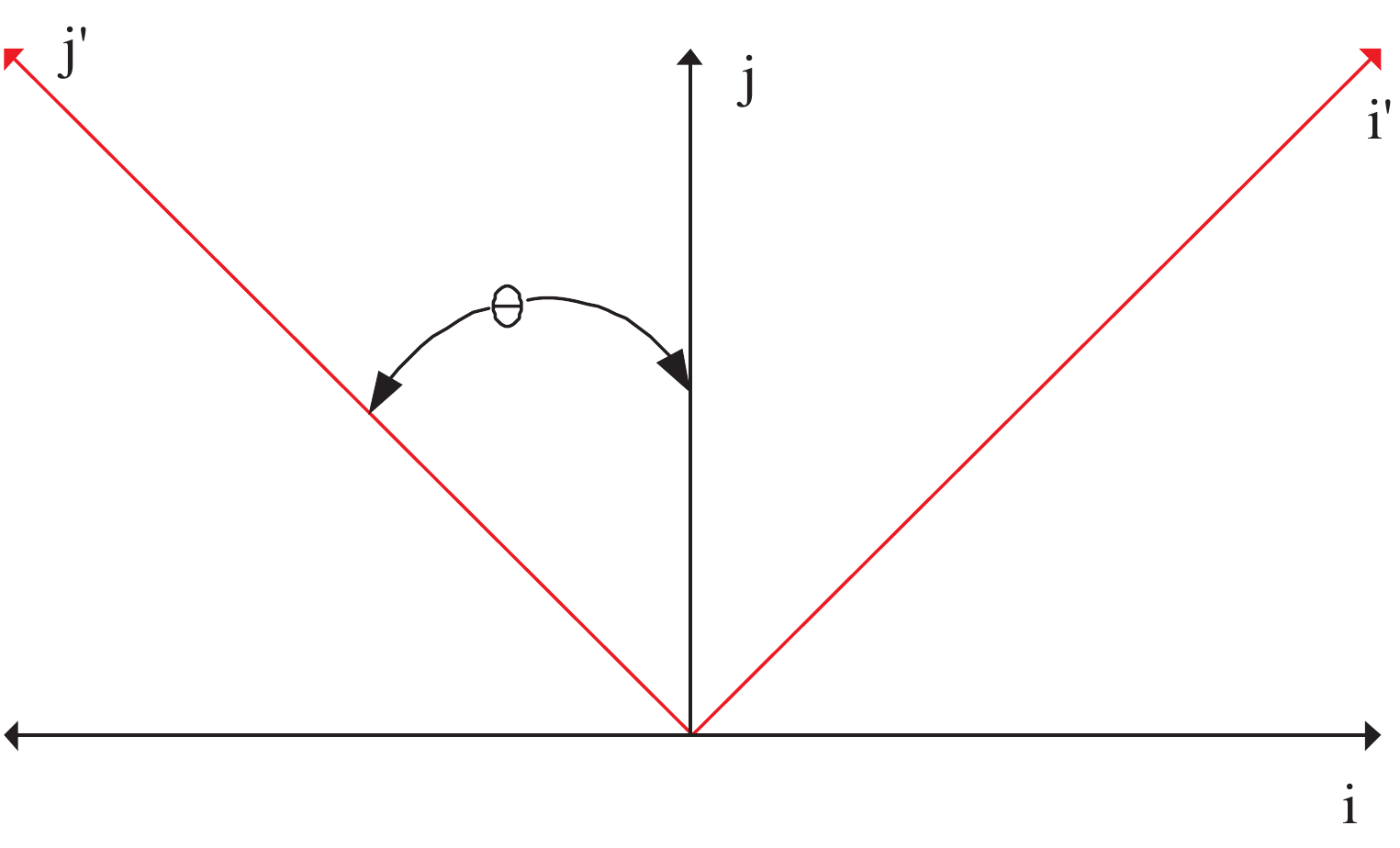}&
\includegraphics[width=7cm]{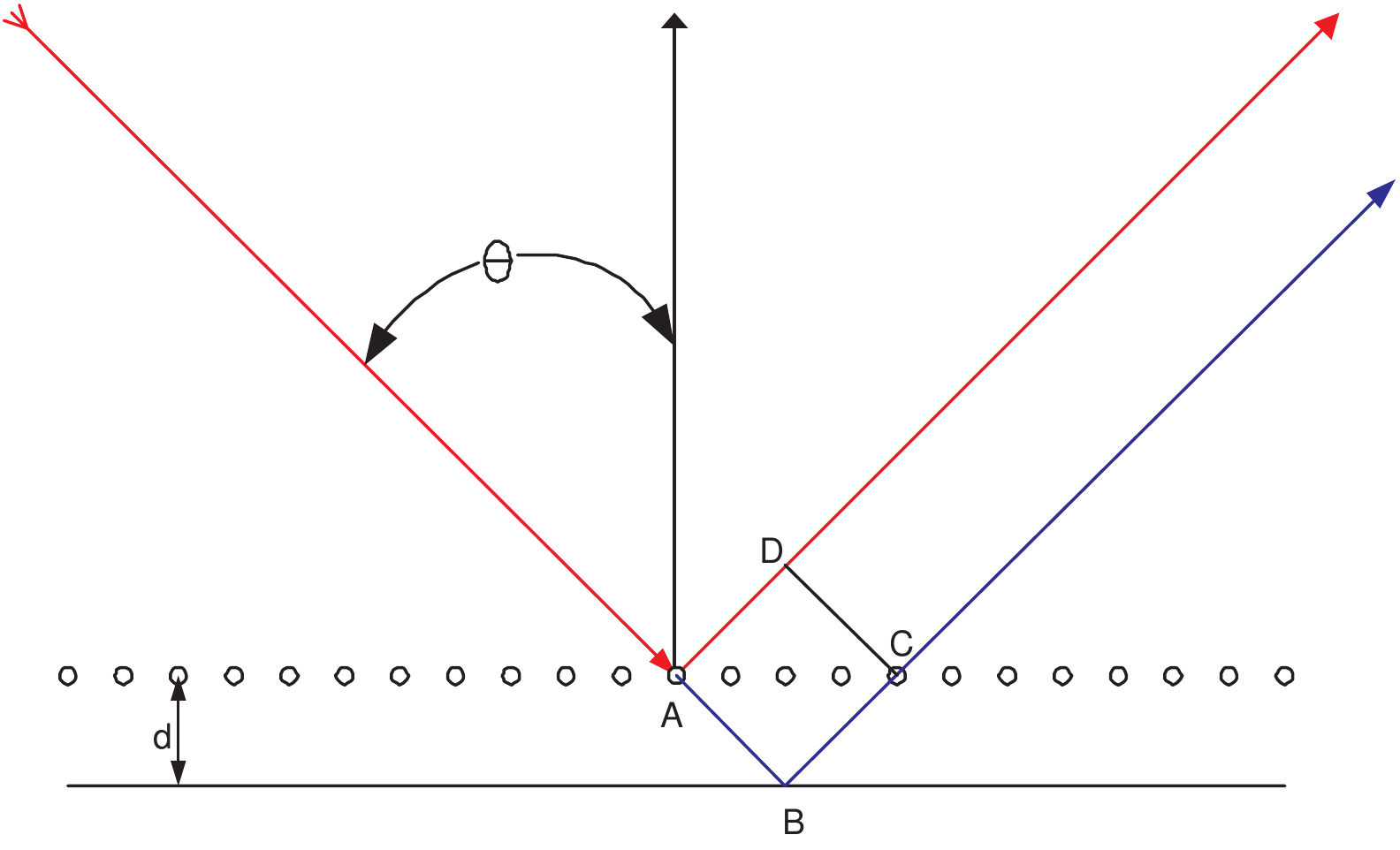}\\
A)  &
B) \\
\end{tabular}
\caption[Basis rotation and wire grid radiation interaction]{$A)$ Incident radiation polarized in the unprimed bases is transformed into the primed bases, a phase shift is added onto the polarization that is perpendicular to the wires and transformed back into the unprimed bases.  The wires are aligned to the $\hat{j}'$ axis. $B)$ Side view of wire grid interaction with the incident radiation and the grid plane mirror spacing shown. \label{fig:WireGridExplained}}
\end{figure}

To make the plane of polarization rotate by $\frac{\pi}{2}$ when the wires are at $45^{\circ}$ the incident radiation that is polarized perpendicular to the wires needs a path length difference from the parallel polarization of $\frac{\lambda}{2}$. From Figure~\ref{fig:WireGridExplained}B the path difference is

\begin{equation}
\begin{array}{l}
  \Delta l=AB+BC-AD, \\
  AB=BC=\frac{d}{\cos(\theta)} ~~\mbox{and}~~ AD=\frac{2 d \sin^{2}(\theta)}{\cos(\theta)}, \\
  \Delta l=\frac{2d}{\cos(\theta)}-\frac{2 d \sin^{2}(\theta)}{\cos(\theta)}=2d \cos(\theta), \\
\end{array}
\label{eqn:WGpathdifference}
\end{equation}
where $\theta$ is the angle of incidence of the radiation. Then for a $\frac{\lambda}{2}$ path difference a wire grid to plane mirror spacing of

\begin{equation}\label{eqn:WGPMspacing}
    d=\frac{\lambda}{4\cos(\theta)}
\end{equation}
gives the appropriate phase shift. Using the B-Machine parameters of $\lambda=.7223$ cm and $\theta=.6632$ gives a spacing of $0.229\mbox{ cm} ~\mbox{or}~ 0.09\mbox{ inches}$.  

\subsection{Polarization Rotator Response}
\begin{figure}[p]
\begin{center}
\includegraphics[width = 13.5cm]{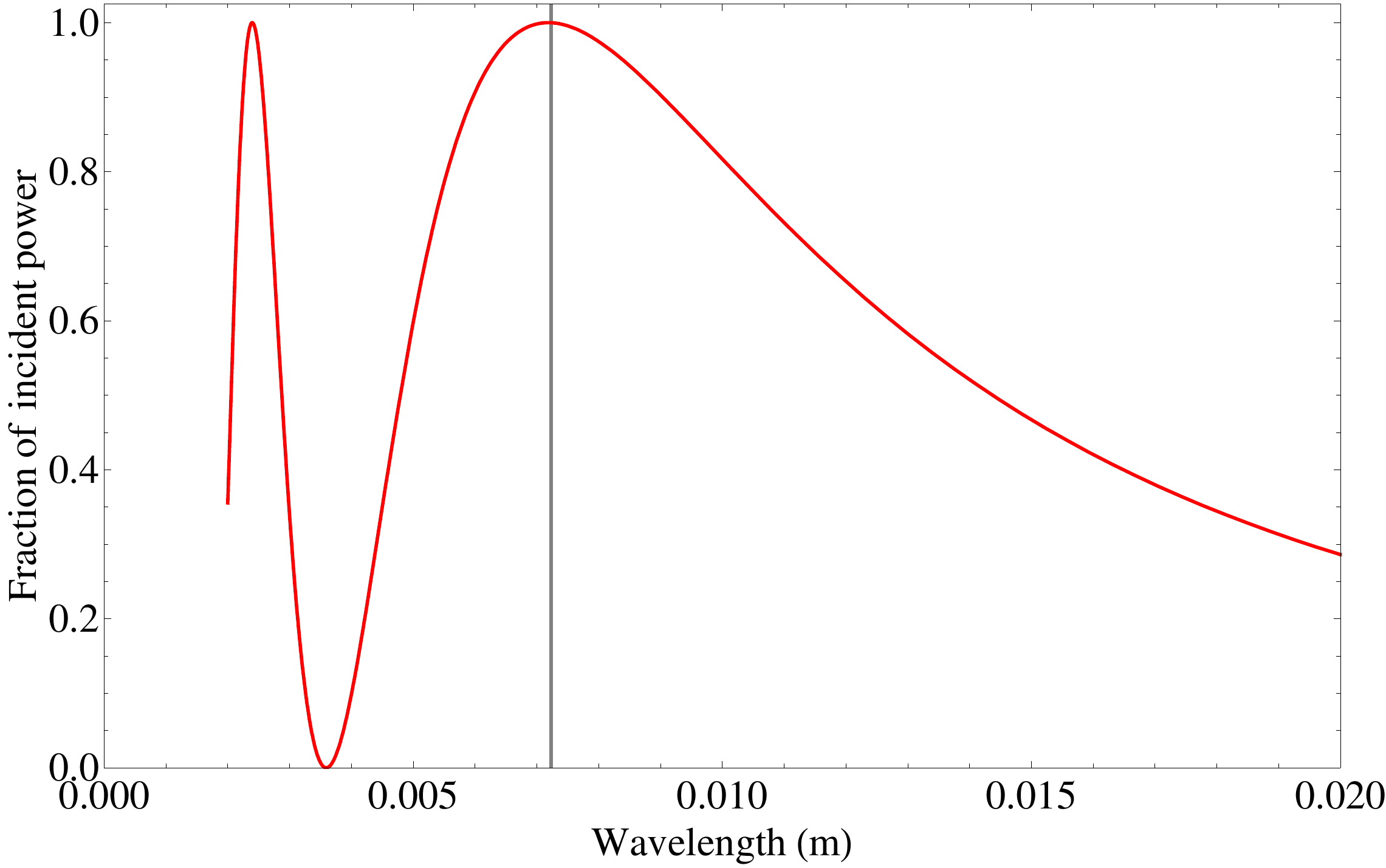}
\caption[Fraction of input power as function of wavelength]{Fractional input power verses wavelength with the black line corresponding to 41.5 GHz using 0.09 inch wire grid to plane mirror spacing. \label{fig:powervslambda}}
\end{center}
\end{figure}

\begin{figure}[p]
\begin{center}
\includegraphics[width = 13.5cm]{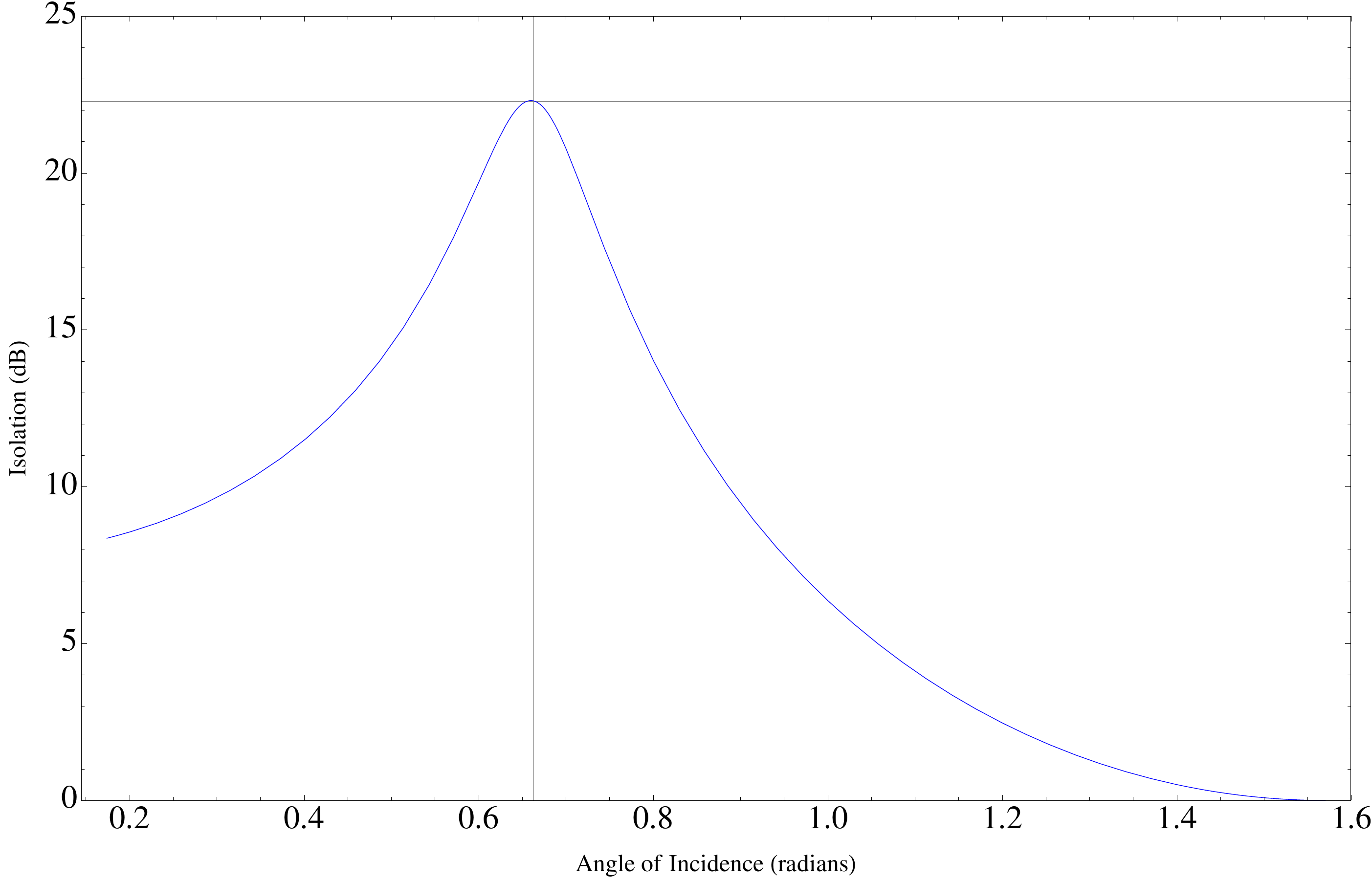}
\caption[Isolation of Q from U as a function of beam divergence]{Isolation of Polarization Rotator as a function of angle of incidence of the horn. The FWHM of the horn is  $18^{\circ}$ and the angle of incidence ranging from $10^{\circ}-90^{\circ}$.  \label{fig:isovsbeam}}
\end{center}
\end{figure}

With a theoretical construction of the operation of the Polarization Rotator in hand, basic questions to test the viability of the technique can be answered. It is necessary for the rotator to work across the B-Machine band-pass and that the maximum polarization rotation occurs for the centeral frequency of the band-pass. Using the fact that the beam makes an angle of $38^{\circ}$ with the normal to the polarization grid and the expected maximum polarization rotation occurs at $45^{\circ}$, a simple plot (Figure~\ref{fig:powervslambda}) shows that a spacing of 0.09 inches gives the best results. With a bandwidth of $16.87\%$  the efficiency of the rotator is calculated to be $99.41\%$ or has an isolation of $22.28$ dB. It is assumed that at $41.5\mathrm{~GHz}$ the rotator is $100\%$ efficient. There is some ambiguity in the definition of bandwidth, for my purposes here percentage bandwidth is defined as

\begin{equation}\label{eqn:bandwidth}
  \beta=\frac{f_{stop}-f_{start}}{f_{bandcenter}},
\end{equation}
where the start and stop frequencies correspond to the 3 dB points of the filter. When the Polarization Rotator has rotated the plane of polarization $90^{\circ}$, the expectation is that the other polarization will have no power in it, but since the rotation is sensitive to frequency, beam size, grid spacing, wire spacing, and angle of incidence some power leaks from one polarization to the other. A measure of this power leakage is the isolation quoted here in decibels (dB). The corrugated feed horns have a FWHM of $18^{\circ}$ which makes the angle of incidence vary from $32.5-50.5^{\circ}$, reducing the isolation a bit more, see Figure~\ref{fig:isovsbeam}.

Convolution of beam width and frequency dependence gives an expected isolation of 20 dB. This is in good agreement with experimental values from sky dips as seen in Chapter~\ref{chap:instrument} Figure~\ref{fig:crosspolarization}.

\subsection{Wire Grid}
\begin{figure}[p]
\begin{center}
\includegraphics[width = 13.5cm]{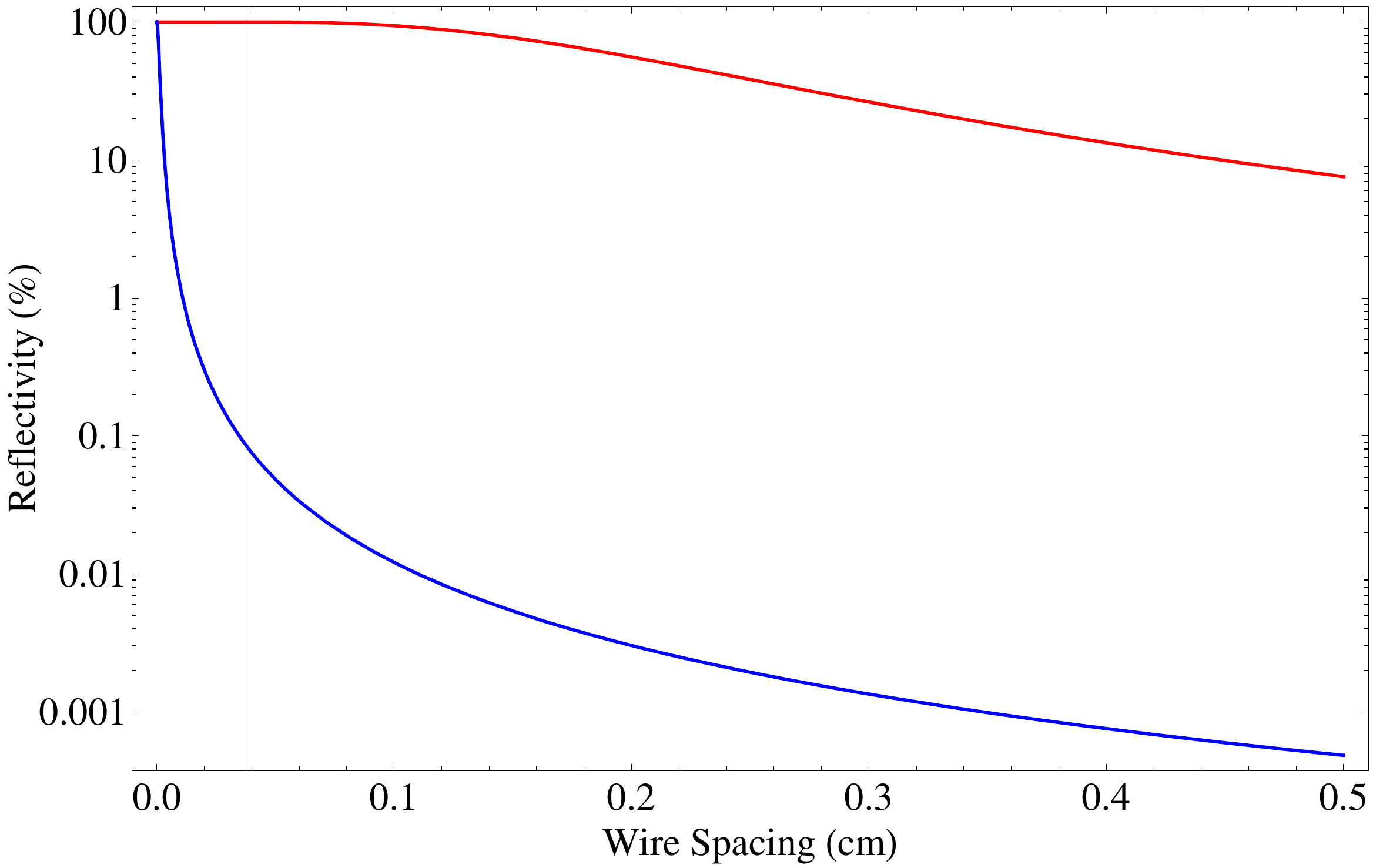}
\caption[Wire grid reflectivity as a function of wire spacing]{Reflectivity of normalized power for radiation polarized parallel (red) to the wires and perpendicular (blue) to the wires. The black vertical line is the central frequency of the B-Machine band-pass (41.5 GHz) and 5 mil (0.127 cm) wide wires with a 15 mil (0.0381 cm) center to center spacing. \label{fig:WGRVsSpacing}}
\end{center}
\end{figure}

The wire grid that creates the polarization splitting can be tuned by adjusting the size of the wires and thier spacing, changing the efficiency of the wire grid as a function of wavelength. B-Machine's band-pass is well known and hence tuning of the wire gird is a straight forward exercise in using Equations~\ref{eqn:rp} and \ref{eqn:rn}. These equations were originally derived by Lamb~\citep{Lamb98} in 1898 and derived again and presented independently by ~\citep{Lesurf90}.

\begin{equation}\label{eqn:rp}
  |r_{p}|^{2}=\left[1+\left(\frac{2S}{\lambda}\right)^{2}\left(ln\left(\frac{S}{\pi d}\right)\right)^{2}\right]^{-1}
\end{equation}

\begin{equation}\label{eqn:rn}
  |r_{p}|^{2}=\frac{(\pi d)^{4}}{(2S\lambda)^{2}[1+(\pi^{2} d^{2})^{2}/(2S\lambda)^{2}]}
\end{equation}
where $d$ is the wire diameter (the calculations assume a wire with a circular cross section) and $S$ is the center to center spacing of wires. The smallest wire width was limited by the fabrication process to be 5 mils ($0.0127$ cm). Seen in Figure~\ref{fig:WGRVsSpacing} the reflectivity of the parallel component is close to $100\%$ at $99.9976\%$, while the reflectivity of the perpendicular component is low at $0.0836\%$.

Several techniques in constructing a durable and robust wire grid were explored before the final usable grid was produced. The first technique was to use a threaded rod (40 pitch) secured in a rigid metal frame and wrap copper wire around end over end. Once wrapped, the wires were secured to the rod with epoxy and one surface of the wires were cut away, leaving the frame with free standing wires. This technique had several problems associated with it. First, it was hard to get the spacing consistent over large frames. Second, the wires were not rigid enough to survive rotating at any reasonable speed. After this technique was abandoned a photo-lithography process to pattern and evaporate wires onto a piece of Polypropylene (Polypro is transparent to microwave radiation)   was pursued.  Again this technique bore little fruit.  The evaporated material was too thin, had poor adhesion to the Polypro and making large patterns with high tolerances with equipment available in the local clean rooms was not possible.  Following these failures several manufacturers of flexible circuit boards were contacted and asked to quote on a 12" diameter wire grid, finally settling on All Flex Inc.\footnote{All Flex Inc., Northfield, MN 55057, www.allflexinc.com}

\section{Effects on Telescope Sensitivity}
Receiver sensitivity can be estimated using the radiometer equation,

\begin{equation}\label{eqn:sensitivity}
   \sigma_{T}=K\left(\frac{T_{\mathrm{sys}}+T_{\mathrm{sky}}}{\sqrt{\Delta\nu\cdot\tau}}\right),
\end{equation}
where $ \sigma_{T}$ is the root-mean-square noise, $T_{\mathrm{sys}}$ is the system noise temperature,
$T_{\mathrm{sky}}$ is the sky antenna temperature, $\Delta\nu$ is the bandwidth, and $\tau$ is the integration time. $K$ is the sensitivity constant and depends on the type of radiometer being used \citep{daywitt89}. For example, an un-differenced receiver will have $K=1$, while a Dicke switched radiometer will have $K=2$.

\subsection{Demodulation Technique}
When calculating the sensitivity of the instrument no correction factor is added for $\frac{1}{f}$ noise. The sensitivity is considered the white noise limit and it is assumed that the $\frac{1}{f}$ noise is taken into account by the sensitivity constant, $K$, in the radiometer equation. Differencing the signal on time scales much faster than the $\frac{1}{f}$ knee is the typical method used to overcome the associated noise. B-Machine's differencing is done by using a lock-in post-processing software tool. The tool written in IDL (see Chapter~\ref{chap:analysis} Subsection~\ref{subsec:IDL}) multiplies the signal by a square wave which oscillates between $+1$ and $-1$. The phase that aligns the square wave with the appropriate polarization is determined by the orientation of the channels horn to the Polarization Rotators wire grid and is generated using the get max phase procedure; see Chapter~\ref{chap:telescopechar} Subsection~\ref{subsec:GetMaxPhase}.

\subsection{Derivation of Sensitivity Constant}
For a standard total power radiometer the sensitivity constant for the Radiometer Equation~\ref{eqn:sensitivity}, is $1$. All other chopping schemes degrade the sensitivity of a given radiometer. In general, the degradation of the sensitivity falls into one of 2 categories. The first is integration time and the second, error propagation (more error accrues on a given reading when you subtract 2 signals with the same error). For B-Machine the sensitivity constant is calculated using a sine wave chopping technique with a square wave demodulation. This is achieved by starting with a square wave chop and a square wave demodulation (Dicke Switched), then multiplying by the efficiency factor between square and sine wave demodulation to get the final answer.

For a standard Dicke Switched Radiometer half the time is spent looking at each of the loads, thus only $\frac{1}{2}$ of the integration time is spent on a given load which degrades the sensitivity by $\sqrt{2}$. Then differencing the 2 signals and adding their error in quadrature yields another factor of $\sqrt{2}$, and multiplying these gives a sensitivity constant of $K=2$. A sine wave demodulation scheme efficiency factor can be derived by looking at a simple example. If a Dicke Switched Radiometer is chopping between a $0$ K and a $1$ K blackbody with a period of $2\pi$ (for simplicity in the sine wave calculations) a signal of $\pi$ K is observed, see Equation~\ref{eqn:SquareInt}, while a similar radiometer that is sine wave chopped will see a signal of $2$ K, see Equation~\ref{eqn:SinInt}. From this the efficiency factor of $Sine\rightarrow Square$ is $\frac{2}{\pi}$.
\begin{equation}\label{eqn:SquareInt}
   \langle T_{Square} \rangle=\int_{0}^{\pi}0dT-\int_{\pi}^{2\pi}1dT=\pi
\end{equation}

\begin{equation}\label{eqn:SinInt}
   \langle T_{Sine} \rangle=\int_{0}^{\pi}\frac{\sin(T)}{2}dT-\int_{\pi}^{2\pi}\frac{\sin(T)}{2}dT=2
\end{equation}
Multiplying the efficiency gives a sensitivity constant of $\pi$ for $\Delta T$. Gaining another factor of $1/2$ from the definition of the $Q$ and $U$,  Stokes parameters
\begin{equation}
  Q=\frac{T_{x}-T_{y}}{2},
\end{equation}

\begin{equation}
  Q=\frac{T_{x'}-T_{y'}}{2},
\end{equation}
 gives the final sensitivity constant as $\frac{\pi}{2}$ for each of the stokes parameters.

\begin{figure}[p]
\begin{center}
\includegraphics[width = 13.5cm]{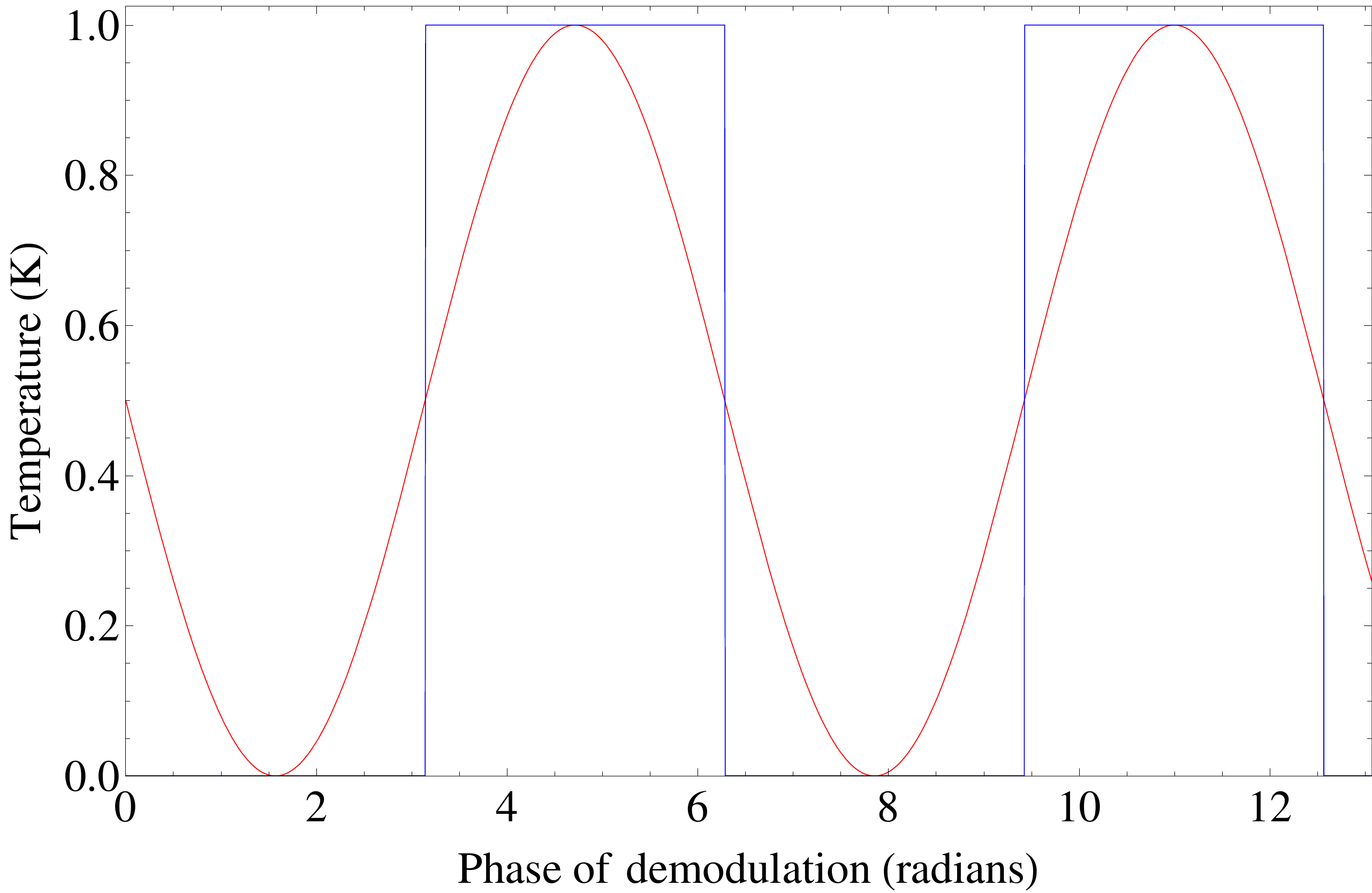}
\caption[Sample square wave and sine wave demodulation wave form]{Square wave (blue) and sine wave (red) demodulation wave forms. \label{fig:SquareVsSin}}
\end{center}
\end{figure}

\subsection{$\frac{1}{f}$ Characteristics}
$1/f$ noise is a signal with a power spectral density that is roughly inversely proportional to the frequency. The noise power spectral density $P(f)$ can be described as,

\begin{equation}\label{eqn:PSD}
    P(f)\sim\sigma^{2}\left[1+\left(\frac{f}{f_{\mathrm{k}}}\right)^{\alpha}\right],
\end{equation}
where $\sigma$ is the white noise component level, $f_{\mathrm{k}}$ denotes the frequency where the white noise and $1/f$ contribute equally to the total noise and is referred to as the knee frequency, and $\alpha$ characterizes the slope of the power spectrum and is typically $\simeq1$. The noise fluctuates the apparent gain of the system, mimicking a real signal on the sky. By differencing the signal on time scales much shorter than the knee frequency the fluctuations can be minimized. For B-Machine a chop rate of 133.6 Hz between polarizations was used with knee frequencies, see Table~\ref{tab:oof}, of $\sim140 \mathrm{ Hz}$ causing some of the $1/f$ noise to pollute the demodulated data. The chop rate was determined based on maximum sampling rates of data acquisition computers and beam smearing effects.

\begin{table}[p]
\begin{center}
\caption[List of $\frac{1}{f}$ Knees Before and After Demodulation]{List of $\frac{1}{f}$ Knees Before and After Demodulation \label{tab:oof}}
\begin{tabular}{|c|c|c|}
  \hline
  Channel & Undifferenced   & Differenced  \\
          & Knee (Hz)       &  Knee (mHz) \\
  \hline
  \hline
   1 & 140.0 & 5.0\\
   \hline
   2 & 143.0 & 4.4 \\
   \hline
   3 & 151.0 & 4.4\\
   \hline
   6 & 135.0 & 5.6\\
  \hline
  \hline
\end{tabular}
\end{center}
\end{table}

\begin{figure}[p]
\begin{center}
\includegraphics[width = 13.5cm]{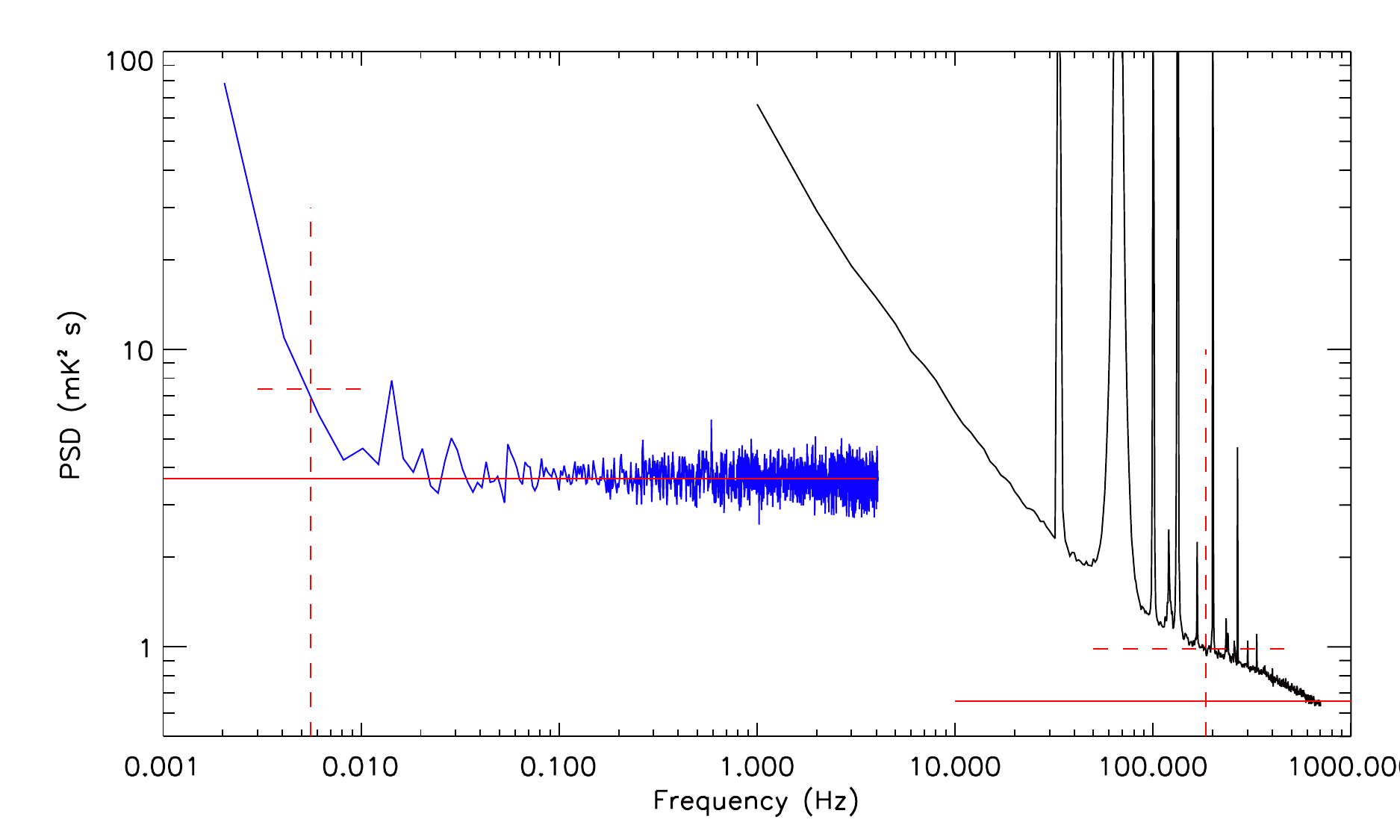}
\caption[Power spectral distribution before/after demodulation]{Power spectral distribution from the central channel of the cold radiometer viewing sky. Black curve is before demodulation and the blue is after demodulation, solid red lines denote noise floor and red dashed lines mark the knee.  \label{fig:PSD}}
\end{center}
\end{figure}

\section{Testing}
The theoretical calculations for the wire grid polarization modulator showed no obvious limitations to stop progress on the development of practical testing. Tests of different reflective materials, including wire grids, were made to determine the reflectivity efficiencies of the different materials compared to a wire grid. As expected Copper and Aluminum plates both had efficiencies over $99\%$. All of the measurements were done by hand, chopping a cold and warm load in reflection off of a plate of the desired material.  For the several wire grids on hand all the efficiencies were above $95\%$. More accurate measurements we not possible due to $\frac{1}{f}$ dominating the errors in the measurements. Small imperfections in the efficiencies only contribute small changes in the calibration constants, for the low levels that were measured the efficiency imperfections were small and corrected for in the final calibration constants, see Chapter~\ref{chap:telescopechar} Subsection~\ref{Sec:Calibration}.  The final grids were not made as of this point, but were eventually measured on the full integrated telescope.

\begin{figure}[p]
\begin{tabular}{cc}
\includegraphics[width=6.5cm,height=6.5cm]{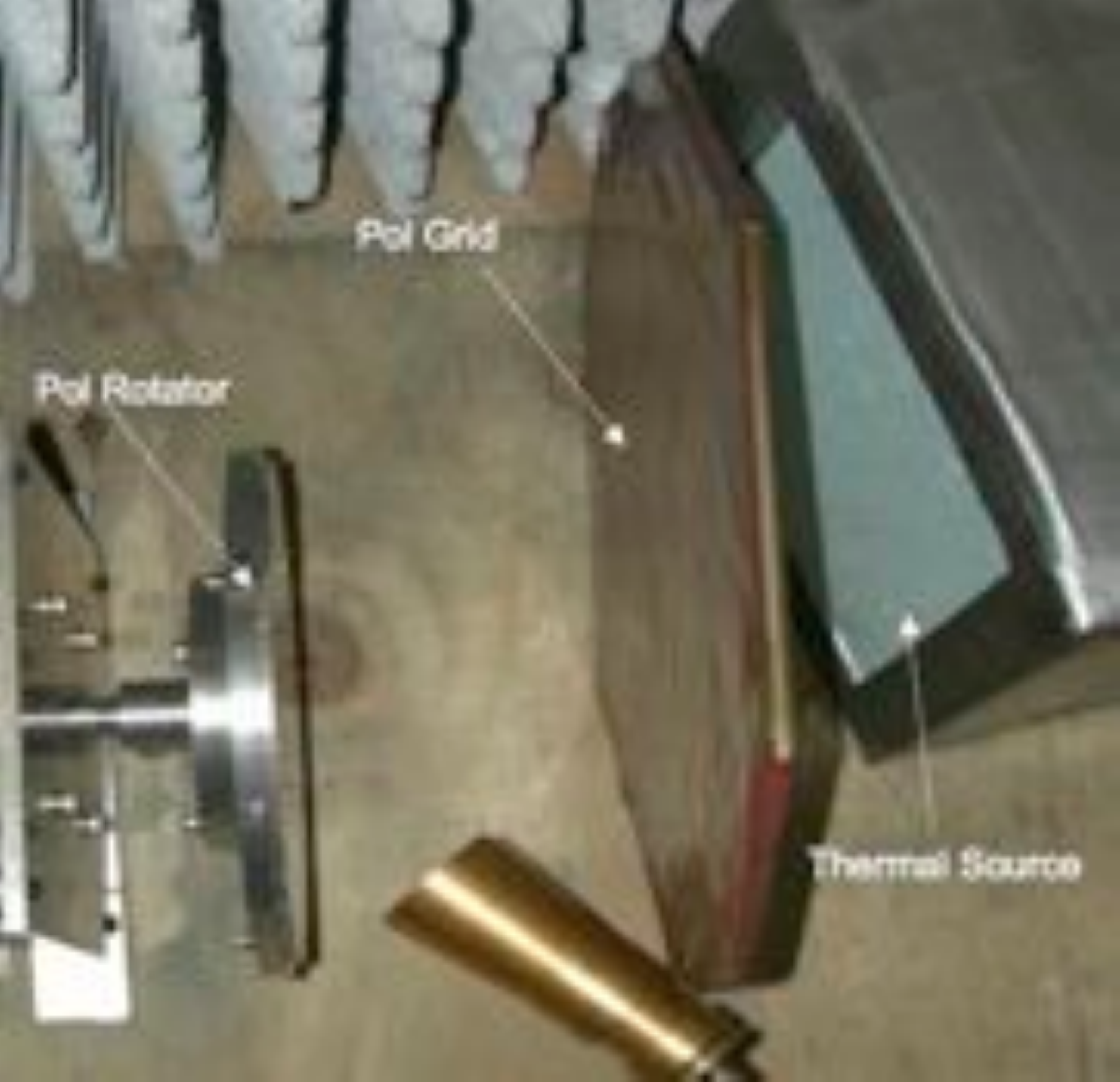}&
\includegraphics[width=6.5cm,height=6.5cm]{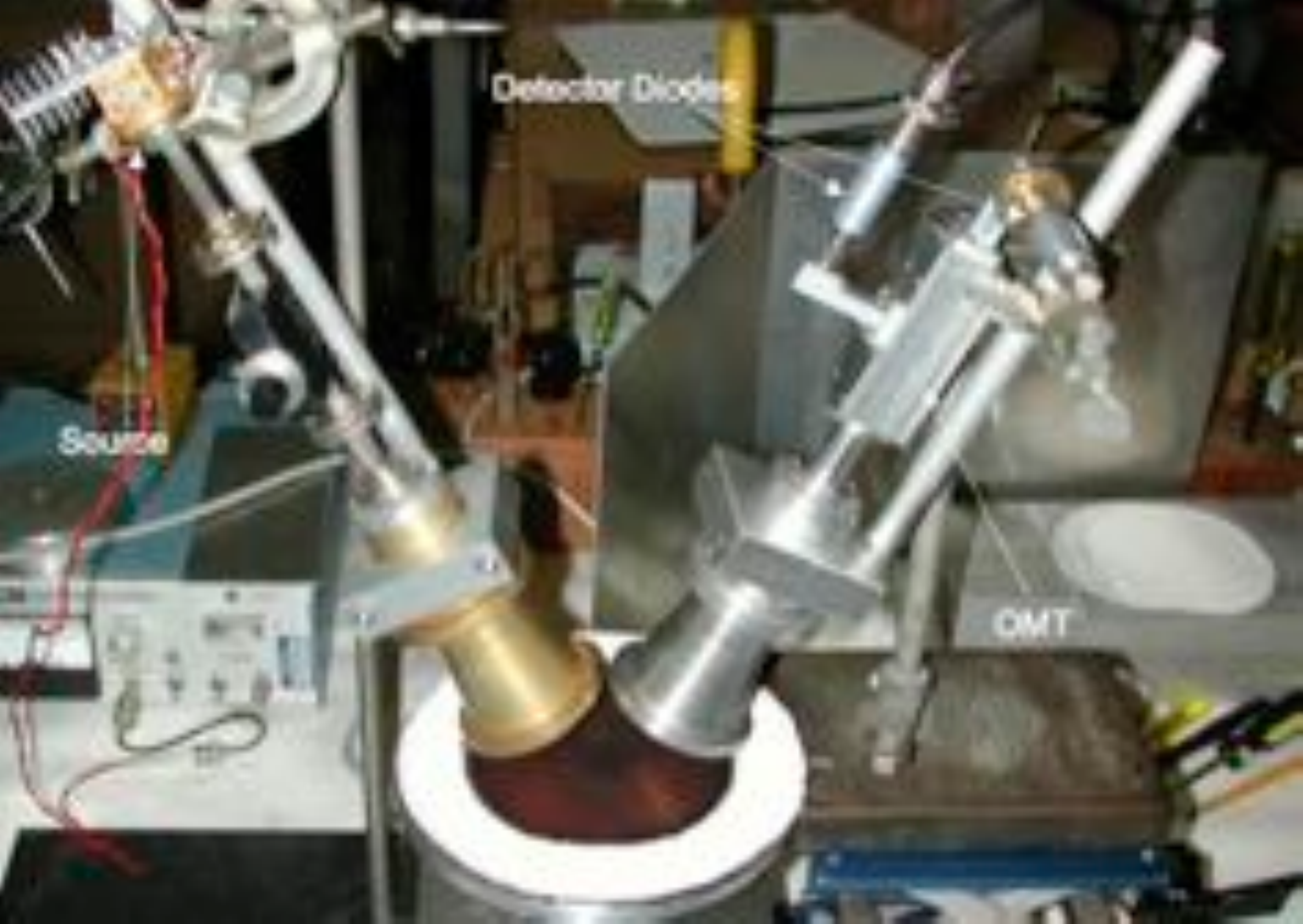}\\
\includegraphics[width=6.5cm,height=6.5cm]{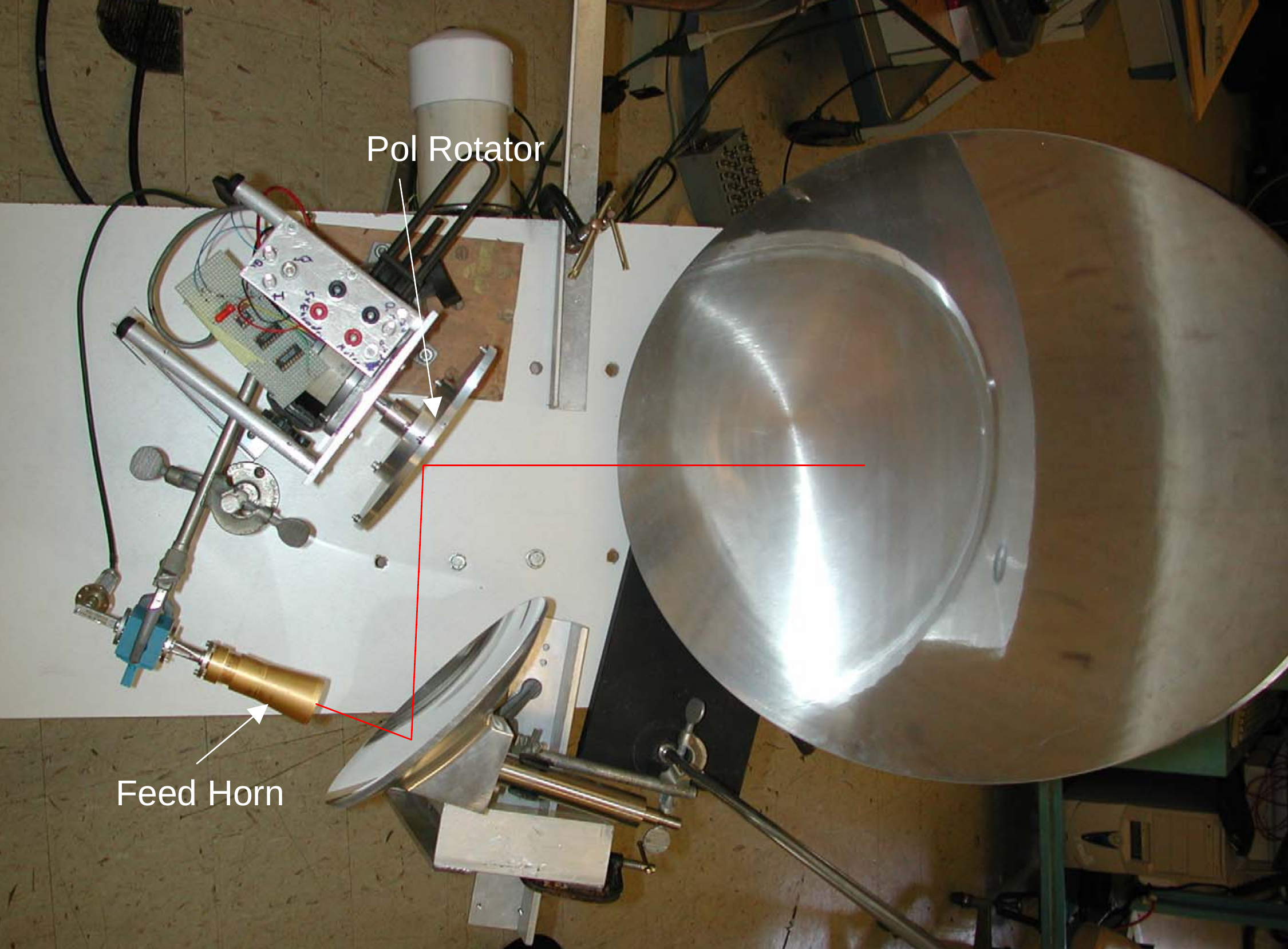}&
\includegraphics[width=6.5cm,height=6.5cm]{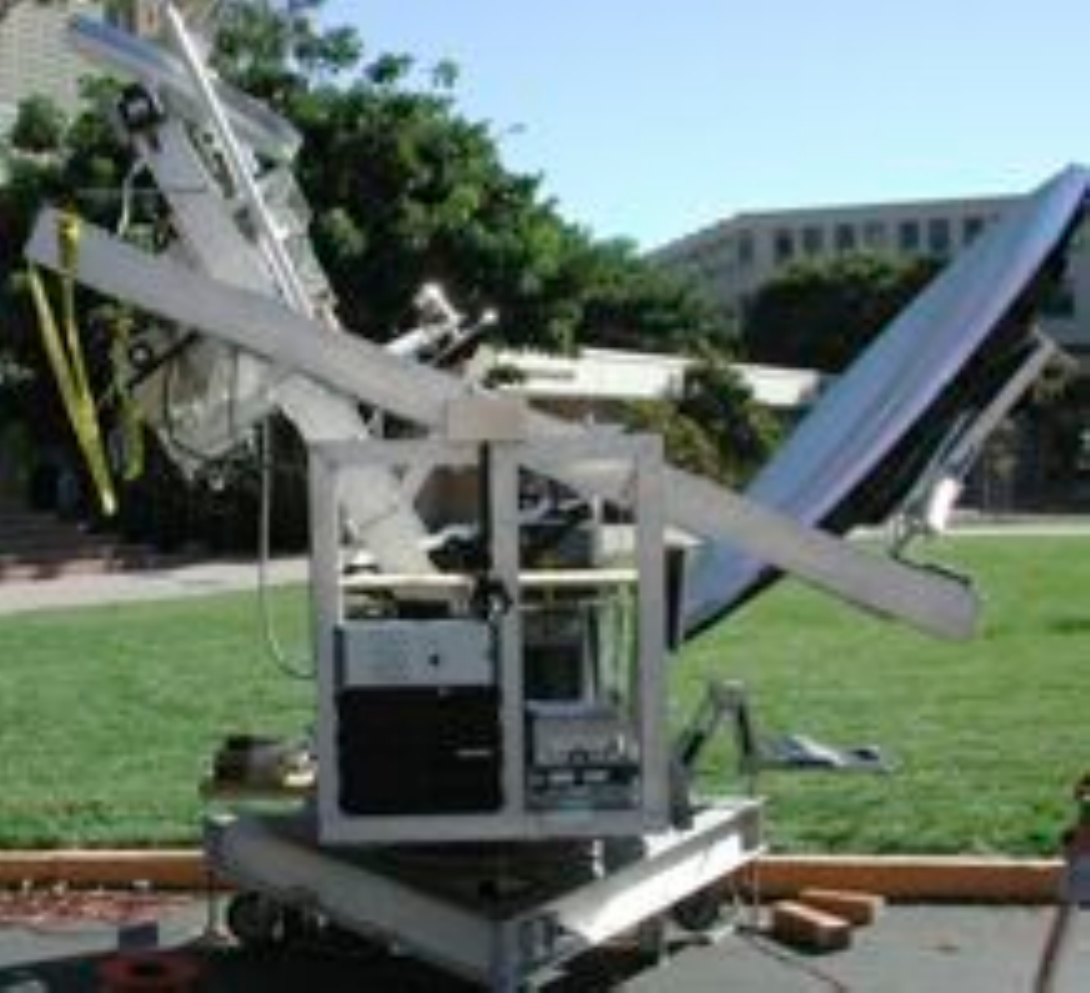}\\
\end{tabular}
\caption[The 4 different testing platforms for the Polarization Rotator]{Polarization Rotator platforms for testing. Top left, Polarization Rotator looking at polarized thermal source. Top right, Gunn diode reflecting off of Polarization Rotator and using an Ortho-Mode Transducer to look at both polarizations simultaneously. Bottom left, small telescope setup with beam path in red, beam coming out of page. Bottom right, integrated telescope looking at sky from East side of Broida Hall at UCSB. \label{fig:PolRotTestPlatforms}}
\end{figure}

\begin{figure}[p]
\begin{center}
\includegraphics[width = 13.5cm]{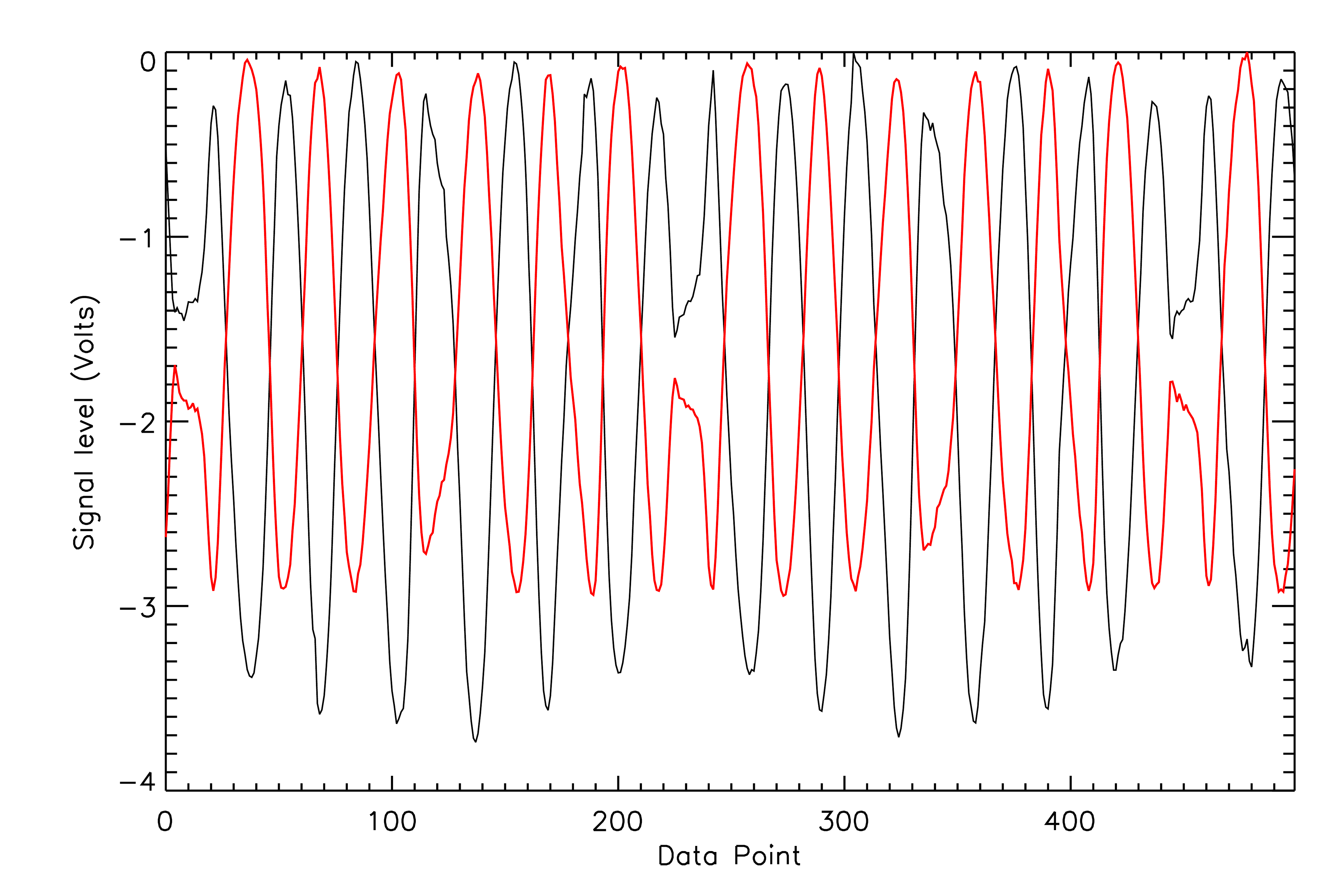}
\caption[Comparison of both polarizations using polarized source and OMT]{Comparison of both polarizations using a polarized source at $\sim45^{\circ}$ from horizontal and OMT. The red line is from the horizontal output of the OMT and black lines from the vertical output of the OMT. \label{fig:OMT}}
\end{center}
\end{figure}

Several small testing platforms, Figure~\ref{fig:powervslambda}, were constructed to make rudimentary measurements. The first platform was a simple Polarization Rotator using a corrugated feed horn and a polarized thermal source. For this setup the Polarization Rotator used a small wire grid that was manufactured by using a standard lift off lithography technique that evaporated a thin layer of copper onto a Polypropylene film producing wires. The second wire grid was purchased as a calibration standard for WMPOL and affixed to a frame. The thermal source, set to $75^{\circ}$C, placed behind the polarizing wire grid transmitted through the wire grid at one polarization and reflected the ambient load for the other polarization giving a polarized signal of roughly $50^{\circ}$C. Rotating the polarizing grid while taking data caused the waveform to shift as expected, confirming that the system could distinguish between both Q and U signals.  The second test setup used a Gunn diode at $41.5$ GHz as a polarized source. In addition to the polarized source an Ortho-Mode Transducer (OMT) was used to observe both polarizations simultaneously, Figure~\ref{fig:OMT}.  The first 2 test setups were relatively easy to construct and yielded data that confirmed the calculations in a rough fashion. This gave us the confidence to invest the time to construct and test a small telescope with focusing optics and a 5" Polarization Rotator.

The small telescope was a modified off-axis Gregorian telescope using a 22 inch primary, 8 inch secondary and a 5" Polarization Rotator. The wire grid for the small telescope was a UCSB manufactured small wire grid. These optics and an $18^{\circ}$ FWHM corrugated feed horn gives a $\sim6^{\circ}$ FWHM beam on the sky. The telescope was used over an extended time period to gather information on the functionality of the Polarization Rotator.  A crude software pipeline to process the data was setup to view the data in the $Q$ and $U$ Stokes parameters.  This presented the first opportunity to verify the $\frac{1}{f}$ characteristics before and after demodulation.  A $\frac{1}{f}$ knee of $150$ Hz before demodulation was reduced to $10$ mHz after demodulation. This platform also allowed many different, not well documented, tests,  things such as placing Eccosorb and other wire grids in the beam path. This yielded some experience and feel for the modulator before an integrated telescope was made. Secure that the Polarization Rotator met minimal standards for a larger telescope, B-Machine was finally constructed. Full tests of B-Machine with the field rotator and detector chains are explored in Chapter~\ref{chap:telescopechar}.

\begin{figure}[p]
\begin{center}
\includegraphics[width = 13.5cm]{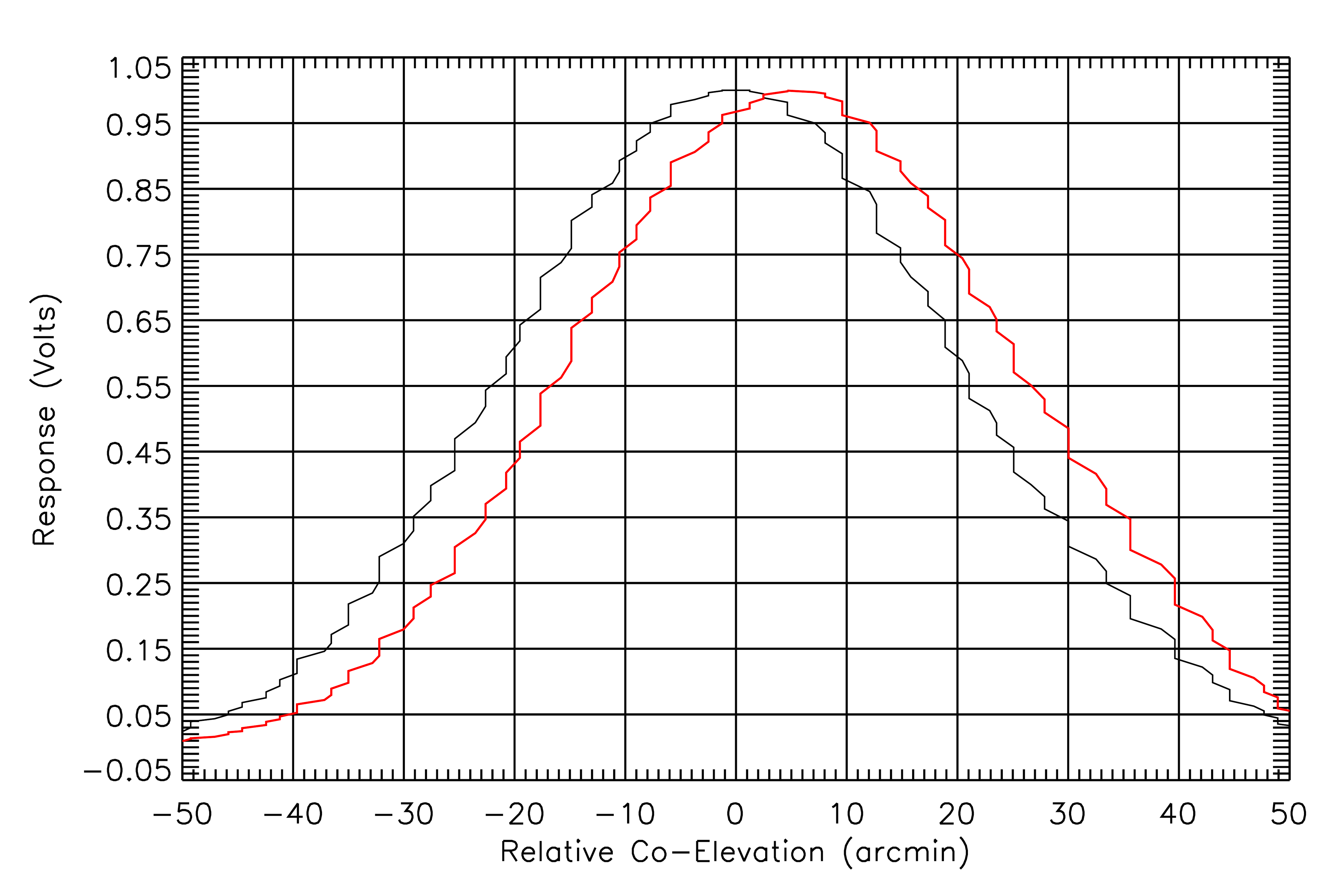}
\caption[Main lobe response from polarized thermal source]{Main lobe response from polarization modulator while viewing a polarized thermal source. Black is reflected off of plane mirror and red is reflection off of wire grid. \label{fig:BeamShift}}
\end{center}
\end{figure}

\begin{figure}[p]
\begin{center}
\includegraphics[width = 13.5cm]{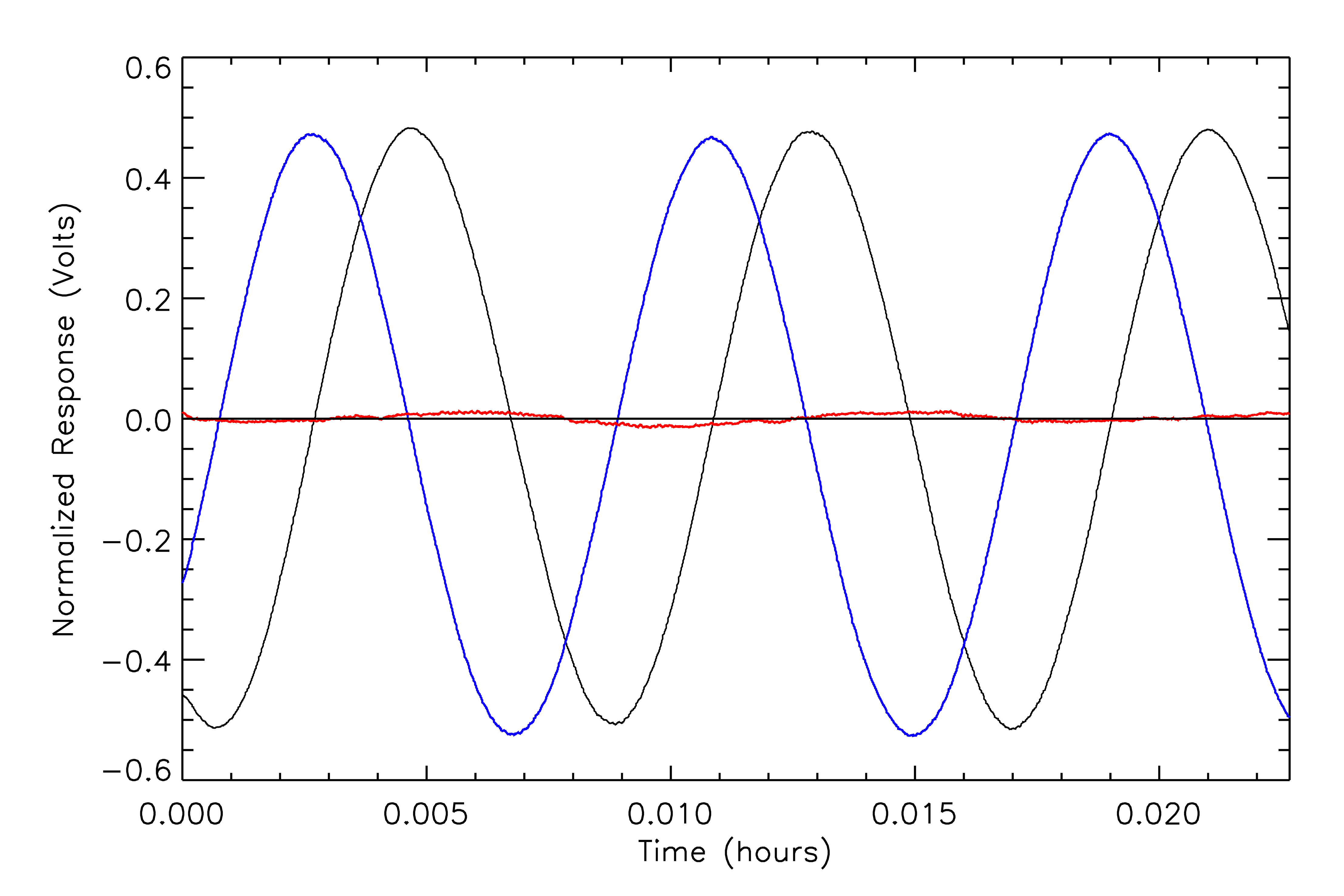}
\caption[Plot of I, Q and U from rotating polarized thermal source]{Radiometer output from small thermal source mounted on the roof of the Bren building rotated several times to show Q (black) and U (blue) signals. Also added is the quadrature sum in red. \label{fig:TQUThermalsource}}
\end{center}
\end{figure} 
\chapter{Telescope Characterization}\label{chap:telescopechar}
The B-Machine telescope was characterized both at UCSB and WMRS Barcroft. During the UCSB characterization phase the focus was mainly on beam shape and noise characteristics of the instrument.  While at WMRS efforts to define calibration constants both in temperature and polarization were explored. At the same time the servo system was exercised to determine its operational fringes.

\section{Beam Characterization} \label{sec:beamcharacterization}
 A full exploration of the beam size and patterns were made to test possible beam shape problems due to the addition of the Polarization Rotator into the Beast \citep{childers05} and WMPOL \citep{levy08} optics.

\subsection{Gaussian Beam Size}
\begin{figure}[p]
\begin{center}
\includegraphics[width = \textwidth]{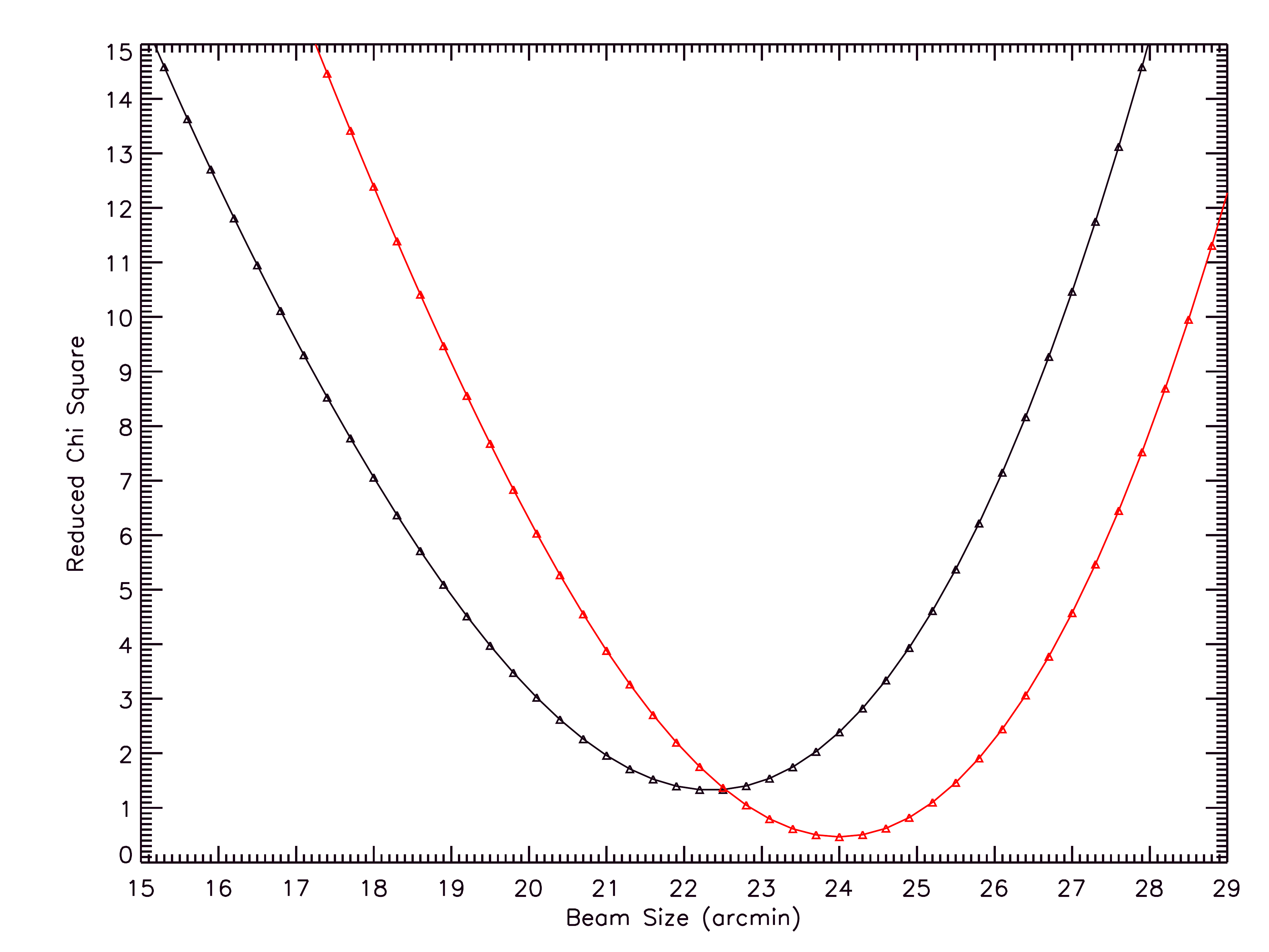}
\end{center}
\caption[Reduced chi-square fit of simulated Moon/beam convolution and Moon scan]{Reduced chi-square fit of simulated Moon/beam convolution and Moon scan. Reduced Chi square minimum corresponds to best fit of simulation with data. The red curve is an off axis horn and black is the central horn. Error estimates appear to be to large for the off axis horn. \label{fig:ChiSquare}}
\end{figure}

\begin{figure}[p]
\begin{center}
\includegraphics[height=17cm]{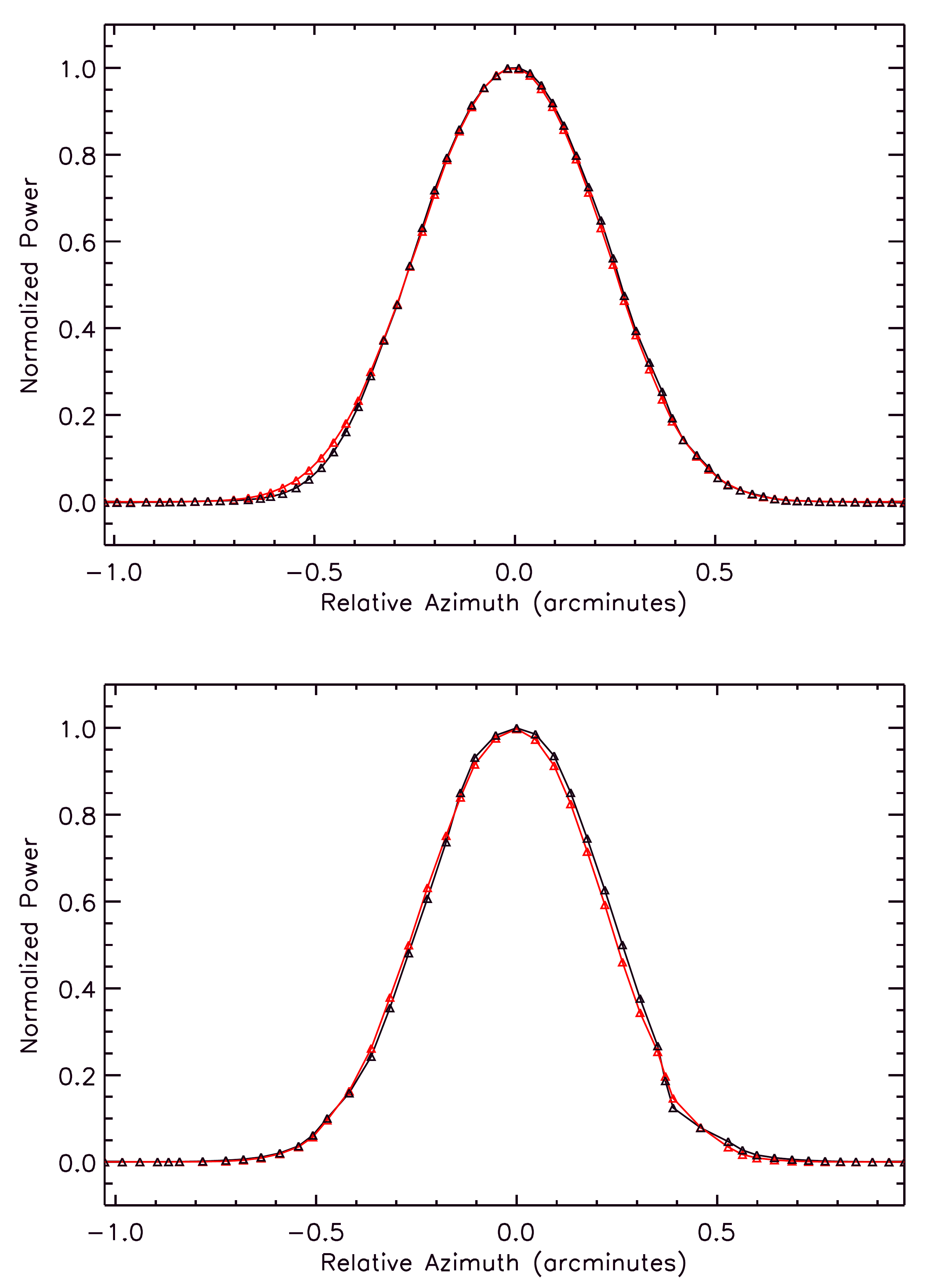}
\end{center}
\caption[Beam size using Moon vs simulated Moon]{Moon data and simulated Moon data using chi square test to determine best fit of convolved simulated Moon and beam. The black points/line are data taken 08/08/2008 and the red points/line are Moon and beam convolved. Top: central horn with best fit beam FWHM $22.2'\pm0.2'$, Bottom: off axis horn with best fit beam FWHM $24.0'\pm0.2'$. \label{fig:BeamSize}}
\end{figure}

\begin{figure}[p]
\begin{center}
\includegraphics[width = \textwidth]{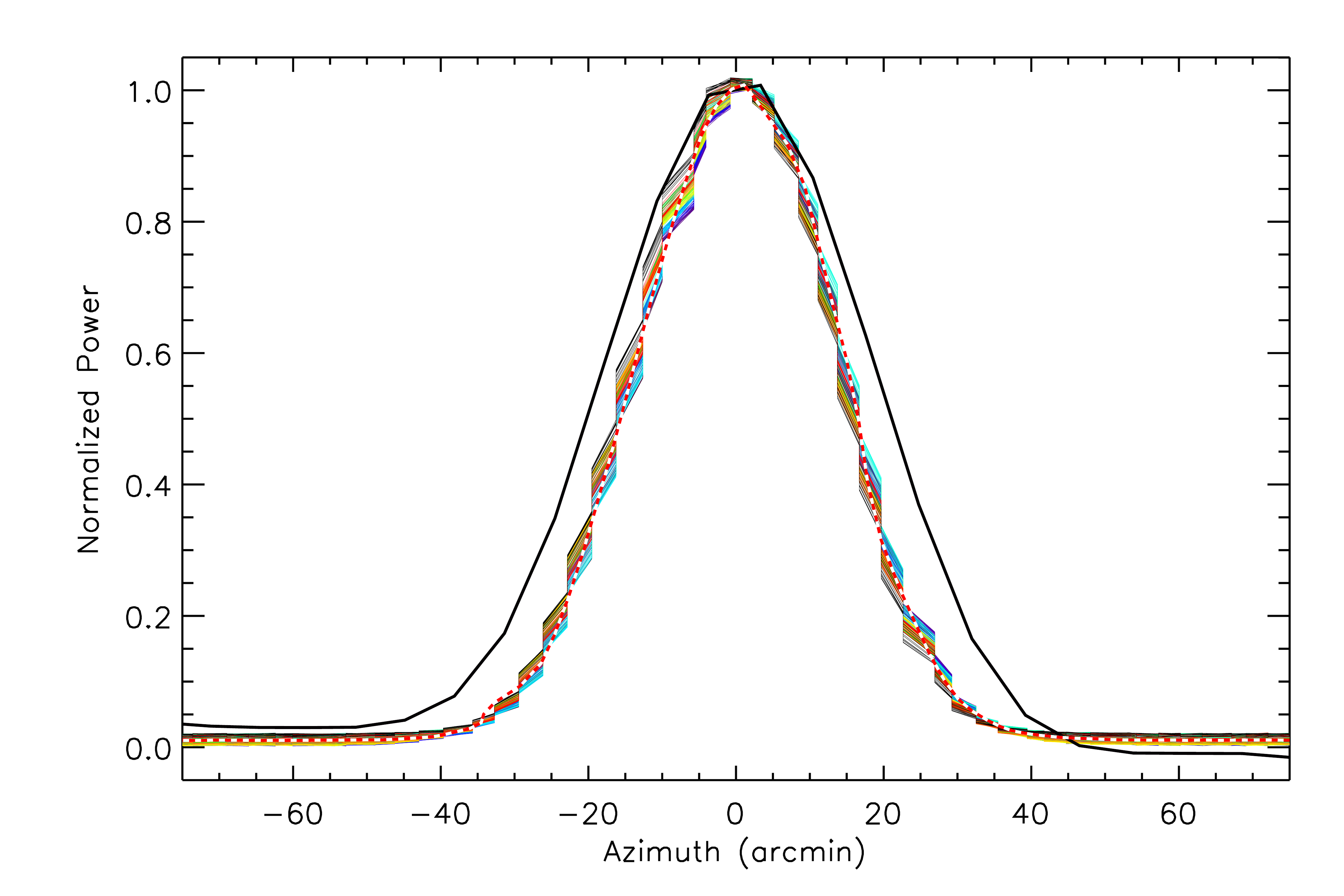}
\caption[Comparison of Beast and B-Machine beam shape]{Thick black line is a Moon scan from the Beast telescope using the same optics as B-Machine. The multi color lines are plots of each sector of the B-Machine Moon scan. The dotted red line is the demodulated Moon scan. \label{fig:BeastBmachineMoon}}
\end{center}
\end{figure}

The beam size was determined using scans of the Moon from August 8th, 2008. The Moon was scanned slowly multiple times giving fine angular resolution and 10-20 crossings in a short time period. The multiple crossings of the Moon made it possible to determine when the beam was centroided.  A simulated temperature map of the Moon was generated using modeling software written by Stephen Keihm~\citep{keihm75}. The Moon simulation used phase angle, frequency ($41\mathrm{~Ghz}$), and polarization angle (zero for temperature maps) relative to the Moon equator to generate accurate temperature maps. The phase angle was determined by the percent of the Moon that was illuminated on August 8th, 2008 ($43\%$).  Simulated maps convolved with different Gaussian beams where checked for goodness of fit with Moon scans using the chi square technique, Figure~\ref{fig:ChiSquare}.  The reduced chi square goodness of fit was generated and the beam size was taken to be where this parameter was minimized. This process was done for the central horn (channel 14) and one of the off axis horns (channel 9), Figure~\ref{fig:BeamSize}, yielding beam FWHM of $22.2'\pm0.2'$ for the central horn and $24.0'\pm0.2'$ for the off axis horn. Beam shapes for the $Q$ stokes parameter were investigated using the Moon scanned data, but inconsistencies in the simulation data (polarization from limbs of Moon not well understood) and issues with saturation made the process untenable. The main source of error in determining the beam size came from the resolution in scanning and the resolution of the simulated Moon maps. Beam sizes from experiments using the same optics have yielded similar results. The Beast Campaign gave FWHM beam size of $23.0'$ and WMPOL $24.0'$ for similar optics. WMPOL used larger corrugated feed horns causing the larger beam size. A direct comparison of a Moon scan done from both B-Machine and Beast telescopes, Figure~\ref{fig:BeastBmachineMoon}, show that a smaller beam from B-Machine is expected. Due to the Polarization Rotator much greater care was taken in aligning the optics for B-Machine than Beast and a slightly smaller (and closer to theoretical size) beam was achieved.

\subsection{Beam Shift}
\begin{figure}[p]
\begin{center}
\includegraphics[width = \textwidth]{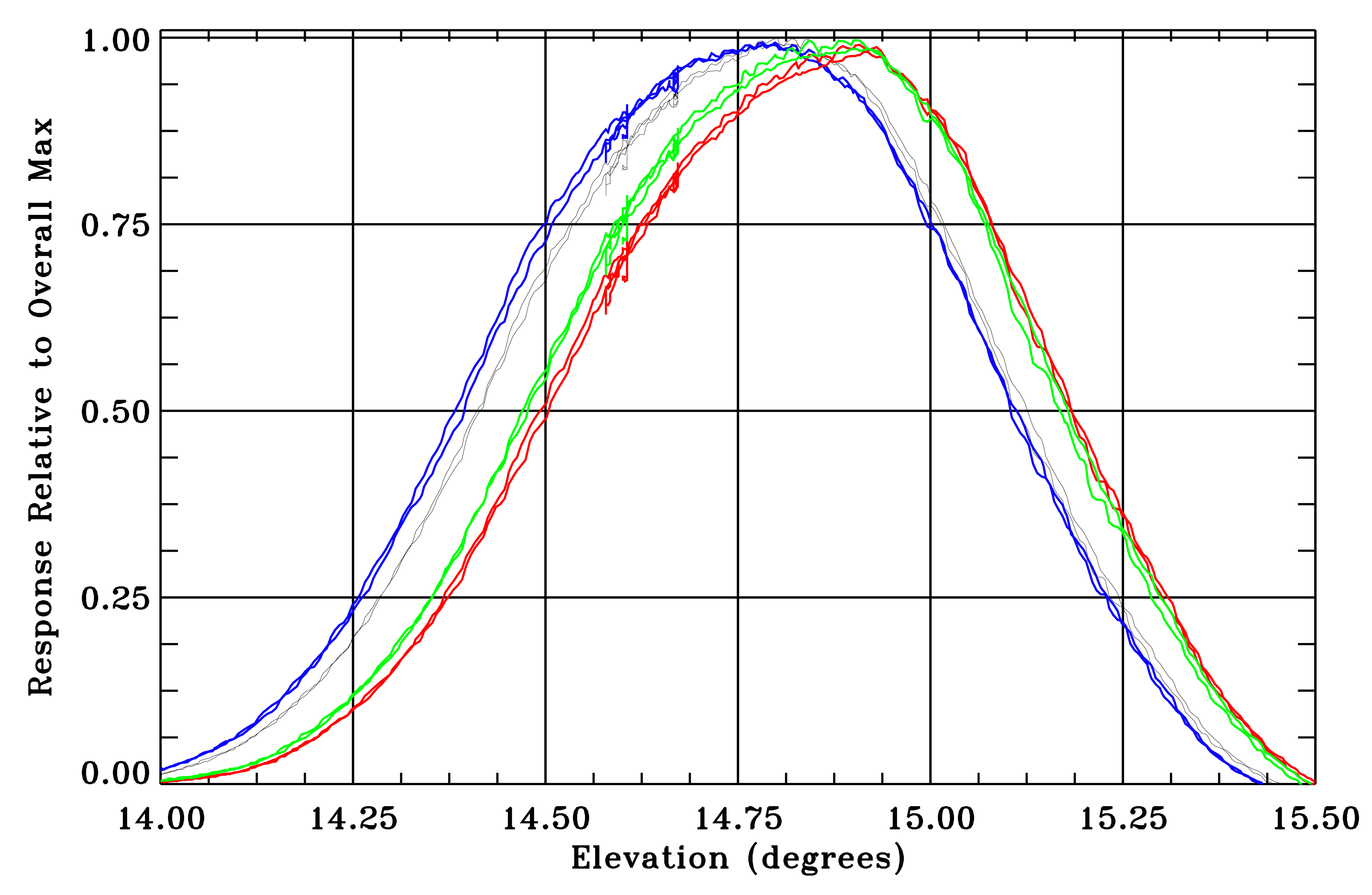}
\end{center}
\caption[Main lobe beam shift 4 max sectors]{Observed beam shift using polarized source,  red and black lines correspond to wires on Polarization Rotator vertical (reflecting off of plane mirror), Green and Red lines wires horizontal (reflecting off of wire grid). Testing done on East side of Broida Hall using the 41.5 GHz source with the polarization aligned with the central horn. \label{fig:MainLobeBeamShift}}
\end{figure}

The central corrugated feed horn observing a polarized aligned source will see maximum signal 4 times per revolution: 2 maxima corresponding to the wires being horizontal (signal reflects off of wires) and 2 corresponding to the wires being vertical (signal reflects off of plane mirror backing plate) relative to the horizon. The additional path length when the wires are vertical causes the beam to shift on the sky.  This is a purely geometric effect and for a $0.09\mbox{ inch}$ wire grid to plane mirror spacing, the calculated beam shift should be $6.5'$. Using the $41.5$ GHz source on the roof of the Bren Institute and plotting the signal as a function of elevation gives an observed beam shift of $6.56'\pm0.17'$, see Figure~\ref{fig:MainLobeBeamShift}. 

\subsection{Full Beam Shapes}
On February 26th, 2008 B-Machine was rolled to the top of the loading ramp on the east side of Broida, the frame was aligned using existing markings from previous observing days. The markings are spray painted circles on the ground that are three different colors for each of the three stabilization feet on the experiment. A 41.5 Ghz source was mounted on the roof of the Bren Institute approximately 150 m from the telescope. Though this is still in the near field, $\frac{D^{2}}{L\lambda}=\frac{2.0^{2}}{150\times0.0072}>1$, the source uses a corrugated feed horn and can be treated like a point source.  Attached to the source were 2 aerowave attenuators and a Direct Read Attenuator (DRA). The Aerowave attenuators had no attenuation on them and the DRA was set at a different attenuation level for each scan, to increase the dynamic range of the beam measurements.
Each polarization, referenced to the horizon as vertical, horizontal or $45^{\circ}$, of the horn was done in the same fashion. Attenuation was added on the DRA, then an elevation scan was taken followed by an azimuth scan. The DRA was adjusted to the next attenuation level (50 dB, 45 dB, 35 dB, 15 dB, or 0 dB) and then Az/El scans were done again. Using multiple attenuation levels allowed for better signal to noise so that the side lobes could be seen out to the 9th or 10th side lobe. Each scan was pieced together by matching the section of the previous scan with the beginning of the uncompressed sections of the next scan. The scans were trimmed when it was clear that the signal to noise was poor. Due to the hand alignment of the source, the $45^{\circ}$ polarization is close but the horizontal and vertical polarizations are off by a couple of degrees.

\begin{figure}[p]
\begin{center}
\includegraphics[width = \textwidth]{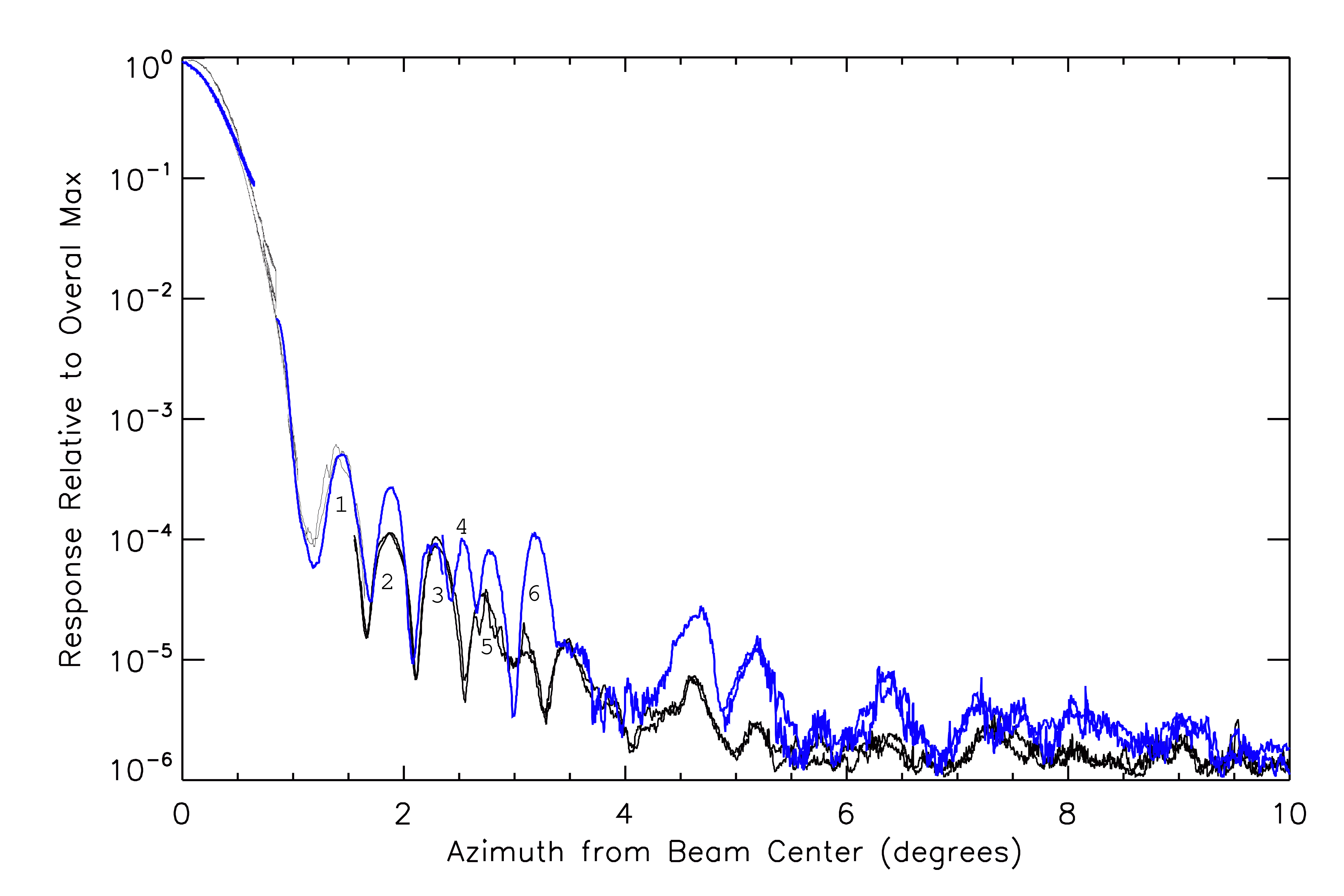}
\caption[Full beam pattern for $T$ source vertical and horizontal]{Full beam pattern of the central horn for $T$ with source Vertical (blue) and Horizontal (black). The variations in beam shape can give rise to spurious $Q$ signals. \label{fig:TVertHor}}
\end{center}
\end{figure}

The side lobe differences of the beam shapes for horizontal and vertical incident polarization can give rise to a spurious $Q$ signal. If an object, for example the sun at 6000 K, is in the 5th and 6th side lobe (in multiple lobes due to angular size) a $1.3\mbox{ mK}$ $Q$ signal is expected.  To confirm the expected value a scan of the sun with multiple crossings was taken. When the data were analyzed the thermal effects overwhelmed the small radiometric effects from the Sun. These effects included heating of baffling, optics, RF chains and surrounding environment. Instead using the source scanned data to get fractional signals by comparing the amplitudes of each side lobe in horizontal and vertical polarizations (see Figure~\ref{fig:TVertHor}) and then looking at the $Q$ signal from the source with polarization at $45^{\circ}$ gives a measure of the beam asymmetries.  With the source at $45^{\circ}$ the expected signal is all $U$.  Comparing the $Q$ signal (measured) with the scans of the source vertical and horizontal to get the expected signal as a fraction of the height of the side lobe estimates of the telescope side lobe pollution were made.  Table~\ref{tab:QoutofTfraction} lists expected and measured numbers for the first 5 side lobes. The 3rd side lobe had some anomalous noise problems and was omitted. The expected numbers represent differences in the side lobe response for the 2 orthogonal polarization states, the measured response includes multiple effects that can't be de-convolved from the side lobe asymmetries. However, the measure of the first side lobe with $0.0\%$ expected conversion (horizontal and vertical side lobe sizes are identical) can give a rough estimate of the other effects such as miss alignment of the source, T to Q conversion from non side lobe effects and U to Q conversion.  A source that was unpolarized and provided the needed signal to noise for all of the side lobes would have separated some of the ambiguities better. Table~\ref{tab:QoutofTfraction} provides a good first estimate of side lobe contamination and with the inclusion of all the effects gives information on how close thermal signals can be to the main lobe before they contaminate the survey. For a possible balloon experiment the galactic plane may present some spurious signals that will need to be modeled for template removal.

\begin{table}[p]
\begin{center}
\caption[Fraction of $Q$ Out of $T$ from Side Lobes of Central Horn]{Fraction of $Q$ Out of $T$ from Side Lobes of Central Horn \label{tab:QoutofTfraction}}
\begin{tabular}{|c|c|c|c|c|}
  \hline
  Side Lobe & Expected & Measured & Error & Fraction of\\
            &          &          &       & Main Lobe \\
  \hline
  \hline
  1 & 0.0$\%$ & 6$\%$ & 2$\%$ & $2.0\pm0.4\cdot10^{-4}$\\
  \hline
  2 & 13$\%$ & 20$\%$ & 2.5$\%$ & $7.5\pm0.2\cdot10^{-5}$\\
  \hline
  3 & 0.0$\%$ & Noise & N/A & $5.0\pm1.4\cdot10^{-5}$\\
  \hline
  4 & 11.0$\%$ & 15$\%$ & 3.0$\%$ & $2.9\pm1.0\cdot10^{-5}$\\
  \hline
  5 & 15.0$\%$ & 20$\%$ & 3.6$\%$ & $2.5\pm0.9\cdot10^{-5}$\\
  \hline
  6 & 22.0$\%$ & 25$\%$ & 2.7$\%$ & $3.0\pm1.0\cdot10^{-5}$\\
  \hline
\end{tabular}
\end{center}
\end{table}

\begin{figure}[p]
\begin{center}
\includegraphics[width = \textwidth]{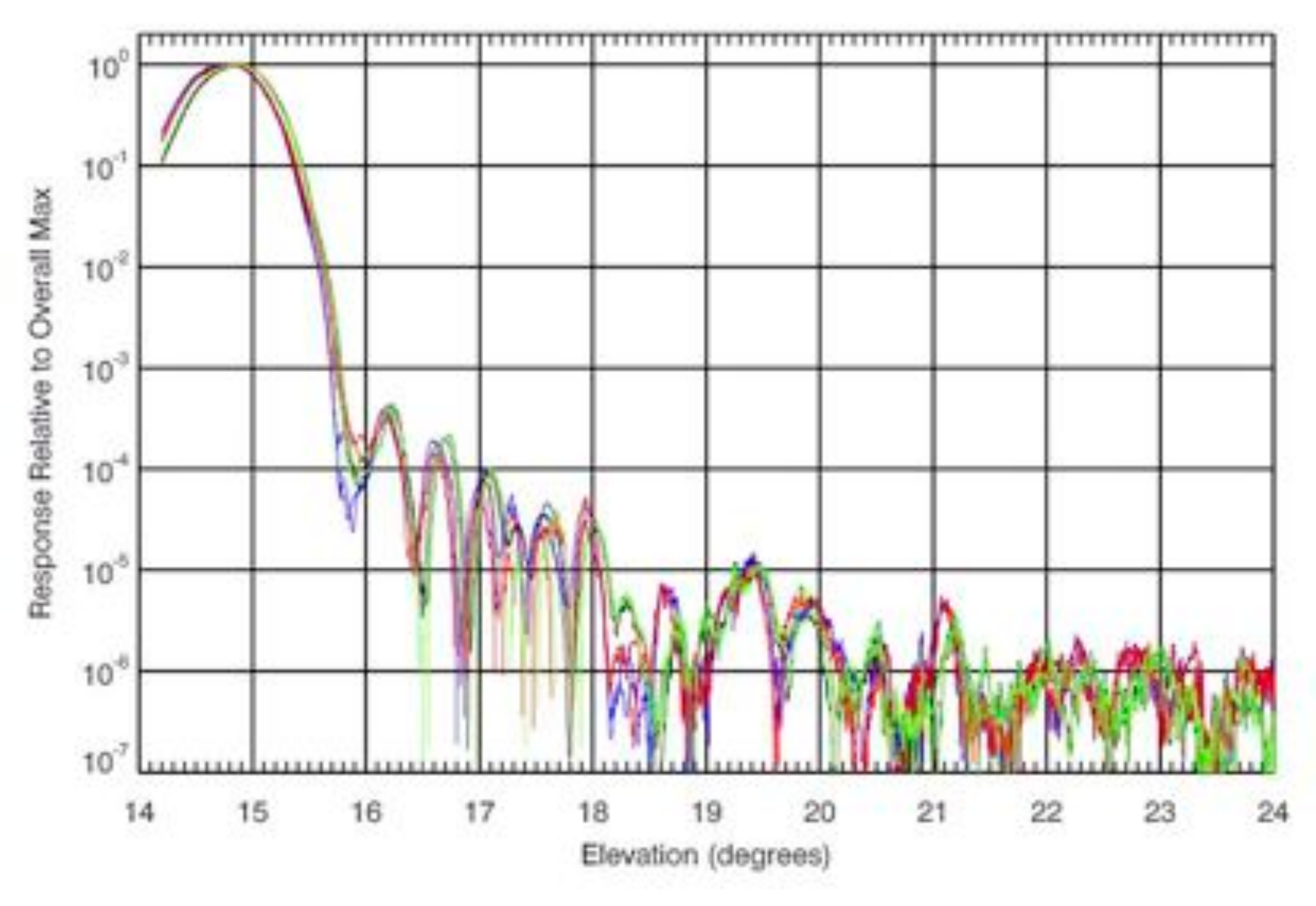}
\caption[Full beam pattern for max sectors source polarization at $45^{\circ}$]{Full beam pattern of the central horn for max sectors with source polarization at $45^{\circ}$.   \label{fig:FullBeam4sect45}}
\end{center}
\end{figure}

\begin{figure}[p]
\begin{center}
\includegraphics[width = \textwidth]{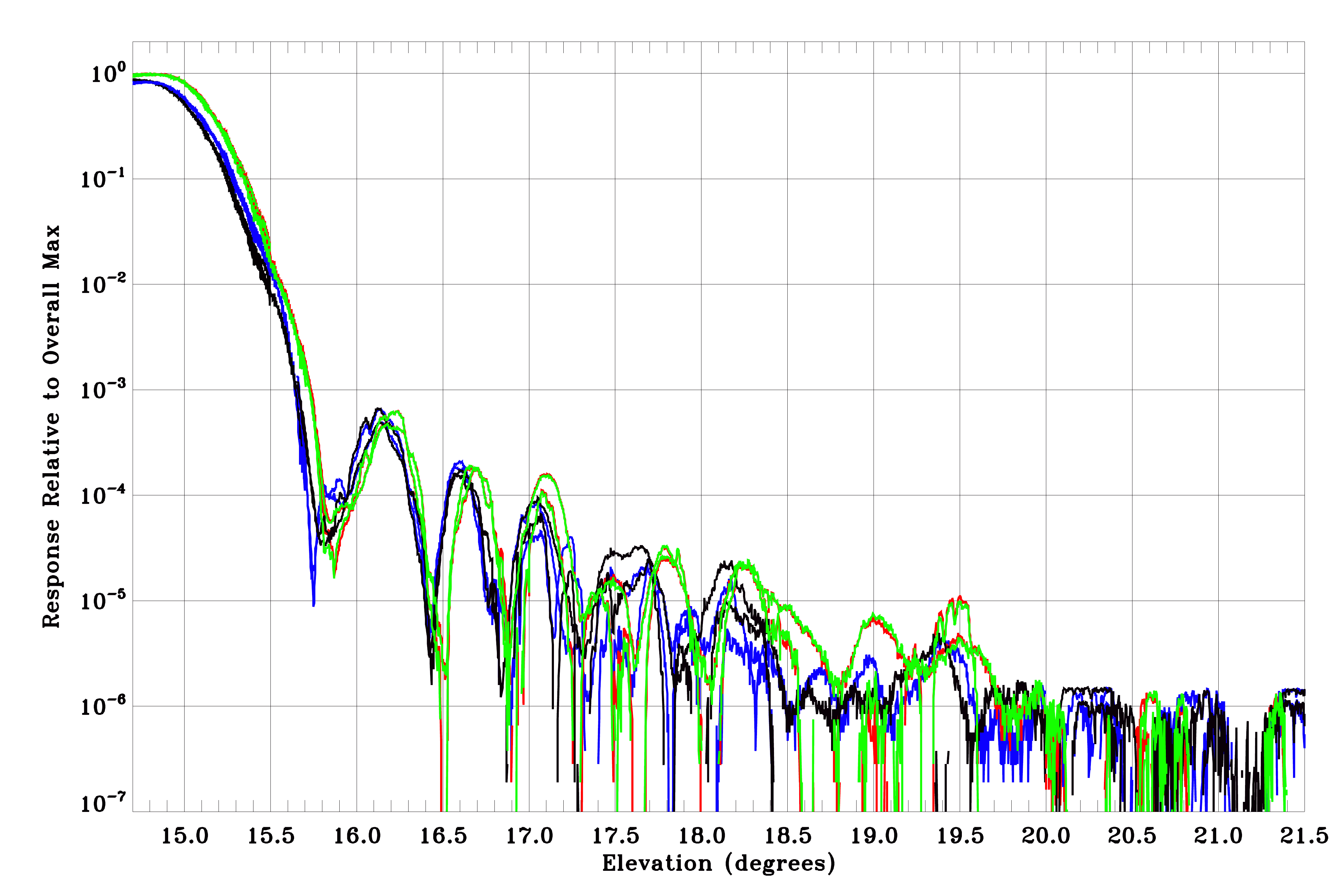}
\caption[Full beam pattern for max sectors source polarization horizontal]{Full beam pattern of the central horn for max sectors with source polarization horizontal. \label{fig:FullBeam4secthor}}
\end{center}
\end{figure}

\begin{figure}[p]
\begin{center}
\includegraphics[width = \textwidth]{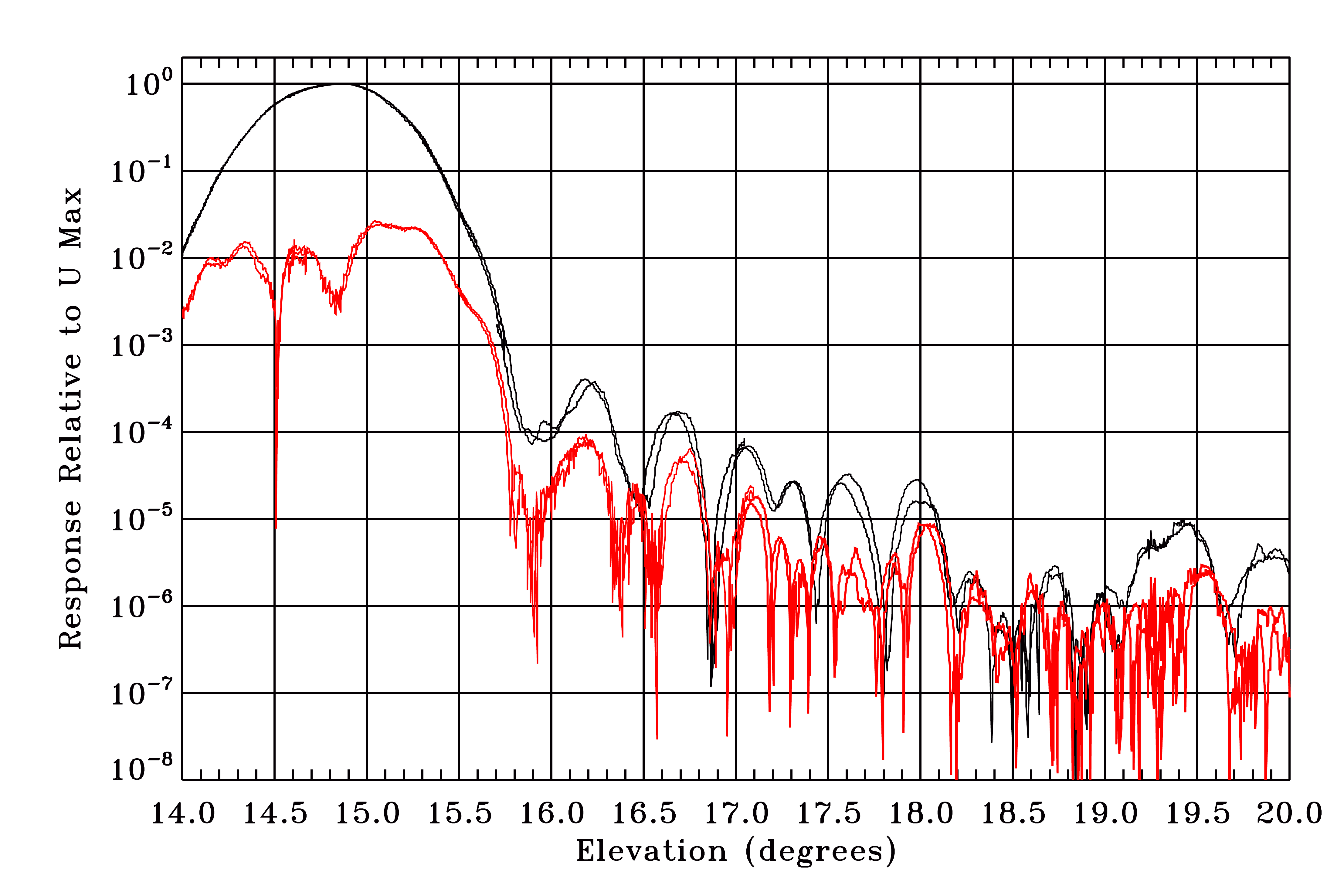}
\caption[Full beam pattern for $Q$ and $U$ source polarization $45^{\circ}$]{Full beam pattern of the central horn for $Q$ (red) and $U$ (black) with source polarization $45^{\circ}$. Ratio of $Q$ to $U$ shows $Q$ to $U$ conversion/isolation and source alignment precision. \label{fig:FullBeamU45}}
\end{center}
\end{figure}

\begin{figure}[p]
\begin{center}
\includegraphics[width = 13.5cm]{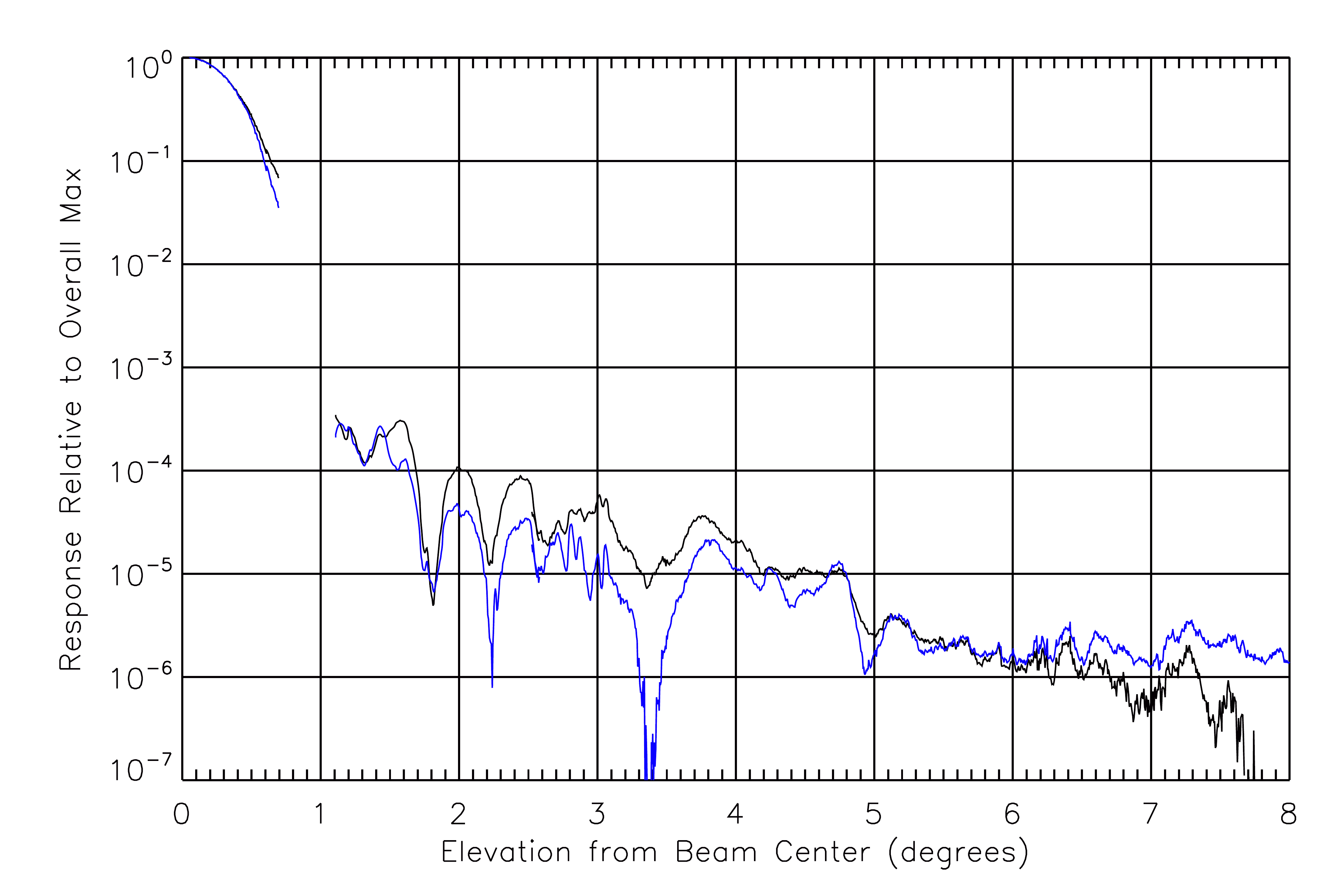}
\caption[Full beam pattern for off axis horn source polarization horizontal]{Full beam pattern for horn C, off axis horn, that is not vertically or horizontally lined up with the central horn. $T$ (black) and $Q$ (blue) with source polarization horizontal. The gap in data where the first side lobe is located is due to lack of on scale data for this region. \label{fig:FullBeamHornCTQ}}
\end{center}
\end{figure}

Full beam maps of the central channel were taken for both azimuth and elevation in T, $Q$, $U$ and of all sectors for the source polarized vertically, horizontally, and $45^{\circ}$.  Each plot was carefully inspected for any indication of sensitivity to signals well off the bore or major differences in the beam shapes.  Any major beam anomalies in $Q$ or $U$ would have presented major problems for this type of chopping technique. All of the beam plots were made and show no unexpected results. A subset of the plots are included here for reference, Figures~\ref{fig:TVertHor} through ~\ref{fig:FullBeamHornCTQ}.

\section{Calibration}\label{Sec:Calibration}
The phrase "calibration" is used to describe the measurement of the instrumental parameters that characterize the three transfer standards (gains) to convert the data stream which is in Volts to temperature, $Q$, or $U$. Each channel has 2 gains, one for the AC channels and the second for the DC channels, they are scaled by the difference in the IF gain between the AC and DC chains, see Chapter~\ref{chap:instrument} Section~\ref{Subsect:datainput}. A full calibration sequence was done at both the beginning and end of the data taking campaign which includes filling the beam with a known cold load, intermediate load, ambient load and sky load, with daily automated calibration sequences ran for all days. Daily calibrations use an automated ambient load and the sky to calibrate.

\subsection{Temperature}
Typically measurements of the gain are done with 2 thermal loads, Eccosorb (AN72) either at room temperature ($\sim300$ K) or soaked in a LN bath ($\sim73.8$ K at $4$ km). They are put in front of the radiometer, just below the extruded polystyrene window, to make sure they fill the beam. With the voltages that correspond to the given temperatures from each RF chains output and,

\begin{center}
\begin{equation}\label{eqn:gainlin}
   G_{0}=\frac{T_{hot}-T_{cold}}{V_{cold}-V_{hot}},
\end{equation}
\end{center}
and

\begin{center}
\begin{equation}\label{eqn:lincal}
   V_{out}=G_{0}\left(T_{sky}+T_{sys}\right),
\end{equation}
\end{center}
to get the gain in Kelvin per Volt. Here $T_{hot}$/$V_{hot}$ is the temperature/voltage ratio of the warm load and $T_{cold}$/$V_{cold}$ is the temperature/voltage ratio of the cold load. B-Machine has roughly $60$ dB of gain and uses a diode with a response of $0.5$ mV per $\mu\mathrm{W}$ observing a $15$ K sky with a system temperature of 45 K gives a diode level of $\sim3$ mV. Using the same calculation for a 300 K load gives a diode level of $\sim19$ mV consistent with the measured values. The Anristu 75K50 Microwave Detector Diodes output voltage is proportional to its incident power from 1 mV to 10 mV.  When outside of this regime the diodes response is nonlinear. For B-Machine the warm load represents a data point outside of the linear response and hence Equation~\ref{eqn:gainlin} cannot be used. As an alternative the detector response can be modeled using,

\begin{center}
\begin{equation}\label{eqn:nonlincal}
   V_{out}=\frac{G_{0}(T_{A}+T_{sys})}{1+bG_{0}(T_{A}+T_{sys})},
\end{equation}
\end{center}
where $V_{out}$ is the output voltage of the radiometer chain, $T_{A}$ is the antenna temperature , $T_{sys}$ is the system temperature, $G_{0}$ is the linear gain response and $b$ is the non-linearity parameter. For the case of a linear receiver $b$ goes to zero and Equation~\ref{eqn:nonlincal} reduces to Equation~\ref{eqn:lincal}.

There were 2 main testing phases for B-Machine in order to get all the necessary parameters so that a fit to Equation~\ref{eqn:nonlincal} could be made. The first phase of the measurements was performed at UCSB. With the detector warm, a test load was used which consisted of a large Aluminum box (no top) insulated with polystyrene sides and a piece of Eccosorb (AN72)\footnote{Emerson $\&$ Cuming Microwave Products} inside. The box was filled with Liquid Nitrogen (LN) so that the top of the Eccosorb was underneath the level of the LN. The test load was positioned directly under the RF window of the dewar and Aluminum walls were erected so that all beam paths ended in the LN. With the detector's beam filled with LN a blue foam load alternated with a room temperature piece of Eccosorb was inserted between the RF window and LN load. In this way the blue foam piece could be well characterized for later calibrations. With the detector warm the gain is low enough that all measurements are well within the linear regime of the diode. All measurements for Table~\ref{tab:BlueFoamChar} are done at the diode (no audio gain involved). Using a similar procedure for getting the blue foam temperature the system temperature was found. The main difference in the procedures is that the detector was cooled and tuned for observing. A compression estimate (based on Beast calibrations) from the diode level readings was used for each channel in getting the system temperature. These estimates turned out to be consistent with the compression numbers given by the model.

\begin{figure}[p]
\begin{center}
\includegraphics[width=17.0cm, angle=90]{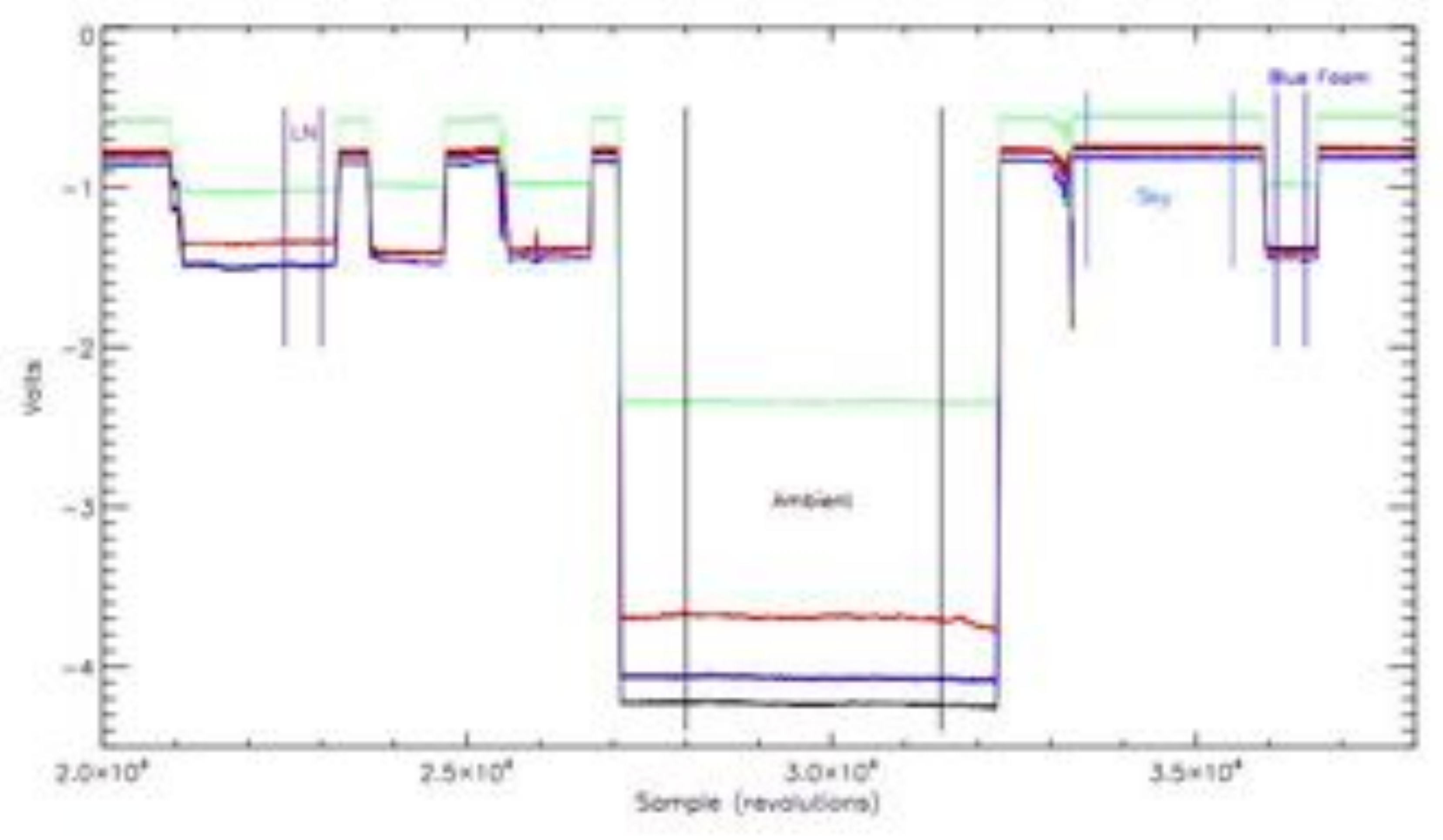}
\caption[Gain calibration sequence 08/07/2008]{Gain calibration sequence (08/07/2008) using LN (73.8 K), Eccosorb (ambient), and Blue Foam (intermediate) loads to calibrate the temperature gain of the system. The telescope was looking at sky when there was no load in the optical path. Each curve is one of the operational channels: red is channel 9, blue is channel 10, green is channel 11, and black is channel 14. \label{fig:gain08072008}}
\end{center}
\end{figure}

\begin{figure}[p]
\begin{center}
\includegraphics[width=17.5cm, angle=90]{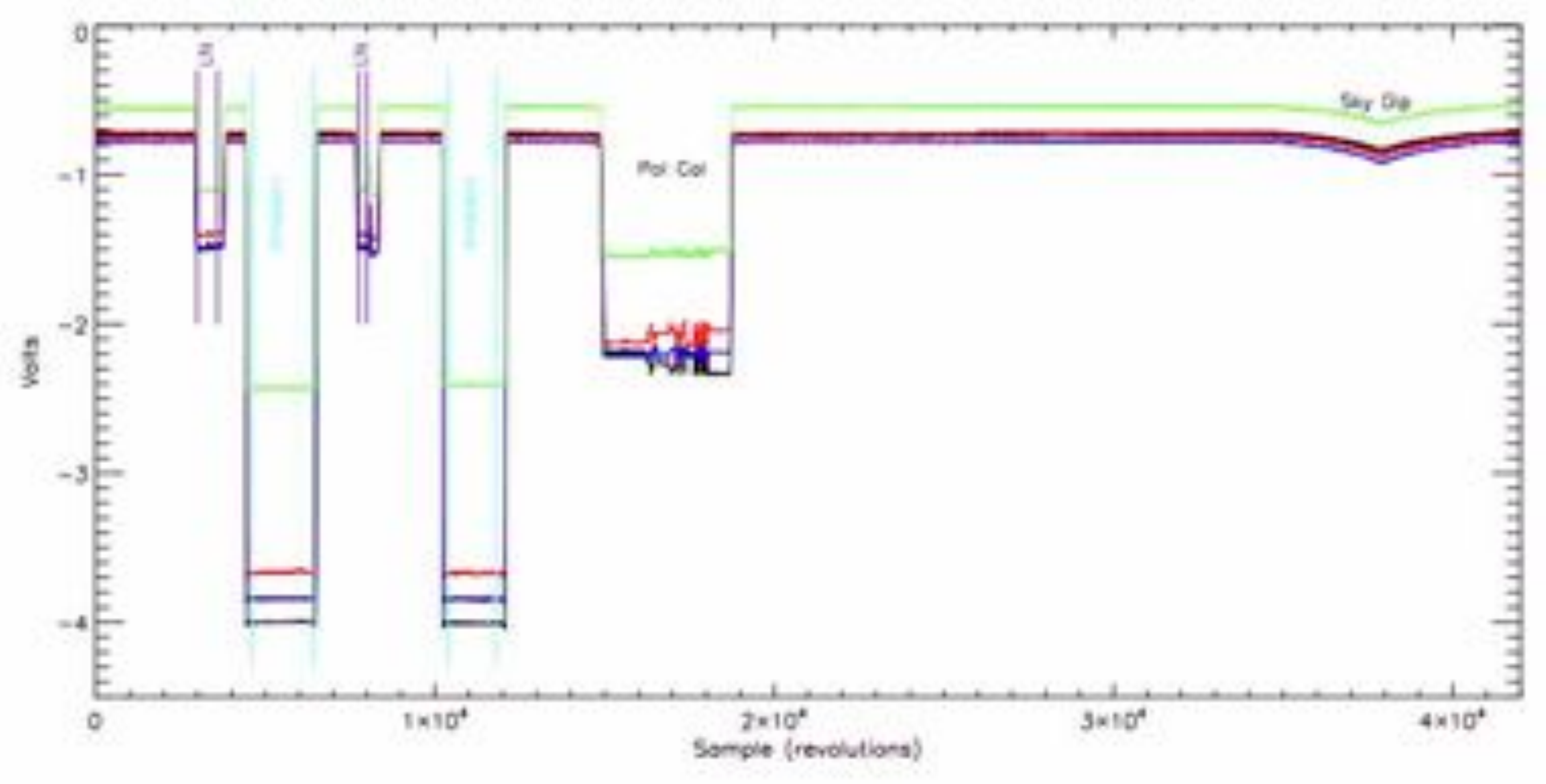}
\caption[Gain calibration sequence 10/14/2008]{Gain calibration sequence (10/14/2008) using LN (73.8 K) and Eccosorb (ambient)loads to calibrate the temperature gain of the system. The telescope was looking at sky when there was no load in the optical path. Each curve is one of the operational channels: red is channel 9, blue is channel 10, green is channel 11, and black is channel 14. \label{fig:gain10142008}}
\end{center}
\end{figure}

\begin{table}[p]
\begin{center}
\caption[Blue Foam Characterization]{Blue Foam Characterization \label{tab:BlueFoamChar}}
\begin{tabular}{|c|c|c|c|c|c|}
  \hline
  Channel & Offset & Hot  &  Cold & Blue Foam & Blue Foam\\
          &  mV & mV & mV & mV & K\\
  \hline
  \hline
  1 / 9  & 1.65 & 6.47  & 4.55  & 4.94  & 44.8 \\
         & 1.70 & 6.62  & 4.67  & 5.08  & 45.8 \\
  \hline
  2 / 10 & 1.23 & 9.44  & 6.48  & 7.08  & 44.2 \\
         & 1.24 & 9.44  & 6.49  & 7.08  & 44.06\\
  \hline
  3 / 11 & 0.89 & 12.01 & 10.46 & 10.82 & 50.63\\
         & 1.07 & 12.46 & 10.87 & 11.23 & 49.36\\
  \hline
  6 / 14 & 0.28 & 2.91  & 2.08  & 2.23  & 39.4 \\
         & 0.29 & 2.88  & 2.07  & 2.22  & 40.3 \\
  \hline
\end{tabular}
\end{center}
\end{table}
Before B-Machine was set to observe (beginning of season) and just before it was shutdown for the winter (end of season) a calibration sequence was done. The first calibration sequence, see Figure~\ref{fig:gain08072008}, consists of a 73.8 K load (LN filled polystyrene cooler that had several sheets of Eccosorb in it) placed in the beam path just below the dewar. This was followed by several tests of calibrated blue foam, white foam, ambient temperature calibrator and polypropylene in several combinations into the beam. A similar sequence was done at the end of the campaign, see Figure~\ref{fig:gain10142008}, modulo no blue foam. The importance of the blue foam calibration was not discovered until after the observing campaign had been finished and in depth data analysis started.

To accurately fit the curve from Equation~\ref{eqn:nonlincal}, voltages and temperatures for each of the thermal loads are found and fit to the model. Ambient temperature and LN temperature can be found from temperature sensors or barometric pressure, respectively and the voltages for these loads from data acquisition readouts. Finding system temperature and added blue foam temperature was determined in previous tests at UCSB. The remaining piece of information to get was the sky temperature which was determined by using sky dips, see Subsubsection~\ref{subsubsec:skytemp}.  Fit parameters have been computed for all of the channels, see Table~\ref{tab:fitparameters}.

\subsubsection{Sky Temperature}\label{subsubsec:skytemp}
\begin{figure}[p]
\includegraphics[width=\textwidth]{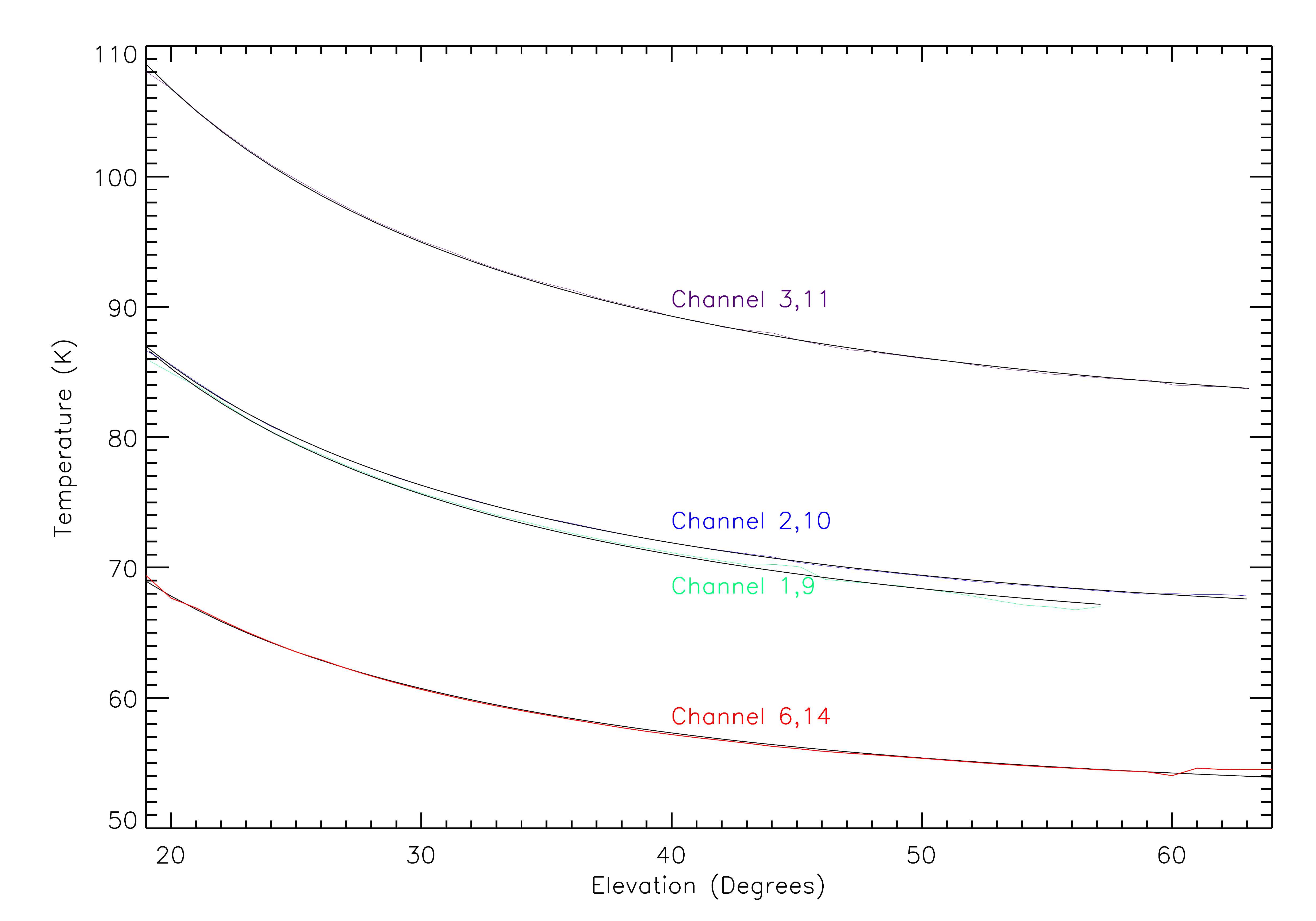}
\caption[Sky dip data with fit from 08/07/2008]{Sky dip data for 08/07/2008: the black line is the fit data and the colored lines are observed data. \label{fig:skydipfit08072008}}
\end{figure}

\begin{figure}[p]
\begin{center}
\includegraphics[width=\textwidth]{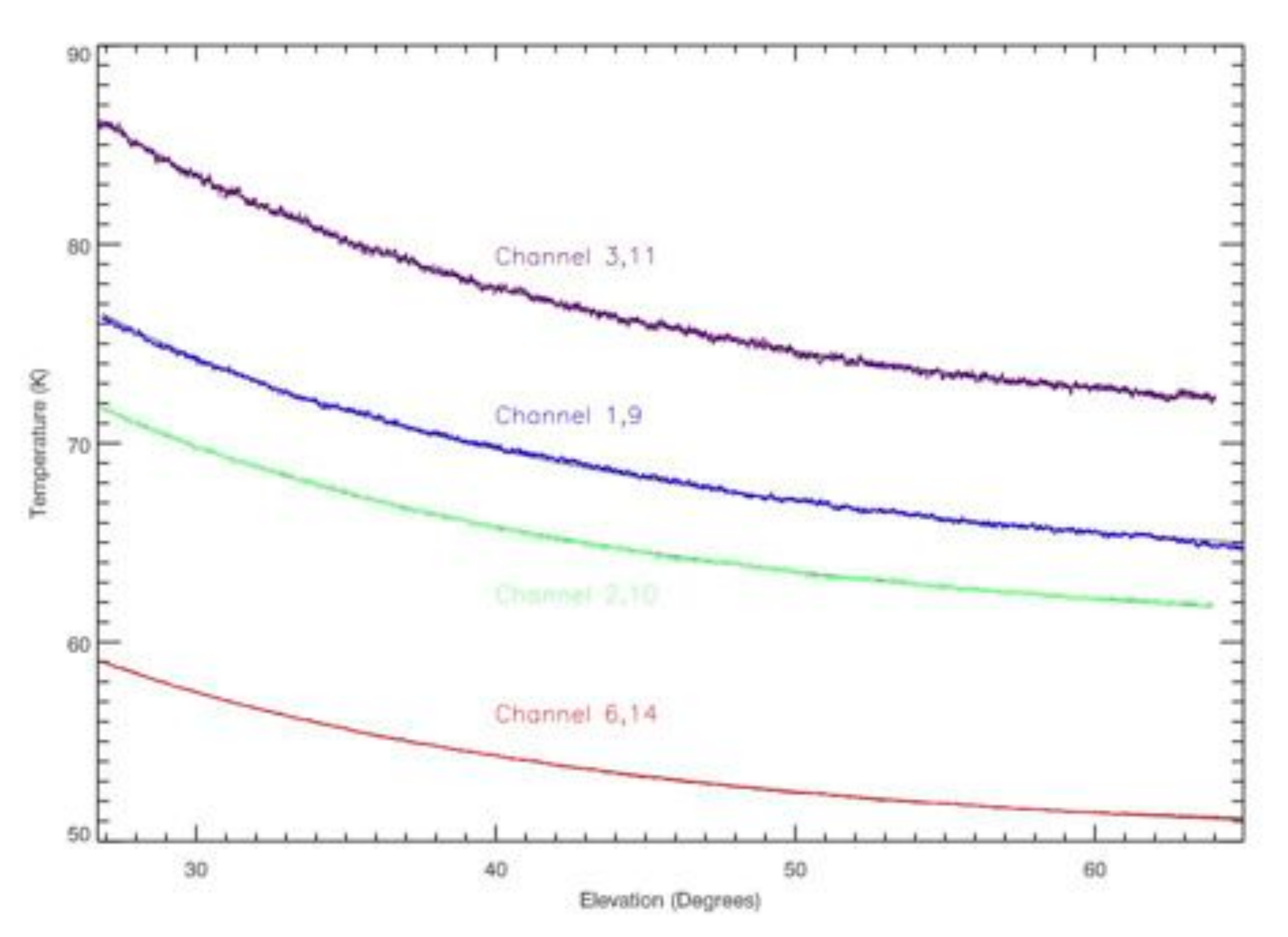}
\caption[Sky dip data with fit from 10/14/2008]{Sky dip data for 10/14/2008: the black line is the fit data and the colored lines are observed data. \label{fig:skydipfit10142008}}
\end{center}
\end{figure}

The sky temperature can be calculated using 2 different methods. First, by using the sky signal between thermal load sources (referred to as the DC method) and second by using a sky dip. The DC method suffers from several critical drawbacks: it is very sensitive to DC voltage drift and $\frac{1}{f}$ noise. By solving Equation~\ref{eqn:zenithtemp} for $T_{Zenith}$ (zenith temperature) and using the approximate gain calculated from thermal loads (estimating compression) gives the zenith sky temperature. A more reliable sky temperature, zenith temperature and rough estimates of the system temperatures can be made using sky dips.  A sky dip is generated by slowly driving the elevation up or down giving a decreasing or increasing signal, respectively, from the change in thickness of the atmosphere. In Figure~\ref{fig:skydipfit08072008} the signal can be seen decreasing as a function of elevation. By fitting this signal to a model of the sky temperature, Equation~\ref{eqn:skytemp}, a zenith temperature and system temperature are found. The sky dip method for getting the sky temperature is less prone to systematic error from DC voltage drift and 1/f noise then the DC method. Scans made just after the calibration sequences can be modeled and fit to Equation~\ref{eqn:skytemp}, see Figures~\ref{fig:skydipfit08072008} and~\ref{fig:skydipfit10142008}. This also turns out to give a good fit to the system temperature. A constant 1.5 K is added to the sky temperature from the integration of the x2 signal, see Subsection~\ref{subsect:emissivity}, by the demodulation technique.

\begin{equation}\label{eqn:zenithtemp}
T_{Zenith}=(T_{LN}-((V_{sky}-V_{LN})*Gain))*\cos(elevation)
\end{equation}
where $T_{LN}$ is the temperature of LN, $V_{sky}$ is the voltage from the sky signal and $V_{LN}$ is the voltage from the LN signal.

\begin{equation}\label{eqn:skytemp}
T=T_{sys}+\frac{T_{Zenith}}{\sin(elevation)}
\end{equation}

\begin{table}[p]
\begin{center}
\caption[Fit Parameters for Calibrations]{Fit Parameters for Calibrations \label{tab:fitparameters}}
\begin{tabular}{|c|c|c|c|c|c|c|c|}
  \hline
  Channel  & Max     &                      &                & Tsys   & Tsys       & $\triangle$T          &    $\triangle Q$ or $\triangle U$\\
  AC/DC    & Phase  &    $G_{0}$    &   $b$    & DC   & Sky Dip &  $\mathrm{mK}/\sqrt{\mathrm{s}}$  & $\mathrm{mK}/\sqrt{\mathrm{s}}$\\
                   &               &                       &              &        (K)            &           (K)            & & \\
   \hline
  1/9 & 23 & 0.0108  & 0.00655 & 53.66 & 54.75& 1.74 & 1.93\\
  \hline
  2/10 & 7 & 0.0119  & 0.00608 & 52.95 & 56.43& 1.72 & 1.91\\
  \hline
  3/11 & 6 & 0.00723 & 0.0506 & 71.19 & 69.55& 2.10 & 2.33\\
  \hline
  6/14 & 14 & 0.0136 & 0.0178 & 40.20 & 43.88& 1.44 & 1.60\\
  \hline
\end{tabular}
\end{center}
\end{table}

\subsubsection{Emissivity}\label{subsect:emissivity}
\begin{figure}
\begin{center}
\includegraphics[width=14.6cm]{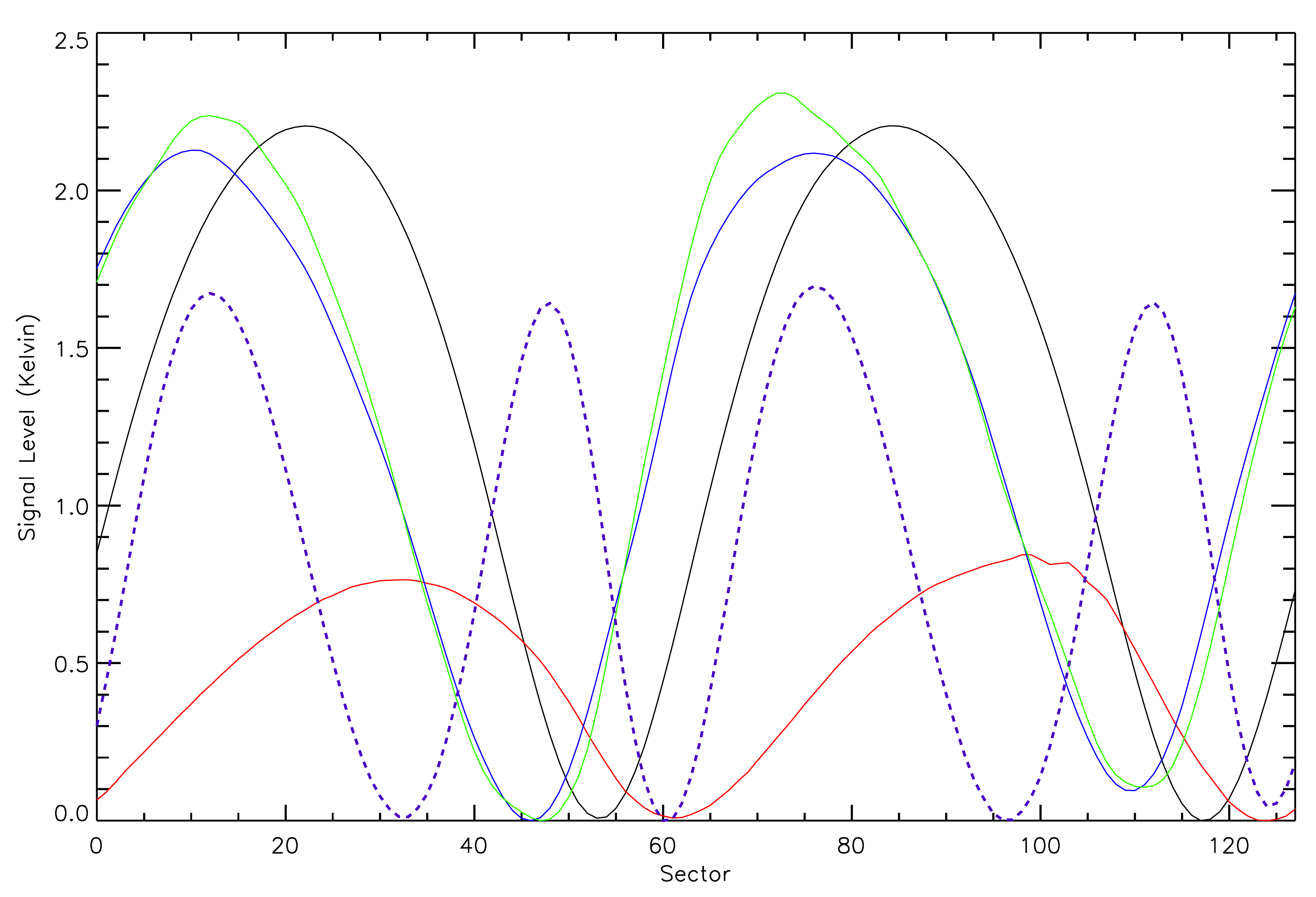}
\caption[Times 2 signal all channels]{Times 2 signal from emissivity difference between wire grid and plane mirror Aluminium plate. Channel 6/14 is black, Channel 1/9 is red, channel 2/10 is blue, channel 3/11 is green, and the thick purple dashed line is a polarized signal on the central channel for periodicity comparison. The demodulation integrates the x2 signal to zero on the AC channels. \label{fig:x2}}
\end{center}
\end{figure}

A difference in the signal level when observing sky between the wire grid and the aluminum plate (plane mirror) causes a synchronous signal that is $\frac{1}{2}$ of the periodicity of the polarization signal, see Figure~\ref{fig:x2}.  This signal is referred to as the x2 signal.  Though the signal level is high the demodulation technique (lock-in amplification) for $Q$ and $U$ is not sensitive to the x2 signal.    The x2 signal is the convolution of several effects including loss of the wire grid, emissivity of the wires, emissivity of the blue foam spacer, and emissivity of the Aluminum plate. All of these effects combine to create an overall level difference in the signal when the feed horns are viewing sky off of the wire grid surface or the plane mirror surface.

Data with no wire grid and blue foam should not exhibit the same added signal. Hence a data set that contains both sky signal with and without the wire grid and blue foam spacer on the Polarization Rotator should yield the emissivity of the wire grid. In addition, a calibration must be performed relatively close to the above sequence to determine the sky temperature. Theoretically the gain is not necessary; all of the calculations could be done in voltage but due to diode compression from the data sets and diode levels not being adjusted for nominal scanning mode, calibration is a must.

To calibrate the data set that viewed the sky with and without the wire grid on, a data set made several hours before that contained a sky dip, a LN load and ambient load in the data was used to get the sky temperature. Also, in the calibrating data set was a section of data with the Polarization Rotator in.  This was necessary due to extreme diode compression in the data set with and without data set. Assuming a moderate compression of 10$\%$ (consistent with previous diode compression calculations) a gain of $86.9\pm2.0\mathrm{~K/V}$ for the sky dip data and using a model to fit the zenith temperature gives a zenith temperature of $22.30\pm0.52^{\circ}\mathrm{C}$ or a sky temperature of $28.98\pm0.68^{\circ}\mathrm{C}$ at $50.3^{\circ}$ from horizon. Utilizing the sky temperature and the LN target the calibration for the data set with and without the wire grid is $30.34\pm0.76\mathrm{~K/V}$.  The signal viewing the sky through the Polarization Rotator with the wire grid is of the form,

\begin{center}
\begin{equation}\label{eqn:Grideff}
   T_{out}=T_{in}+T_{amb}(1-\epsilon_{grid}).
\end{equation}
\end{center}
Using the section of the data that is looking at the sky with no wire grid gives the input temperature and the section of data looking at the sky through the wire grid averaged over all sectors gives the output temperature. Solving for $\epsilon$ yields an efficiency of $99.33\pm0.20\%$ using the temperature sensor on the calibrator for the ambient temperature. Also available is an estimate of the efficiency of the polarization calibrator $80.0\pm2.0\%$. This number is not an estimate of the emissivity but rather the reflectivity of the Eccosorb, how well the piecemeal wire grids are put together, and the non-uniformity of the grids used. The compression at the ambient temperature side is dominating the error of these measurements.

\subsection{Polarization}
\begin{figure}
\begin{center}
\includegraphics[width = \textwidth]{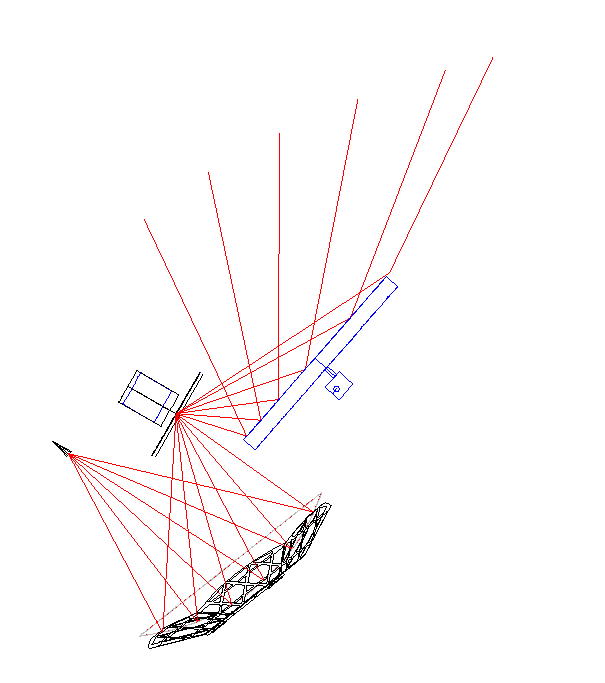}
\caption[Optical design of B-Machine Polarization Calibrator]{Beam paths for the central horn with Polarization Calibrator in. \label{fig:PolCal}}
\end{center}
\end{figure}

Converting the signal from voltage to temperature is not the optimal way to run this experiment because of the chopping method. Converting the AC channels into the $Q$ and $U$ Stokes parameters is the preferable mode of operation. The basic idea of getting a polarization calibration is to put a known polarized signal in front of the detector and then demodulate that section of data. The ratio of the demodulated voltage to the known Q or U signal (Q is used for this test) is the calibration constant. Following the temperature calibration sequence with LN on 08/07/2008 and 10/14/2008 the Polarization Calibrator was lowered into the beam and rotated.  The Polarization Calibrator is a large round piece of Hexcel covered by several layers of Eccosorb, AN72, covered by a thin layer of Styrofoam. On top of the Styrofoam is a piecemeal wire grid. This wire grid consists of several panels of wire grids that didn't pass the quality control for the Polarization Rotator and were visually aligned and attached together. Finally, a large Styrofoam cover is put over the whole assembly and taped down to the Hexcel for thermal stability and mechanical strength. The Polarization Calibrator is mounted to a frame that rotates in and out of the optical path by a bushing system that lets it rotate about its center, see Figure~\ref{fig:PolCal}. A potentiometer is attached to the end of the bushing system to measure the relative position of the calibrator and the voltage is stored on channel 15 of the DAQ data files. The voltage is calibrated to get the exact position of the wires from the max phase measurement (see Section~\ref{subsec:GetMaxPhase}).

The orientation of the wheel dictates the polarization angle and the difference between the sky (zenith) and ambient (Eccosorb) temperatures give the amplitude of the signal. Knowing the rough orientation of the wires on the Polarization Calibrator makes it possible to determine the matching Stokes parameter. 

A data set with polarization calibrator in was taken and a section of demodulated data were found that represented a signal that was all $Q$ for the given channels. The revolution that corresponds to the max signal was used to find the minimum (sky zenith) and maximum (Eccosorb or ambient) voltage in the time ordered data (TOD). The temperature difference divided by 2 gives the corresponding $Q$ or $U$ Stokes parameter that the level 1 file for this revolution should give. Taking the ratio of the Stokes parameter and the voltage gives the gain that each channel should be multiplied by to get the corresponding Stokes parameter ($Q$ or $U$).

\begin{equation}\label{eqn:Stokes}
   Q=\frac{T_{x}-T_{y}}{2}=\frac{f_{gain}(V_{eccco})-f_{gain}(V_{sky})}{2},
\end{equation}
Where $f_{gain}(\mathrm{~V})$ is the gain model which gives a temperature for the given voltage. See Table~\ref{tab:Polcal08072008} through Table~\ref{tab:Polcalall1014} for the relevant data for the polarization calibrations.

\begin{table}[p]
\caption{Polarization Calibration File Information 08/07/2008 \label{tab:Polcal08072008}}
\begin{center}
\begin{tabular}{|c|c|c|c|c|c|c|}
  \hline
  Channel & Rev Num & Max (V) & Min (V) & $\Delta$ T (K) & $\Delta$ V (V)& Pol Gain \\
  \hline
  \hline
  1/9 & 1462640 & -1.029 & -3.400 & 226.49 &  2.371 & 166.22 \\
  \hline
  2/10 & 1462487 & -1.086 & -3.688 & 224.69 & 2.602 & 148.47 \\
  \hline
  3/11 & 1461193 & -0.746 & -2.230 & 240.94 & 1.484 & 279.93 \\
  \hline
  6/14 & 1462484 & -0.939 & -3.750 & 225.804 & 2.911 & 139.284\\
  \hline
\end{tabular}
\end{center}
\end{table}
A couple of points need to be made about Table~\ref{tab:polcalall}. First, the temperature calibration model previously derived was used. Secondly, another method of getting the polarization calibration might be to use the voltages to get an absolute temperature for both the min and max voltages and then calculate the Stokes parameter using these temperatures. The reason that this method wasn't used is it is very sensitive to any offset voltages that might have changed since the calibration sequences. The polarization calibration takes into account both the inefficiencies of the polarization calibrator and the wire grid. In addition, a small effect due to the non-symmetric wave form of the polarization signal is buried in the analysis. This effect was explored in some detail and by averaging the demodulation over a full revolution of the Polarization Rotator it can be removed.

\begin{table}[p]
\caption{Calibration Numbers Using All Peaks and Averages for Data Taken on 08/07/2008 \label{tab:polcalall}}
\begin{center}
\begin{tabular}{|c|c|c|c|c|}
  \hline
  Stokes & Channel 9 & Channel 10 & Channel 11 & Channel 14 \\
  \hline
  \hline
  +Q   &   166.574   &   148.385    &  278.396    &  141.609 \\
  \hline
  -Q   &   165.054   &   148.408 &     280.452   &   142.110 \\
  \hline
  +U   &   163.671   &   148.029   &   278.060   &   141.580 \\
  \hline
  -U   &   167.597     & 148.064    &  279.176   &   141.262 \\
  \hline
  Average & 165.72  &   148.22 &  279.00   &    141.64  \\
  \hline
\end{tabular}
\end{center}
\end{table}

\begin{table}
\caption{Calibration Numbers Using All Peaks and Averages for Data Taken on 10/14/2008 \label{tab:Polcalall1014}}
\begin{center}
\begin{tabular}{|c|c|c|c|c|}
  \hline
  Stokes & Channel 9 & Channel 10 & Channel 11 & Channel 14 \\
  \hline
  \hline
  +Q  & 150.000    &  140.843   &   224.678     & 139.915\\
  \hline
  -Q   &   149.191   &   140.244   &   226.366    &  140.473 \\
  \hline
  +U   &  145.991   &   139.341    &  223.443  &    139.888 \\
  \hline
  -U & 150.260   &   138.980   &   224.293   &   139.718 \\
  \hline
  Average & 148.86 &  140.10  &  224.70  &  140.00   \\
  \hline
\end{tabular}
\end{center}
\end{table}
As a simple test to check that the polarization calibration is consistent, a data set that contained a signal that was all $Q$ from the get max phase data, see Subsection~\ref{subsec:GetMaxPhase}, was normalized and shifted so that a signal that is roughly $0.2$ K peak to peak, was demodulated and a $Q$ signal of $106$ mK was seen. From the definition of $Q$ a polarized signal of $0.2$ K should give a $100$ mK $Q$ signal. The waveform of the Polarization Rotator doesn't yield itself to this type of analysis due to its asymmetries, but it is clear that the calibration for the polarization is at worst $6\%$ off.

\subsubsection{Getting Max Phase}\label{subsec:GetMaxPhase}
Each of the feed horns is coupled via circular to rectangular waveguide transition to the input of a Low Noise Amplifier (LNA). Rectangular waveguides will only carry radiation polarized perpendicular to its long dimension and each of the horns is arbitrarily orientated with respect to the central horn. Some effort was made to align the central horn so that the polarization was parallel to the horizon, but this was a gross alignment and was not used as a standard. To calibrate the horns alignment the telescope required a polarized signal of known polarization with a large signal to noise. When discussing local sources $Q$ is a source polarized vertically or horizontally and $U$ is a source polarized at $45^{\circ}$. Once a signal of say all Q is observed the signal is demodulated with a phase shift for the demodulation square wave of 0 to 31, the shift that yields the maximum signal is the max phase. The max phase corresponds to the number of sectors the Polarization Rotator must rotate from the zero position (the point at which the index pulse of the encoder triggers the sector rest and data sample) before the wires are lined up with the polarization of a given horn. The wire grid is subdivided into 128 sectors or 32 counts per polarization rotation.

To get the max phase for B-Machine, a person was sent to the north ridge of Mount Barcroft with a 41.5 GHz source mounted on a tripod. The source had 2 Aerowave uncalibrated attenuators and one calibrated Direct Read Attenuator (DRA). The source was leveled using a digital level, that is accurate up to $0.01^{\circ}$, to better than $0.1^{\circ}$ and pointed roughly at B-Machine. The source was setup such that when it is leveled on the tripod the emitted polarization is all $Q$. The signal level was adjusted with the use of the DRA so that it was on scale at its maximum for the DC channels.  B-Machine then scanned the source taking care that the central horn (channels 6 and 14) saw the source and the signal was on scale. Following the central channel scan the polarization calibrator was placed into the beam and rotated several times for calibration of the other channels. This sequence was done at the beginning and end of the data taking campaign (08/07/2008 and 10/14/2008).  Several Polarization Calibration sequences were also done at random intervals in the data taking campaign to confirm relative phase.

It was noticed after taking a look at the max phase data from the first day of calibration that a max phase of 13 or 14 could be achieved depending on the revolution number used for demodulation. This was caused by the source not being on the peak of Mt. Barcroft so the side lobes of the beam were seeing the mountain. Also, better care aligning the source would have made the measurement more accurate. For the second max phase calibration sequence much more care was taken into leveling and placement of the source, placing the source on the peak of Mt. Barcroft. These numbers were not dependent on the revolution for the second max phase calibration. Ultimately, a max phase of 14 for the central channel was used, based on the second calibration sequence and parts of the first.

Once the max phase for the central channel is established a calibration of the Polarization Rotator is easily made. The polarization angle of the Polarization Calibrator as a function of the output voltage on channel 15 can be used to get the max phase of the remaining channels. Data from the rotation of the Polarization Calibrator is demodulated and the revolution number corresponding to the maximum signal for channel 14 (all of the AC channels are not on scale) is the revolution number that corresponds to a signal that is all Q (in the telescope frame). This coincides with the wires being vertical on the Polarization Calibrator.  Finding the phase that demodulates to the maximum value for each channel from this revolution of data gives the max phase.

\begin{table}[p]
\begin{center}
\caption[Maximum Phase and System Temperature]{Maximum Phase and System Temperature \label{tab:MaxPhaseTsys}}
\begin{tabular}{|c|c|c|c|c|}
  \hline
  Chan  & Chan  & Max   & Tsys DC        & Tsys      \\
   AC   &   DC  & Phase & Measurements   & sky dips  \\
        &       &       & (K)            & (K)      \\
  \hline
  \hline
  1 & 9 & 23 & 53.66 & 54.75 \\
  \hline
  2 & 10 & 7 & 52.95 & 56.43 \\
  \hline
  3 & 11 & 6 & 71.19 & 69.55 \\
  \hline
  6 & 14 & 14 & 40.20 & 43.88 \\
  \hline
\end{tabular}
\end{center}
\end{table}

\subsection{Error}
\begin{figure}[p]
\begin{center}
\includegraphics[width=14.6cm]{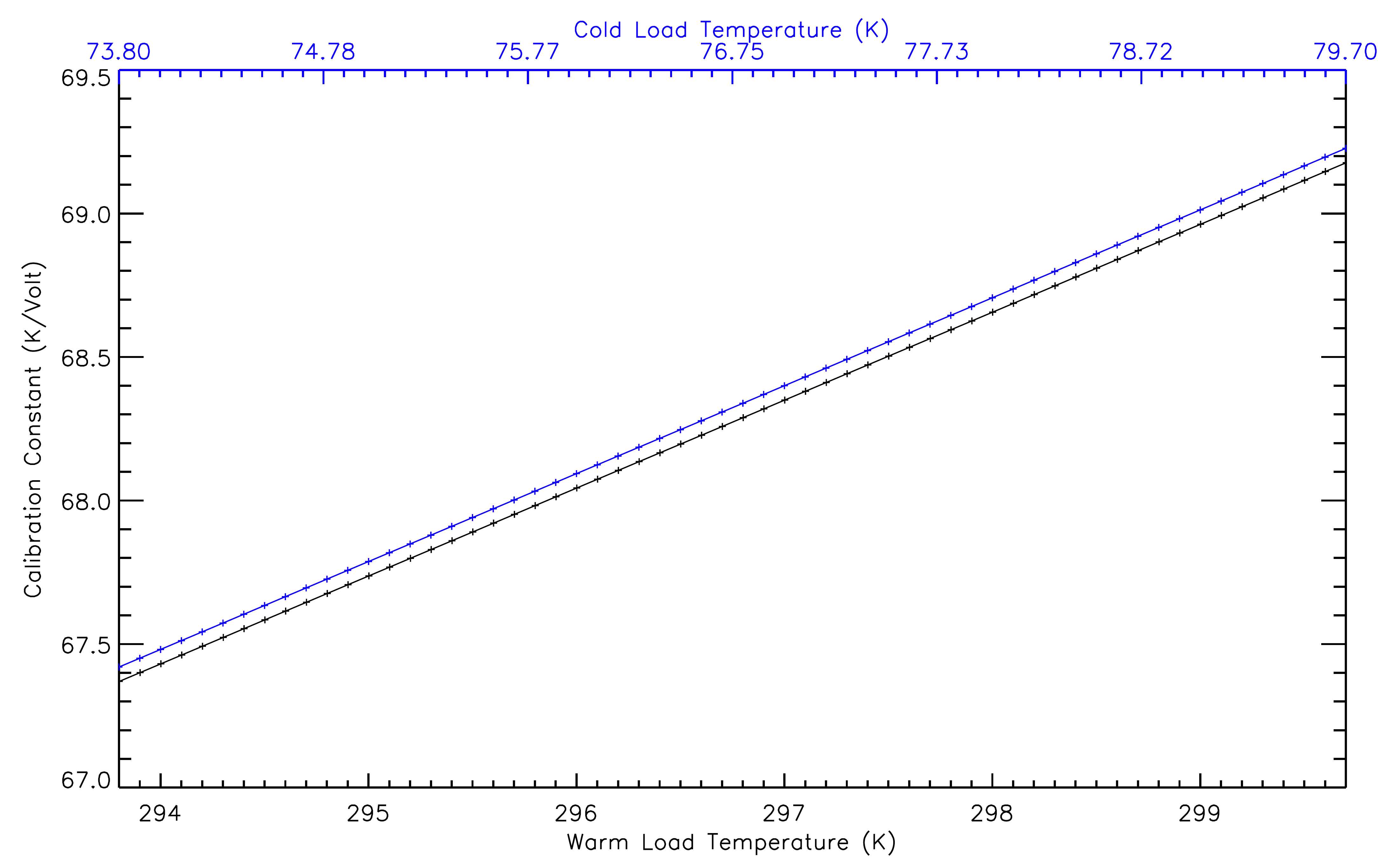}
\caption[Gain changes from temperature variations in load]{Gain of channel 14 for small variations in load temperatures. The cold load calibrations have been offset by $0.5$ K/V so that the lines are distinguishable from each other. \label{fig:gainvstemp}}
\end{center}
\end{figure}

The errors in calibration will be propagated throughout the experiment to the final answer. It is important to understand the sources of error and thier impact on the calibration constants. The uncertainty based on the measurements can be quantified by looking at the variance of each of the values and combining them in the appropriate fashion, Equation~\ref{eqn:gainerror}. 
\begin{center}
\begin{equation}\label{eqn:gainerror}
   \Delta G^{2}=G(\frac{\sigma_{Th}^{2}}{T_{h}^{2}}+\frac{\sigma_{Tc}^{2}}{T_{c}^{2}}+\frac{\sigma_{Vh}^{2}}{V_{h}^{2}}+\frac{\sigma_{Vc}^{2}}{V_{c}^{2}})
\end{equation}
\end{center}
The standard deviation for each of the voltage values can be calculated from the data sets. The standard deviations in the temperature are a bit more difficult to quantify. Variations in the thermal conductivity of the Eccosorb loads, radiative loading, and small variations in the surface temperature can all lead to errors between the RF signal and the projected temperature of the loads. For the warm load a separation of statistical and systematic uncertainty is possible, since a temperature sensor is placed directly on the load. Looking at the data from the warm load temperature read out over a $2.2$ minute interval the temperature varied by $0.5$ K and the voltage from the radiometer varied by $0.03$ V ($\sim1.5$ K). This indicates that the surface of the warm load was radiatively cooling faster than the read out was changing. Using the extreme of this measurement of $\pm0.5$ K leads to a change of $0.5\%$ in the gain constant. Figure~\ref{fig:gainvstemp} shows how the calibration constant changes as a function of temperature for the warm and cold loads.  The data that were used to get the warm load voltages was much shorter ($\sim$30 s) so this will safely over estimate the error in the warm load temperature. No readout was possible for the cold load temperature, hence an estimation of the deviation will need to be made. Using the entire section of cold load data and looking at the decay in the signal, gives an estimate of the change in cold load temperature as a function of time. The cold load section represents about $52$ seconds of data and shows a change in temperature of roughly $0.5$ K. The data that are used for the calibration sequence is $\frac{1}{4}$ of the cold load data, a small section in the middle of the run.  This suggests that the deviation in temperature will be dominated by systematic unknowns such as beam filling, thermalization of Eccosorb, and reflectivity at the surface. A conservative estimate of the deviation from the temperature of LN at altitude ($\sim73.78$ K) is $0.3$ K. The error for the cold load won't be symmetric around the boiling point temperature, because the thermal load is much more likely to be warmer in this case than colder. The temperature of LN at altitude is found using the "Thermophysical Properties of Nitrogen" webpage from NIST. The results from the error calculations are all summarized in Table~\ref{tab:DCgains}.

\begin{table}[p]
\begin{center}
\caption[Voltages, Temperatures and Standard Deviations for Gain Measurements]{Voltages, Temperatures and Standard Deviations for Gain Measurements \label{tab:DCgains}}
\begin{tabular}{|c|c|c|c|c||c|c|c|c|}
  \hline
  &\multicolumn{4}{c||}{08-07-2008}&\multicolumn{4}{c|}{10-14-2008}\\
  \hline
  Channel & 9 & 10 & 11 & 14 & 9 & 10 & 11 & 14 \\
  \hline
  \hline
  $V_{hot}$ & 3.91 & 4.47 & 2.58 & 4.66 & 3.88 & 4.23 & 3.01 & 4.40\\
  \hline
  $\sigma_{Vhot}$ & 0.011 & 0.011 & 0.008 & 0.008 & 0.008 & 0.008 & 0.008 & 0.005 \\
  \hline
  $V_{cold}$ & 1.35 & 1.49 & 1.03 & 1.49 & 1.41& 1.49 & 1.11 & 1.50\\
  \hline
  $\sigma_{Vcold}$ & 0.0069 & 0.007 & 0.006 & 0.007 & 0.003 & 0.003 & 0.004 & 0.002 \\
  \hline
  $T_{hot}$ & 296.8 & 296.8 & 296.8 & 296.8 & 273.46 & 273.46 &273.46 & 273.46\\
  \hline
  $\sigma_{Thot}^{*}$ & 0.144 & 0.144 & 0.144 & 0.144 & 0.08 & 0.08 & 0.08 & 0.08 \\
  \hline
  $\sigma_{Thot}$ & 0.5 & 0.5 & 0.5 & 0.5 & 0.5 & 0.5 & 0.5 & 0.5 \\
  \hline
  $T_{cold}$ & 73.8 & 73.8 & 73.8 & 73.8& 73.8 & 73.8 & 73.8 & 73.8\\
  \hline
  $\sigma_{Tcold}$ & 0.25 & 0.25 & 0.25 & 0.25& 0.25 & 0.25 & 0.25 & 0.25 \\
  \hline
  Error$\pm$ & 0.48 & 0.41 & 0.71 & 0.35  & 0.43 & 0.38 & 0.67 & 0.33\\
  \hline
  Cal & 87.23 & 74.60 & 118.47 & 69.09 & 80.25 & 72.56 & 105.20 & 68.62 \\
  \hline
\end{tabular}
\end{center}
\end{table}
All voltages are in Volts, temperatures are in Kelvin and gains are in Kelvin per Volt. The 2 deviations in the warm load temperature have been added in quadrature to get the standard deviation for the final calculation. The first warm load standard deviation is statistical while the second is systematic.

\section{General Telescope Properties}
In addition to RF and beam characterization other properties of the telescope were quantified for observation.

\subsection{Servo System}
The original controls and programs for controlling the telescope were written for a telescope that scanned between 2 fixed azimuths \citep{levy08}. While B-Machine is set up to continuously scan in azimuth, the necessity for raster ability was not deemed important in the original design. While attempting to raster scan a small source for a full 2-D beam map it was found that the drive system was not finely tuned enough to allow for raster scanning. As a result all point sources both terrestrial and extra solar were observed by fixed elevation scans while slowly scanning between 2 azimuths letting the source drift through (frequently referred to as a drift scan).

\subsection{Scan Strategy} \label{subsec:scanstrategy}
B-Machine needed to balance sample rate limitations, beam smearing effects and beam overlap in the nominal scan strategy for day to day operation. To measure the maximum sample rate, the rate of the Polarization Rotator was slowly increased (rate of rotator proportional to the input voltage) until missed samples in the data were spotted.  A final rotation rate of the primary table was determined once a maximal rotation rate of $33.4$ Hz was determined. Using a $70$ s scan rate gives approximately 9 samples per beam in azimuth. In addition the sky will drift approximately $17.5'$ at the worst point per rotation giving 2 samples per beam in elevation. This allows for maximal sky coverage with minimal missing sky area between rotations. Ultimately the rotation rate of B-Machine is held hostage by the upper limit of the data acquisition system.

\subsection{Thermopile}
B-Machine was also equipped with a Thermopile device\footnote{Dexter research, Inc. M5 Thermopile Detector}, which produces a voltage proportional to the difference in temperature of its housing vs. observed temperature. The thermopile has a large field of view and consequently a long Aluminum tube was placed on the end to narrow its field of view from $78^{\circ} \mathrm{~to~} \sim20^{\circ}$. Final analysis of this data will not be addressed in any detail here and will be combined with a data set that is being taken at WMRS with additional Infra-Red sensors.

\chapter{Data Analysis} \label{chap:analysis}
The ambition of any good experiment is to transform its data into results that agree with the initial investigative expectations. For B-Machine, as per all other CMB experiments, the goal is to produce angular power spectra, sky maps and cosmological parameters. By selecting the best data and using modern computing skills and tools that are currently available to the CMB community, B-Machine's data has been changed into just that.

\section{Interactive Data Language} \label{subsec:IDL}
The majority of the data analysis and reduction is done using the Interactive Data Language (IDL)\footnote{ITT Visual Information Solutions www.ittvis.com}.  IDL has been a key developmental tool for CMB analysis and has thousands of routines and functions written by various experts in the field of astronomy, CMB research, and astrophysics. In addition all maps that are generated in IDL use the HEALpix scheme~\citep{gorski99}.  At UCSB all significant data analysis uses codes developed throughout the CMB community and also written locally by Peter Meinhold, Rodrigo Leonardi, John Staren, Mike Seifert, and Brian Williams.

\section{Data Selection}
One hurdle to overcome with large data sets is the selection of data with the appropriate noise levels and signal levels. B-Machine has a set of automated calibrations and was meant to be rotating in azimuth once every $\sim70$ s. If these parameters are not being met then the data needs to be stamped as not in standard operating mode. Further, each channel needs to be checked for weather, DC noise level, and DC levels just to name a few. The manner in which this is done was developed by Rodrigo Leonardi when he was processing the WMPOL (\citep{levy08}) data. It is referred to as the compact data set or just CDS for short. The CDS pipeline is a set of IDL routines that divide the data into $\sim70$ s sections and generate statistics for each section; a description of each binned variable is described in Table~\ref{tab:cdsdatafields}. Only channels 1, 2, 3, and 6 were used for statistics on the AC channels and 9, 10, 11, and 14 for the DC channels.

\begin{table}[p]
\begin{center}
\caption[Description of CDS Variables]{Description of CDS Variables \label{tab:cdsdatafields}}
\begin{tabular}{|l|l|}
    \hline
    Variable name & Description \\
    \hline
    \hline
    AVERAGEAC & Average value for AC channels.\\
    \hline
    AVERAGEDC &  Average value for DC channels for I.\\
    \hline
    AVERAGEIR & Average Value of the IR sensor.\\
    \hline
    CENTERCHANTEMP & Temperature of the central horn.\\
    \hline
    COLDHEADTEMP & Temperature of the cold head.\\
    \hline
    FILE & File that 70 s segment came from.\\
    \hline
    FRAMETEMP & Temperature of frame sensor 10 K/V.\\
    \hline
    GAIN & RF gain of each channel in K/V.\\
    \hline
    JULIANDAY & Julian day of first sample in data segment.\\
    \hline
    MAXDC & Maximum DC value for DC channels.\\
    \hline
    MEANAZ & Average azimuth reading for data Segment.\\
    \hline
    MEANEL & Average elevation reading for data Segment.\\
    \hline
    MINDC & Minimum DC value for DC channels.\\
    \hline
    NOBS & Number of samples for 70 s data segment.\\
    \hline
    PEAK2PEAK & Difference of min/max values for AC channels.\\
    \hline
    POLCALPOSITION & Averaged voltage of Polarization Calibrator.\\
    \hline
    PRIMARYTEMP & Temperature of the primary mirror 10 K/V.\\
    \hline
    SIGMAAC & Standard deviation of AC channels.\\
    \hline
    SIGMAAZ & Standard deviation of azimuth.\\
    \hline
    SIGMACENTCHANTEMP & Standard deviation of temperature for\\
                        & the central cryogenic amplifier.\\
    \hline
    SIGMACOLDHEADTEMP & Standard deviation of cryo head temperature.\\
    \hline
    SIGMAEL & Standard deviation of elevation.\\
    \hline
    SIGMAFRAMETEMP & Standard deviation of frame temperature.\\
    \hline
    SIGMAPRIMARYTEMP & Standard deviation of primary mirror\\
                     & temperature.\\
    \hline
    SIGMATILTTEMP & Standard deviation of clinometer temperature.\\
    \hline
    SIGMATILTX & Standard deviation of tilt in the X direction.\\
    \hline
    SIGMATILTY & Standard deviation of tilt in the Y direction.\\
    \hline
    STATUS & Average value of status word.\\
    \hline
    TILTTEMP & Temperature of clinometer $100^{\circ}\mathrm{C}/\mathrm{V}$.\\
    \hline
    TILTX & Average X tilt $0.407792 ^{\circ} /\mathrm{V}$ at $15.8^{\circ}\mathrm{C}$.\\
    \hline
    TILTY & Average Y tilt $0.408163 ^{\circ} /\mathrm{V}$ at$15.6^{\circ}\mathrm{C}$.\\
    \hline
    TIME & The time of the first and last data points.\\
    \hline
    WHITENOISEAC & Average white noise value for AC channels.\\
    \hline
\end{tabular}
\end{center}
\end{table}

\begin{figure}[p]
\begin{tabular}{cc}
\includegraphics[width=7cm,height=6cm]{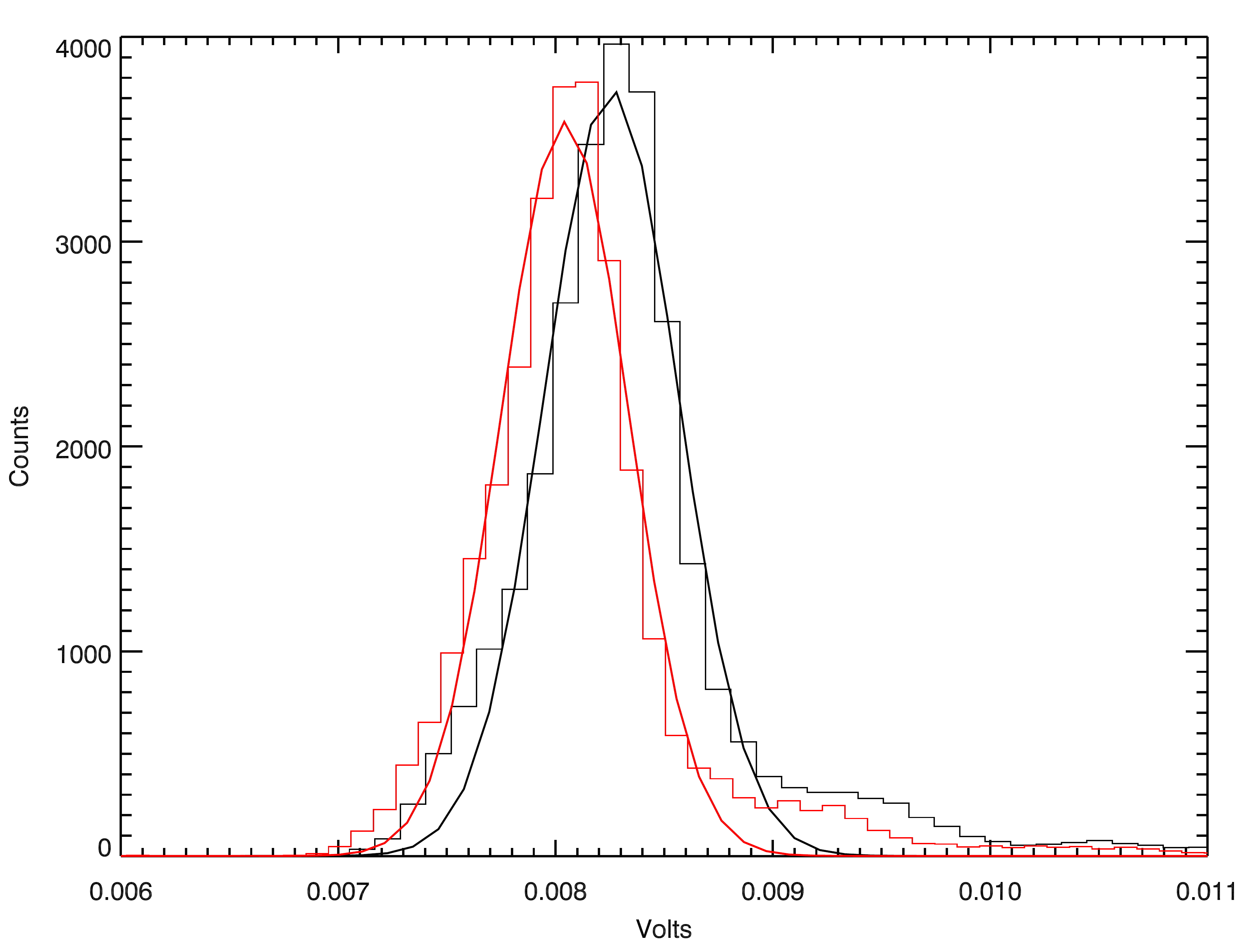}&
\includegraphics[width=7cm,height=6cm]{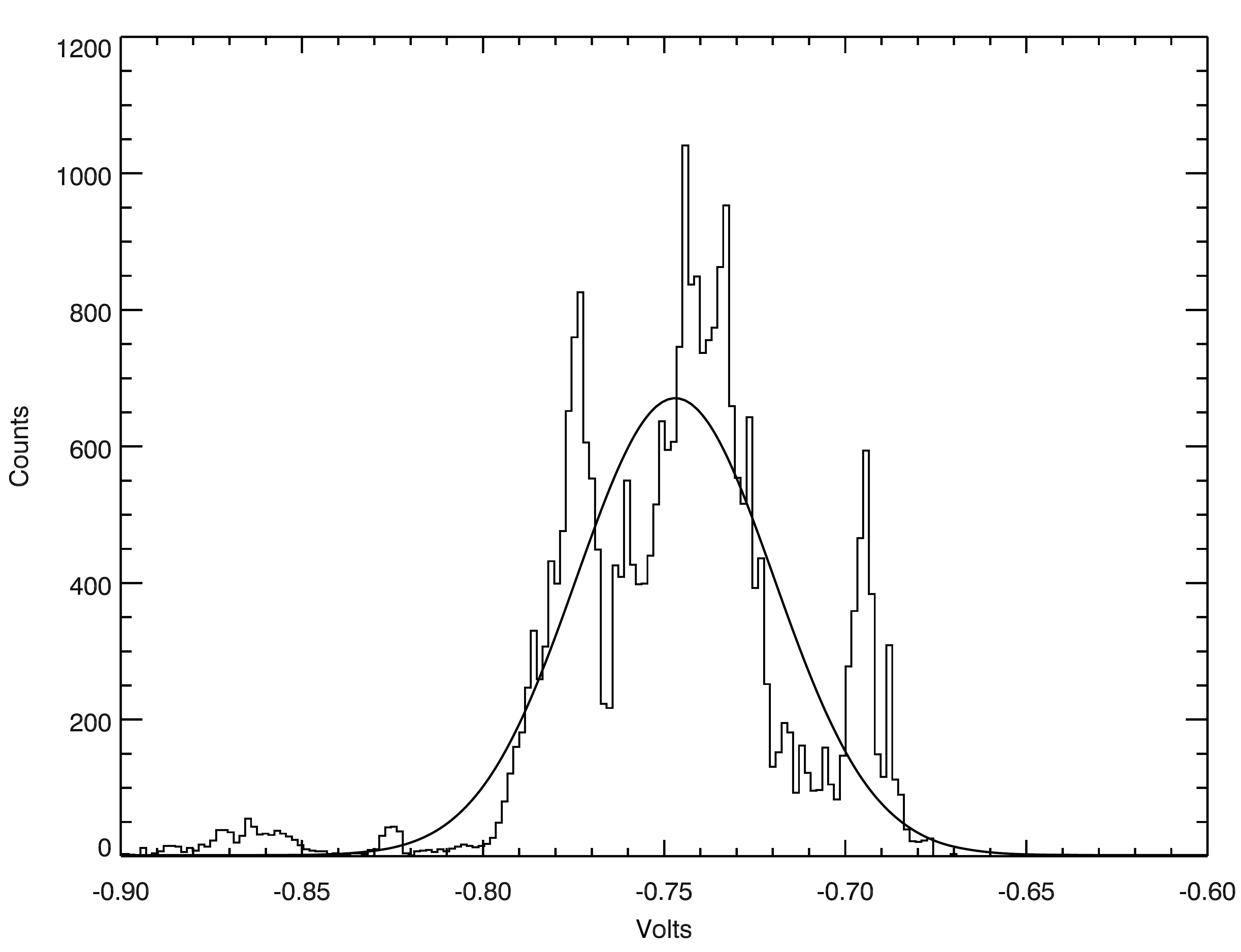}\\
\includegraphics[width=7cm,height=6cm]{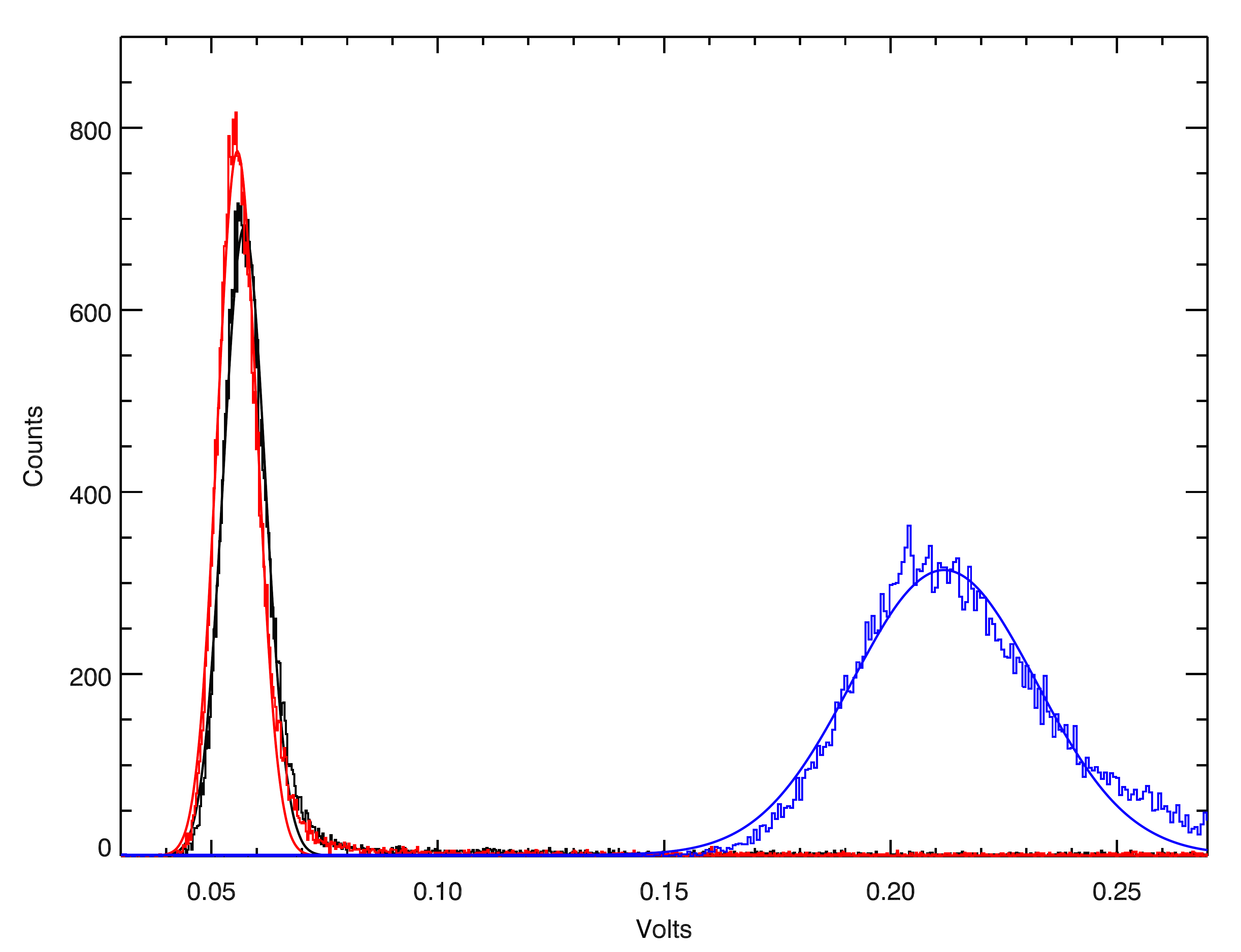}&
\includegraphics[width=7cm,height=6cm]{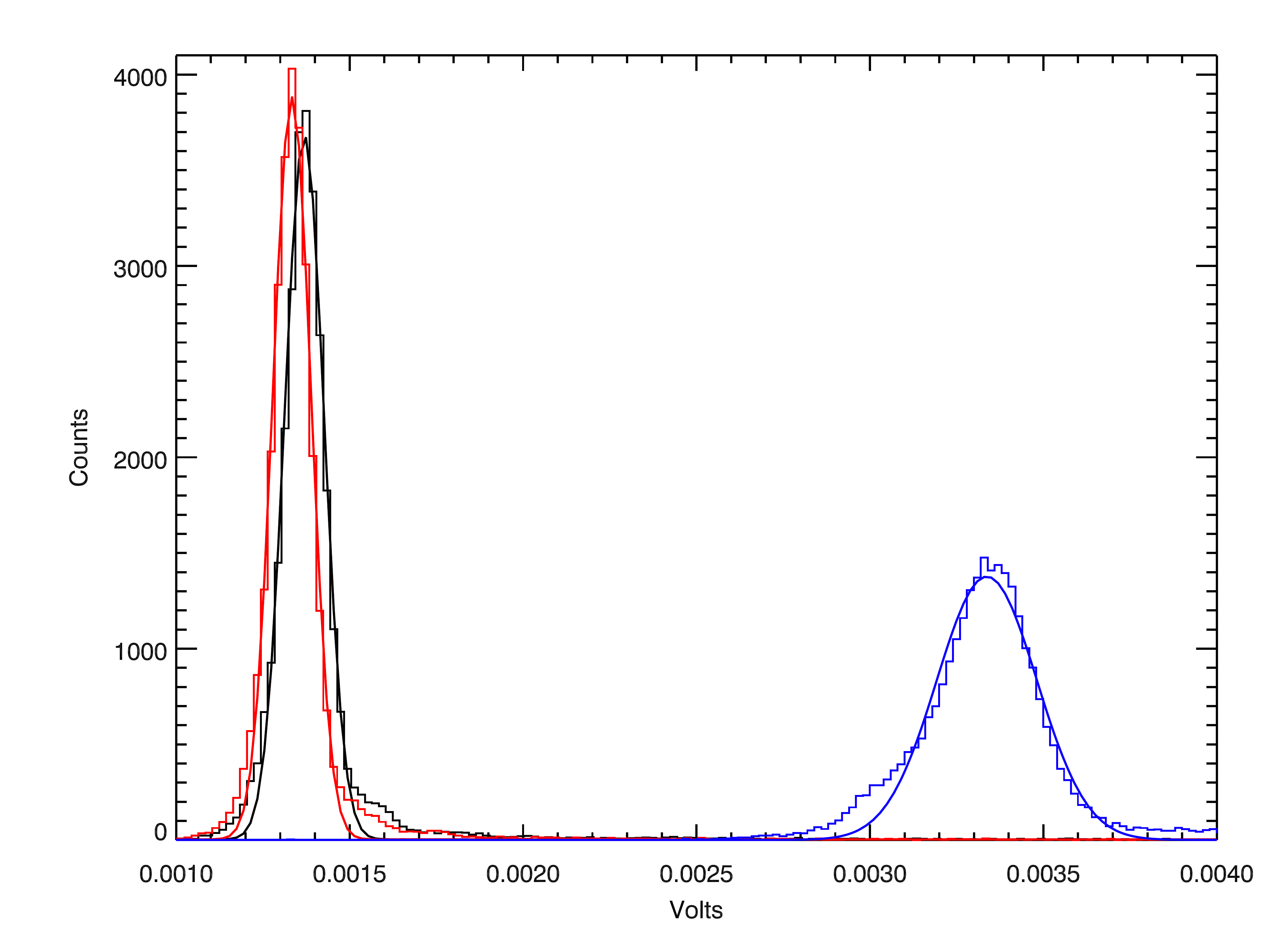}\\
\end{tabular}
\caption[Example data cut histograms with gaussian fits included]{Example data cut histograms with gaussian fits, all data from the central channel 6/14. Top left, is the standard deviation of Q(black) and U(red). Top right, is the average DC value, the odd shape of this curve represents changing sky temperatures over the campaign. Bottom left, is the peak to peak values of T (blue), Q (black), and U (red). Bottom right, is the white noise level of T (blue) , Q (black) and U (red). \label{fig:datacuthistograms}}
\end{figure}

The CDS was generated for all data taken at WMRS, including all calibration runs, Jupiter and Tau A scans as well as all specialized sky scanning strategies (NCP scans and stationary data taking). The final data cutting parameters can be seen in Table~\ref{tab:cdsdataselection}. The basic technique for data cutting is straightforward. A histogram for all data is generated for a set of parameters, some sample histograms can be seen in Figure~\ref{fig:datacuthistograms}. The mean value of the gaussian fit is the cut parameter and the standard deviation of the fit is the unit of measure for cutting. A very conservative cut ($2\sigma$)was used to generate a batch of data labeled level 2 data. The list of parameters that yields the best cutting information with the fewest fields is in Table~\ref{tab:cdsdataselection}. These data were uploaded to the Nersc site where the analysis was performed.

\begin{table}[p]
\begin{center}
\caption[CDS Data Selection Parameters]{CDS Data Selection Parameters \label{tab:cdsdataselection}}
\begin{tabular}{|l|l|}
  \hline
  Variable name & Description \\
  \hline
  \hline
  $\mathrm{SigmaAC}\_\mathrm{chan}\#$ & Standard Deviation of AC channels \\
  \hline
  MinDC & Minimum DC value\\
  \hline
  MaxDC & Maximum DC value\\
  \hline
  Peak2Peak & Peak to Peak value of AC channels \\
  \hline
  Nobs & Number of samples in a given bin \\
  \hline
  MeanEl & Mean elevation \\
  \hline
  Status & Average value of status word for nominal operation\\
  \hline
  $\mathrm{WhiteNoiseAC}\_\mathrm{chan}\#$ & White noise level for AC channels\\
  \hline
\end{tabular}
\end{center}
\end{table}

\section{Pointing Reconstruction} \label{sec:pointreconstruct}
Currently the largest problem presented in reconstructing maps from the data collected is accuracy of the pointing reconstruction. The majority of the pointing corrections were found using Moon or Tau A crossings, see Chapter~\ref{chap:instrument} Section~\ref{sec:pointing}. A list of days with Moon crossings is in Table~\ref{tab:mooncrossings}, which also includes the angular diameter and the phase of the Moon in a percentage of full.  The phase of the Moon was used during beam characterization to account for the thermal profile of the moon depending on its illumination, see Chapter~\ref{chap:telescopechar} Section~\ref{sec:beamcharacterization}.

\begin{table}[p]
\begin{center}
\caption[List of Moon Crossings]{List of Moon Crossings \label{tab:mooncrossings}}
\begin{tabular}{|c|c|c|c|c|}
  \hline
  Day & Time of      & Time of          &  Ang Diameter & Percentage \\
      & Crossing (UT) &  Crossing (UT) & (arcseconds)  & of Full\\
  \hline
  \hline
  $8/17/2008$ & 7:56-8:45 &  & 1870 & $97\%$\\
  \hline
  $8/18/2008$ & 7:25-8:35 &  & 1890 & $96\%$\\
  \hline
  $8/19/2008$ & 7:20-8:30 & 11:15-12:10 & 1906 & $91\%$\\
  \hline
  $8/20/2008$ & 7:40-8:20 & 12:40-13:25 & 1920 & $83\%$\\
  \hline
  $8/21/2008$ & 8:00-8:50 & 13:55-14:40 & 1930 & $74\%$\\
  \hline
   $8/22/2008$ & 8:40-9:40 & 15:15-15:50 & 1950 & $64\%$\\
  \hline
   $8/23/2008$ & 9:20-9:55 &  & 1950 & $50\%$\\
  \hline
   $8/26/2008$ & 12:15-12:55 &  & 1960 & $20\%$\\
  \hline
   $8/27/2008$ & 13:25-14:10 & & 1960 & $10\%$\\
  \hline
   $8/28/2008$ & 14:35-15:20 & & 1955 & $5\%$\\
  \hline
   $8/28/2008$ & 15:50-16:15 &  & 1940 & $1\%$\\
  \hline
   $9/14/2008$ & 5:35-8:15 &  & 1890 & $100\%$\\
  \hline
   $9/15/2008$ & 5:35-6:45 & 8:50-9:55 & 1915 & $100\%$\\
  \hline
  $9/16/2008$ & 5:50-6:40 & 10:35-11:10 & 1930 & $97\%$\\
  \hline
  $9/18/2008$ & 6:40-7:20 & 13:00-13:35 & 1960 & $86\%$\\
  \hline
  $9/20/2008$ & 8:10-8:50 & 15:20-16:00 & 1960 & $66\%$\\
  \hline
  $9/21/2008$ & 9:10-9:45 & 16:20-16:45 & 1960 & $58\%$\\
  \hline
  $9/24/2008$ & 12:25-13:10 &  & 1940 & $22\%$\\
  \hline
  $9/26/2008$ & 14:45-15:25 &  & 1910 & $7\%$\\
  \hline
  $10/14/2008$ & 9:25-10:10 &  & 1940 & $100\%$\\
  \hline
\end{tabular}
\end{center}
\end{table}

\begin{figure}[p]
\includegraphics[width=13.5cm]{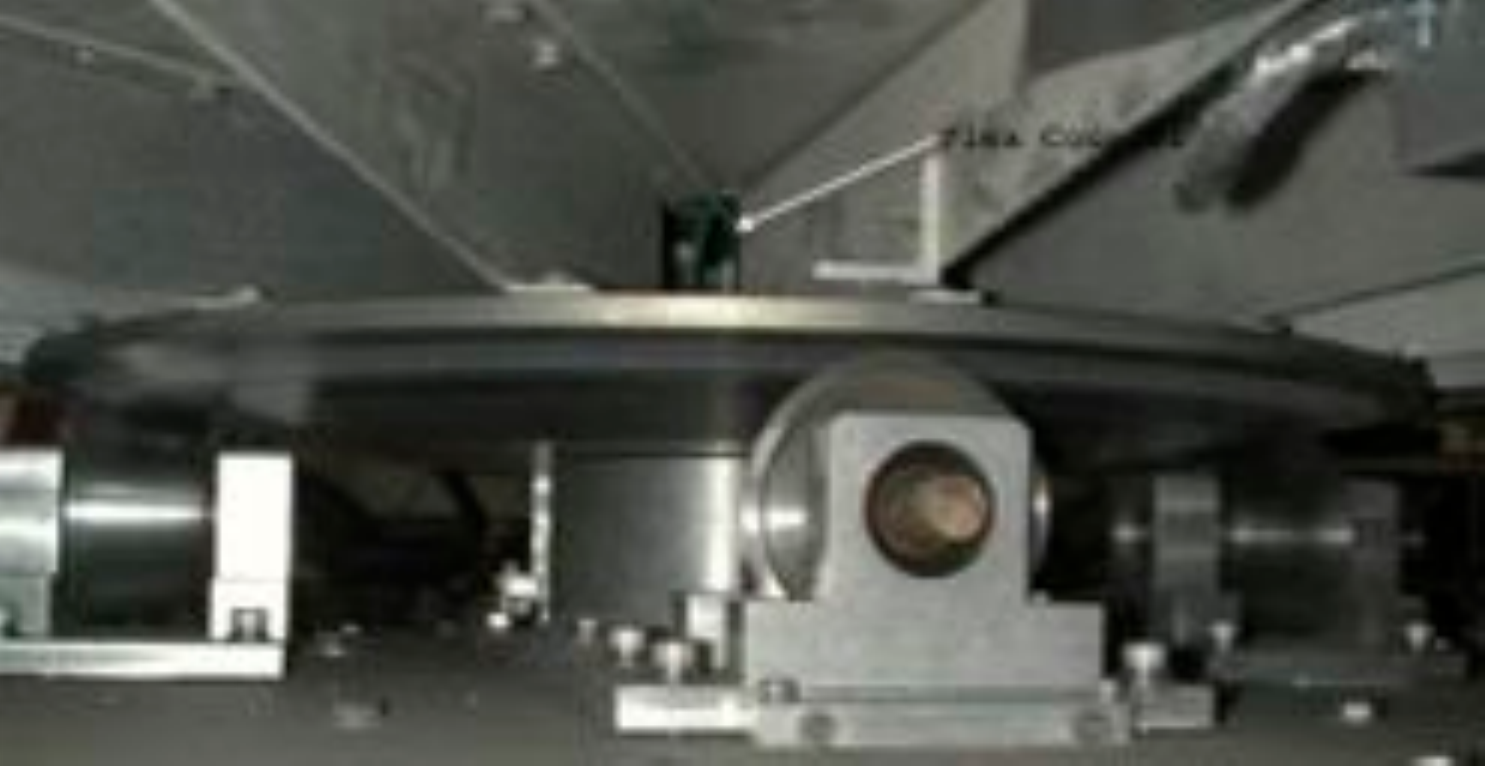}
\caption[Flex coupler for B-Machines azimuth pointing]{Image of the underside of B-Machines rotating table. Three of the support drive cones are visible in the image. Also barely distinguishable is the Flex Coupler (black) between 2 of the support ribs and criss-crossed by green wires. \label{fig:flexcoupler}}
\end{figure}

It now appears that a loose Helical Beam Shaft Coupler (Flex Coupler, see Figure~\ref{fig:flexcoupler}) is the culprit of the pointing problems. The Flex Coupler is used to minimize the torque on the Gurley absolute encoder from shaft misalignment. If the stationary shaft that is secured to the underside of B-Machine is slightly misaligned with the absolute encoder that is affixed to the rotating table, small torque variations will cause unusual wearing in the encoder bearings. This can lead to inaccuracies in the position readout that are gradual and very difficult to characterize. A Flex Coupler is placed between the 2 shafts to maintain precise shaft alignment. After thermally cycling for several weeks of observing and testing the Flex Coupler gradually loosened to a point where it would slip slightly ($\sim1-2^{\circ}$) when starting rotation or changing direction. For nominal observing with Moon crossings the data can be salvaged by reconstructing the centroid of the Moon, using this positional offset to adjust all the data for that day. Days with no Moon crossings are more difficult to reconstruct and further efforts to correct were needed. Each of the days that didn't contain Moon crossings were evaluated for Tau A crossings. Of the 15 days with no Moon crossings 9 of these days were salvaged using Tau A. Comparing the Tau A and Moon crossings gave consistent results to within a beam size for 2 days that contained both crossings.

\begin{figure}[p]
\begin{tabular}{cc}
\includegraphics[width=7cm,height=6cm]{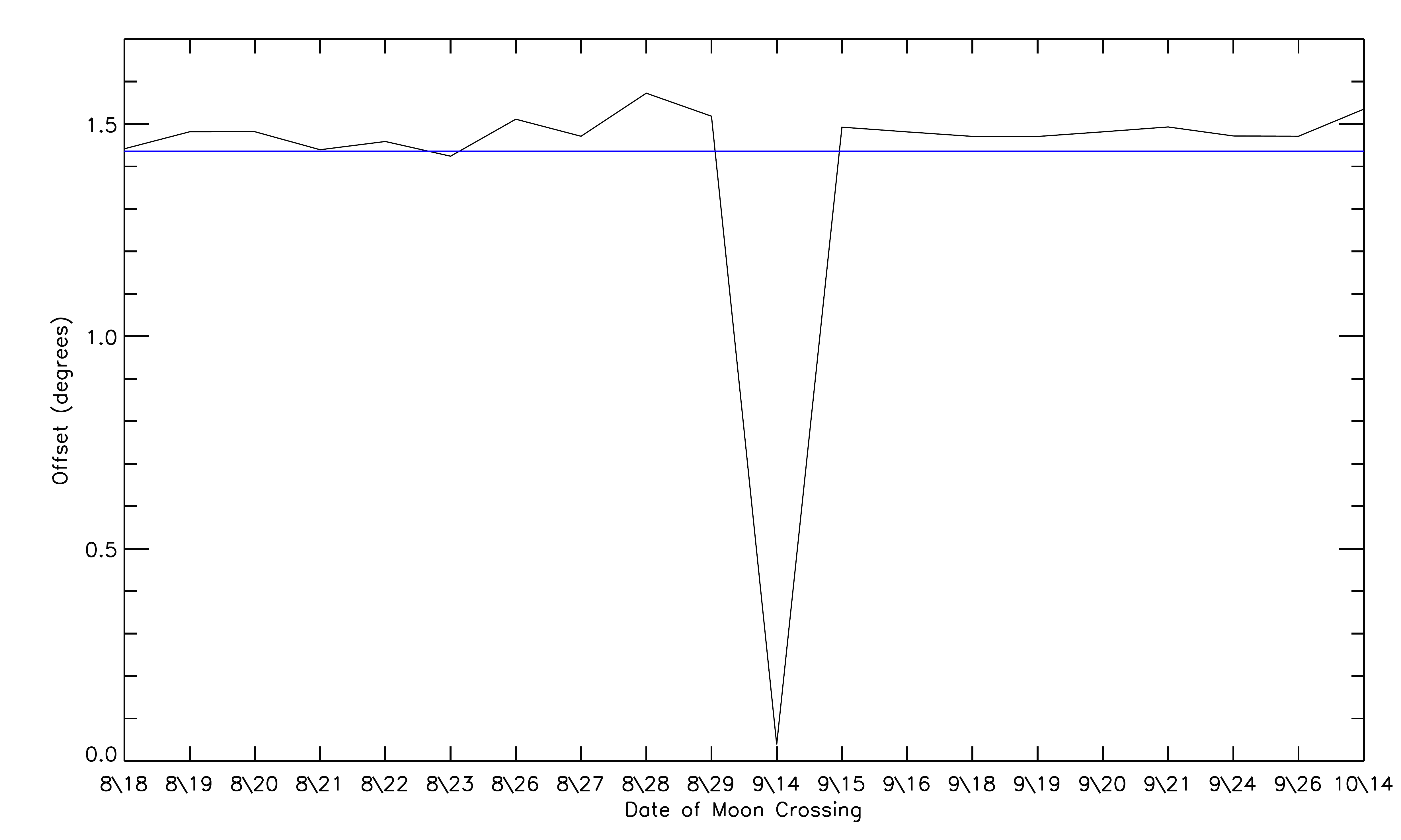}&
\includegraphics[width=7cm,height=6cm]{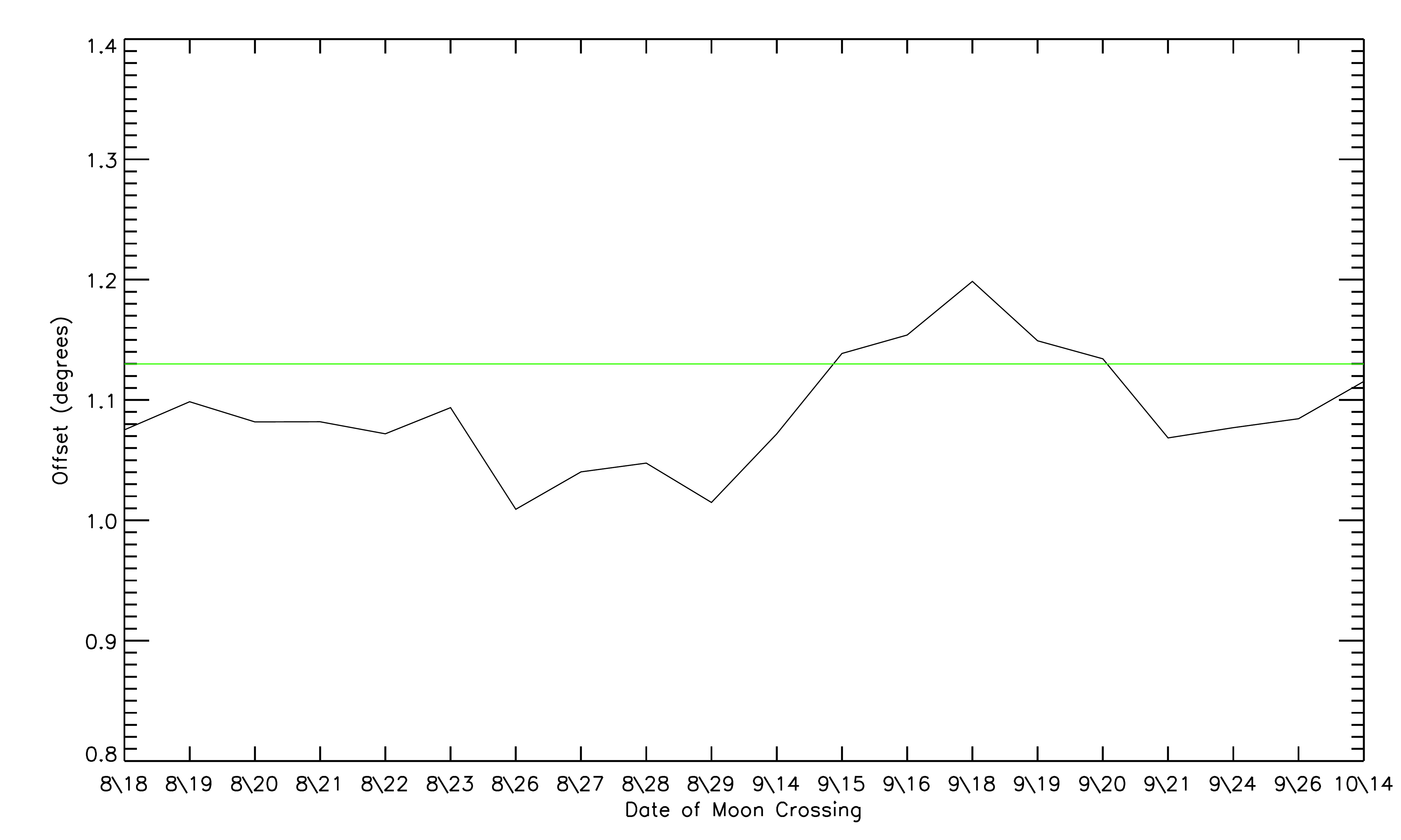}\\
\includegraphics[width=7cm,height=6cm]{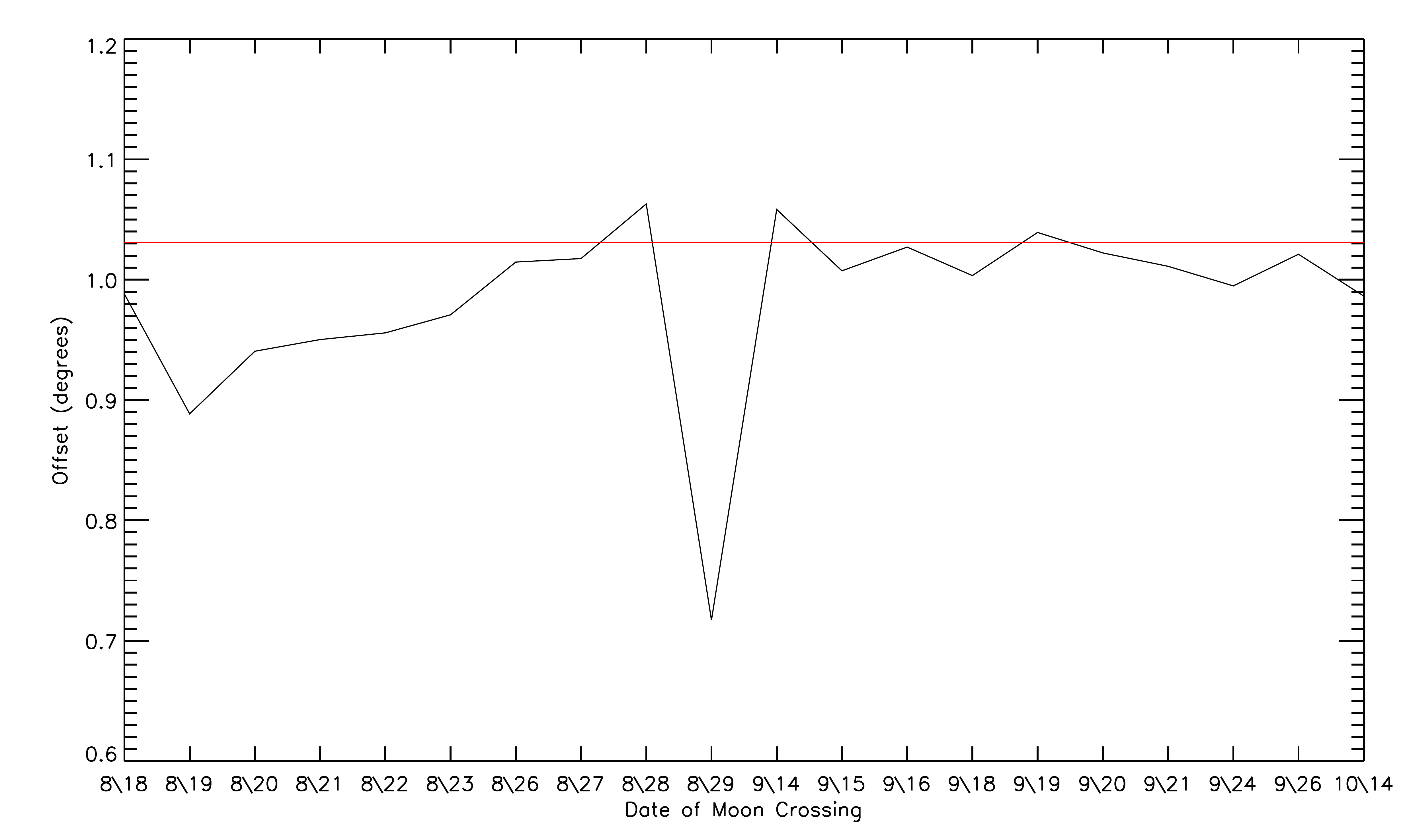}&
\includegraphics[width=7cm,height=6cm]{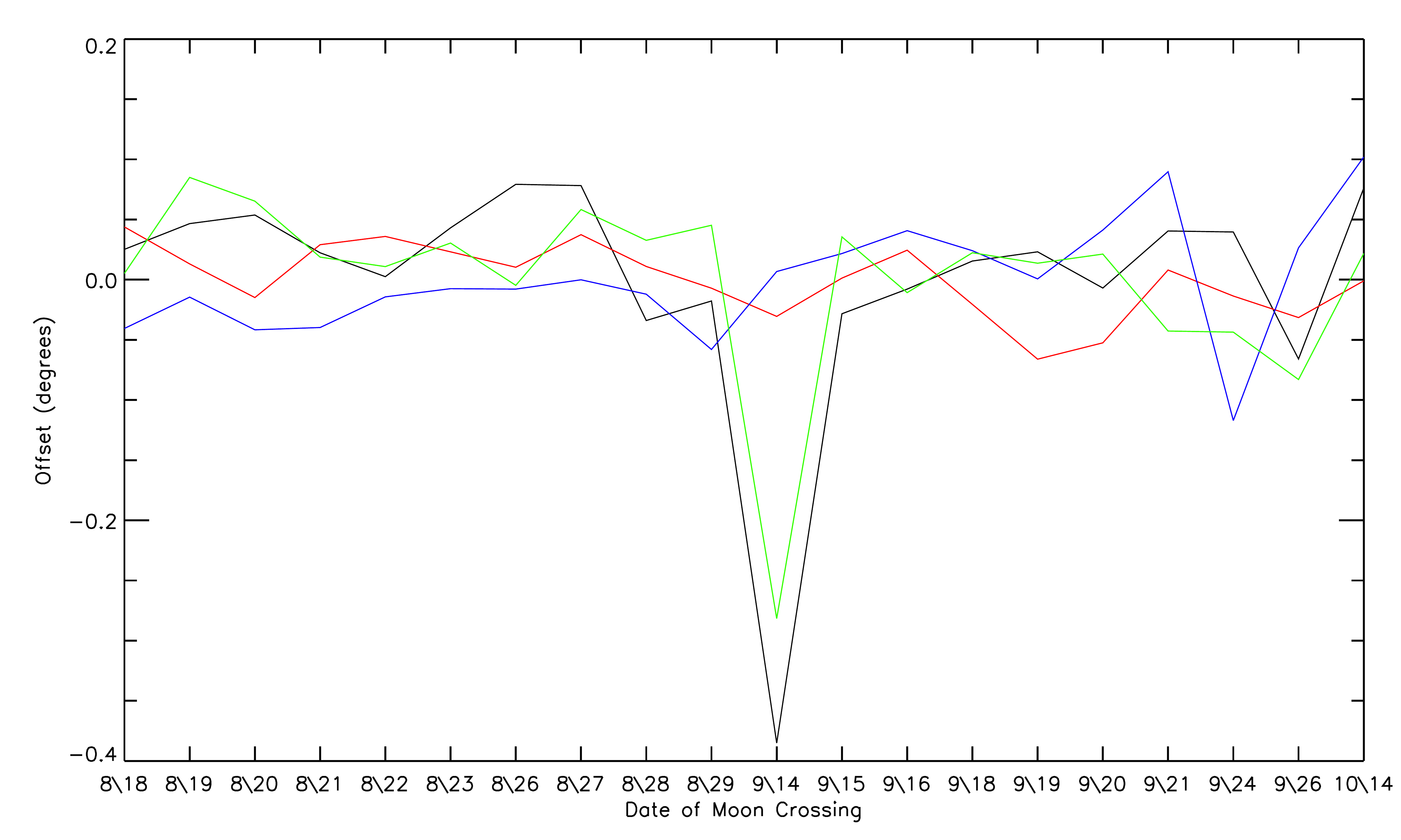}\\
\end{tabular}
\caption[Pointing offsets for each channel compared to central channel]{Pointing offsets for each channel compared to the central channel. Top left, is channel 1, Top right is channel 2, Bottom left channel 3, and bottom right is the elevation comparison of all channels. The straight lines in the first 3 plots are the expected offset from horns position compared to central horn. For the elevation comparison the offset was removed. The large spike for channel 1 is a problem in the Moon comparison code due to some small clouds in the data and is not a real artifact of the pointing. \label{fig:moonoffsetcompare}}
\end{figure}

Pointing comparisons, see Figure~\ref{fig:moonoffsetcompare}, of days with Moon crossings show the pointing reconstruction to be reliable. Also, in Figure~\ref{fig:moonoffsetcompare} is a comparison of the elevation pointing which is consistent as well. The corrections for the elevation from the Moon pointing data were sub-beam size and reinforces the notion that the only problem with the pointing data is the Flex Coupler and nothing more complicated.

\section{Point Sources} \label{sec:pointsources}
An effort to observe point sources to confirm calibrations, pointing and facilitate comparisons to other experiments was done several times during the observing campaign. Table~\ref{tab:radiosources} gives a list of possible sources; all the sources for this table use Jupiter as an absolute standard to calibrate all of the other sources and bands. All of the reference information is outside of our band-pass making it necessary to extrapolate up to our frequency of 41.5 GHz using the spectral index quoted in \citet{page07} and
\begin{center}
\begin{equation}\label{eqn:specindex}
   S\propto \nu^{\beta},
\end{equation}
\end{center}
where $\beta$ is the spectral index and S is the flux. In addition to temperature calibrations, polarization calibrations are also necessary, but a standard candle in polarization presents fewer opportunities to observe and less accurate data for cross calibrating experiments. Only 2 of the sources in Table ~\ref{tab:radiosources} were observable and sufficiently bright to hope to observe on a single drift scan. Tau A, known more commonly as the Crab Nebula, and Jupiter were pursued. Due to the pointing difficulties only Tau A was observed. It turns out that all of the drift scans for Jupiter were slightly off on azimuth and thus no observations were made.

\begin{table}[p]
\begin{center}
\caption[Radio Source Data]{Radio Source Data \citep{Hafez08,page07}}
\begin{tabular}{|c|c|c|c|c|c|}
  \hline
  Radio  &   RA    &   DEC & Flux Density & Spectral  & Freq \\
       Source         &        &       &     (Jy)     &            Index    & (GHz)\\
  \hline
  \hline
  Cas A      & $23^h23^m26^s$ &  $58^{\circ}48'$    & 182.0 & $-0.69$  & 33\\
  \hline
  Cyg A      & $19^h59^m28^s$ & $40^{\circ}44'2''$   & 36.4  & $-1.21$  & 33\\
  \hline
  Tau A (M1) & $ 5^h35^m 4^s$ & $22^{\circ}01'1''$  & 299.2 & $-0.23$  & 40.4\\
  \hline
  NGC7027    & $21^h 7^m 2^s$ & $42^{\circ}14'1''$  & 5.39  & $-0.119$ & 33\\
  \hline
  Hydra A    & $ 9^h18^m 6^s$ & $-12^{\circ}65'4''$ & 0.127 & $0.19$   & 15\\
  \hline
  Jupiter    &                &                     & 146.6 & $0.248$  & 33\\
  \hline
  Venus      &                &                     & 460.3 & $-0.278$ & 33\\
  \hline
  Saturn     &                &                     & 140.5 & $0.00$   & 33 \\
  \hline
\end{tabular}
\end{center}
\end{table}

\begin{table}[p]
\begin{center}
\caption[Radio Source Brightness and Flux Extrapolated to 41.5 GHz]{Radio Source Brightness and Flux Extrapolated to 41.5 GHz \citep{Hafez08} \label{tab:radiosources}}
\begin{tabular}{|c|c|c|c|c|c|}
  \hline
  Radio  & Planet & Flux Density & T Brightness & Effective Ta & Angular\\
    Source &   & (Jy) & (K) & (mK) & Size ($''$)\\
  \hline
  \hline
    & Jupiter &  & 157 & 172 & 42-47\\
  \hline
   & Venus &  & 431.61 & 86.6 & 17\\
  \hline
   & Saturn &  & 140.50 & 71.4 & 12\\
  \hline
  Cas A & & 155.1 &  & 82.5 & 5 \\
  \hline
  Cyg A & & 27.58 &  & 14.67 & 0.03\\
  \hline
  Tau A & & 281.3 & & 112.7 & 6$'$x4$'$\\
  \hline
  NGC7023 & & 5.25 &  &  2.793 & 18\\
  \hline
  Hydra A & & 0.154 &  & 0.0819 & 0.0001 \\
  \hline
\end{tabular}
\end{center}
\end{table}

\subsection{Converting Flux Units to Temperature Units}\label{subsect:flux2temp}
Point source measurements are typically quoted in the radio astronomy flux unit Jansky (Jy). This unit is the total flux incident on a telescope for a given source, and conversion of this flux unit to antenna temperature is instrument dependent and requires knowledge of the telescope for conversion.  The flux from a radio telescope can be converted to antenna temperature by,

\begin{center}
\begin{equation}\label{eqn:Flux}
   Flux(\frac{W}{m^2\cdot Hz})=\frac{k\times Ta}{Ae},
\end{equation}
\end{center}
where Ae is the effective receiving area, Ta is the antenna temperature, and \textit{k} is Boltzmann's constant. We use

\begin{center}
\begin{equation}\label{eqn:effectivearea}
   \lambda^2=Solid Angle \times Ae,
\end{equation}
\end{center}
to get the effective receiving area, where $\lambda$ is the center frequency of the band-pass and the solid angle is the integrated gaussian over the sphere for each horn. Using B-Machine's FWHM of $22.2'\pm0.2'$ for the central horn and $24.0'\pm0.2'$ for the off axis horns (see Section~\ref{sec:beamcharacterization}), we obtain for a $1.0$ mK Ta,

\begin{center}
\begin{equation}\label{eqn:conversionconstant}
  Flux(\frac{\mathrm{W}}{\mathrm{m}^2\cdot \mathrm{Hz}})=\frac{1.38\cdot10^{-23}\times 1\cdot 10^{-3}}{1.106 \mathrm{m}^2}=1.248 \cdot 10^{-27}\frac{\mathrm{W}}{\mathrm{m}^2 \cdot \mathrm{Hz}}=1.248 \frac{\mathrm{Jy}}{\mathrm{mK}},
\end{equation}
\end{center} 
where $1 \mathrm{ Jy} = 10^{-26}~\frac{\mathrm{W}}{\mathrm{m}^2 \cdot \mathrm{Hz}}$. However, standard nomenclature for radio source flux is to quote total flux, which includes both polarizations, while B-Machine measures only one polarization and calibrates with a blackbody. The conversion is doubled, to get $2.496~\frac{\mathrm{Jy}}{\mathrm{mK}}$ for an unpolarized source from the central horn, see Table~\ref{tab:jytomk} for the off axis horns. As a check to test the reliability of the calculation WMAP data was used.  The WMAP~\citep{page07} observations of Tau A show a $\sim72$ mK signal and using the published WMAP instrument characteristics  and Equations~\ref{eqn:Flux} and~\ref{eqn:effectivearea} gives $\sim74$ mK. 

\begin{table}[p]
\begin{center}
\caption[Jansky to Kelvin Conversion Constants]{Conversion Constants for Jansky's to Kelvin \label{tab:jytomk}}
\begin{tabular}{|c|c|c|}
  \hline
  Constant& $22.2'$ & $24.0'$ \\
  \hline
  \hline
   Solid Angle & $4.725\cdot10^{-5}$ sr     & $5.523\cdot10^{-5}$ sr \\
  \hline
   Ae          & $1.106~\mathrm{m}^{2}$              & $0.946~\mathrm{m}^{2}$ \\
  \hline
   Ta Radio    & $1.25\pm.03~\frac{\mathrm{Jy}}{\mathrm{mK}}$ & $1.450\pm.035~\frac{\mathrm{Jy}}{\mathrm{mK}}$ \\
  \hline
   Ta          & $2.50\pm.06~\frac{\mathrm{Jy}}{\mathrm{mK}}$ & $2.90\pm.07~\frac{\mathrm{Jy}}{\mathrm{mK}}$ \\
  \hline
\end{tabular}
\end{center}
\end{table}

\subsection{Tau A}
Tau A was used for pointing reconstruction, calibration, and functionality tests. For pointing reconstruction, Tau A was located in each of the daily temperature maps and the day's pointing was adjusted so that Tau A appeared in the appropriate pixels, see Section~\ref{sec:maps}. For calibration and functionality testing, drift scans were used due to the raster scan limitation. Tau A is sufficiently bright that it can be seen over the $1/f$ noise in the temperature channels with minimal data analysis. There were 5 drift scans at different elevations (5 sections) with each of the scans having 7-10 crossings each. Each crossing that had a threshold voltage above a certain level (different for each channel) was used for pointing calibration. The azimuth and elevation for each point was compared to the expected azimuth and elevation of Tau A and the average difference for all crossings of one drift scan is considered the offset for that scan, see Table~\ref{tab:tauaoffsets}. The azimuth and elevation for Tau A was found by using the time stamp in the level 1 data and the known right ascension and declination to convert to azimuth and elevation.

\begin{table}[p]
\begin{center}
\caption[Tau A Offsets for Each of the 5 Drift Scans.]{Tau A Offsets for Each of the 5 Fixed Elevation Drift Scans \label{tab:tauaoffsets}}
\begin{tabular}{|c|c|c|c|c|c|c|}
  \hline
      & \multicolumn{2}{|c|}{Expected}&\multicolumn{2}{|c|}{Actual}&\multicolumn{2}{|c|}{Offset}\\
  \hline
    Pass  &  Azimuth  & Elevation &  Azimuth & Elevation & Co-El &  Azimuth \\
               & (degrees) & (degrees) & (degrees)& (degrees) &   (degrees)  & (degrees)\\
  \hline
  \hline
   1 & 79.333 & 23.165 & 81.810 & 23.003 & -2.277 & 0.168 \\
  \hline
   2 & 84.323 & 30.234 & 86.770 & 30.001 & -2.114 & 0.233 \\
  \hline
   3 & 93.451 & 42.259 & 95.866 & 42.007 & -1.787 & 0.252 \\
  \hline
   4 & 100.78 & 50.328 & 103.06 & 49.995 & -1.454 & 0.333 \\
  \hline
   5 & 110.31 & 58.347 & 112.48 & 58.008 & -1.136 &  0.340\\
  \hline
\end{tabular}
\end{center}
\end{table}

\begin{figure}[p]
\begin{tabular}{cc}
\includegraphics[width=7cm,height=6cm]{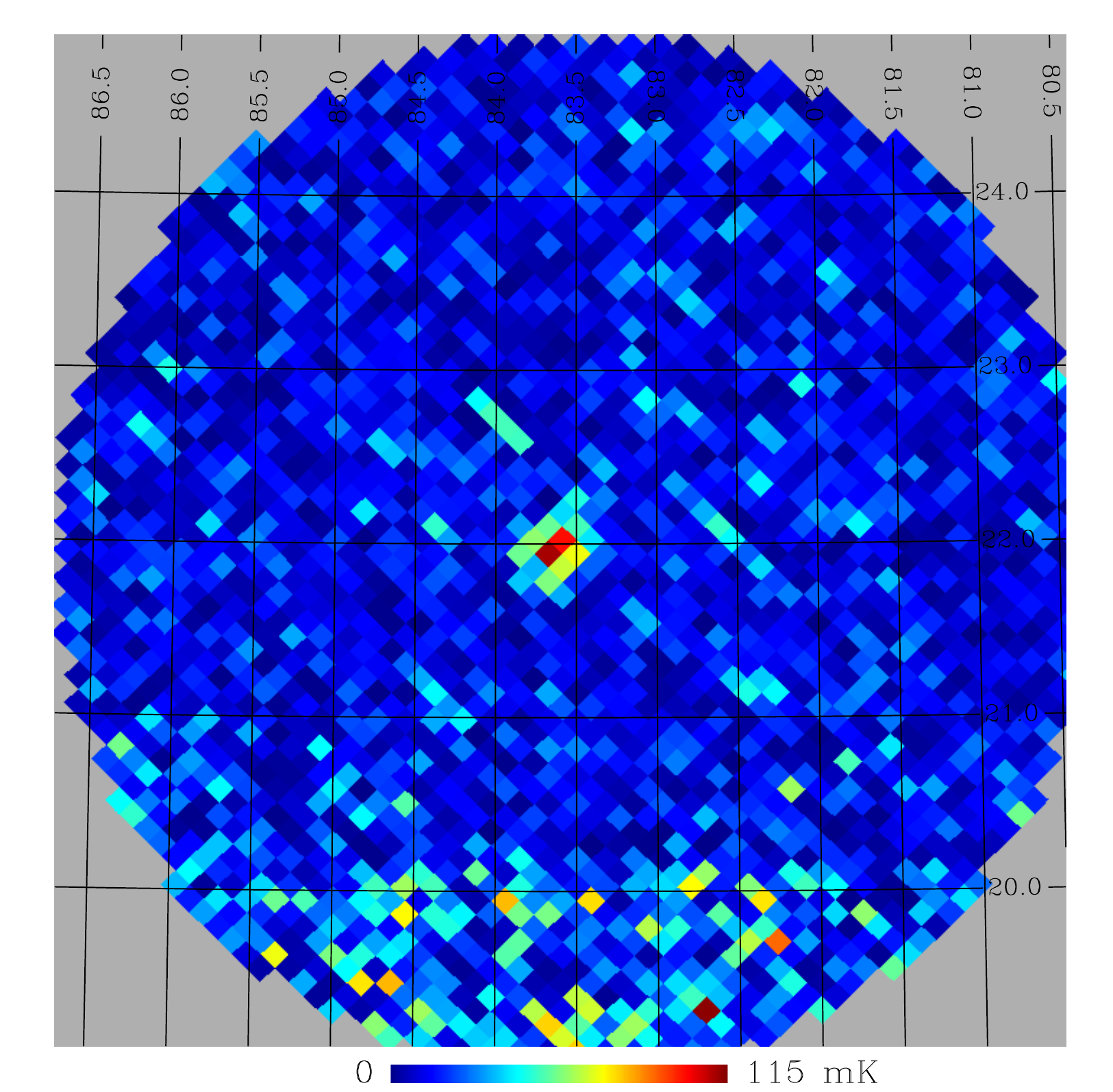}&
\includegraphics[width=7cm,height=6cm]{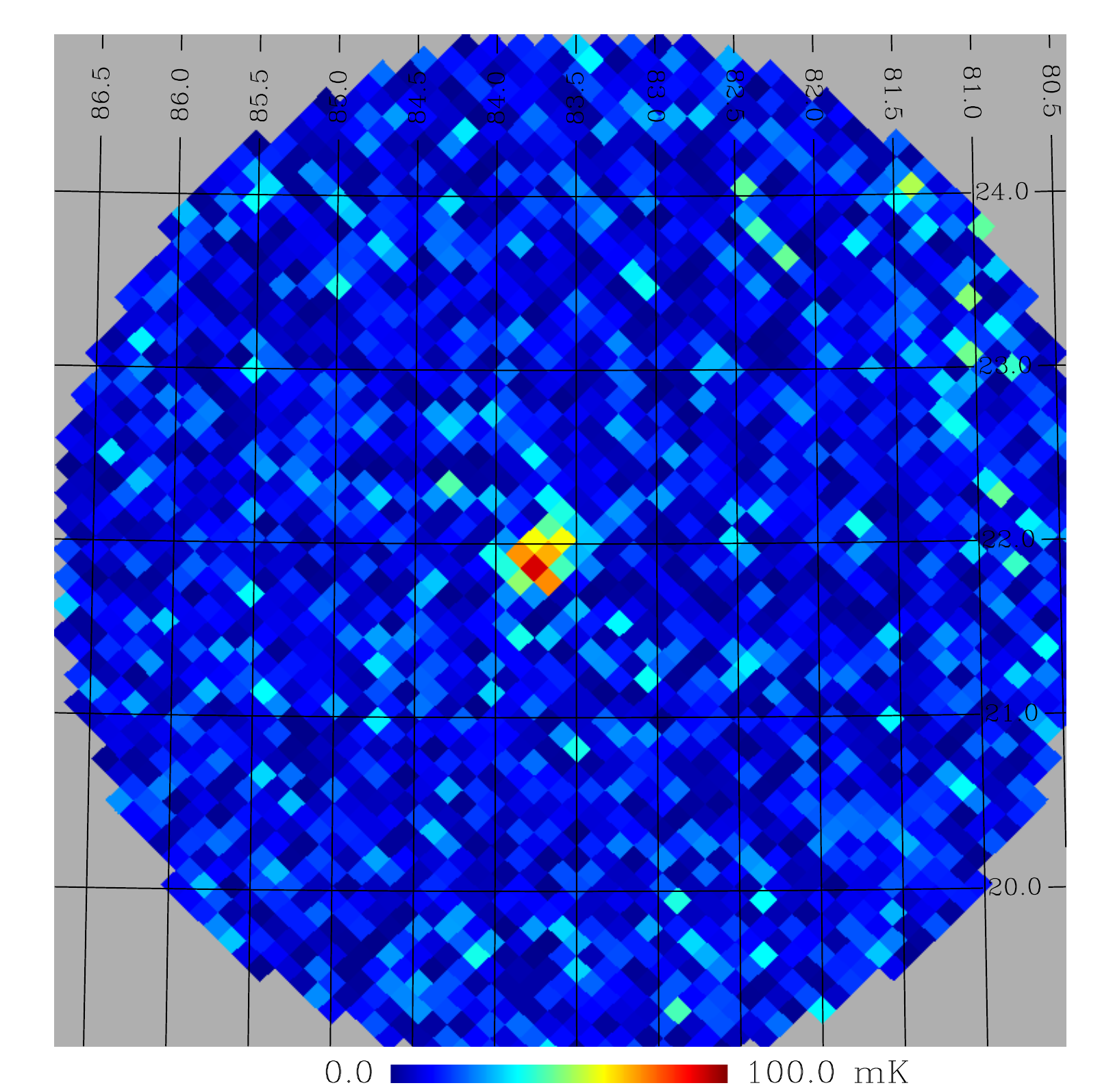}\\
\includegraphics[width=7cm,height=6cm]{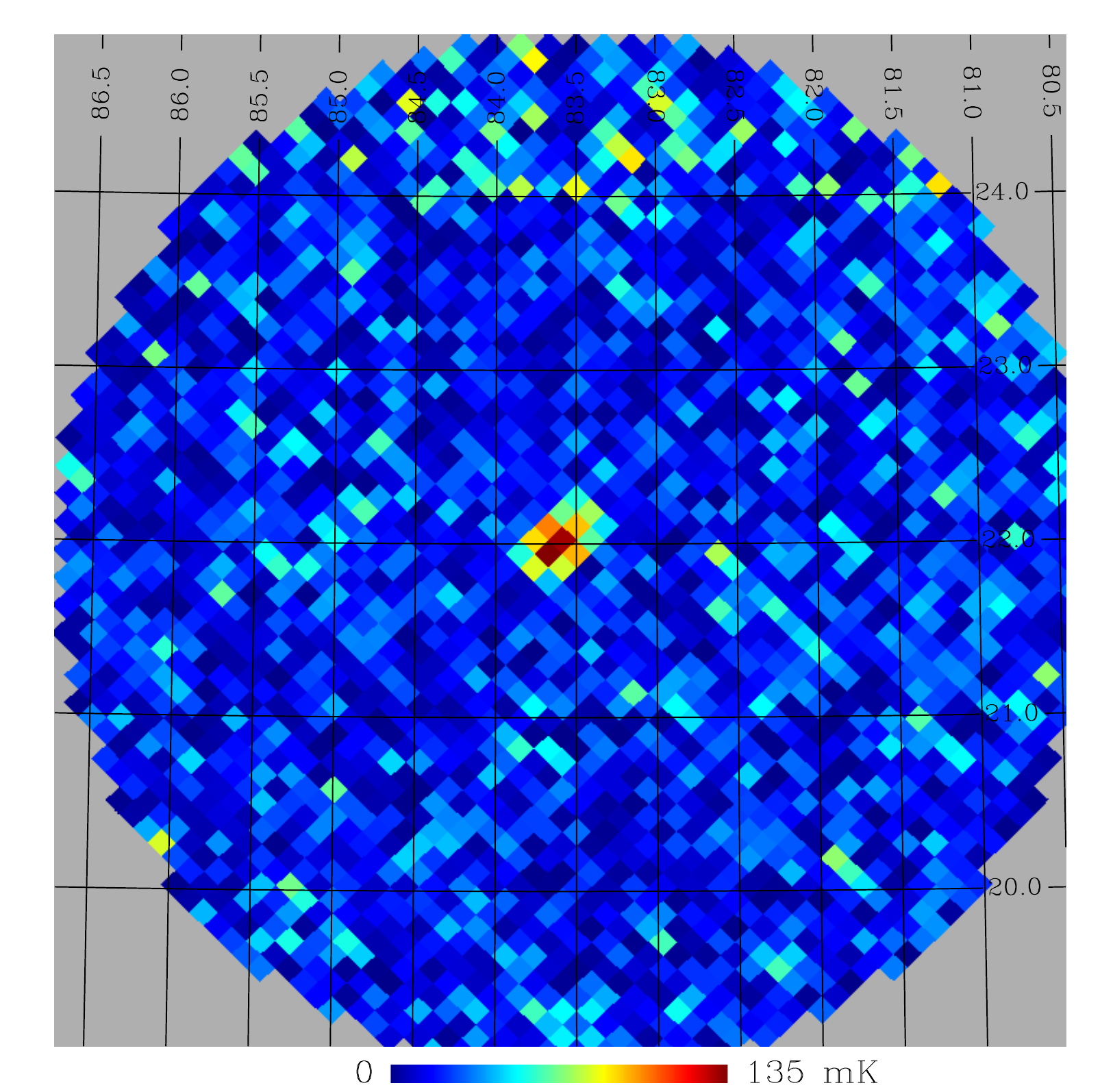}&
\includegraphics[width=7cm,height=6cm]{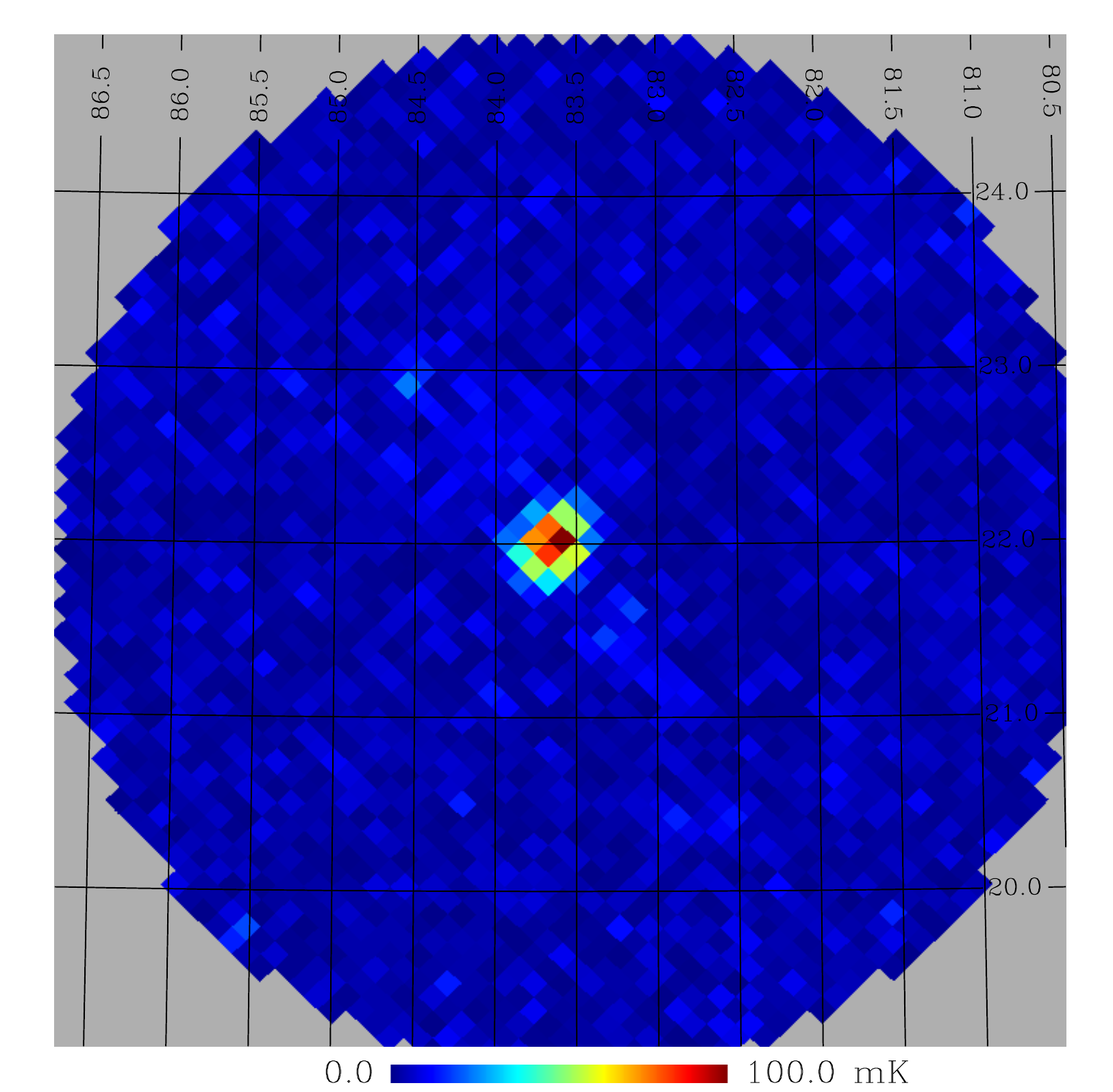}\\
\end{tabular}
\caption[Tau A temperature maps nside=512]{Tau A temperature maps with nside=512. Top left, channel 1, Top right channel 2, Bottom left channel 3, and bottom right channel 6. Each map is centered on Tau A (RA: $83.6332^{\circ}$ DEC: $22.015^{\circ}$) and represents $28.27$ square degrees. The peak temperatures are $111.0\pm11.0\mathrm{~mK}$, $91.9\pm10.5\mathrm{~mK}$, $134.4\pm15.0\mathrm{~mK}$ and $100.0\pm6.0\mathrm{~mK}$, respectively. \label{fig:tauatempmaps}}
\end{figure}

\begin{figure}[p]
\begin{tabular}{cc}
\includegraphics[width=7cm,height=6cm]{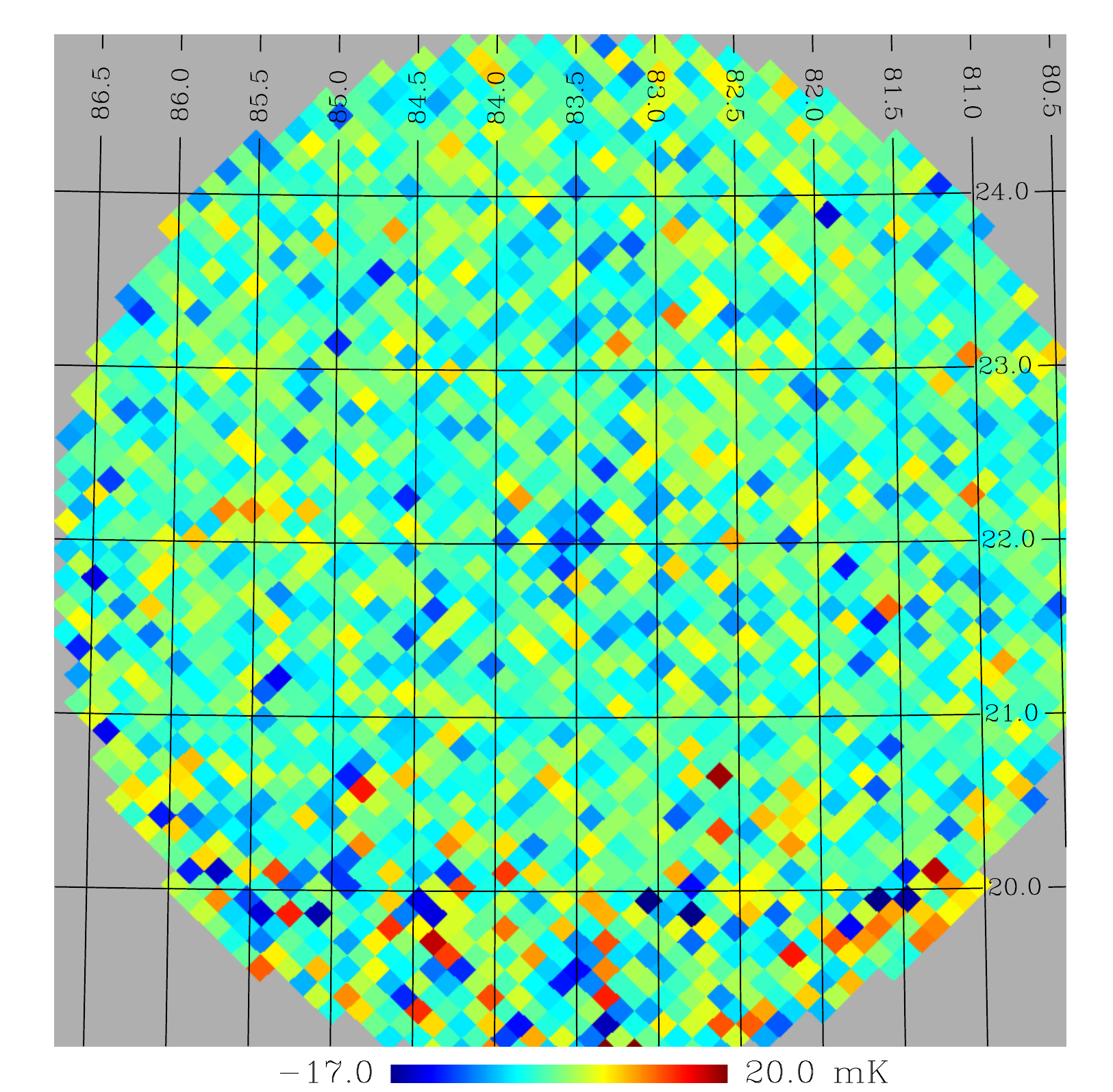}&
\includegraphics[width=7cm,height=6cm]{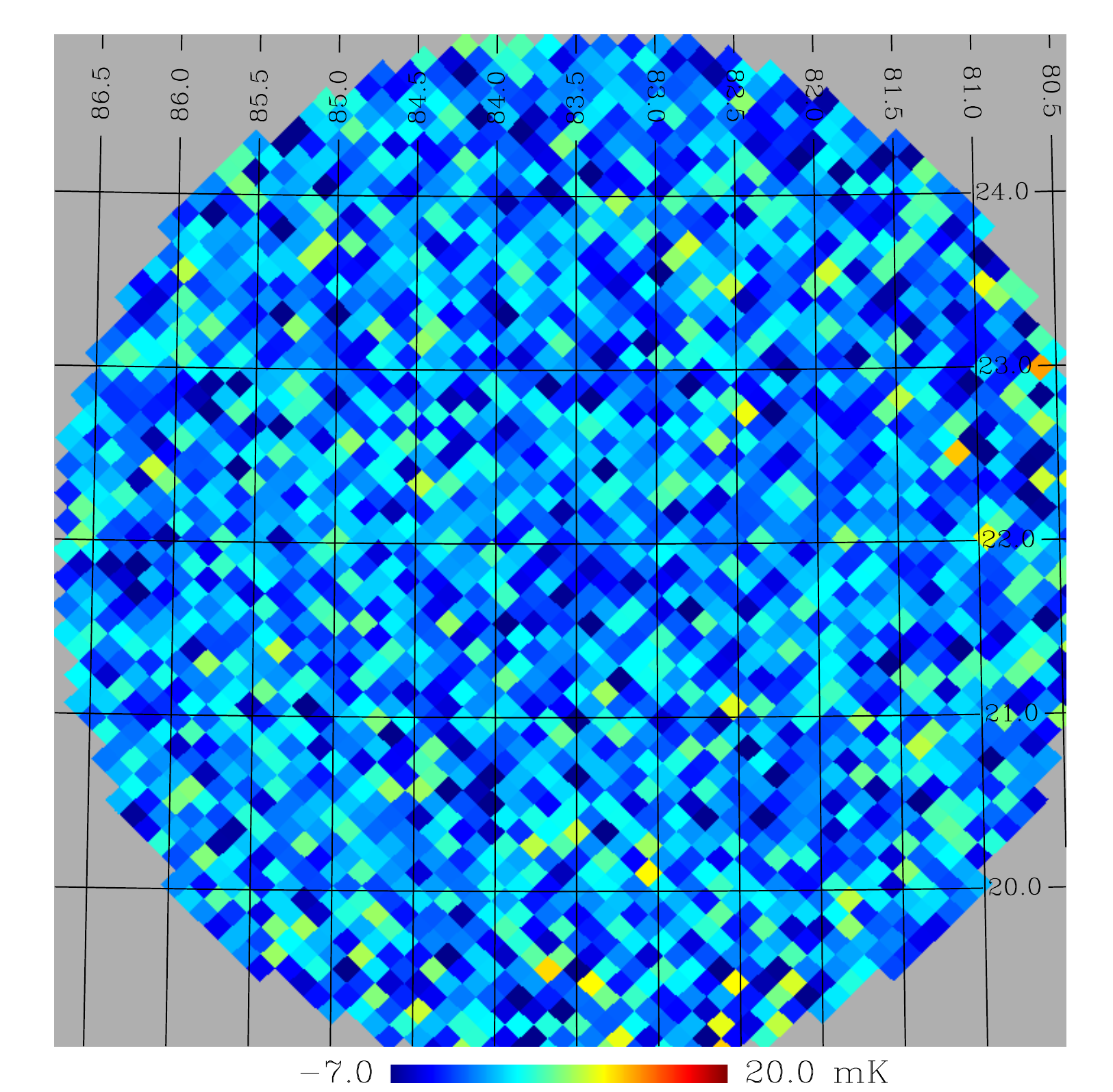}\\
\includegraphics[width=7cm,height=6cm]{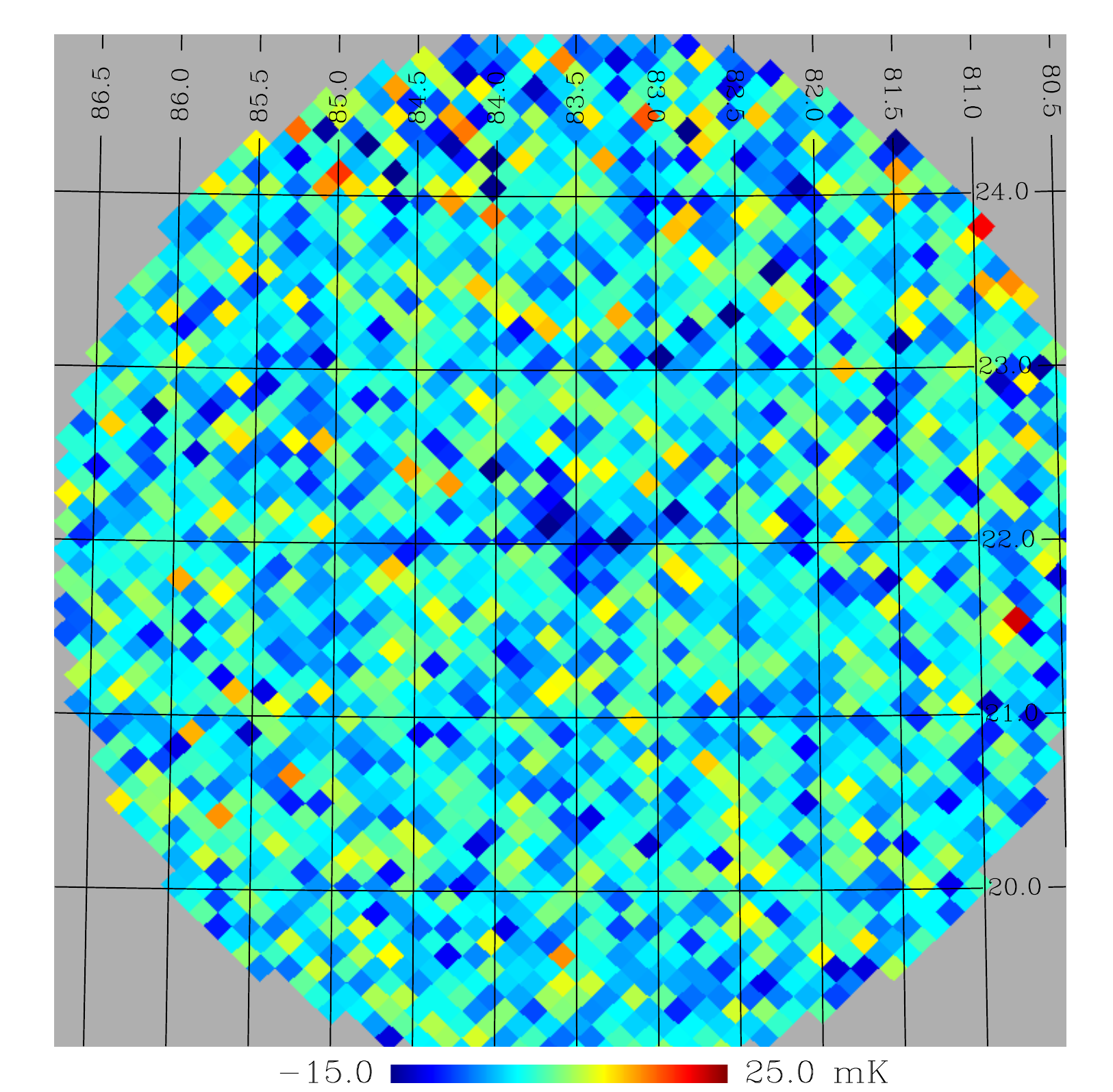}&
\includegraphics[width=7cm,height=6cm]{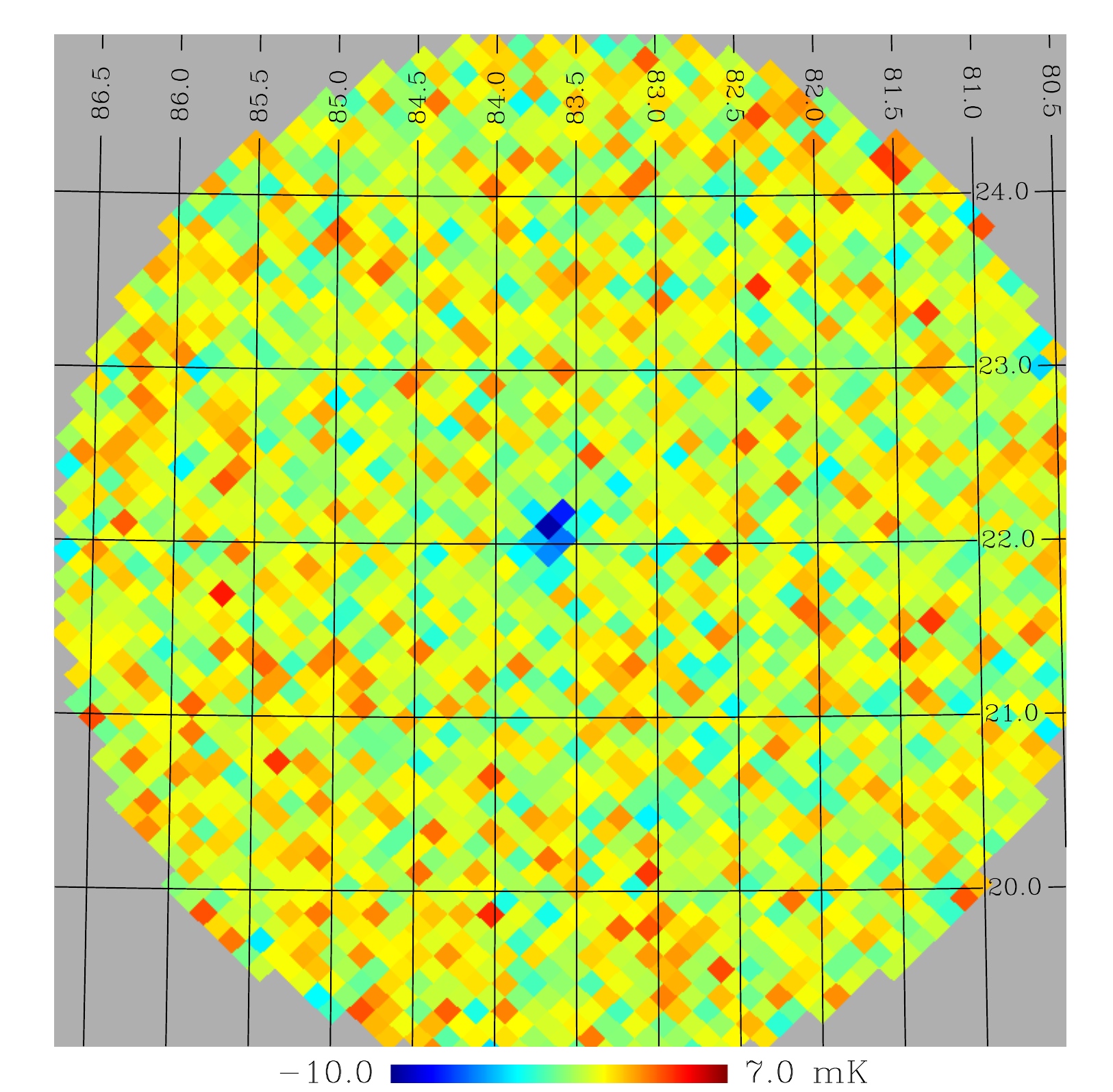}\\
\end{tabular}
\caption[Tau A $Q$ maps]{Tau A $Q$ maps with nside=512. Top left, channel 1, Top right channel 2, Bottom left channel 3, and bottom right channel 6. Each map is centered on Tau A (RA: $83.6332^{\circ}$ DEC: $22.015^{\circ}$) and represents $28.27$ square degrees. The peak temperatures are $-12.6\pm6.0\mathrm{~mK}$, $-8.5\pm4.0\mathrm{~mK}$, $-18.0\pm22.0\mathrm{~mK}$ and $9.8\pm1.2\mathrm{~mK}$, respectively.\label{fig:tauaqmaps}}
\end{figure}

\begin{figure}[p]
\begin{tabular}{cc}
\includegraphics[width=7cm,height=6cm]{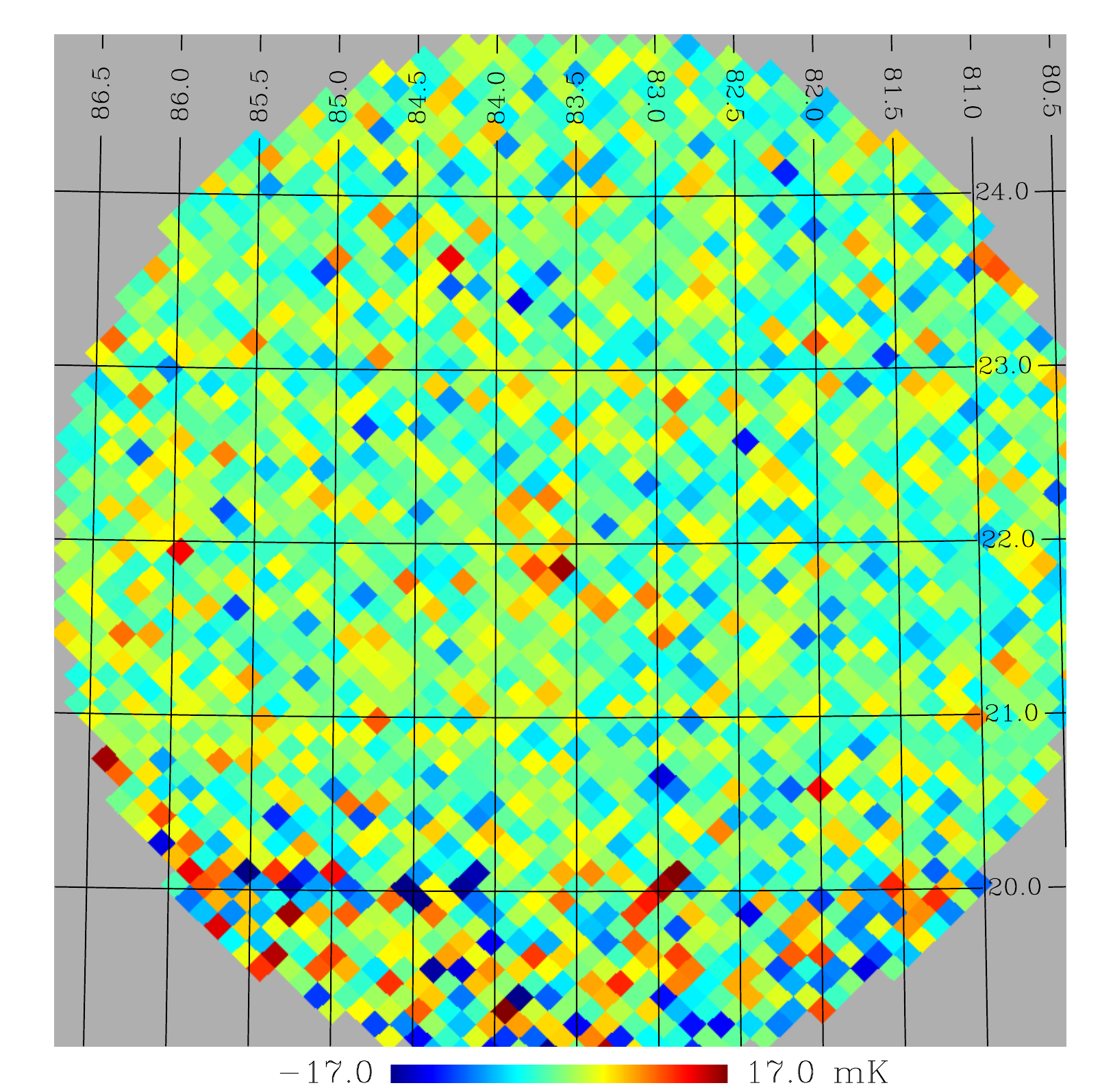}&
\includegraphics[width=7cm,height=6cm]{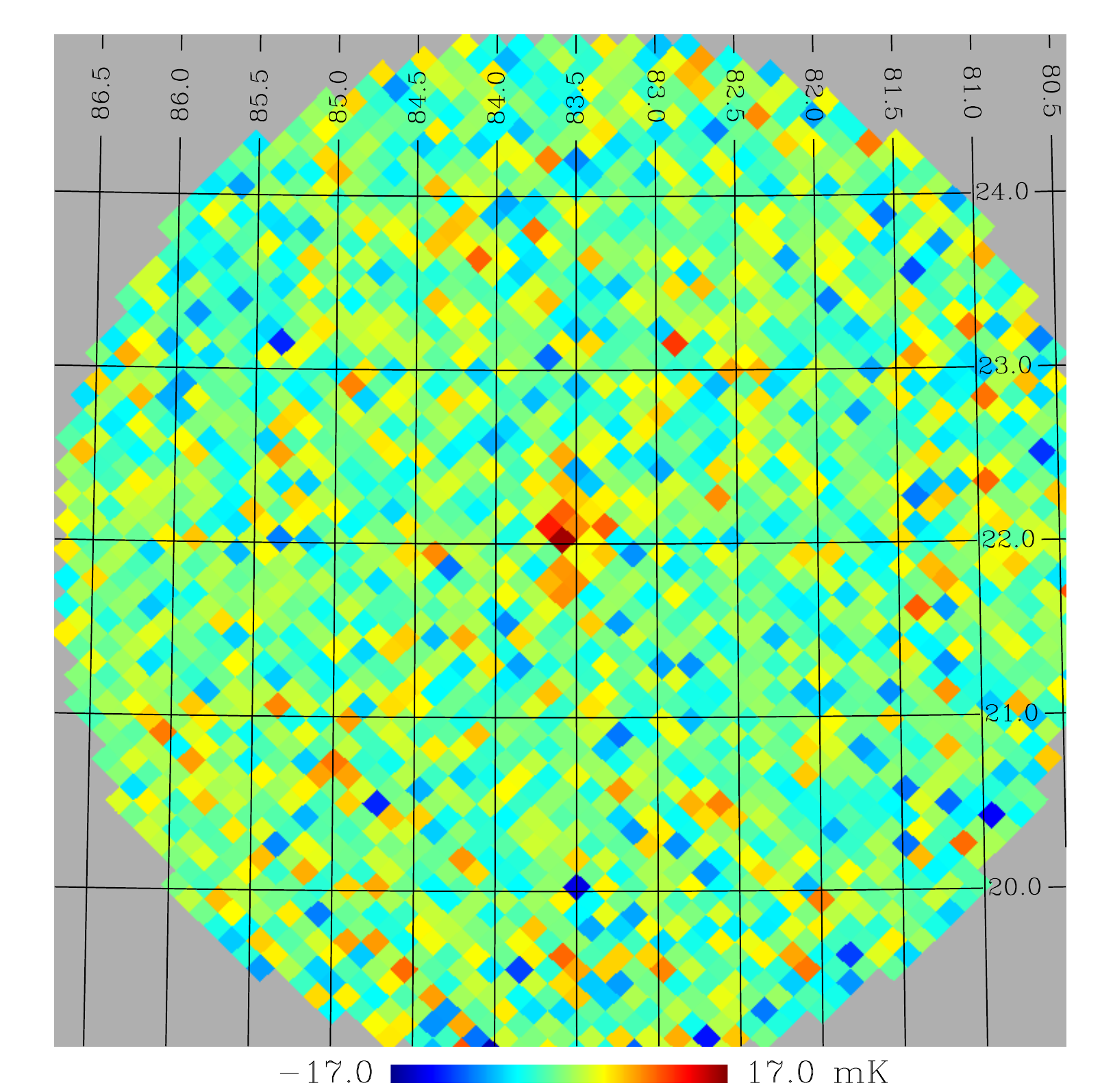}\\
\includegraphics[width=7cm,height=6cm]{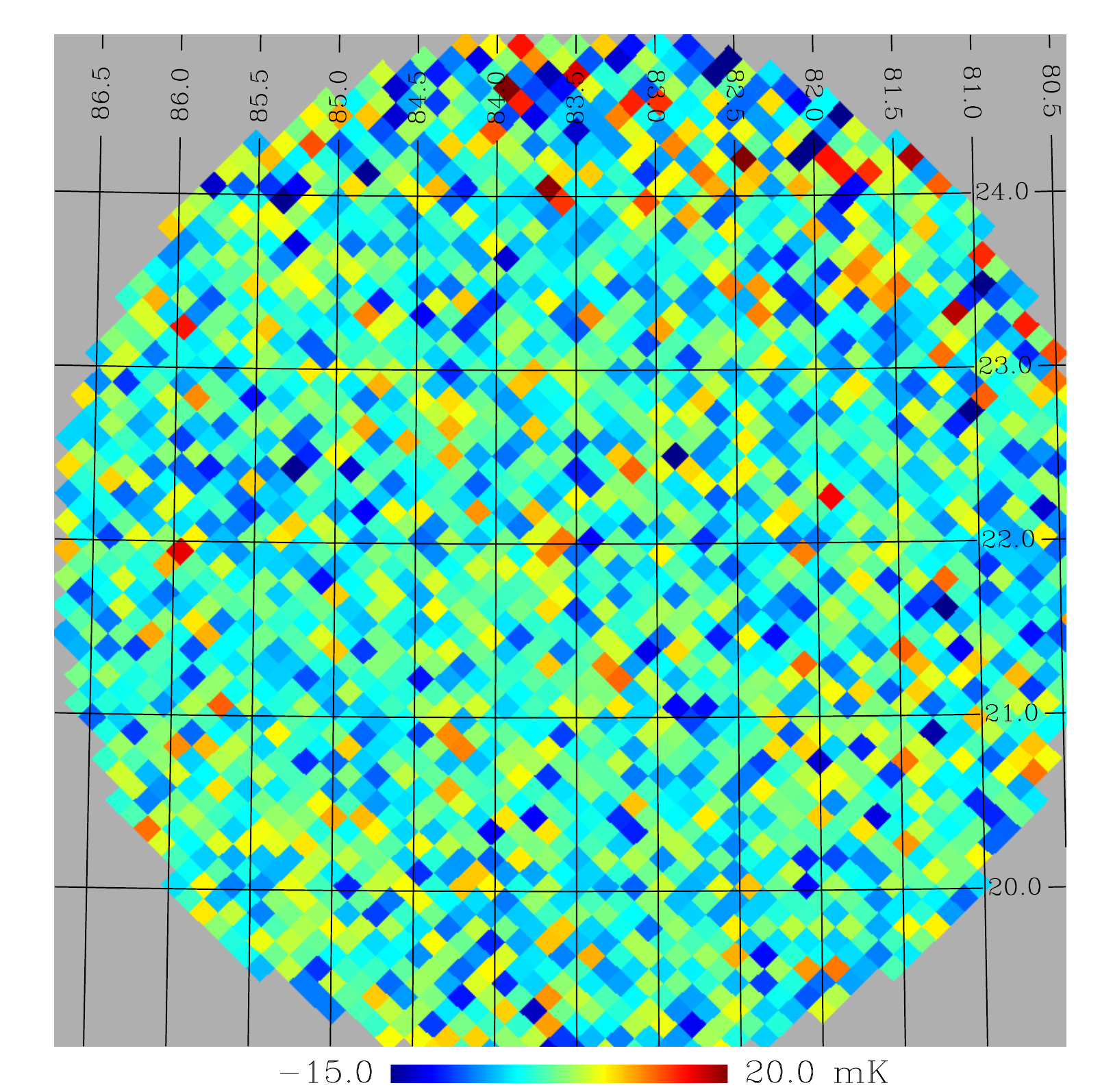}&
\includegraphics[width=7cm,height=6cm]{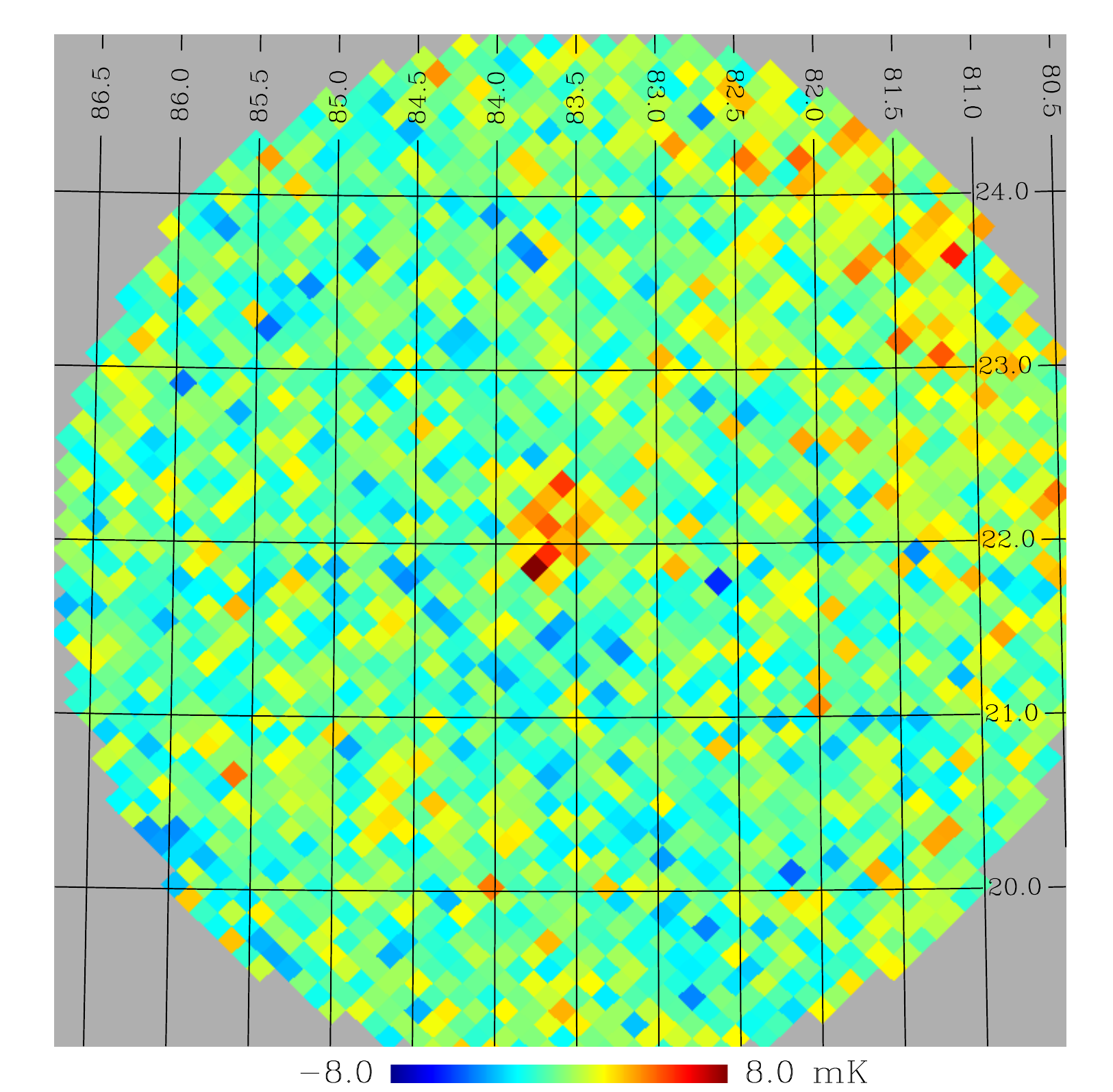}\\
\end{tabular}
\caption[Tau A $U$ maps]{Tau A $U$ maps with nside=512. Top left, channel 1, Top right channel 2, Bottom left channel 3, and bottom right channel 6. Each map is centered on Tau A (RA: $83.6332^{\circ}$ DEC: $22.015^{\circ}$) and represents $28.27$ square degrees. The peak temperatures are $12.4\pm6.0\mathrm{~mK}$, $12.5\pm4.0\mathrm{~mK}$, $15.0\pm22.0\mathrm{~mK}$ and $-8.5\pm1.2\mathrm{~mK}$, respectively.\label{fig:tauaumaps}}
\end{figure}

In Figures~\ref{fig:tauatempmaps},~\ref{fig:tauaqmaps}, and~\ref{fig:tauaumaps} maps of T, Q, and U respectively are presented. The temperature maps are consistent with the expected temperatures from the flux measurements from the WMAP data, with the exception of channel 3. Channel 3 has been consistently problematic, with much higher noise levels and calibration constants that are a factor of 2 higher then the other 3 channels. Channel 3 has a high gate current indicating a leaky gate which contributes significantly to its higher noise temperature.  An additional note on the temperature measurement is that the noise levels are dominated by the $\frac{1}{f}$ noise from each channel. B-Machine was designed as a polarimeter and it is surprising that Tau A is visible in the temperature maps. The $Q$ and $U$ maps both consistently show more signal than expected. The full result of the Tau A drift scans are in Table~\ref{tab:tauatemps}. There are 2 sources of error contributing to these figures. First, the pixelization scheme changes the beam size in a non-gaussian fashion, which in turn causes the conversion from flux to temperature units to differ slightly from the derived formula. Secondly, the beam shift as a function of position of the Polarization Rotator is contributing a polarized signal that would not otherwise be present.  As the telescope scans across the source the beam is shifting in elevation, and each sector of the rotator sees a slightly different part of the sky. For example, if the horizontal sector is centered on Tau A then the vertical sector won't be, causing an erroneous polarization signal. This effect is only apparent due to the fact that a point source is being observed.

\begin{figure}[p]
\begin{tabular}{cc}
\includegraphics[width=\textwidth]{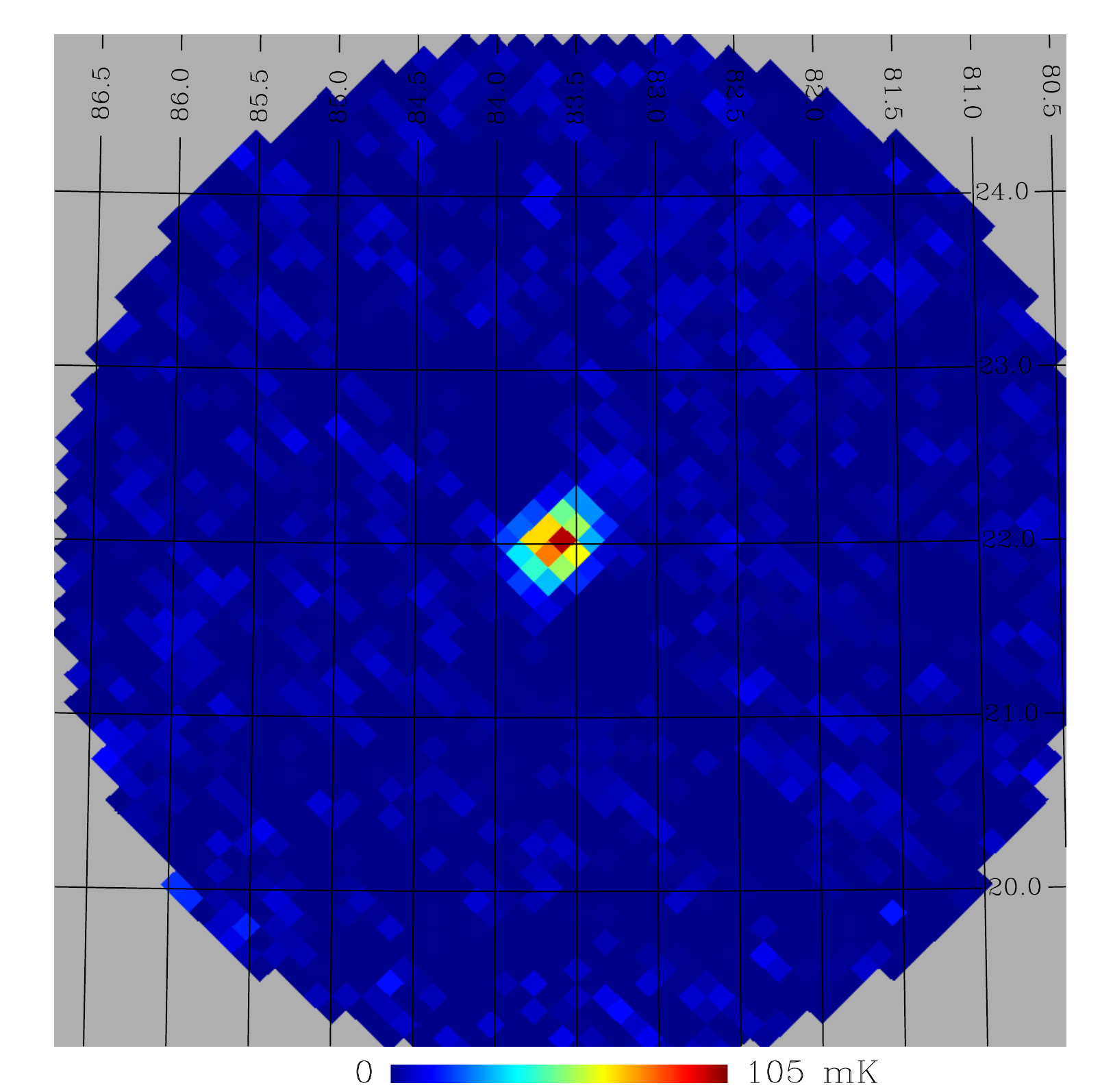}
\end{tabular}
\caption[Combined Tau A temperature map]{Tau A temperature map with nside=512.  All channels were combined using sigma weighting. The map is centered on Tau A (RA: $83.6332^{\circ}$ DEC: $22.015^{\circ}$) and represents $28.27$ square degrees. The centroid temperature of $99.8\pm4.0\mathrm{~mK}$. \label{fig:tauatempsummap}}
\end{figure}

\begin{figure}[p]
\begin{tabular}{cc}
\includegraphics[width=\textwidth]{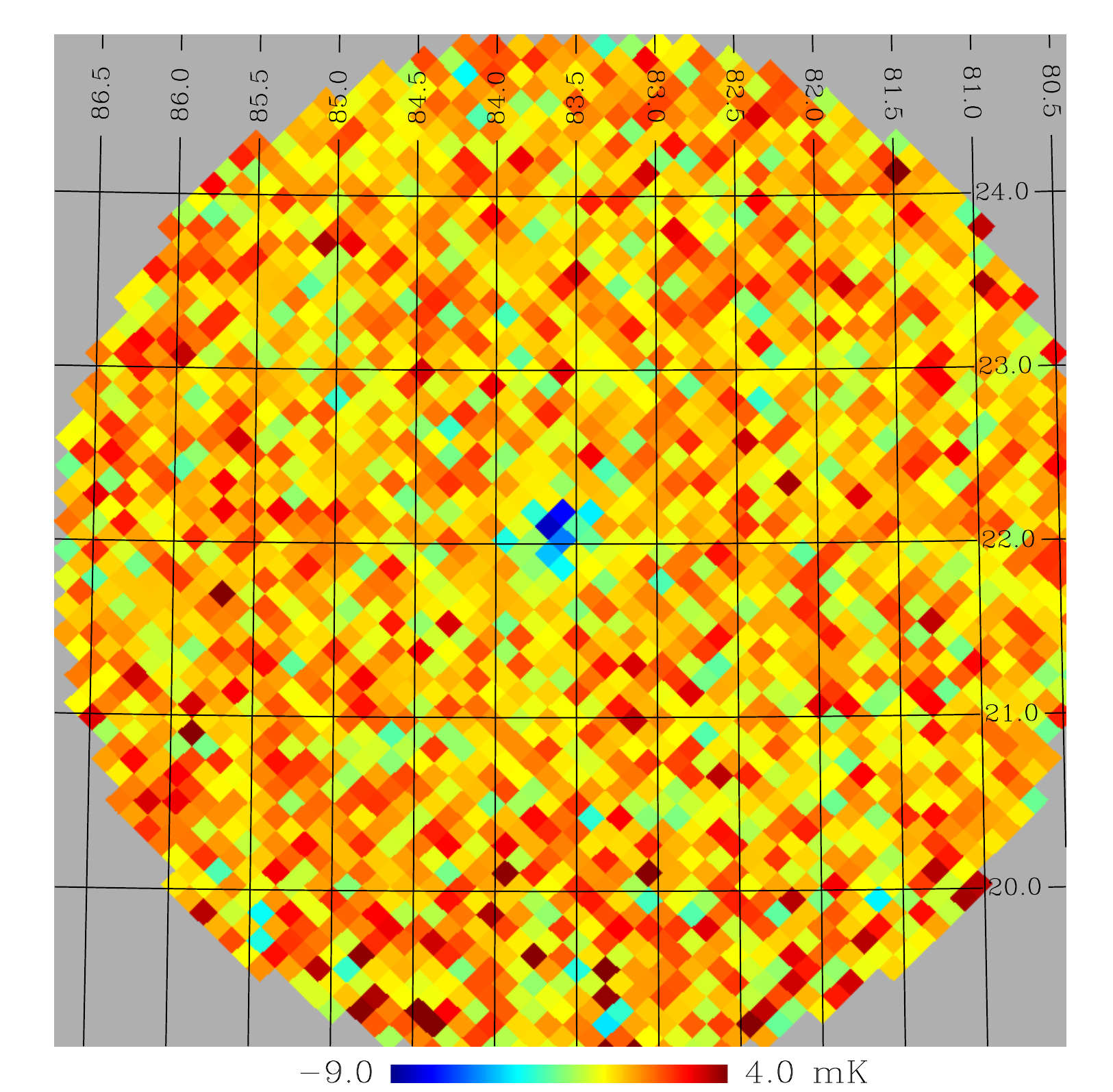}
\end{tabular}
\caption[Combined Tau A $Q$ map]{Tau A $Q$ map with nside=512.  All channels were combined using sigma weighting. The map is centered on Tau A (RA: $83.6332^{\circ}$ DEC: $22.015^{\circ}$) and represents $28.27$ square degrees, with a centroid $Q$ of  $-8.2\pm3.0\mathrm{~mK}$. \label{fig:tauaqsummap}}
\end{figure}

\begin{figure}[p]
\begin{tabular}{cc}
\includegraphics[width=\textwidth]{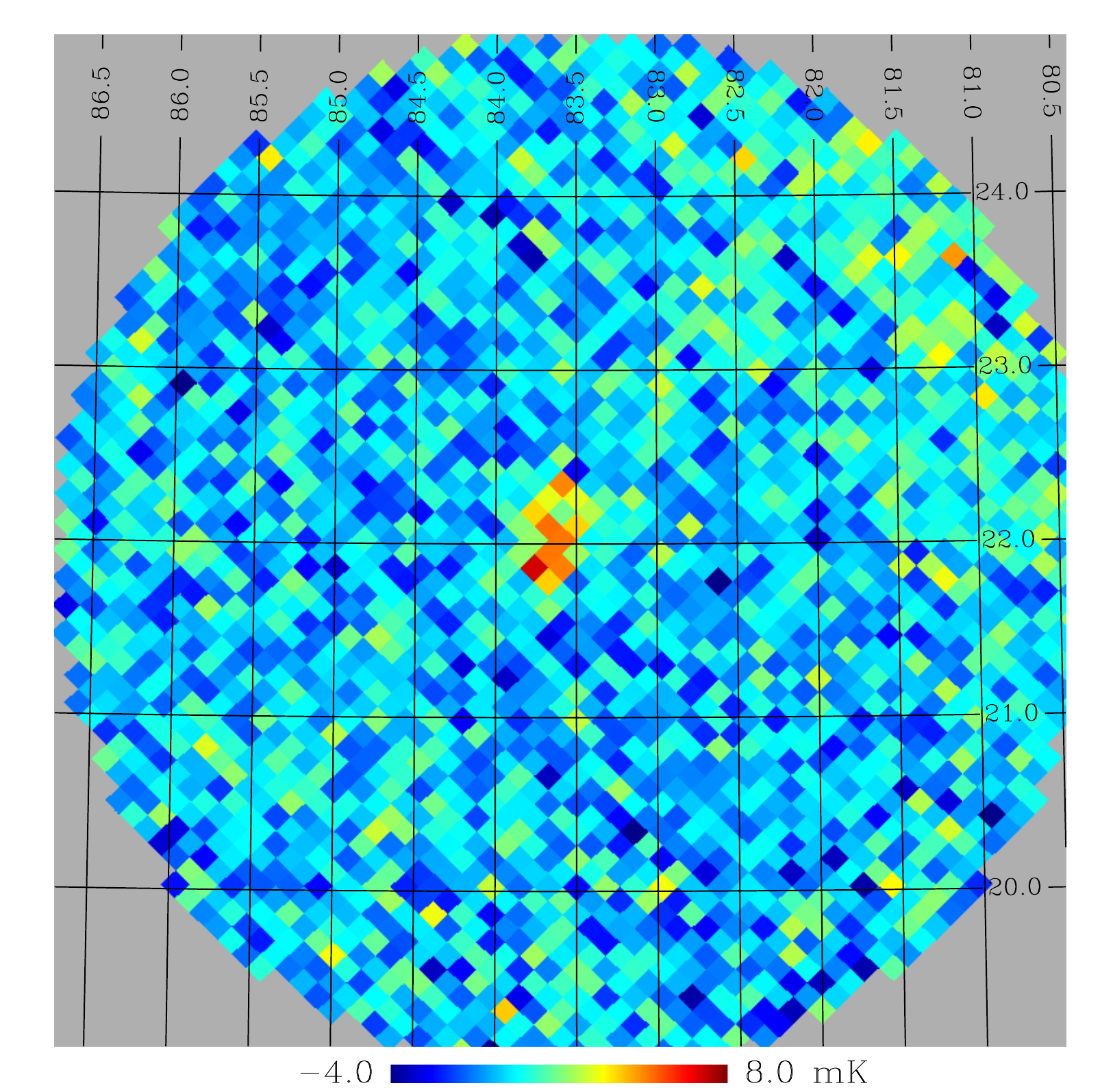}
\end{tabular}
\caption[Combined Tau A $U$ map]{Tau A $U$ map with nside=512.  All channels were combined using sigma weighting. The map is centered on Tau A (RA: $83.6332^{\circ}$ DEC: $22.015^{\circ}$) and represents $28.27$ square degrees, with a centroid $U$ of $7.1\pm3.0\mathrm{~mK}$. \label{fig:tauausummap}}
\end{figure}

\begin{figure}[p]
\begin{tabular}{cc}
\includegraphics[width=7cm,height=6cm]{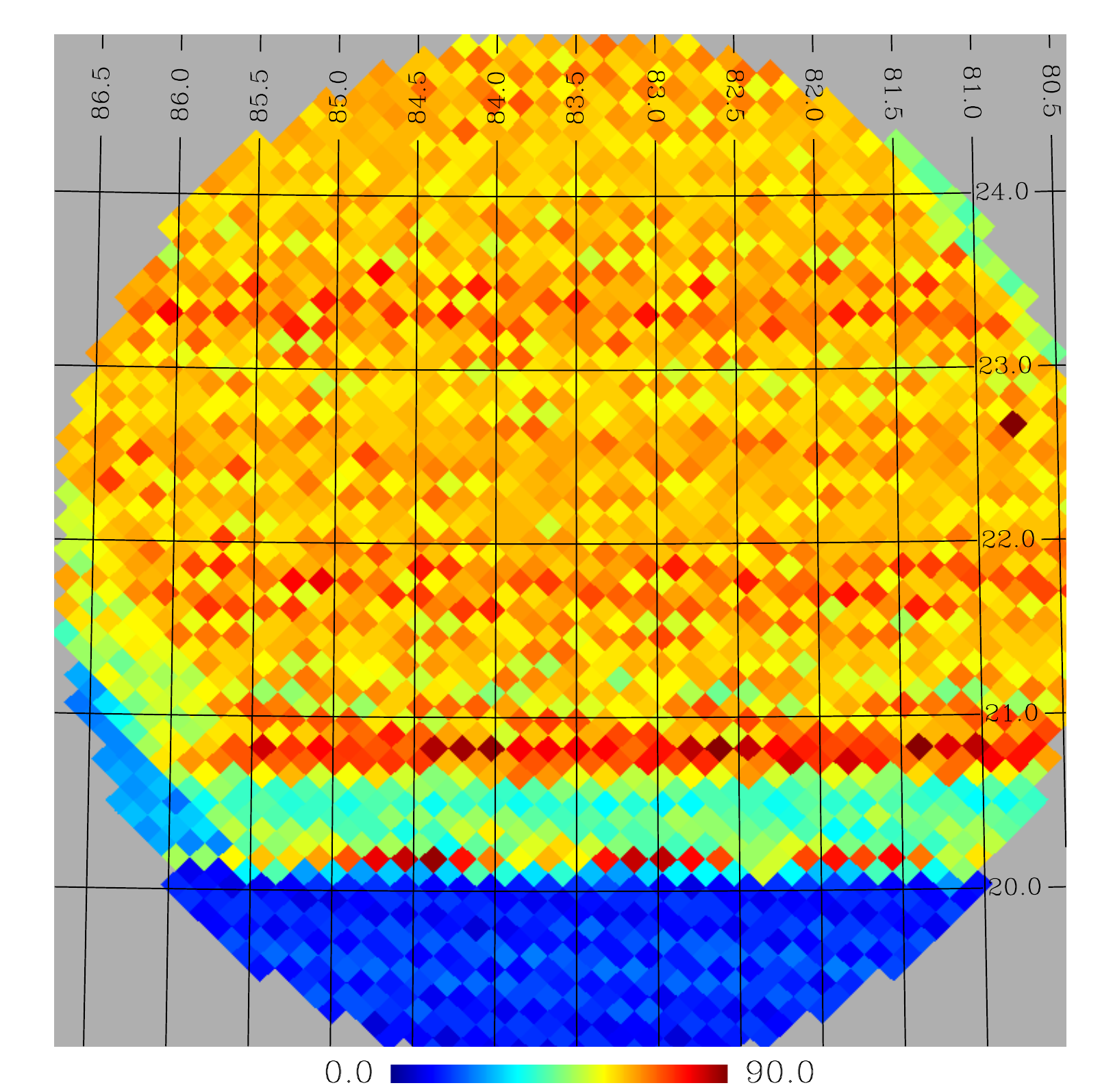}&
\includegraphics[width=7cm,height=6cm]{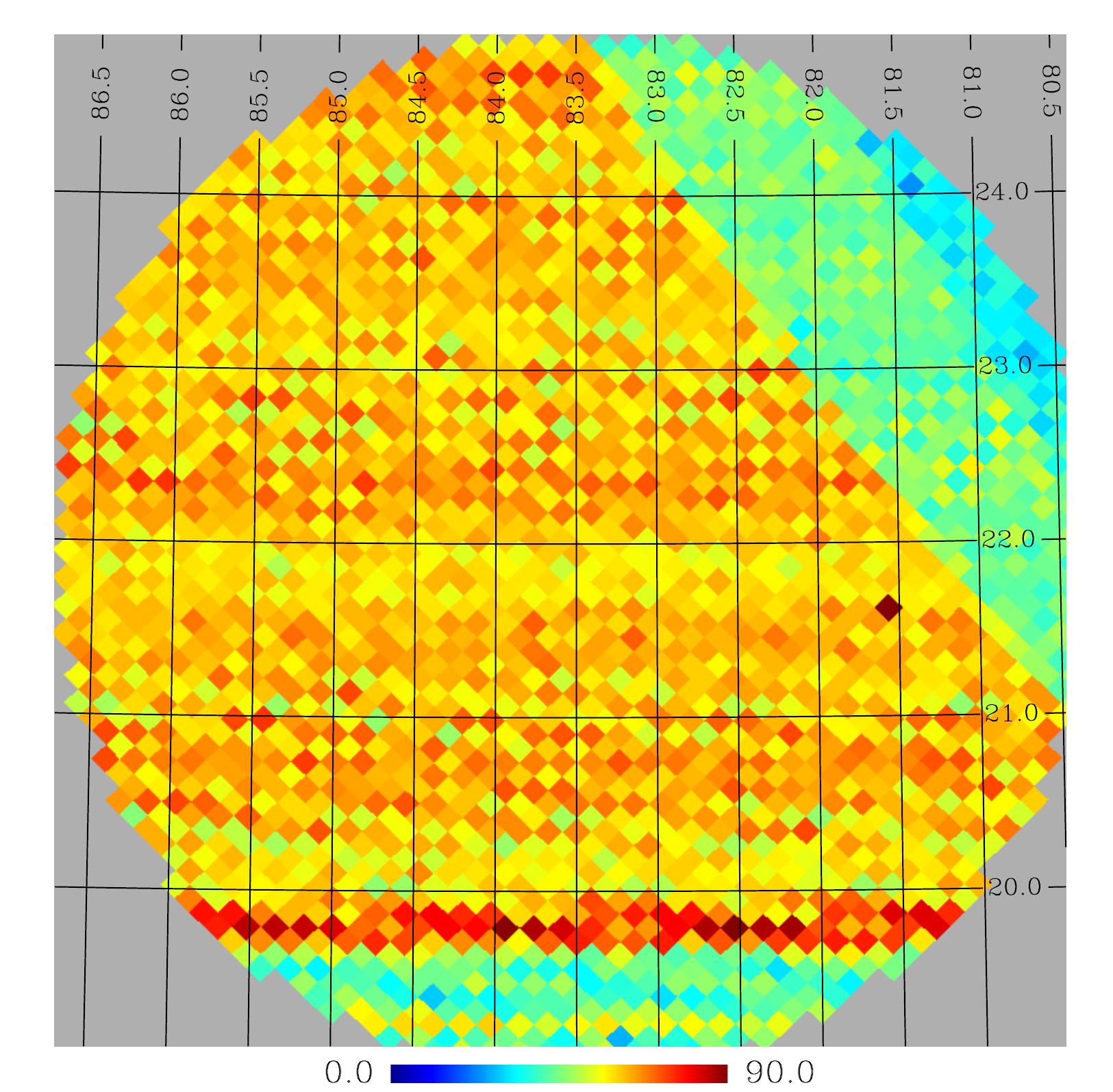}\\
\includegraphics[width=7cm,height=6cm]{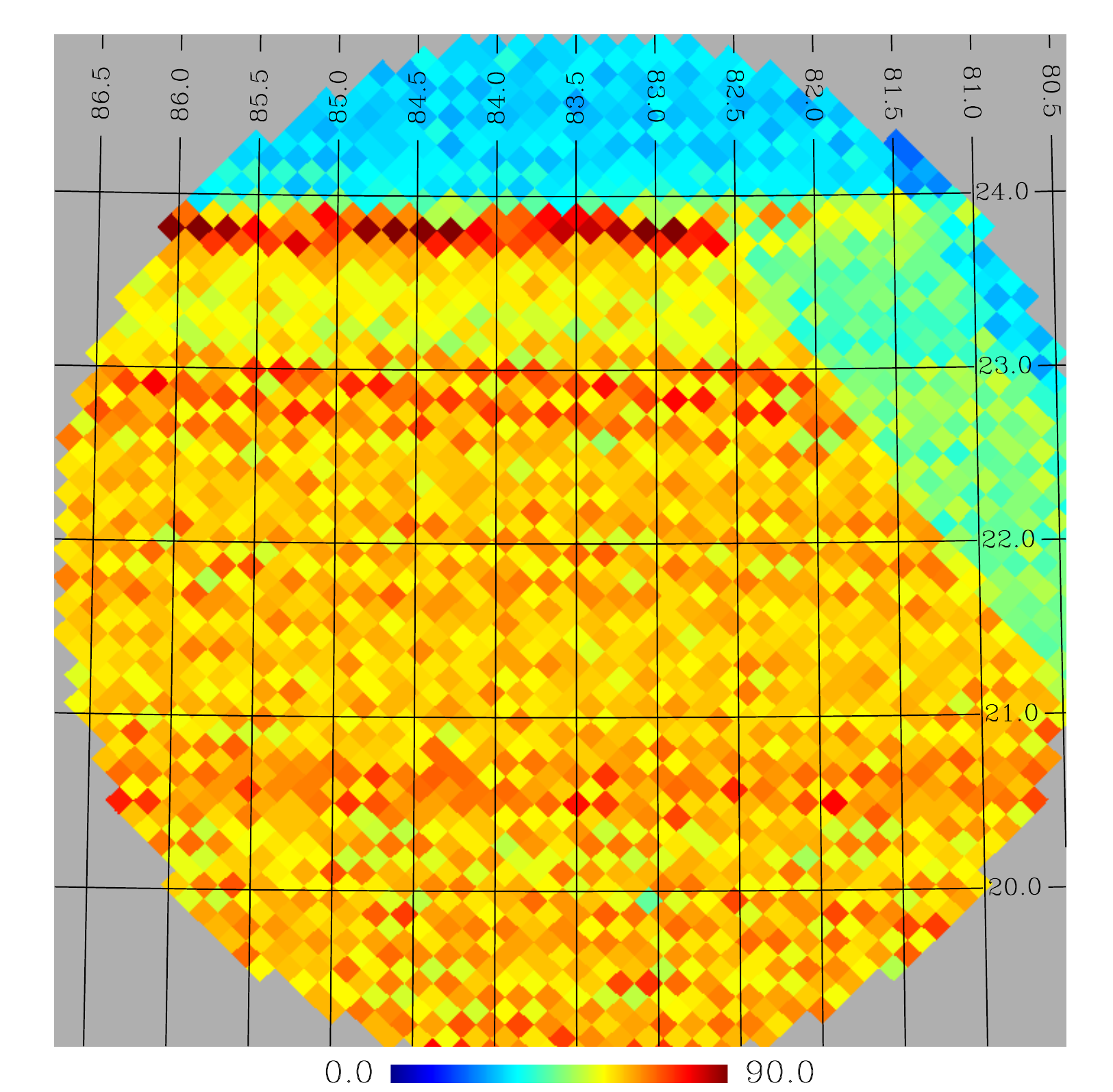}&
\includegraphics[width=7cm,height=6cm]{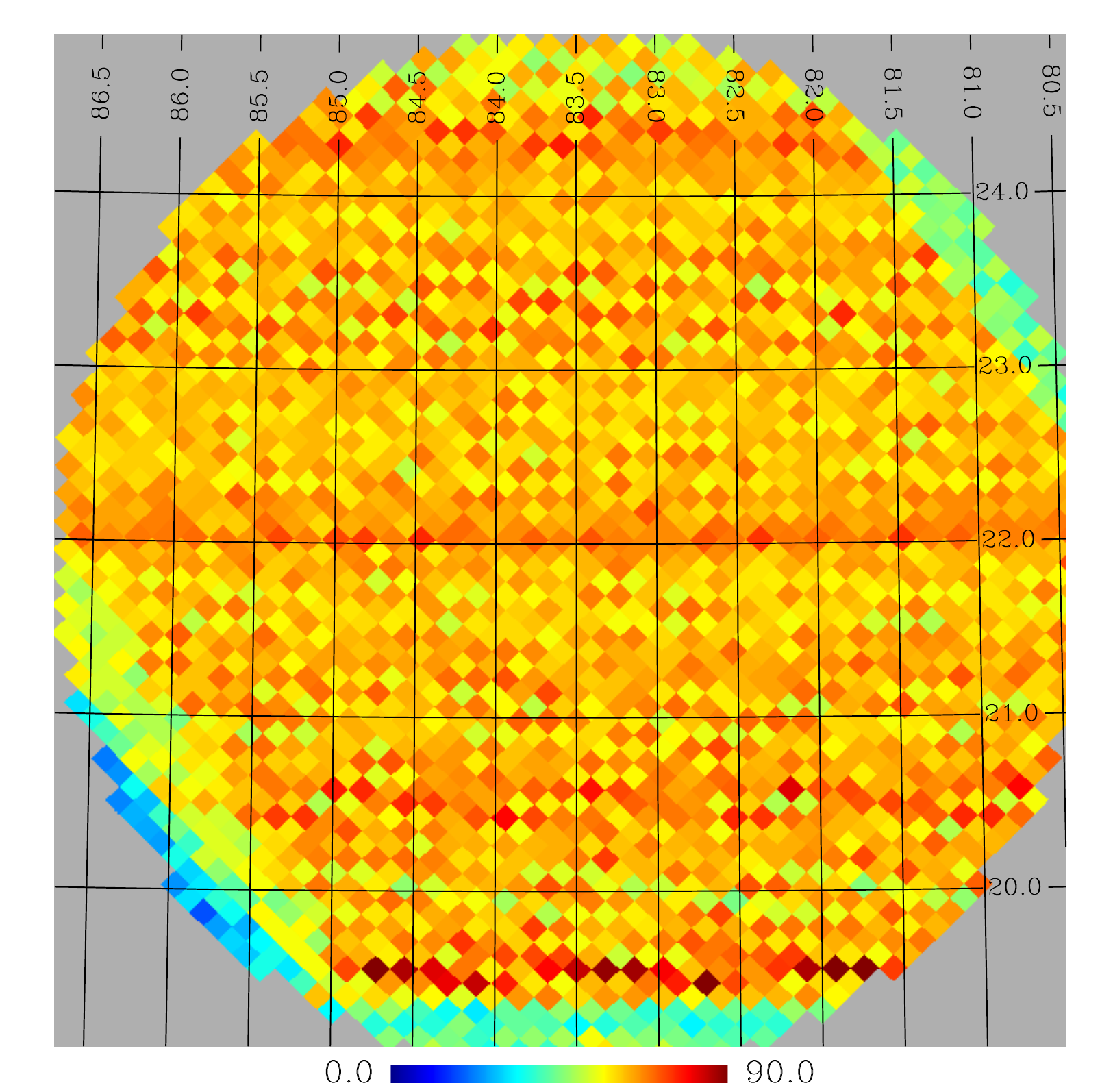}\\
\end{tabular}
\caption[Tau A number of samples per bin maps]{Tau A number of samples per bin maps. Top left, channel 1, top right channel 2, bottom left channel 3, and bottom right channel 6. Each map is centered on Tau A (RA: $83.6332^{\circ}$ DEC: $22.015^{\circ}$) and represents $28.27$ square degrees. The average samples per bin are 62.7 (1.88 s), 61.7 (1.85 s), 61.9 (1.85 s) and 63.0 (1.88 s), respectively. The regions with low counts have been excluded from the averages. \label{fig:tauanobsmap}}
\end{figure}

The pixel size for the maps (nside=512) was selected based on beam shift, number of pixels per beam ($\sim3$), and total number of pixels. This last criterion was necessary due to computational limits on memory usage imposed by IDL.

\begin{table}[p]
\begin{center}
\caption[Tau A Stokes Parameters at 41.5 GHz]{Tau A Stokes Parameters at 41.5 GHz \label{tab:tauatemps}}
\begin{tabular}{|c|c|c|c|}
  \hline
  Channel &  Flux Density & Temperature & Observed Temperature \\
          &    (Jy)       &      (mK)   &      (mK)             \\
  \hline
  \hline
  \multicolumn{4}{|c|}{Temperature}\\
  \hline
  1 & $281\pm6$ &  $98\pm6$ & $111.0\pm11$      \\
  \hline
  2 & $281\pm6$ & $98\pm6$ & $103.4\pm10.5$      \\
  \hline
  3 & $281\pm6$ &  $98\pm6$ & $124.9\pm15.0$     \\
  \hline
  6 & $281\pm6$ &  $110\pm6$ & $95.2\pm6.0$     \\
  \hline
  Sum &         &  $101\pm3$  & $99.8\pm4.6$ \\
  \hline
  \multicolumn{4}{|c|}{Q Stokes parameter}\\
  \hline
  1 & $-18.5\pm3.0$ &  $-6.7\pm1.3$ & $-10.8\pm6.0$      \\
  \hline
  2 & $-18.5\pm3.0$ &  $-6.7\pm1.3$ & $-6.8\pm4.0$      \\
  \hline
  3 & $-18.5\pm3.0$ &  $-6.7\pm1.3$ & $-14.7\pm22.0$     \\
  \hline
  6 &$ -18.5\pm3.0$ &  $-7.4\pm1.3$ & $-9.4\pm1.2$     \\
  \hline
  Sum &         &  $-6.9\pm0.7$  & $-8.2\pm3.0$\\
  \hline
  \multicolumn{4}{|c|}{U Stokes parameter}\\
  \hline
  1 & $0.5\pm3.0$ &  $0.2\pm1.1$ & $16.0\pm6.0$      \\
  \hline
  2 & $0.5\pm3.0$ &  $0.2\pm1.1$ & $15.8\pm4.0$      \\
  \hline
  3 & $0.5\pm3.0$ &  $0.2\pm1.1$ & $11.2\pm22.0$     \\
  \hline
  6 & $0.5\pm3.0$ &  $0.2\pm1.1$ & $8.1\pm1.2$     \\
  \hline
  Sum &         &  $0.2\pm0.6$  & $7.1\pm3.0$ \\
  \hline
\end{tabular}
\end{center}
\end{table}

\section{Maps} \label{sec:maps}
B-Machine spent 81 days at White Mountain from arrival with the experiment in the truck to final shut down of the facility due to weather and funding constraints. Of the 81 possible days, 40 of these were used for observing. Of the remaining 41 days 8 were used for the initial reconstruction and calibration of B-Machine, 12 of the days B-Machine was down due to mechanical issues, and the final balance was from weather.  Of the 40 observing days 4 were used for a dedicated NCP scan which was not usable due to pointing issues and 29 were salvaged from the pointing debacle, see Section~\ref{sec:pointreconstruct}, for map generation. A full observing day consisted of 14 hours, with the time being limited by the Sun's height in the sky. During the final shut down sequence the telescope was allowed to operate during the day to test if Sun crossings would damage the instrument.  After inspecting the telescope after a Sun crossing it was determined that operation during day light hours was possible, though thermal cycling would certainly lead to data analysis issues.

Of the 29 days of usable data each channel for each day was cut based on several values generated out of the CDS arrays. In Table~\ref{tab:sparameters} the final selection parameters for each of the channels can be seen.

\begin{table}[p]
\begin{center}
\caption[CDS Data Selection Values]{CDS Data Selection Values, Values are in Volts Unless Otherwise Stated \label{tab:sparameters}}
\begin{tabular}{|c|c|c|c|c|}
  \hline
  Selection Field &  Channel &   I   &   Q   & U  \\
  \hline
  \hline
  AC Sigma       &          &          &          &    \\
  \hline
                  &  1       &  0.11   & 0.021     &  0.022  \\
  \hline
                  &  2       &  0.11   &  0.0186   & 0.0203  \\
  \hline
                  &  3       &  0.11   & 0.0156    &  0.0164  \\
  \hline
                  &  6       &  0.11   & 0.00918   &  0.00892  \\
  \hline
  Peak to Peak    &          &         &           &    \\
  \hline
                  &  1       &  0.72   & 0.149     &  0.161  \\
  \hline
                  &  2       &  0.78   &  0.142   & 0.155  \\
  \hline
                  &  3       &  0.70   & 0.124    &  0.126  \\
  \hline
                  &  6       &  0.30   & 0.0701   &  0.0680  \\
  \hline
  White Noise     &          &         &           &    \\
  \hline
                  &  1       &  0.012   & 0.00336  &  0.00366  \\
  \hline
                  &  2       &  0.013  &  0.00322 & 0.00352  \\
  \hline
                  &  3       &  0.010  & 0.00266  &  0.00281  \\
  \hline
                  &  6       &  0.004  & 0.00154  &  0.00150 \\
  \hline
  Min DC Value    &          &         &          &    \\
  \hline
                  &  1      &  -0.825  &          &   \\
  \hline
                  &  2      &  -0.922  &          &   \\
  \hline
                  &  3      &  -0.638  &          &   \\
  \hline
                  &  6      &  -0.829  &    &   \\
  \hline
  Max DC Value    &          &         &          &    \\
  \hline
                  &  1      &  -0.627  &          &   \\
  \hline
                  &  2      &  -0.703  &          &   \\
  \hline
                  &  3      &  -0.473  &          &   \\
  \hline
                  &  6      &  -0.665  &    &   \\
  \hline
  Number of Samples &  All   &  3000   &  counts         &  \\
  \hline
  Status            &  All   &  35   &        &  \\
  \hline
  Max Elevation     &  All   &  46   &   degrees        &  \\
   \hline
\end{tabular}
\end{center}
\end{table}

\begin{figure}[p]
\begin{center}
\includegraphics[width=9.5cm ]{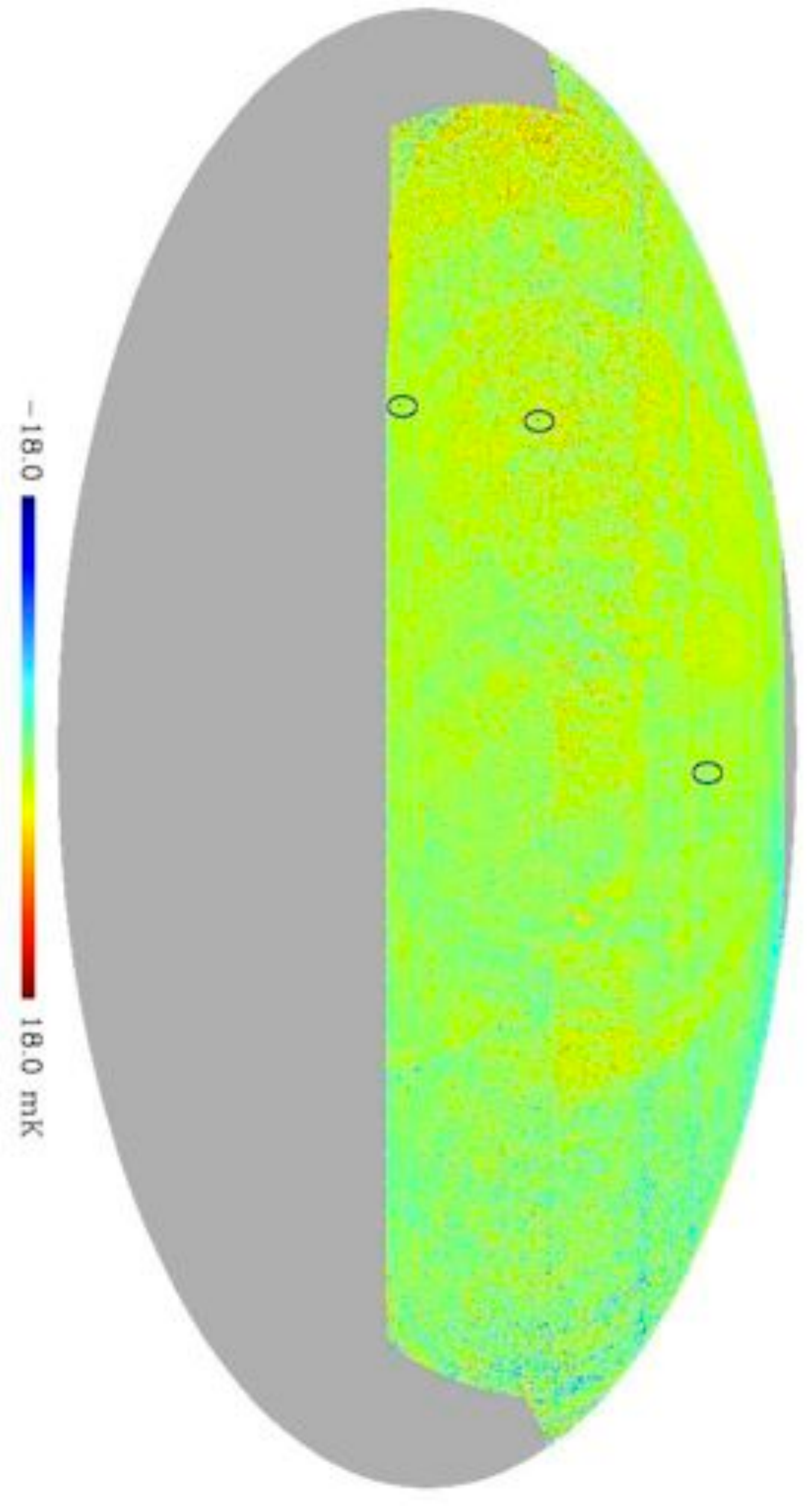}
\caption[Temperature sky map]{Temperature sky map for all days and channels. The three circles are observed point sources  Cygnus A, Tau A, and M42 from the top down. The Galaxy can be seen as a blue line on the right side of the figure. The map represents a $53.07\%$ sky coverage, with an average integration time of 20.9 seconds per pixel and is in ecliptic coordinates.\label{fig:tempmapall}}
\end{center}
\end{figure}

\begin{figure}[p]
\begin{center}
\includegraphics[width=10cm ]{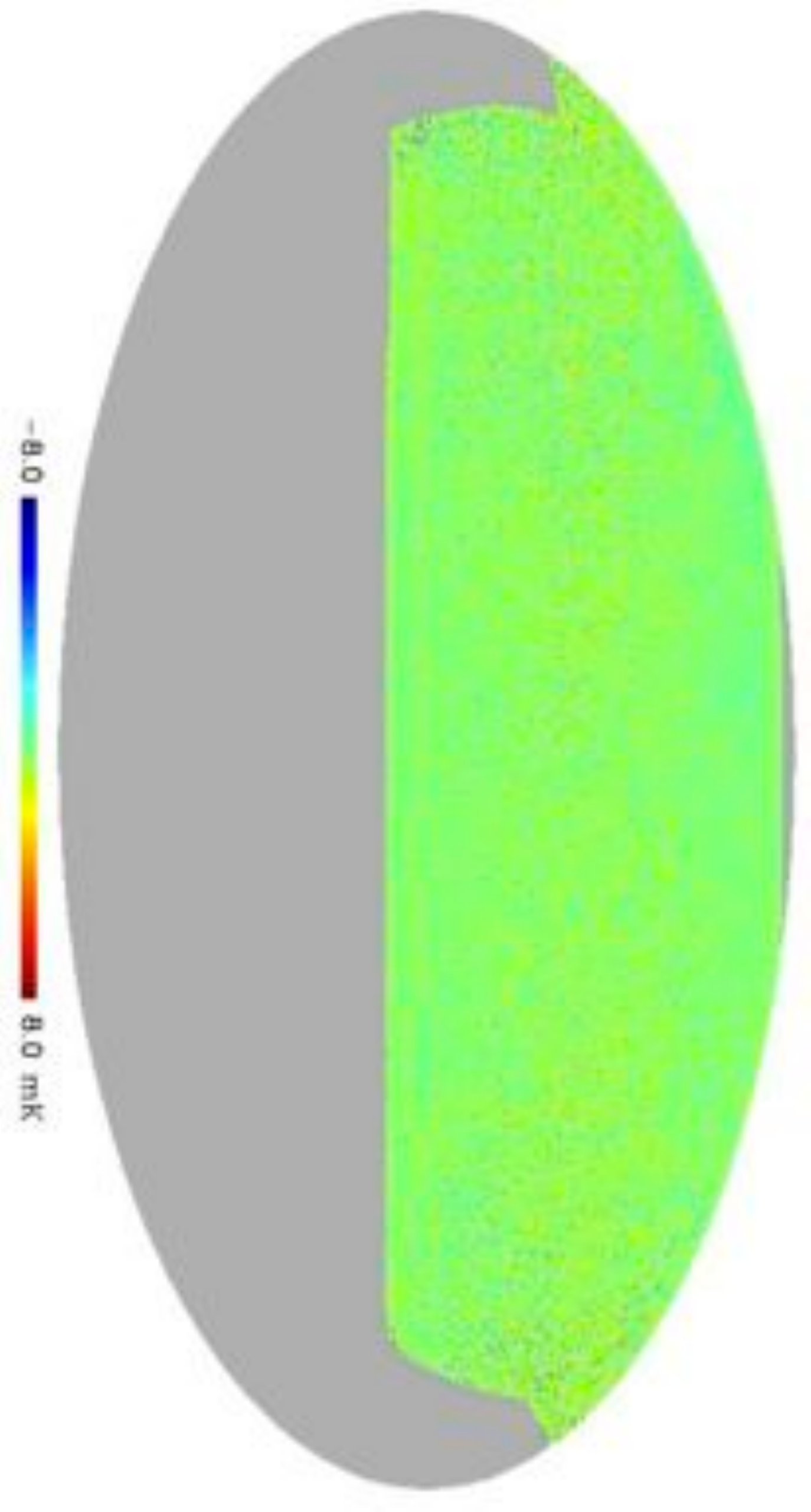}
\caption[Q sky map]{Q sky map. This map is featureless and represents a white noise dominated signal. \label{fig:Qmapall}}
\end{center}
\end{figure}

\begin{figure}[p]
\begin{center}
\includegraphics[width=10cm ]{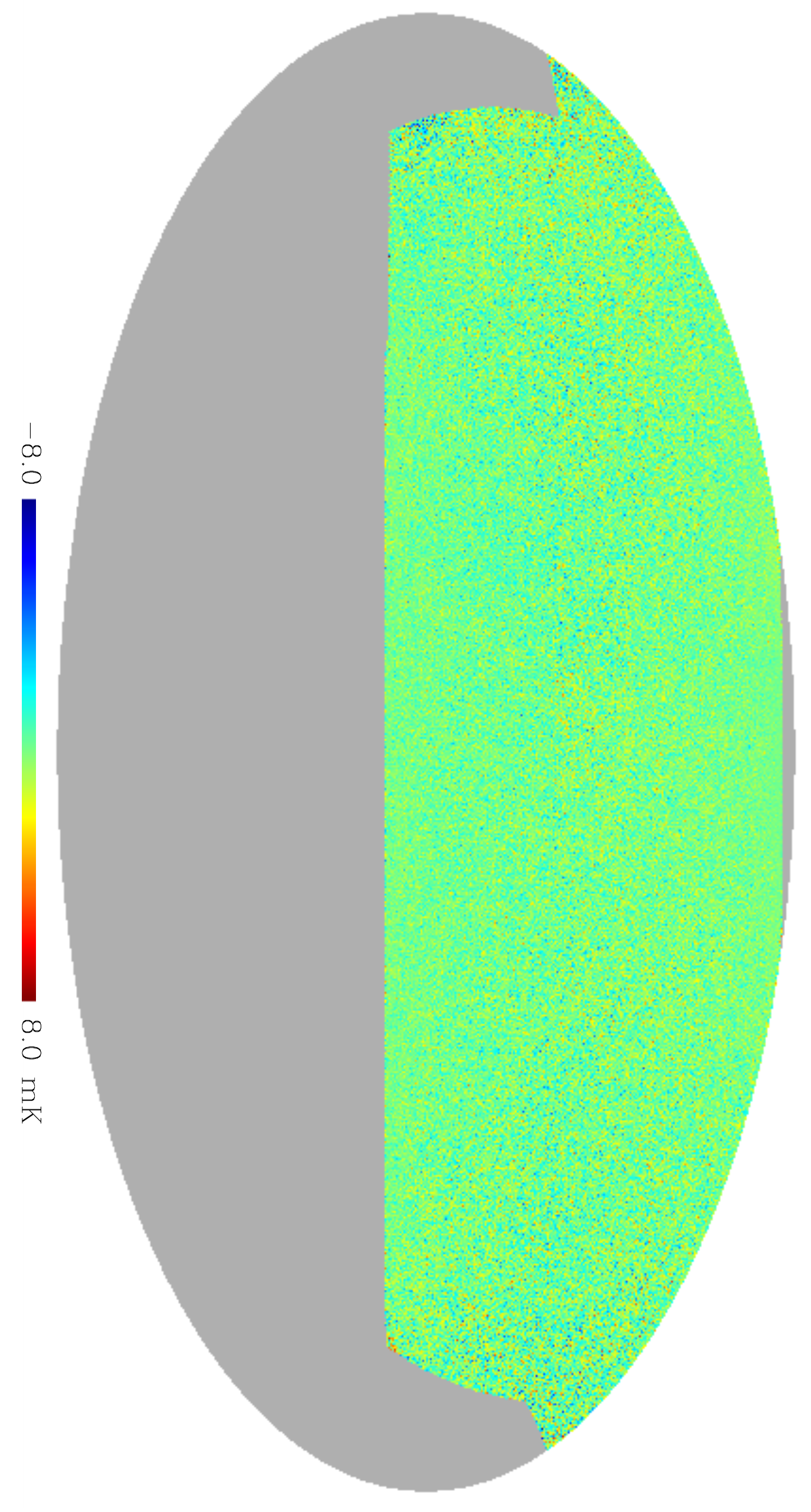}
\caption[U sky map]{U sky map. This map is featureless and represents a white noise dominated signal. \label{fig:Umapall}}
\end{center}
\end{figure}

\begin{figure}[p]
\begin{center}
\includegraphics[width=10cm ]{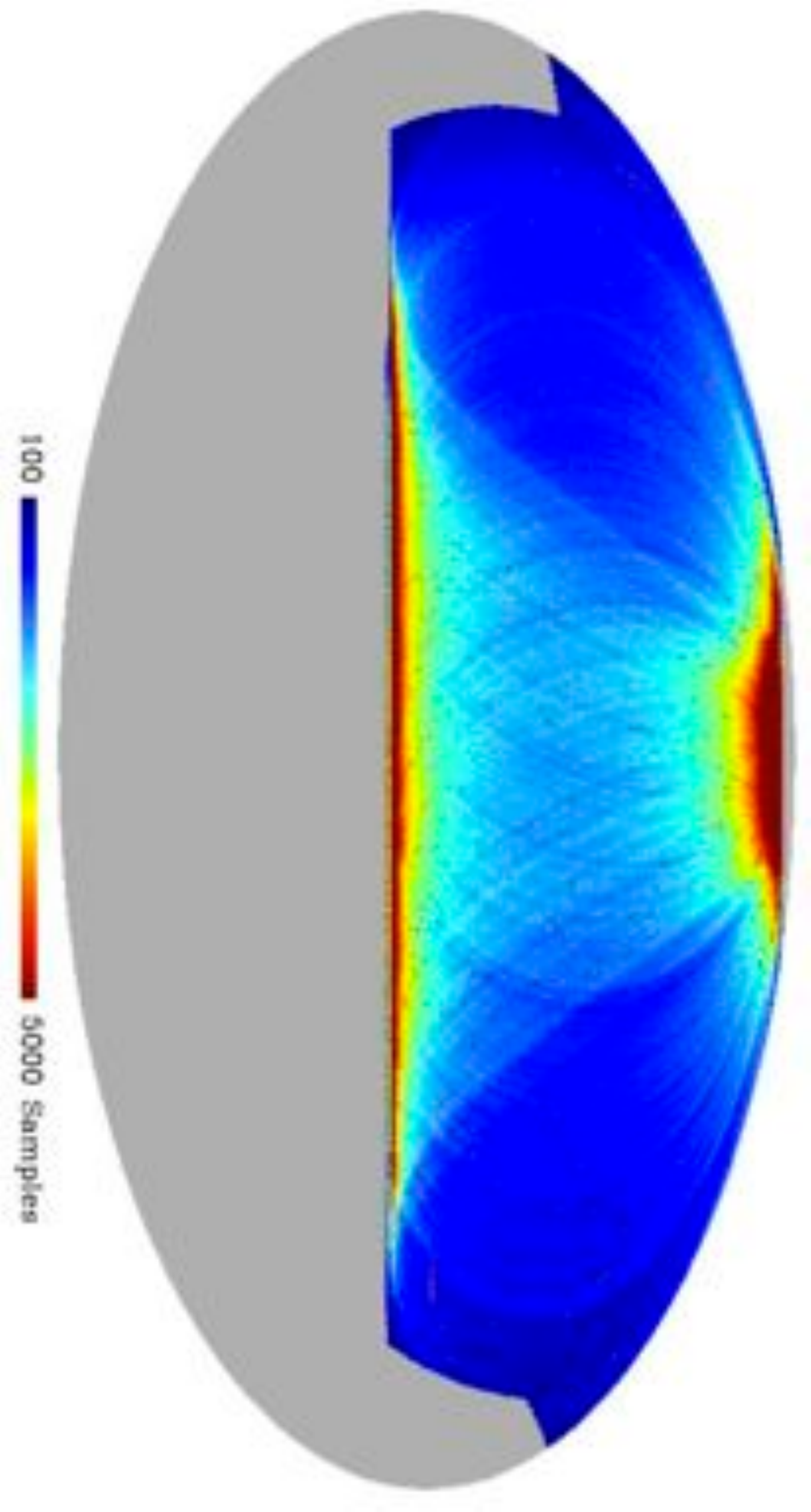}
\caption[Number of observations per bin sky map]{Number of observations per bin map. The features at the top and bottom represent the edges of scans.\label{fig:nobsmapall}}
\end{center}
\end{figure}

\begin{figure}[p]
\begin{tabular}{cc}
\includegraphics[width=7cm,height=6cm]{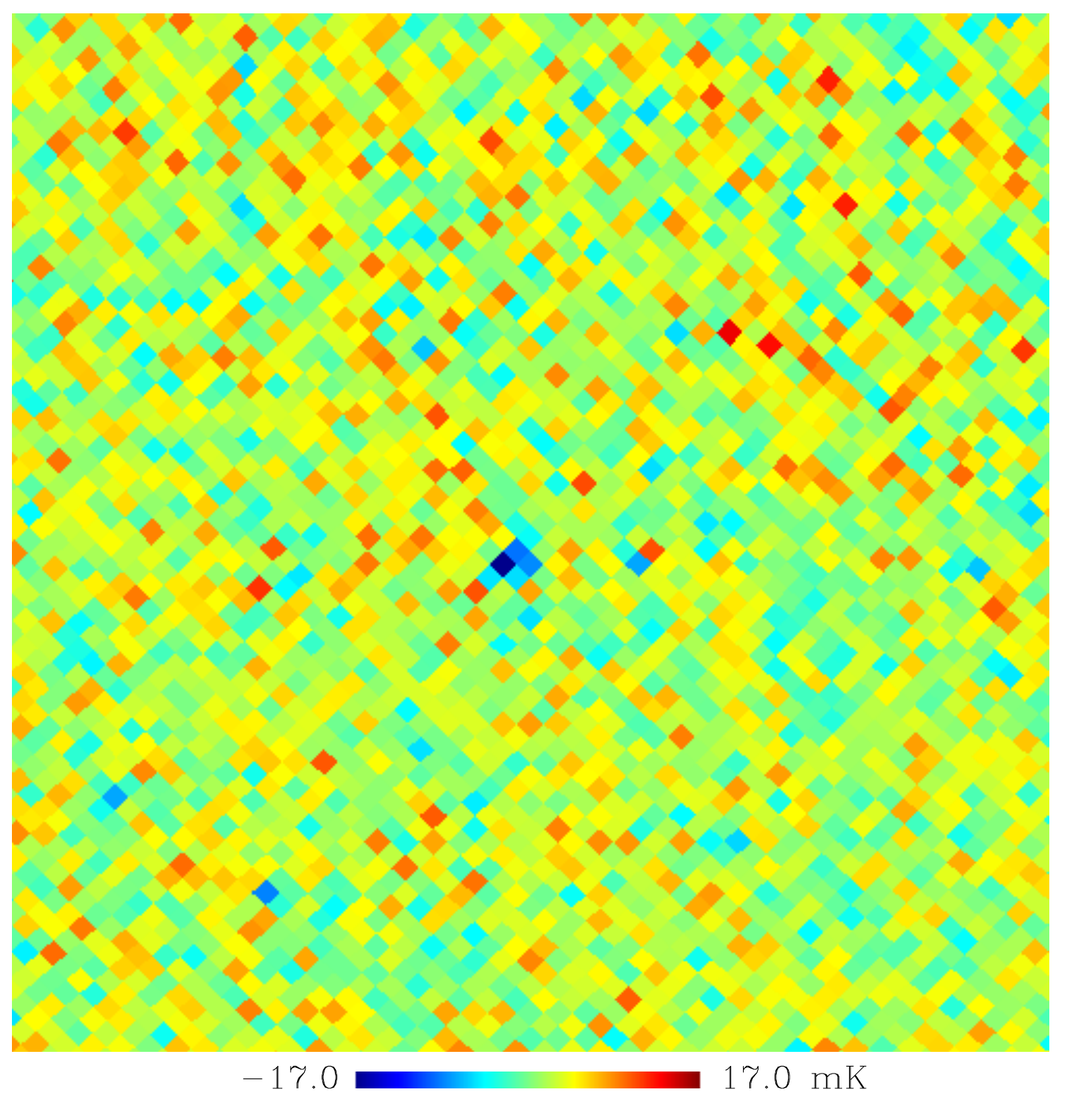}&
\includegraphics[width=7cm,height=6cm]{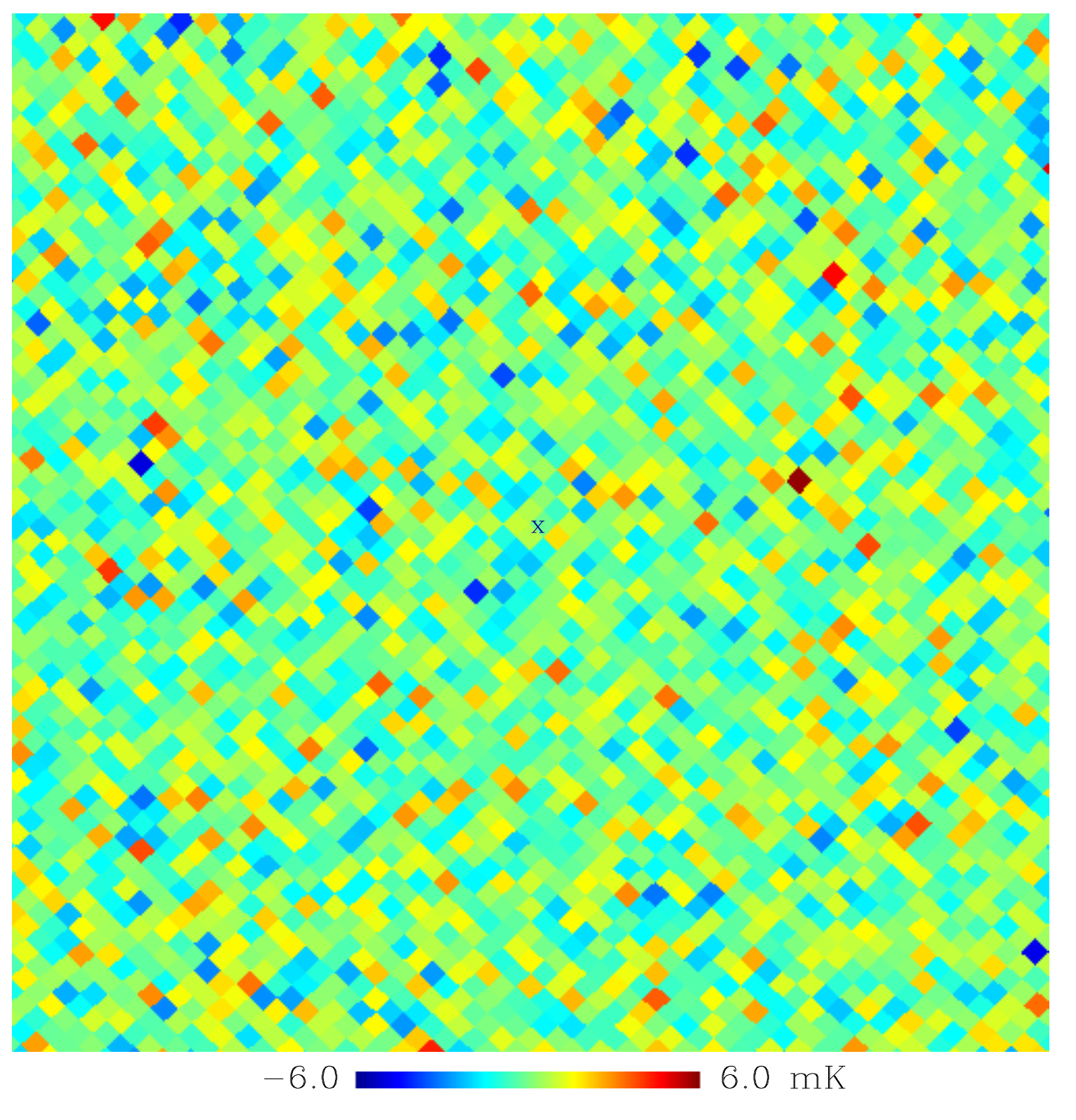}\\
\includegraphics[width=7cm,height=6cm]{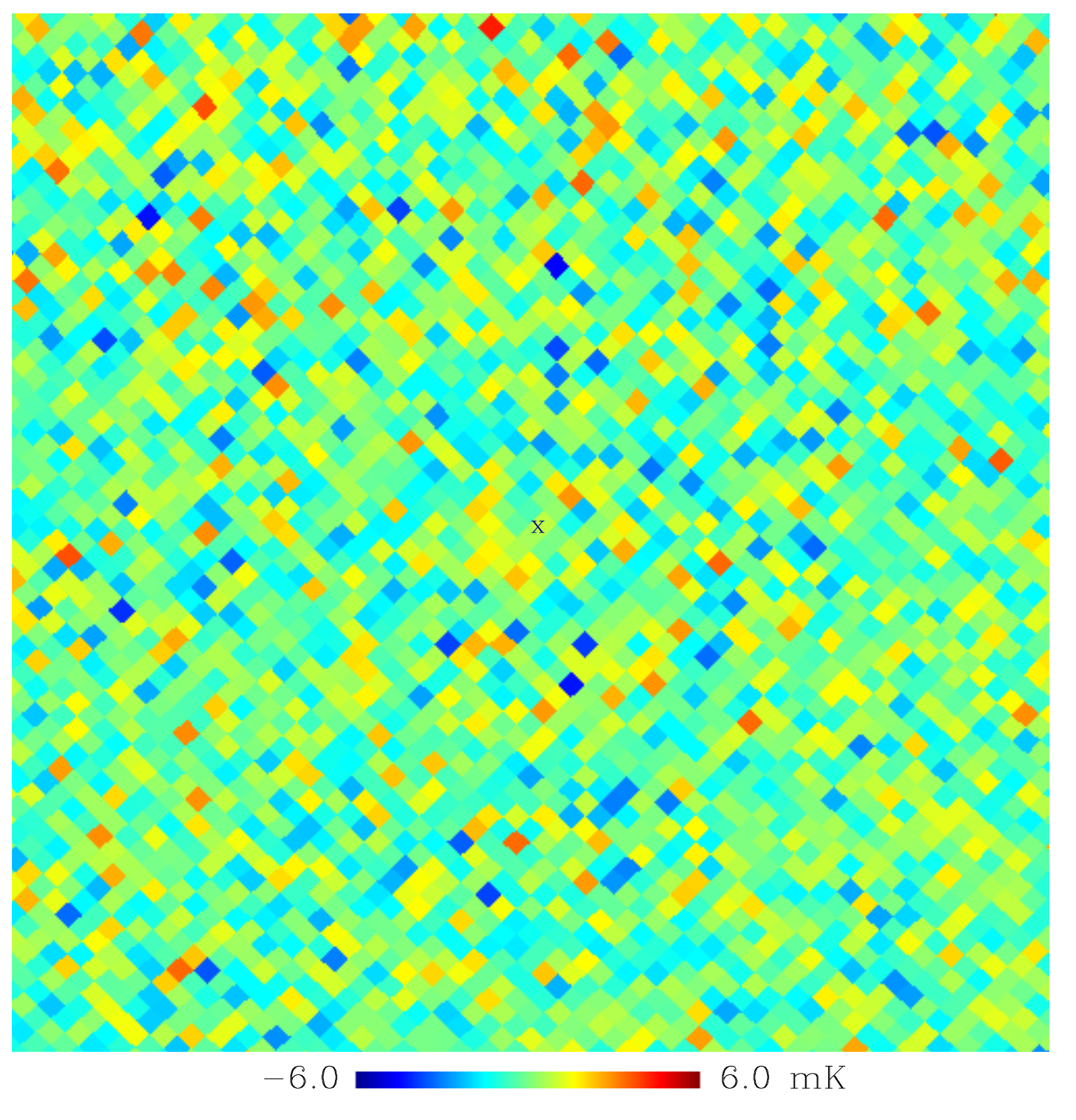}&
\includegraphics[width=7cm,height=6cm]{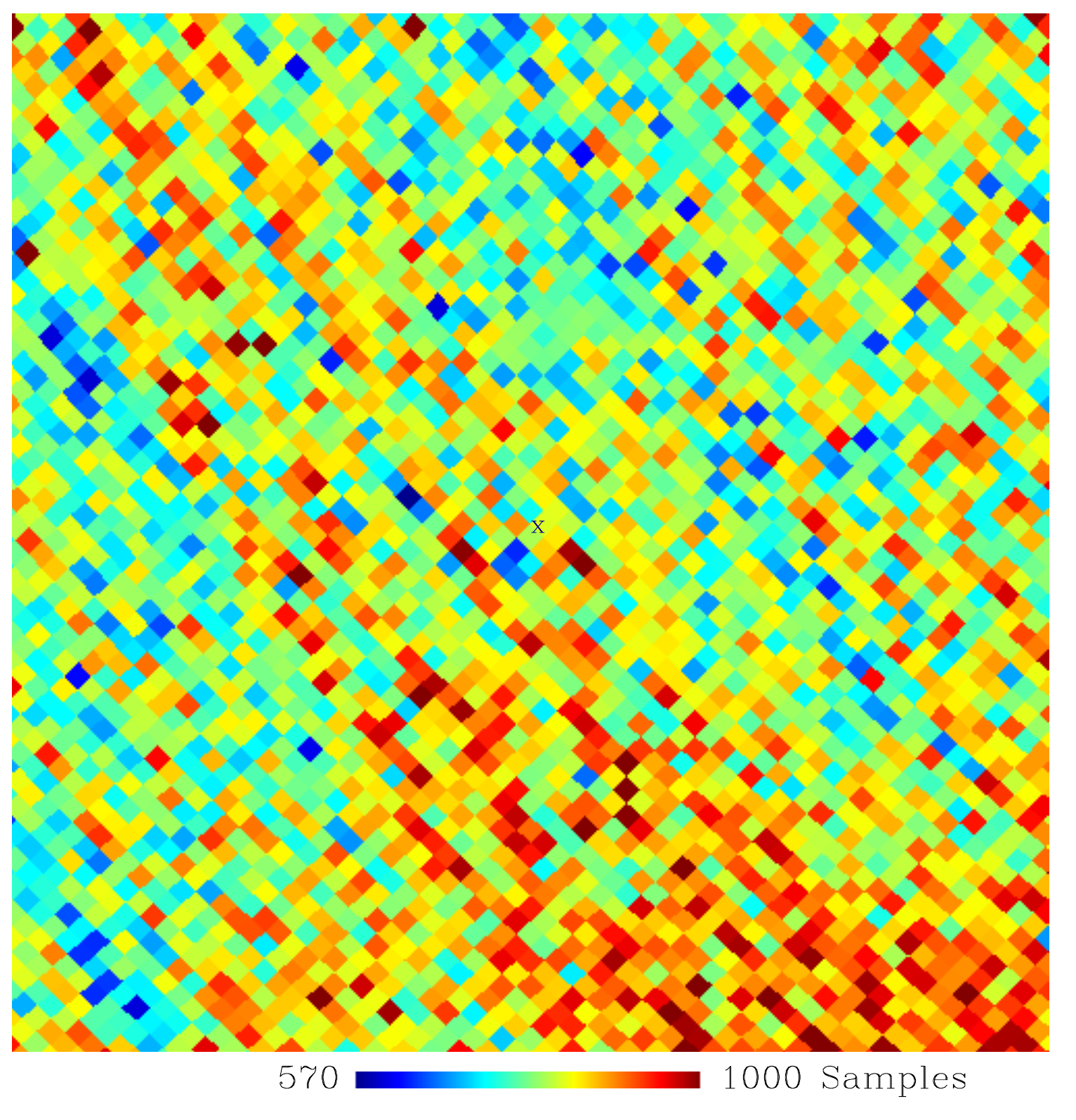}\\
\end{tabular}
\caption[Zoom in of Tau A region of full map]{Zoom in of Tau A region of full maps for all Stokes parameters and number of sampled bins. Top left, Temperature, top right, $Q$, bottom left, $U$, bottom right, number of binned samples. The small "x" is the position of Tau A.\label{fig:tauazoommap}}
\end{figure}

The final values were determined by using 2 sigma cuts on gaussian fits to the data histograms. Bulk cuts were made based on Healpix position. These cuts can be seen on the maps as rectangular cut outs on the leading and trailing edges of the maps.  These cuts correspond to the start and stop of data for each day and are consistent with thermal cycling of optical components. In Figures~\ref{fig:tempmapall} through~\ref{fig:tauazoommap} maps of all 3 Stokes parameters can be seen with all channels and days combined.

\section{Angular Power Spectrum}
\begin{figure}[p]
\begin{center}
\includegraphics[width=\textwidth ]{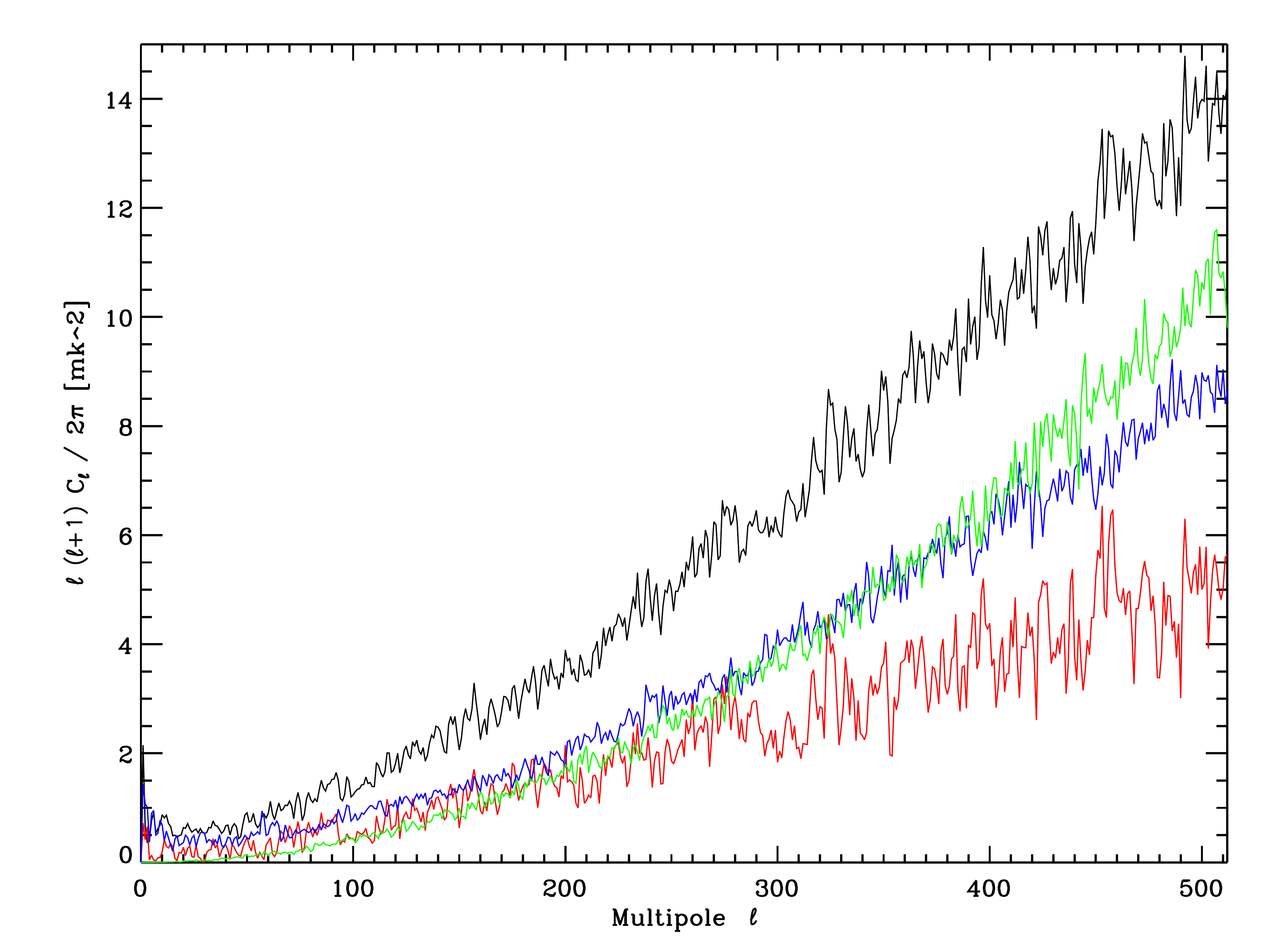}
\caption[TT Angular Power Spectrum]{TT Angular Power Spectrum with the three main point sources and the galaxy removed before spectrum generation. Black is Jack knifed data, blue is B-Machines entire data set, red is the differenced data, and green is a white noise spectrum.\label{fig:TTPowerSpec}}
\end{center}
\end{figure}

\begin{figure}[p]
\begin{center}
\includegraphics[width=\textwidth ]{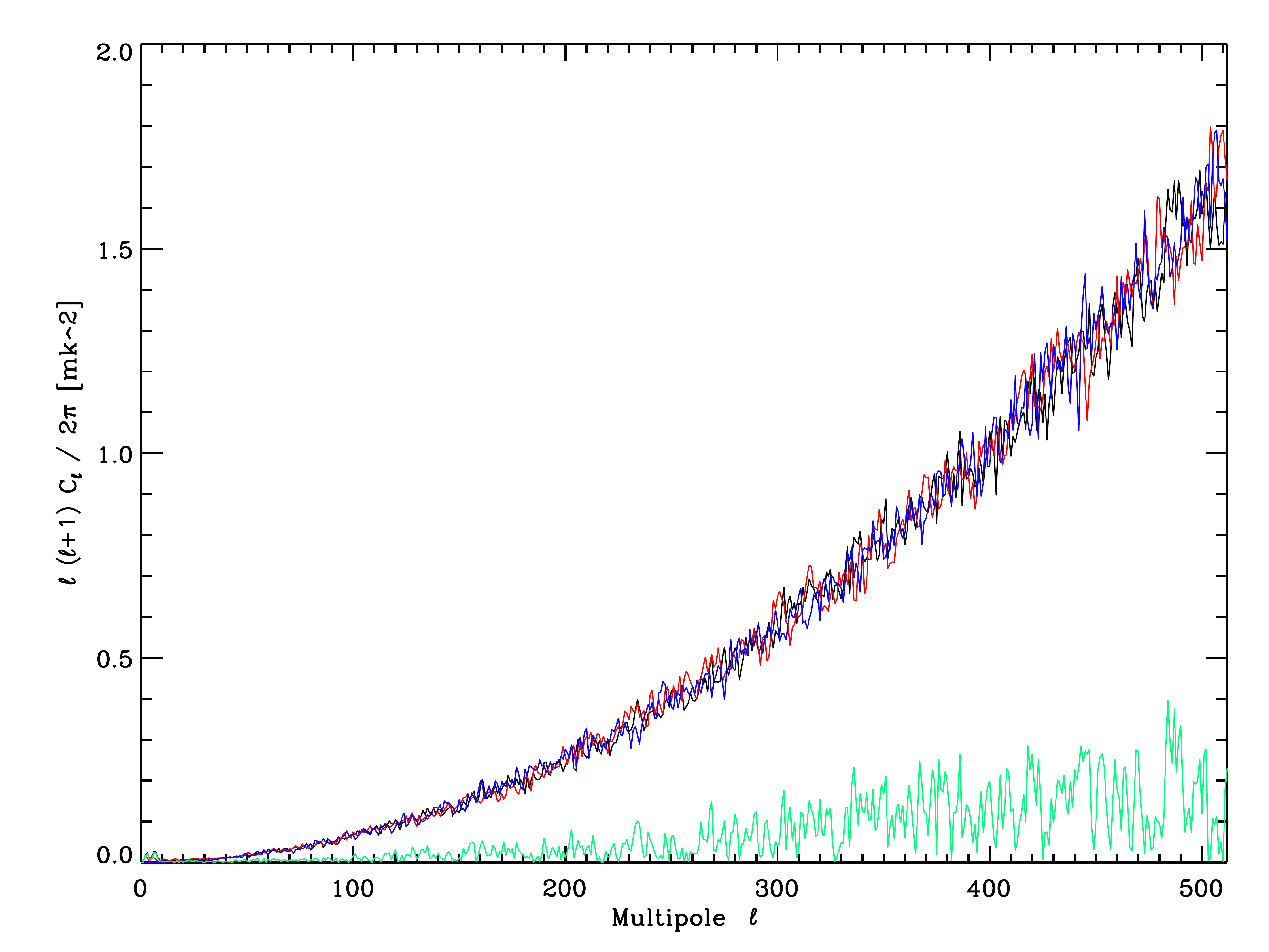}
\caption[EE Angular Power Spectrum]{EE Angular Power Spectrum in black and a white noise spectrum in blue. The differenced Jack knifed data is in green and consistent with no cosmological signal. \label{fig:EEPowerSpec}}
\end{center}
\end{figure}

\begin{figure}[p]
\begin{center}
\includegraphics[width=\textwidth ]{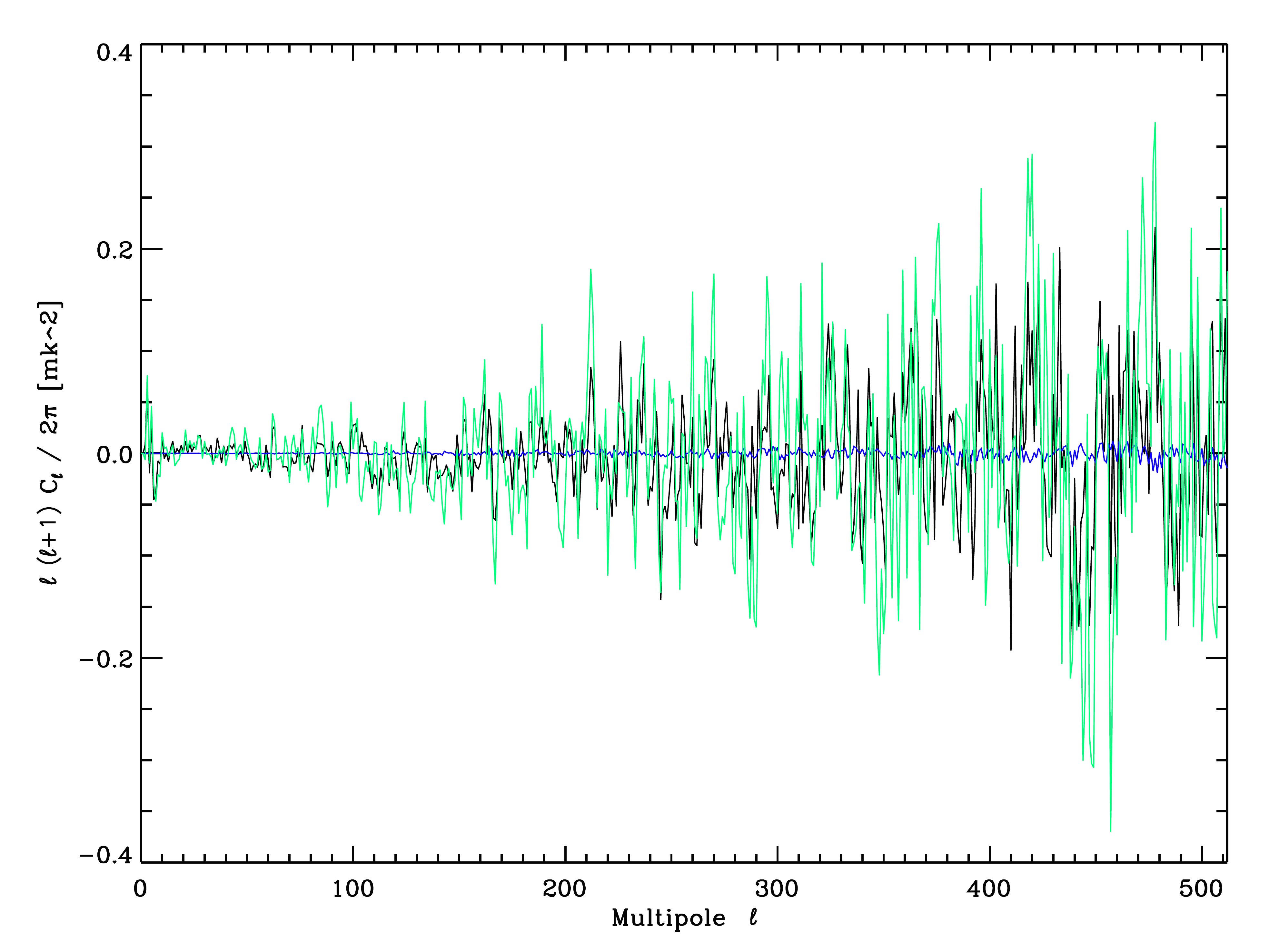}
\caption[TE Angular Power Spectrum]{TE Angular Power Spectrum in black and a white noise spectrum in blue. The differenced Jack knifed data is in green and consistent with no cosmological signal. \label{fig:TEPowerSpec}}
\end{center}
\end{figure}

Preliminary angular power spectra are consistent with noise dominated measurements. For accurate power spectra de-stripping codes, such as Madam 3.5, need to be used to remove the obvious stripping in the current temperature maps, see Figure~\ref{fig:tempmapall}.  Included in Figures~\ref{fig:TTPowerSpec} through Figures~\ref{fig:TEPowerSpec} are power spectra for TT, EE, and TE. In addition, power spectra for a random white noise map with standard deviation scaled to that of the data set for the given Stokes parameter are shown.  A Jack knife difference was taken for first order instrument noise removal from the maps.  The process of Jack knifing the data consisted of splitting the data set into 2 roughly equal length data sets, subtracting the maps, and generating power spectra from the differenced map.

\begin{figure}[p]
\begin{center}
\includegraphics[width=\textwidth ]{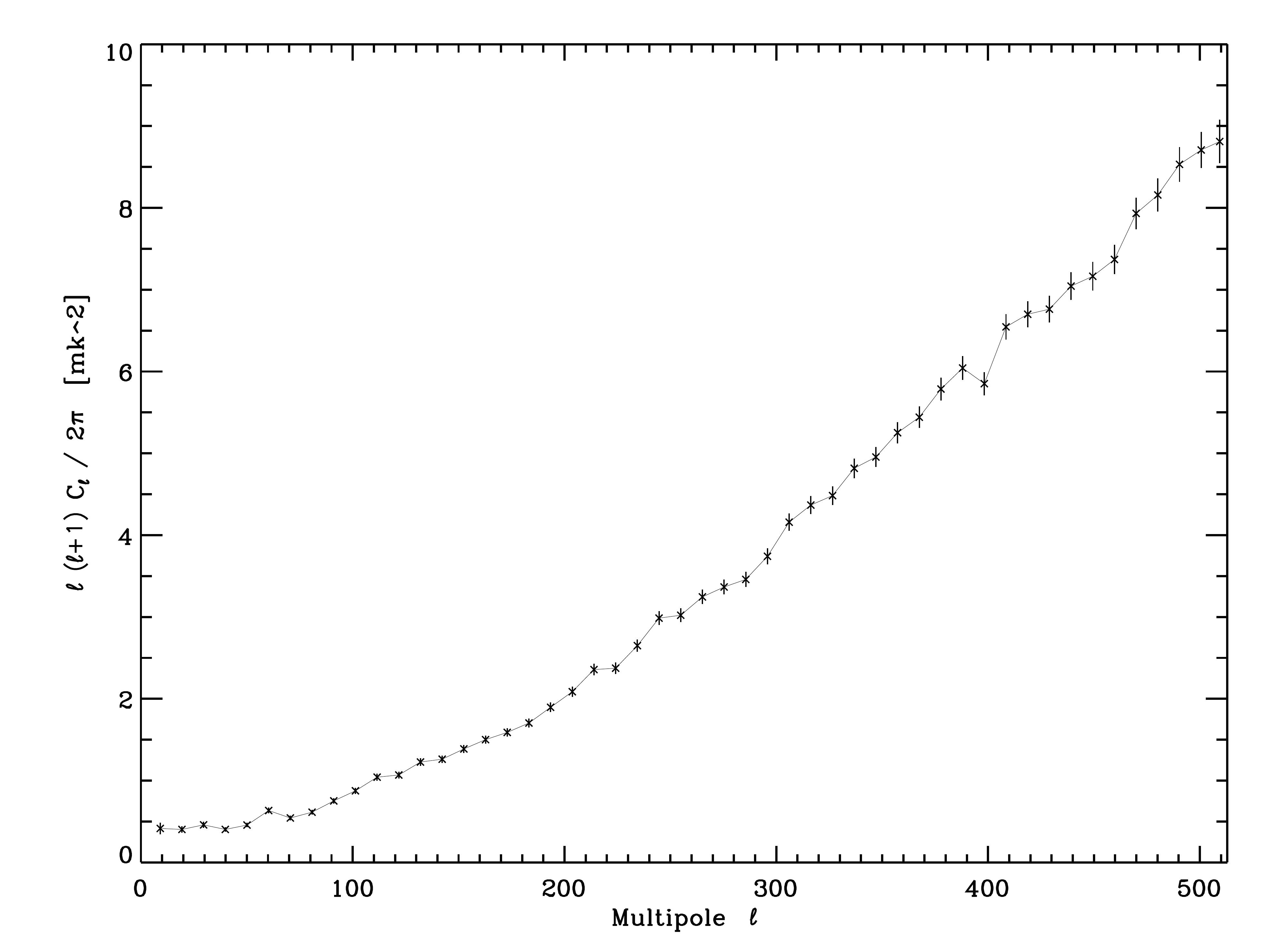}
\caption[TT Angular Power Spectrum with estimate errors]{TT Angular Power Spectrum including estimates of errors. The shape of the power spectrum is clearly indicative of a white noise dominated power spectrum.\label{fig:TTPowerSpecErr}}
\end{center}
\end{figure}

\begin{figure}[p]
\begin{center}
\includegraphics[width=\textwidth]{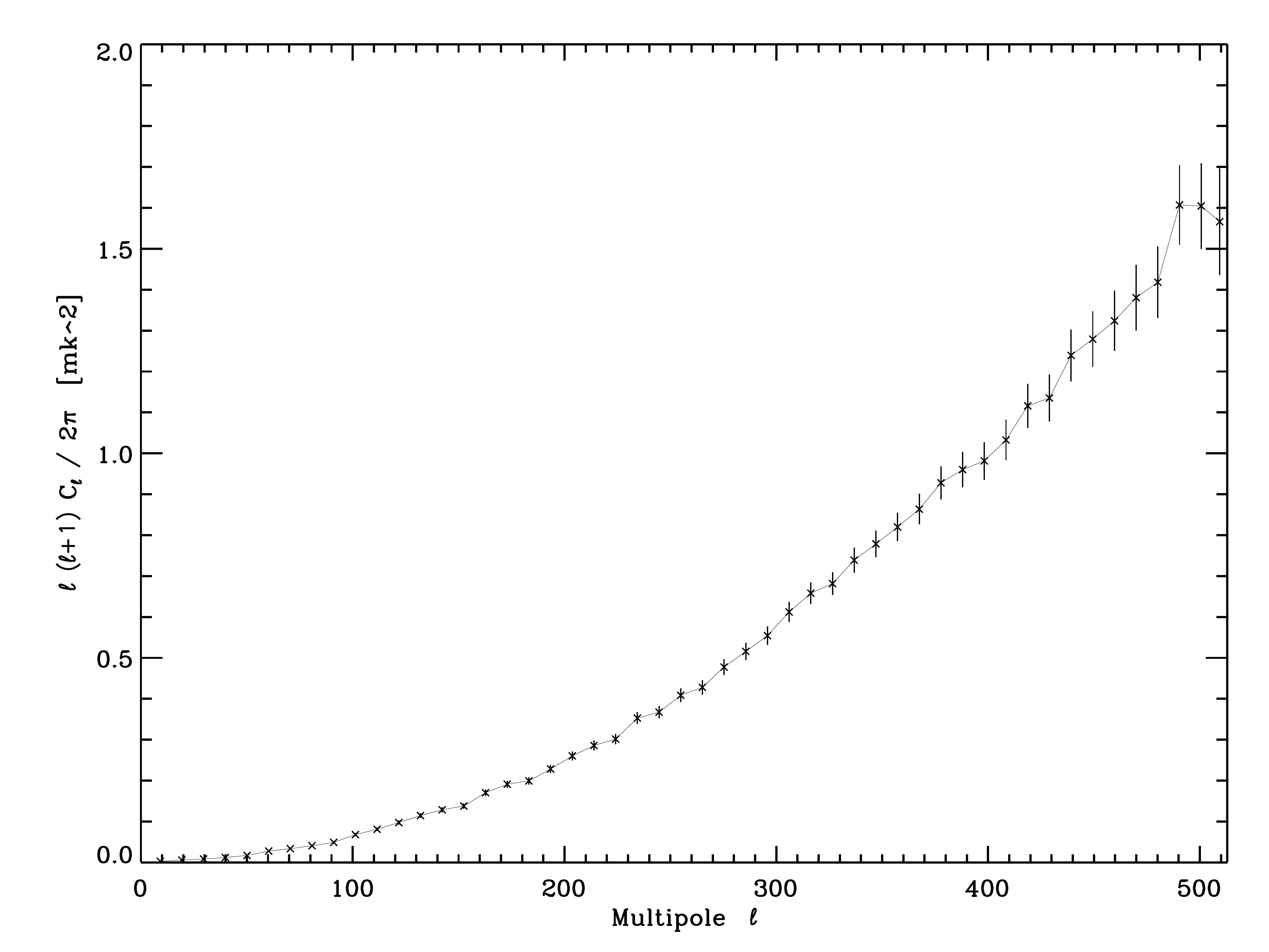}
\caption[EE Angular Power Spectrum with estimate errors]{EE Angular Power Spectrum including estimates of error.  The shape of the power spectrum is clearly indicative of a white noise dominated power spectrum. \label{fig:EEPowerSpecErr}}
\end{center}
\end{figure}

\begin{figure}[p]
\begin{center}
\includegraphics[width=\textwidth]{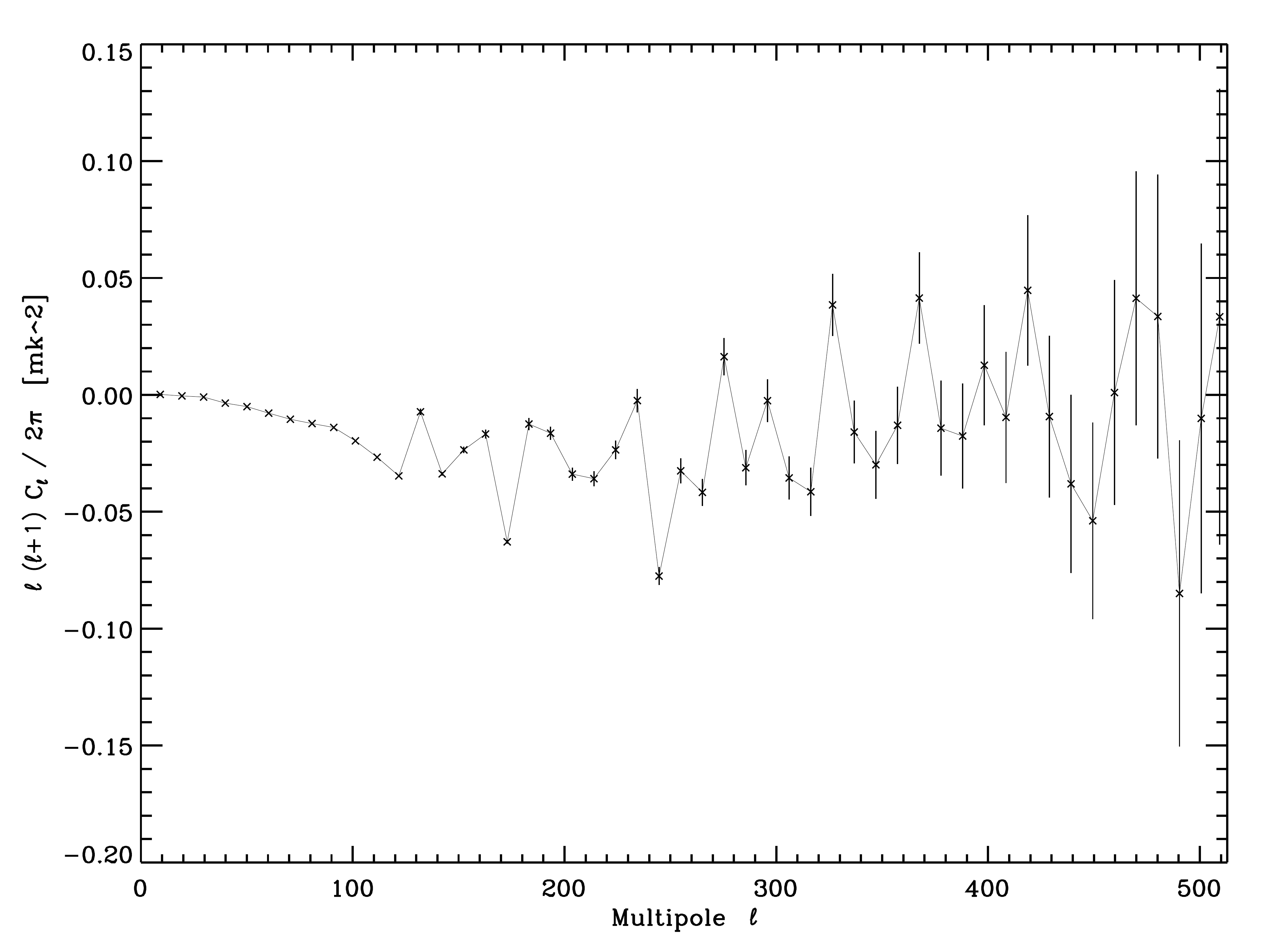}
\caption[TE Angular Power Spectrum with estimate errors]{TE Angular Power Spectrum including estimates of errors. This curve is consistent with no cosmological signal.\label{fig:TEPowerSpecErr}}
\end{center}
\end{figure}

The TT spectrum  differs from a white noise spectrum for 2 main reasons. First, its noise is dominated by $\frac{1}{f}$ noise and secondly the maps are not de-stripped. This accounts for the slope difference of the 2 curves in Figure~\ref{fig:TTPowerSpec}.  The polarization map is consistent with a white noise dominated map as expected by our sensitivity given the limited integration time and large sky coverage, see Chapter~\ref{chap:telescopechar} Table~\ref{tab:MaxPhaseTsys}.  A first attempt to estimate the power spectrum uncertainties was made using an analytic tool developed by Lloyd Knox \citep{knox96}.  Plots of the three interesting power spectra have been generated with error bars in Figures~\ref{fig:TTPowerSpecErr} through ~\ref{fig:TEPowerSpecErr}. Using the errors from the analytical model a goodness of fit test reveals that the TE power spectrum is consistent with the null hypothesis, reduced chi square is $0.65$.

\chapter{Conclusion} \label{chap:conclusion}
Though B-Machine didn't yield grand results, it did achieve the goals it set out to meet. B-Machine was fielded and operated in the summer of 2008 at White Mountain Research Station for 2 main reasons: first, to test the polarization rotator chopping strategy and  second, to collect data to generate maps and power spectra of the CMB polarization. Jack knife analysis of the the current sky maps show that the map noise integrates down with added integration time and no features become apparent to the tenths of mK level.  This demonstrates that no major systematics are polluting the data stream in an odd fashion.  Moreover, rebinning of the maps into nside=64 reduces the RMS noise of the map from $1.0\mathrm{~mK}$ to  $0.20\mathrm{~mK}$ increasing the sensitivity at low l's at the price of high l sensitivity, yielding yet another avenue for continued analysis.  It seems clear that the B-Machine platform was effective as both a test platform for larger experiments (arrays of multiple telescopes for B-mode observations and balloon borne experiments for foreground observations) and a polarimeter for observing the CMB.

Though several problems were encounter when fielding the telescope these problems are easy to solve for the next observing campaign and the next year of data promises to yield a minimum of twice as much data. The addition of point source observations will give a very rich data set for future analysis.  While future observations are certainly not guaranteed for B-Machine, it is clear at frequencies below 50 GHz ground based observations can provide excellent results on par with that of satellites, see Table~\ref{tab:compare}.  With a very stable $\frac{1}{f}$ knee frequency,  B-Machine is a gold mine of data waiting to be taken advantage of.  Drawing your attention to row 12 of Table~\ref{tab:compare}, it becomes clear that with a minimal financial and man power investment low frequency observations can be made from the ground and compete with satellites. The reduction in the $f_{\mathrm{knee}}$  frequency in Table~\ref{tab:compare} is done with a linear fit to the data before differencing and is a data analysis technique. In addition, the numbers quoted in Table~\ref{tab:compare} don't take into account that B-Machine style radiometers have a fundamental advantage over both WMAP and Planck-LFI style radiometers in that they are designed as polarimeters and measure $Q$ and $U$ directly through the same RF path. 

\begin{table}[p]
\begin{center}
\caption[Comparison of WMAP, Planck,  Future B-Machine, and COFE]{Comparison of B-Machine, WMAP \citep{jarosikwmap03}, Planck-LFI \citep{meinhold09}, Future B-Machine, and Cofe \citep{leonardi05} \label{tab:compare}}
\begin{tabular}{|l|c|c|c|c|c|}
  \hline
   &  B-Machine                    & WMAP              & Planck & Future    & COFE \\
   &                                          & Satellite           & LFI & B-machine & \\
  \hline
  \hline   
  Center Freq. (GHz)         			    & 41.5     &  41.0  &  44.0   & 40.0 & 20 \\
  \hline 
  T$_{sys}$ (K)         			                       & 54        &      59  &  30       & 30 & 10 \\
  \hline
  T$_{sky}$ (K)                                                     & 16        &     2.7  &  2.7      & 16 & 2.7 \\
  \hline   
  $f_{\mathrm{knee}}$ (mHz)		              & 5          &      4    &  30      &  2.8 & 2.8 \\
  \hline
  Number of Detectors 		                        & 4         &         4  &  8      & 8    & 6\\
  \hline
  Angular Res. (arcmin)                                     & 22.2    &    30.6  &  30    & 22.2 & 42.0\\
  \hline
  $\triangle$T (mK $\sqrt{\mathrm{s}}$)                   & 1.8      &  0.90   & 0.439  & 0.80 & 0.318\\
  \hline
  $\triangle$P (mK $\sqrt{\mathrm{s}}$)                      & 1.37          &  0.90   & 0.439  & 0.61 & 0.227\\
  \hline
  Aggr. $\triangle$P (mK $\sqrt{\mathrm{s}}$)  &     0.685        &  0.450   & 0.155  & 0.215 & 0.094\\
  \hline
  \multicolumn{6}{|c|} {$\sim1\sigma$ sensitivities for 7 months of observations$^{\star}$}\\
  \hline
  Observing Efficiency                                               & 0.375 &  1.00 & 1.00   &  0.50 &  0.50\\
  \hline
   Aggr. $\triangle$P /Pixel ($\mu K$)$^{\dag}$& 452.0$^{\ddag}$   &   67.0   & 25.7  &  48.0 & 17.0 \\
  \hline
  Sky coverage ($\%$)                               & 53.1            & 100     & 100 & 60 & 60 \\
  \hline
  \multicolumn{6}{l}{$~~\dag$ Pixel Size $0.5^{\circ} \mathrm{~x~} 0.5^{\circ}$}\\
  \multicolumn{6}{l}{$~~\ddag$ From actual B-Machine data set of 40 days}\\
  \multicolumn{6}{l}{$~~\star$ For similar observations B-Machine achieves 165 $\mu K$ Aggr. $\triangle$P Per Pixel }\\
  
\end{tabular}
\end{center}
\end{table}

\section{The Future}
With some minor contributions in time and money B-Machine could get a data set that would rival even that of the Planck Satellite Mission.  By changing the current front end LNA's with more modern lower noise LNA's and populate the remaining 4 horns with the same LNA's,  the current system temperature would be reduced by a factor of 2, doubling the integration time. B-Machine has the capability to have 8 feed horns and is currently equipped with the hardware and software to run all 8 horns. Drop in replacements for the data acquisition boards would allow for faster DAQ rates reducing the addition of $\frac{1}{f}$ noise from the LNA's and the addition of a 5 point calibration sequence on a monthly basis would improve the performance of the telescope. The sensitivity for all of these upgrades can be seen in the Future B-Machine column of Table~\ref{tab:compare}.

\appendix
\chapter{Blackbody Temperature to Antenna Temperature} \label{app:blackbody}
\begin{figure}[p]
\begin{center}
\includegraphics[width = 13.5cm]{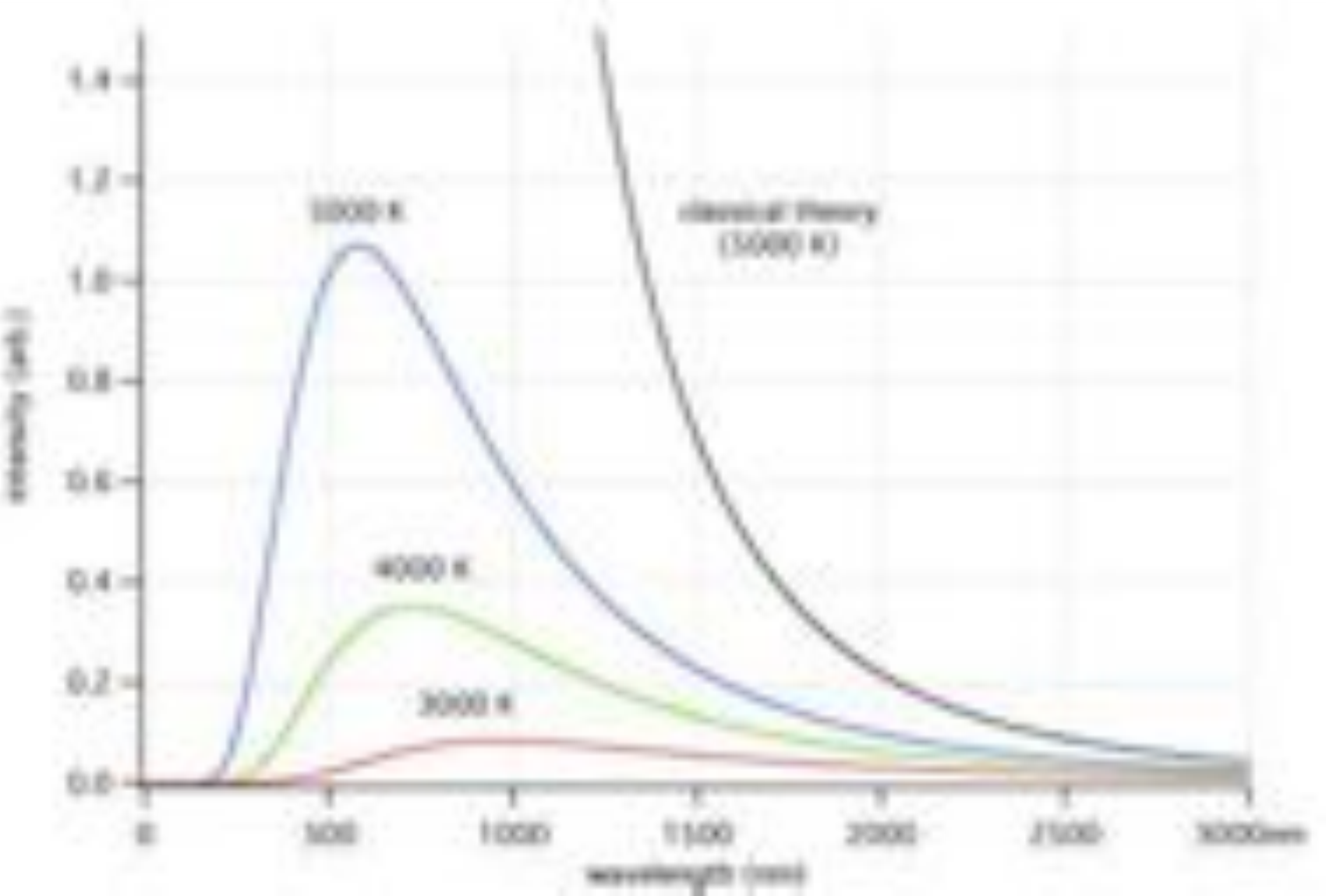}
\caption[Curves of blackbody radiators at different temperatures]{Curves of blackbody radiators at different temperatures retrieved may 28, 2009 from $\mathrm{http://en.wikipedia.org} / \mathrm{wiki} / \mathrm{Black\_body}$. \label{fig:Blackbody}}
\end{center}
\end{figure}

Blackbody radiation refers to an idealized object or system which absorbs all radiation incident upon it and re-radiates the energy which is characteristic of the radiating system only, not dependent upon the type of radiation which is incident upon it, see Figure~\ref{fig:Blackbody}. The brightness of radiation from a black body is given by Planck's law,
\begin{equation}
 B_{\nu}(T) =\frac{2h\nu^3}{c^2}\frac{1}{\mathrm{e}^{h\nu /kT}-1}
\end{equation}
where $h$ is Planck's constant, $\nu$ is the frequency, $c$ is the speed of light, $k$ is Boltzmann's constant, $T$ is the physical temperature and B$_{\nu}$ is the surface brightness in Watts $\mathrm{m}^{-2}$ $\mathrm{Hz}^{-1}$ $\mathrm{sr}^{-1}$. Typically a transfer standard (gain/calibration) is generated based on the input flux of a telescope versus output voltage level using beam filling objects that radiate as black bodies at different temperatures, see Chapter~\ref{chap:telescopechar} Section~\ref{Sec:Calibration}.
A brightness temperature is what is strived for when observing, but antenna temperature is what is measured. Antenna temperature is the convolution of the brightness temperature with the beam of the instrument and is given by,
\begin{equation}
 T_A =\frac{1}{\Omega_A}\int \int T_s(\theta, \phi)P_n(\theta,\phi)\mathrm{d}\Omega,
\end{equation}
where $T_s(\theta,\phi)$ is the source temperature and $P_n(\theta,\phi)$ is the normalized antenna pattern. For very small sources with uniform temperatures this reduces to,
\begin{equation}
T_A =\frac{\Omega_s}{\Omega_A}T_s,
\end{equation}
where $\Omega_s$ and $\Omega_A$ are the source and antenna areas respectively. For large sources that fill the beam (but are small compared to the entire sky) and an antenna pattern that does not have significant side lobe contribution then
\begin{equation}
 T_A \simeq T_s.
\end{equation}

\clearpage \ssp

\addcontentsline{toc}{chapter}{Bibliogrpahy}

\bibliographystyle{plainnat}
\bibliography{bdwbiblio}

\clearpage

\end{document}